\documentclass[aps,prb,twocolumn,showpacs,10pt,floatfix,superscriptaddress,floatfix]{revtex4-1}
\usepackage{epsf}
\usepackage{bm}
\usepackage{hyperref}
\usepackage{amsfonts}
\usepackage{amssymb}
\usepackage{amsmath,mathtools}
\usepackage{array}
\usepackage{enumerate,dsfont}
\usepackage{dcolumn,multirow}
\usepackage[utf8]{inputenc}
\usepackage{latexsym}
\usepackage{xcolor}
\newcommand{\vk}{\vec k}
\newcommand{\ve}{\vec e}

\newcommand{\vR}{\vec R}

\newcommand{\vx}{\vec x}

\newcommand{\ZZ}{\mathbb{Z}}
\newcommand{\zz}{z}

\newcommand{\sign}{{\rm sign}\,}

\newcommand{\nootimes}{}

\usepackage{ bbold }
\newcommand{\id}{\textbf{1}}
\newcommand{\gen}{\mathrm{e}}
\newcommand{\idn}{0}

\newcommand{\SI}{\text{SI}}

\newcommand{\BS}{\text{BS}}
\newcommand{\BSz}{\text{BL}}
\newcommand{\AI}{\text{AI}}

\newcommand{\XX}{\{ \vk_{\rm s} \}}
\newcommand{\vkslist}{\vk_{\rm s}}

\newcommand{\be}{\begin{equation}}
\newcommand{\ee}{\end{equation}}

\renewcommand{\vec}[1]{\mathbf{#1}}

\newcommand{\classD}{{\rm D}}
\newcommand{\classA}{{\rm A}}
\newcommand{\classAII}{{\rm AII}}
\newcommand{\classDIII}{{\rm DIII}}
\newcommand{\classBDI}{{\rm BDI}}
\newcommand{\classC}{{\rm C}}
\newcommand{\classCI}{{\rm CI}}
\newcommand{\classCII}{{\rm CII}}

\begin{document}

\title{Symmetry-based indicators for topological Bogoliubov-de Gennes Hamiltonians}

\author{Max Geier}
\affiliation{Dahlem Center for Complex Quantum Systems and Physics Department, Freie Universit\"at Berlin, Arnimallee 14, 14195 Berlin, Germany}
\author{Piet W. Brouwer}
\affiliation{Dahlem Center for Complex Quantum Systems and Physics Department, Freie Universit\"at Berlin, Arnimallee 14, 14195 Berlin, Germany}
\author{Luka Trifunovic}
\affiliation{Dahlem Center for Complex Quantum Systems and Physics Department, Freie Universit\"at Berlin, Arnimallee 14, 14195 Berlin, Germany}
\affiliation{RIKEN Center for Emergent Matter Science, Wako, Saitama 351-0198, Japan}
\affiliation{Department of Physics, University of Zurich, Winterthurerstrasse 190, 8057 Zurich, Switzerland}
\date{\today}

\begin{abstract}
We develop a systematic approach for constructing symmetry-based indicators of a topological classification for superconducting systems. The topological invariants constructed in this work form a complete set of symmetry-based indicators that can be computed from knowledge of the Bogoliubov-de Gennes Hamiltonian on high-symmetry points in Brillouin zone. After excluding topological invariants corresponding to the phases without boundary signatures, we arrive at natural generalization of symmetry-based indicators [H.  C. Po, A. Vishwanath, and H. Watanabe, Nature Comm. {\bf 8}, 50 (2017)] to Hamiltonians of Bogoliubov-de Gennes type.
\end{abstract}
\maketitle

\section{Introduction}

Although topological phases of matter have been known for four decades now,
starting with the discovery of the quantized Hall effect,~\cite{klitzing1980}
the study of topological phases and phase transitions became central to quantum
condensed matter physics only in the early 2000s, after the theoretical
proposals for topological superconducting phases~\cite{kitaev2001,ivanov2001} 
and the
quantum spin Hall effect~\cite{kane2005b,kane2005} and the subsequent
experimental observation of these phases of matter.~\cite{mourik2012,bruene2011}
These theoretical and experimental developments paved the way for the complete
classification of all possible topological phases of single-particle systems,
protected by either time-reversal, particle-hole or sublattice symmetries ---
the so-called ``tenfold way classification''\cite{altland1997} ---, which, reflecting its
periodicity as a function of dimensionality and honoring its fundamental
importance to the field, was coined ``periodic table of topological
phases''.~\cite{kitaev2009,schnyder2009}

In addition to the presence or absence of the fundamental non-spatial
symmetries that define the tenfold-way classes, real materials have crystalline
symmetries. The combination of topology and crystalline symmetries leads to an
exceedingly rich set of ``topological crystalline phases''.~\cite{fu2011}
Unlike non-crystalline tenfold-way topological phases, for which a nontrivial
topology of the bulk is always associated with a unique anomalous boundary
signature, topological crystalline phases may come with a variety of possible
boundary signatures. These include the protected existence of anomalous
boundary states on all boundaries --- a ``first-order'' topological phase, for
which the crystalline symmetry is not essential for the protection of the
nontrivial topology ---, the appearance of higher-order boundary states on
hinges or corners of a
crystal,~\cite{parameswaran2017,schindler2018,peng2017,langbehn2017,song2017,fang2018,ezawa2018,shapourian2017,zhu2018,yan2018,wang2018,wang2018b,khalaf2018,khalaf2018b,nobuyuki2018}
or even the complete absence of protected boundary states. The latter scenario
applies to ``atomic-limit phases'', in which the electronic states can be
continuously deformed to a collection of localized orbitals, while preserving
all relevant symmetries as discussed by Po~\textit{et al.}~\cite{po2017} and
Bradlyn~\textit{et al.}~\cite{bradlyn2017}. In such cases, the presence of the
crystalline symmetry may form an ``obstruction'' that prevents from different
arrangements of localized orbitals to be continuously connected to each
other,~\footnote{Although atomic-limit phases have no protected boundary
states, they may have protected fractional charges at ends or corners, see,
{\it e.g.}, Refs.~\onlinecite{lau2016,benalcazar2017,benalcazar2017b}.}
allowing for the existence of multiple topologically-distinct atomic-limit
phases.~\cite{turner2012,lau2015,lau2016,benalcazar2017,benalcazar2017b,vanmiert2017,rhim2017}

Although the classification of topological crystalline phases can be considered
largely under
control~\cite{shiozaki2014,shiozaki2016,chiu2016,trifunovic2019,cornfeld2019,shiozaki2019classification,huang2017,thorngren2018}
(for initial partial classifications results, see Refs.\
\onlinecite{fang2012,fang2013,chiu2013,jadaun2013,morimoto2013,teo2013,benalcazar2014,liu2014b,alexandradinata2014,zhang2013b,slager2013,dong2016,kruthoff2017})
the explicit computation of topological invariants for a given band structure
is often computationally expensive. Explicit expressions for the invariants
need not always be readily available, since a full classification does not
always come with explicit expressions for topological invariants.  A practical
--- but partial --- solution to this problem is the use of a set of
easy-to-compute ``symmetry-based indicators'', which, when nonzero, are a
sufficient indicator of a nontrivial topology of the bulk band
structure.~\cite{po2017} The Fu-Kane criterion,
which links the existence of a strong~\footnote{In this work, the term ``strong
topological phase'' is reserved for a phase that remains topologically
non-trivial after the translational symmetry is broken. Topological crystalline phases that do not rely on the crystalline symmetry for their protection will be referred to as ``first-order'', corresponding to the dimension of their boundary signature.} topological insulator
phase in an inversion-symmetric crystal to the parity of the number of occupied
bands with odd-inversion-parity at the high-symmetry momenta, is an example of
such a symmetry-based indicator.~\cite{fu2007b} For normal-state insulating
phases, symmetry-based indicators were constructed for the complete set of
point group symmetries in two and three
dimensions,~\cite{po2017,bradlyn2017,song2018,song2018b} taking into account
the order of the boundary states.~\cite{khalaf2018,zhang2019} Like the classifying groups of topological phases, symmetry-based
indicators of topological phases have a group structure, the group operation
being the direct sum ``$\oplus$'' of representative
Hamiltonians.~\footnote{Although Refs.~\onlinecite{po2017,bradlyn2017} have equivalent definitions of ``atomic-limit phases'', Bradlyn~\textit{et  al.}~\cite{bradlyn2017} consider band labels that lack the full group structure --- one can perform addition of the band labels, but not their
subtraction. Subsequent careful comparison revealed that some
phases without protected boundary states are topologically equivalent to
a ``difference'' of atomic-limit phases, {\it i.e.}, they can only be deformed to an atomic-limit phase after addition of topologically trivial bands, see H.\ C.\ Po, H.\ Watanabe, and A.\ Vishwanath, Phys.\ Rev.\ Lett.\ {\bf 121}, 126402 (2018). Such phases were called ``fragile topological insulators''. They are trivial in the classifying scheme of Po~\textit{et  al.}~\cite{po2017}, while Bradlyn~\textit{et al.}~\cite{bradlyn2017} label them as non-trivial. The extension of the definition of atomic-limit insulators to superconductors discussed in the present work allows one to define ``fragile topological superconductors'' in analogous way, although this analogy is not being pursued here.}

The general strategy underlying the construction of symmetry-based indicators
for a  Hamiltonian $h(\vk)$ is to replace the topological classification of the
matrix-valued function $h(\vk)$ by the simpler problem of the topological
characterization of the hermitian matrices $h(\vk_{\rm s})$ at a selected set
of ``high-symmetry'' points $\{ \vk_{\rm s} \}$ in the Brillouin
zone.~\cite{po2017} The topological characterization of the hermitian matrices
$h(\vk_{\rm s})$ may be considered a set of ``topological band labels'', the
calculation of which is
considerably easier to obtain than the topological classification of the full
functions $h(\vk)$. The group $\SI$ of symmetry-based indicators then follows
by ``dividing out'' all combinations of topological band labels that correspond
to atomic phases and imposing a set of compatibility constraints derived from
the topological classification of the matrices $h(\vk)$ at lower-symmetry
points in the Brillouin zone.\cite{po2017,bradlyn2017} This procedure ensures
that only topological phases with a nontrivial boundary signature have
nontrivial symmetry-based indicators.

Despite their enormous computational advantage,\cite{bradlyn2017}
symmetry-based indicators are not guaranteed to give complete classification
information: There exist topologically nontrivial phases with protected
anomalous boundary signatures, but trivial symmetry-based indicators.  An
example is the quantized Hall effect in the absence of any crystalline
symmetries, for which no symmetry-based indicators exist ({\it i.e.}, the group
$\SI$ is trivial), in spite of the existence of topological phases with nonzero
Chern number. 

In the present article, we extend the construction of symmetry-based indicators
to superconductors, which, on the mean-field level, are described by
Hamiltonians $H(\vk)$ of Bogoliubov-de Gennes (BdG)
type.~\cite{gorkov1958,degennes1966} We explicitly construct the topological
band labels and the symmetry-based indicators for selected point groups using
the complete topological classification of the BdG Hamiltonian $H(\vk_{\rm s})$
at high-symmetry momenta $\vk_{\rm s}$ and compare the symmetry-based
indicators with a full classification of topological phases with nontrivial
boundary signature. Such a comparison gives information to what extent the
symmetry-based indicators can be used as a proxy for a complete classification.
Depending on the crystalline symmetries considered, we find that certain
aspects of the wealth of boundary signatures available to superconducting
phases --- Majorana modes at surfaces or hinges, zero-energy Majorana bound
states at corners --- are reflected in the symmetry-based indicators, but not
all. 

Recently, a number of articles appeared in the literature that also consider
the construction of symmetry-based indicators for superconducting phases. Based
on a general analysis of the principles underlying the construction of
symmetry-based indicators, Ono and Watanabe~\cite{ono2018} arrive at the
conclusion that the sets of symmetry-based indicators that describe
superconducting phases and the underlying normal-state phases are essentially
the same. Our construction of symmetry-based indicators, which is based on the
full classification of ``zero-dimensional'' Hamiltonians at high-symmetry
momenta,\cite{shiozaki2018}
 shows that this statement needs to be corrected to the extent that we
successfully put to use a topological band label based on the Pfaffian of a
BdG-type Hamiltonian $H(\vk_{\rm s})$ at high-symmetry momenta $\vk_{\rm s}$,
which has no counterpart in the normal state. A different approach to the problem of symmetry-based indicators in
superconductors was taken in a later article by these authors, together
with Yanase,~\cite{ono2019b} as well as by Skurativska, Neupert, and
Fischer,~\cite{skurativska2019} who consider the ``weak-pairing limit''
(superconducting order parameter $\Delta$ much smaller than energy
scales typical for the normal-state band structure) and derive a
classification of superconducting phases that is based on the
topological classification of the normal state and the symmetry of the
superconducting order parameter. An approach that uses the normal-state Hamiltonian as its sole
input has the practical advantage that the symmetry-based indicators can be
calculated from the vast body of band-structure knowledge available for normal
phases.  Furthermore, Skurativska {\it et al.} argue that there is no
alternative to such an approach, because an atomic limit can not be defined on
the level of the BdG Hamiltonian.~\cite{skurativska2019} We arrive at a
different conclusion, showing that there exists a consistent definition of an
``atomic-limit superconductor'' as an ``array'' of zero-dimensional
superconductors.~\cite{trifunovic2019} In the weak-pairing limit, the symmetry-based indicators derived here can be expressed in terms of the normal part of the Hamiltonian only, so
that in that limit our approach offers the same computational advantages
as the approaches that rely on the weak-pairing limit at the outset. Very recently, an article by Ono, Po, and Watanabe appeared,\cite{ono2019c} which bases its symmetry indicators on the full BdG Hamiltonian, be it without the Pfaffian band labels, and has a definition of an atomic limit that is consistent with ours. A recent article by Shiozaki also reports the construction of symmetry-based indicators for Hamiltonians of BdG-type and which has results very similar to ours.~\cite{shiozaki2019SI}

The remaining part of this article is organized as follows: In
Sec.~\ref{sec:symmetry_classification_orderparameter} we discuss the symmetry
of the superconducting order parameter and the set of data that needs to be
specified in order to define a ``topological class'' for a crystalline
superconductor. In Sec.~\ref{sec:0d} we discuss the classification and
topological invariants of zero-dimensional Hamiltonians. This discussion is the
cornerstone for defining topological invariants at high-symmetry momenta in
Brillouin zone. We present our main result, a method to construct and calculate
symmetry-based indicators for superconductors, in Sec.~\ref{sec:SI}.
Sections~\ref{sec:examples}--\ref{sec:examples3} contain examples for various crystalline
symmetries compatible with a square or cubic lattice structure, for which we
present a detailed calculation of symmetry-based indicators and relate this to
higher-order boundary phenomenology. We conclude in Sec.~\ref{sec:conclusion}.
The appendices contain additional examples as well as fully worked-out
classifications of anomalous boundary states for crystalline symmetry groups
not readily available in the literature.

\section{Superconductors with crystalline symmetries}
\label{sec:symmetry_classification_orderparameter}
The mean-field theory of superconductors uses an effective non-interacting
description with a Hamiltonian of the Bogoliubov-de Gennes (BdG)
form~\cite{gorkov1958,degennes1966}
\begin{equation}
  H(\vk) = \begin{pmatrix} h(\vk) & \Delta(\vk) \\ \Delta^{\dagger}(\vk) & -h^*(-\vk) \end{pmatrix}, \label{eq:HBdG}
\end{equation}
where the normal-state Hamiltonian $h(\vk)$ is hermitian and the
superconducting order parameter $\Delta(\vk) = -\Delta(-\vk)^{\rm T}$ is
antisymmetric. The $2 \times 2$ block structure describes particle and hole
degrees of freedom. The special choice of the blocks in Eq.~(\ref{eq:HBdG}) is
equivalent to imposing that $H(\vk)$ is antisymmetric under particle-hole
conjugation ${\cal P} = \tau_1 K$,
\begin{equation}
  H(\vk) = -\tau_1 H(-\vk)^* \tau_1,
  \label{eq:HBdGP}
\end{equation}
where $\tau_1$ is a Pauli matrix acting within particle-hole space and $K$ is
complex conjugation. (Note that $\vk \to -\vk$ under complex conjugation.) In
addition to the BdG structure, we assume that the system obeys translation
invariance --- which is what allows us to use the Fourier language in
Eqs.~(\ref{eq:HBdG}) and (\ref{eq:HBdGP}) in the first place --- and that it
has additional symmetries described by the point group $G$. We restrict
ourselves to symmorphic symmetries, for which the unit cell can be chosen in
such a way that it is left invariant under $G$. For the initial discussion we
focus on systems without time-reversal symmetry or other antiunitary
symmetries.

Requiring that the normal part $h(\vk)$ be symmetric under $G$ implies that
there exists a projective representation $u(g)$ for $g \in G$ such that
\begin{equation}
  h(\vk) = u(g) h(g \vk) u^{\dagger}(g).
  \label{eq:htransform}
\end{equation}
Note that the unitary matrix $u(g)$ does not depend on $\vk$, as the symmetry
group $G$ acts within the unit cell. The representation $u(g)$ is projective,
because the transformation rule (\ref{eq:htransform}) determines $u(g)$ up to a
phase factor only. In general, a consistent choice of these phase factors is
possible up to a sign only, the sign ambiguities being captured by the ``factor
system''
\begin{equation}
  \{\zz_{g,h} = u(g h)^{-1} u(g) u(h) = \pm 1\, \mbox{for}\, g,\, h \in G \}
  \label{eq:factorsystem}
\end{equation}
of the projective representation. Two realizations $u_1(g)$ and $u_2(g)$ that
have the same factor system may still differ by a one-dimensional
representation $\Theta(g)$ of $G$ with trivial factor system. Although the
mathematical structure of the theory allows many non-equivalent factor groups
for the representation $u(g)$, for physical systems only two factor groups are
relevant: The trivial one, which applies to spinless particles, and the
nontrivial factor system associated with the spinful particles. The
representation $\Theta(g)$, which describes the difference between two
representations with the same factor system, always has the trivial factor
group.

The canonical form for the representation $u(g)$ is
\begin{equation}
  u(g) = \mbox{diag}\,[r_1(g)\otimes \openone_{N_1},\ldots,r_n(g) \otimes \openone_{N_n}],
  \label{eq:ugeneral}
\end{equation}
where the $r_{\alpha}(g)$ are irreducible representations ("irreps") of $G$ (with the appropriate factor system) and $\openone_{N_{\alpha}}$ the $N_{\alpha} \times N_{\alpha}$ unit matrix. The dimension of $h(\vk)$, corresponding to the total number of orbitals in the unit cell, is $\sum_{\alpha} d_{\alpha} N_{\alpha}$, where $d_{\alpha}$ is the dimension of the irreducible presentation $\alpha$. For notational simplicity, we choose to set $N_1 = N_2 = \ldots = N_n \equiv N$.

The general symmetry constraint for the full BdG Hamiltonian $H(\vk)$ is obtained by allowing different realizations $u_{1,2}(g)$ of the transformation matrices for the particle and hole degrees of freedom. Such a symmetry constraint is compatible with the group operation of $G$ if $u_1(g)$ and $u_2(g)$ have the same factor system, {\it i.e.}, if there exists a one-dimensional representation $\Theta(g)$ of the symmetry group $G$ with trivial factor system such that
\begin{equation}
  u_1(g) \equiv u(g) = u_2(g) \Theta(g).
\end{equation}
For the BdG Hamiltonian $H(\vk)$ one then finds the general symmetry constraint
\begin{equation}
  H(\vk) = U(g) H(g \vk) U(g)^{\dagger},
  \label{eq:HBdGconstraint}
\end{equation}
with
\begin{equation}
  U(g) = \begin{pmatrix} u(g) & 0 \\ 0 & u(g)^* \Theta(g) \end{pmatrix}.
%  H(g \vk) \begin{pmatrix} u_1(g)^{\dagger} & 0 \\ 0 & u_2(g)^{\rm T} \end{pmatrix}.
  \label{eq:HBdGconstraint2}
\end{equation}
For the superconducting order parameter $\Delta(\vk)$, Eqs.\ (\ref{eq:HBdGconstraint}) and (\ref{eq:HBdGconstraint2}) imply that
\begin{equation}
  \Delta(\vk) = u(g) \Delta(g \vk) u(g)^{\rm T} \Theta(g)^*,
  \label{eq:Deltaconstraint}
\end{equation}
{\it i.e.}, $\Delta(\vk)$ transforms under a one-dimensional representation of the group $G$.\cite{sigrist1991} Whereas the projective representation $u(g)$ is a property of the normal phase, the additional phase factor $\Theta(g)$ exists by virtue of the superconducting order only. 
Together, the representation $u$ and the one-dimensional representation $\Theta$ fully determine the symmetry class of the Bogoliubov-de Gennes Hamiltonian (\ref{eq:HBdG}).

Alternatively, Eqs.~(\ref{eq:HBdGconstraint}) and (\ref{eq:HBdGconstraint2})
may be recast in the form of an algebraic relation between the elements $g$ of
the symmetry group $G$ and the particle-hole conjugation operation ${\cal P}$,
\begin{equation}
  g {\cal P} = \Theta(g) {\cal P} g.
\end{equation}
This is the formulation used in
Refs.~\onlinecite{chiu2013,morimoto2013,shiozaki2014,chiu2016,trifunovic2019,geier2018},
which considered the full classification of BdG Hamiltonians in the presence of
a single order-two spatial symmetry ${\cal S}$ and discriminated between the
cases in which ${\cal S}$ commutes or anticommutes with the Fermi constraint
${\cal P}$.

The presence of time-reversal symmetry and/or spin-rotation symmetry does not
change the above considerations, provided these commute with the crystalline
symmetries. Time-reversal symmetry imposes the additional constraints\cite{altland1997}
\begin{equation}
  h(\vk) = \sigma_2 h(-\vk)^* \sigma_2, \ \
  \Delta(\vk) = \sigma_2 \Delta(-\vk)^* \sigma_2,
  \label{eq:TRSconstraints}
\end{equation}
where $\sigma_2$ is a Pauli matrix that acts in spin space. With time-reversal
symmetry, the one-dimensional representation $\Theta(g)$ of $G$ that
characterizes the superconducting state must be real. In the presence of spin
rotation symmetry, one arrives at an effective description in terms of a
BdG-type Hamiltonian of the form~(\ref{eq:HBdG}), but with a symmetric order
parameter $\Delta(\vk) = \Delta(-\vk)^{\rm T}$, so that $H(\vk)$ is
antisymmetric under the antiunitary operation $\tau_2 K$, which plays the role
of an effective particle-hole symmetry. If time-reversal symmetry is present
additionally, $H(\vk)$ is symmetric under complex conjugation $K$, which plays
the role of an effective time-reversal operation. Note that while ${\cal P}$
and ${\cal T}$ square to $1$ and $-1$, respectively, in the spinful case, the
effective particle-hole conjugation and time-reversal operations square to $-1$
and $1$, respectively, if spin-rotation symmetry is present. Similarly, in the presence of spin-rotation symmetry, the effective representation of crystalline symmetry operations is changed into the corresponding ``spinless'' type.

\begin{table}
\begin{tabular*}{\columnwidth}{c @{\extracolsep{\fill}} ccc} \hline\hline
  & $\Gamma$ & $\id$ & ${\cal I}_\pi$
  \\ \hline 
  $r_{+}$ & $A_{g}$ & 1 & 1 \\
  $r_{-}$ & $A_{u}$ & 1 & -1 \\ \hline\hline
\end{tabular*}
\caption{Irreducible representations of the point group $C_{i}$. The second
column lists the standard crystallographic notation for the representations,
the first column gives the notation used in the text.}
\label{tab:rep_Ci}
\end{table}

The general framework described here is best illustrated using examples. As a
first example, we consider a system with inversion symmetry. In this case the
group $G = C_i = \{ \id, {\cal I} \}$, where $\id$ is the identity and ${\cal
I}$ is the inversion operation. Choosing a basis that has a well-defined parity
under inversion, one finds that one may represent $G$ by
\begin{equation}
  u(g) = \mbox{diag}\,[r_+(g),r_-(g)],\label{eq:upm_inversion} 
\end{equation}
where $r_{\pm}(g)$ are the two irreducible representations of $G$, see
Table~\ref{tab:rep_Ci}, which are both one-dimensional. Specifically,
Eq.~(\ref{eq:upm_inversion}) reads
\begin{equation}
  u(\id) = \rho_0,\ \
  u({\cal I}) = \rho_3,
\end{equation}
where $\rho_3$ is a Pauli matrix that acts in the space of even/odd-parity
states and $\rho_0$ is the $2 \times 2$ unit matrix. From the two
one-dimensional representations of $G$ we also find two possibilities for the
representation $U(g)$ for the BdG Hamiltonian $H(\vk)$. Choosing the trivial
one-dimensional representation $\Theta(g) = r_+(g)$ in
Eq.~(\ref{eq:HBdGconstraint}) (the $A_g$ representation, see
Table~\ref{tab:rep_Ci}), we find that the BdG Hamiltonian $H(\vk)$ obeys the
symmetry constraint
\begin{equation}
  H(\vk) = (\rho_3 \nootimes \tau_0) H(-\vk) (\rho_3 \nootimes \tau_0),
  \label{eq:umm_inversion}
\end{equation}
where $\tau_0$ is the identity matrix in particle-hole space and $\rho_3$ is a
Pauli matrix that acts in the space of even/odd-parity states. For the blocks
$h(\vk)$ and $\Delta(\vk)$ this implies
\begin{equation}
  h(\vk) = \rho_3 h(-\vk) \rho_3,\ \
  \Delta(\vk) = \rho_3 \Delta(-\vk) \rho_3,
\end{equation}
{\it i.e.}, the order parameter is even under inversion. Alternatively,
choosing the nontrivial one-dimensional representation $\Theta(g) = u_-(g)$ in
Eq.~(\ref{eq:HBdGconstraint}) (the $A_u$ representation), we find that the BdG
Hamiltonian $H(\vk)$ satisfies
\begin{equation}
  H(\vk) = (\rho_3 \nootimes \tau_3) H(-\vk) (\rho_3 \nootimes \tau_3).
\end{equation}
In this case the blocks $h(\vk)$ and $\Delta(\vk)$ satisfy the constraint
\begin{equation}
  h(\vk) = \rho_3 h(-\vk) \rho_3,\ \
  \Delta(\vk) = -\rho_3 \Delta(-\vk) \rho_3,
\end{equation}
so that the order parameter is odd under inversion. The two different
transformation rules of the order parameter $\Delta(\vk)$ under inversion imply
two different algebraic relations between the inversion operation ${\cal I}$
and the Fermi constraint ${\cal P}$: If $\Delta(\vk)$ is even under ${\cal I}$,
${\cal I}$ and ${\cal P}$ commute, whereas ${\cal I}$ and ${\cal P}$
anticommute if $\Delta(\vk)$ is odd under inversion.

\begin{table}
\begin{tabular*}{\columnwidth}{c @{\extracolsep{\fill}} ccccc} \hline\hline
	& $\Gamma$ & $\id$ & ${\cal R}_{\pi/2}$ & ${\cal R}^2_{\pi/2}$ & ${\cal R}^3_{\pi/2}$
  \\ \hline 
  $r_{0}$ & $A$ & $1$ & $1$ & $1$ & $1$ \\
  $r_{\pi/2}$ & $^2 E$ & $1$ & $i$ & $-1$ & $-i$ \\
  $r_{\pi}$ & $B$ & $1$ & $-1$ & $1$ & $-1$ \\
  $r_{3 \pi/2}$ & $^1 E$ & $1$ & $-i$ & $-1$ & $i$ \\ \hline
  $r_{\pi/4}$  & $^1 \overline{E}_1$ & $1$ & $e^{i \pi/4}$ & $i$ & $e^{3 \pi i/4}$ \\
  $r_{3\pi/4}$ & $^2 \overline{E}_2$ & $1$ & $e^{3 \pi i/4}$ & $-i$ & $e^{i \pi/4}$ \\
  $r_{5\pi/4}$ & $^1 \overline{E}_2$ & $1$ & $e^{5 \pi i/4}$ & $i$ & $e^{7 \pi i/4}$ \\
  $r_{7\pi/4}$ & $^2 \overline{E}_1$ & $1$ & $e^{7 \pi i/4}$ & $-i$ & $e^{5 \pi i/4}$ \\ \hline\hline
\end{tabular*}
\caption{Spinless representations (top four rows) and one-dimensional spinful representations (bottom four rows) of the point group $C_{4}$. The second column lists the standard crystallographic notation for the representations, the first column gives the notation used in the text.}
\label{tab:rep_C4}
\end{table}

As a second example, we consider a fourfold rotation around a fixed axis for a
system of spinful particles. We write $G = C_4 = \{ \id, {\cal
R}_{\pi/2}^{\vphantom{1}}, {\cal R}_{\pi/2}^2, {\cal R}_{\pi/2}^3 \}$, where
the generator ${\cal R}_{\pi/2} \equiv {\cal R}$ denotes a clockwise rotation
by $\pi/2$. In this case one has ${\cal R}^4 = -\id$, which corresponds to the
factor system 
\begin{equation}
  \zz_{{\cal R}^k,{\cal R}^l} = \left\{ \begin{array}{ll}
  1 & \mbox{if $0 \le k + l < 4$}, \\
 -1 & \mbox{if $4 \le k+l < 8$}. \end{array} \right.
\end{equation}
Choosing basis states with well-defined angular momentum $j=\frac{1}{2}$,
$\frac{3}{2}$, $\frac{5}{2}$, and $\frac{7}{2}$ (defined modulo $4$), we find
the spinful representation
\begin{equation}
  u(g) = \mbox{diag}\,[r_{\pi/4}(g),r_{3\pi/4}(g),r_{5\pi/4}(g),r_{7\pi/4}(g)],
  \label{eq:ug_diag_C4}
\end{equation}
see Table~\ref{tab:rep_C4}.

The transformation rule for the BdG Hamiltonian $H(\vk)$ requires the choice of
a a one-dimensional spinless representation $\Theta$ of $G$. There are four of
those, and we denote these as $r_{\theta}$ with $\theta = 0$, $\pi/2$, $\pi$,
and $3\pi/2$, see Table \ref{tab:rep_C4}. Proceeding as before, we find four
possible transformation rules for the BdG Hamiltonian $H(\vk)$ under a $\pi/2$
rotation ${\cal R}$,
\begin{equation}
  H(\vk) = U_{\theta}({\cal R}) H({\cal R} \vk) U_{\theta}({\cal R})^{\dagger},
\end{equation}
with
\begin{equation}
  U_{\theta}({\cal R}) = 
  \begin{pmatrix} u({\cal R}) & 0 \\ 0 & u({\cal R})^* e^{i \theta} \end{pmatrix}.
%  H({\cal R}\vk) \begin{pmatrix} u_0 & 0 \\ 0 & u_0 e^{i \theta} \end{pmatrix}^{\dagger}.
  \label{eq:Ufourfold}
\end{equation}
Alternatively, the transformation rules for the blocks $h(\vk)$ and $\Delta(\vk)$ are
\begin{align}
  h(\vk) =&\, u({\cal R}) h({\cal R} \vk) u({\cal R})^{\dagger},\nonumber \\
  \Delta(\vk) =&\, e^{i \theta} u({\cal R}) \Delta({\cal R} \vk) u({\cal R})^{\rm T},
\end{align}
which corresponds to a superconducting order parameter with finite angular
momentum.

The time-reversal symmetry operation is represented by $U({\cal T}) K$,
where the definition of $U({\cal T})$ in the angular momentum basis is
\begin{equation}
	U({\cal T})= 
  \begin{pmatrix} 0 & 0 & 0 & -i\\
                             0 & 0 & -i& 0 \\
                             0 & i & 0 & 0 \\
                             i & 0 & 0 & 0 \end{pmatrix}.
\end{equation}
Time-reversal commutes with ${\cal P}$ and satisfies the property ${\cal T}^2 =
-1$.  Time-reversal symmetry imposes the additional
conditions~(\ref{eq:TRSconstraints}) on the blocks $h(\vk)$ and $\Delta(\vk)$.
One easily verifies that these additional conditions allow for a nonzero
superconducting order parameter $\Delta(\vk)$ for the real one-dimensional
representations corresponding to $\theta=0$ or $\theta=\pi$ only.

As a third example, we consider spinful fermions in a system with two
perpendicular mirror symmetries, $G = C_{2v}$. We write the point group
elements as $G = \{ \id, {\cal M}_x, {\cal M}_y, {\cal R}_\pi \}$, where ${\cal
M}_{x}$ and ${\cal M}_{x}$ are mirror reflections in planes perpendicular to
the $x$ and $y$ axis, respectively, and ${\cal R}_{\pi} = {\cal M}_y {\cal
M}_x$ is a rotation around the $z$ axis by $\pi$. For spinful particles the
factor system is nontrivial with ${\cal M}_x^2 = {\cal M}_y^2 = {\cal
R}_{\pi}^2 = -1$. In that case there is only one (projective) representation
$u$ of $G$ (up to unitary transformations), which is two dimensional, see Table
\ref{tab:rep_C2v}. Superconducting phases are characterized by one-dimensional
representations of $G$ with a trivial factor system. There are four of these,
denoted $r_{\sigma_x,\sigma_y}(g)$, with $\sigma_{x,y} = \pm 1$, see Table
\ref{tab:rep_C2v}. Each of the one-dimensional representations of $G$
corresponds to a distinct symmetry of the superconducting order parameter
$\Delta$. For example, for the one-dimensional representation $r_{--}$, the
order parameter is odd under mirror reflections in the $yz$ and $xz$ planes and
even under a twofold rotation around the $z$ axis.
\begin{table}
\begin{tabular*}{\columnwidth}{c @{\extracolsep{\fill}} ccccc}
  \hline\hline
  & $\Gamma$ & $\id$ & ${\cal R}_\pi$ & ${\cal M}_x$ & ${\cal M}_y$
  \\ \hline 
  $r_{++}$ & $A_{1}$ & 1 & 1 & 1 & 1 \\
  $r_{--}$ & $A_{2}$ & 1 & 1 & -1 & -1 \\
  $r_{-+}$ & $B_{1}$ & 1 & -1 & 1 & -1 \\
  $r_{+-}$ & $B_{2}$ & 1 & -1 & -1 & 1 \\ \hline
  $u$ & $E$ & $\sigma_0$ & $i \sigma_3$ & $i \sigma_1$ & $i \sigma_2$ \\
  \hline\hline
\end{tabular*}
\caption{Spinful representation (bottom row) and one-dimensional spinless representations (top four rows) of the point group $C_{2v}$. The representation $u$ is unique up to unitary transformations. The second column lists the standard crystallographic notation for the representations, the first column gives the notation used in the text.}
\label{tab:rep_C2v}
\end{table}

\section{Classification of zero-dimensional Hamiltonians}
\label{sec:0d}
The cornerstone of the construction of symmetry-based indicators is the
classification of zero-dimensional Hamiltonians with additional unitary
symmetries. Concretely, the classification problem is that of a Hamiltonian $H$
with or without time-reversal symmetry, particle-hole antisymmetry, or chiral
antisymmetry, and with additional unitary symmetries specified by the group
$G$. The presence or absence of time-reversal symmetry, particle-hole
antisymmetry, and chiral antisymmetry determines the tenfold-way
(Altland-Zirnbauer) class and is indicated by the triple\cite{altland1997}
\begin{equation}
  \eta = (\eta_{\cal T},\eta_{\cal P},\eta_{\cal C}).
  \label{eq:etadef}
\end{equation}
Here $\eta_{\cal T} = {\cal T}^2 = \pm 1$ if time-reversal symmetry is present
and $\eta_{\cal T} = 0$ if time-reversal symmetry is absent. Similarly,
$\eta_{\cal P} = {\cal P}^2 = \pm 1$ or $0$ if particle-hole antisymmetry is
present or absent and $\eta_{\cal C} = {\cal C}^2 = 1$ or $0$ in the presence
or absence of chiral antisymmetry, respectively. Table \ref{tab:1} summarizes
the tenfold way classification and lists both the notation using the triple
$(\eta_{\cal T},\eta_{\cal P},\eta_{\cal C})$ and the Cartan labeling. If the
context allows, we will use the Cartan label instead of the triple $\eta =
(\eta_{\cal T},\eta_{\cal P},\eta_{\cal C})$. 

\begin{table}
  \begin{tabular*}{\columnwidth}{c @{\extracolsep{\fill}} ccc}
    \hline\hline
    $\eta = (\eta_{\cal T},\eta_{\cal P},\eta_{\cal C})$ & 
    Cartan & $\mathfrak{K}[0]$ & $\nu$\\ \hline
    $(0,0,0)$ & A      & $\ZZ$ & $\mathfrak{N}$ \\
    $(0,0,1)$ & AIII & $0$ & - \\ \hline
    $(1,0,0)$ & AI     & $\ZZ$ & $\mathfrak{N}$\\
    $(1,1,1)$ & BDI    & $\ZZ_2$ & $\mathfrak{p}$\\
    $(0,1,0)$ & D      & $\ZZ_2$ & $\mathfrak{p}$\\
    $(-1,1,1)$ & DIII   & $0$ & - \\
    $(-1,0,0)$ & AII    & $\ZZ$ & $\mathfrak{N}$\\
    $(-1,-1,1)$ & CII    & $0$ & -\\
    $(0,-1,0)$ & C     & $0$ & -\\
    $(1,-1,1)$ & CI    & $0$ & -\\ \hline \hline
  \end{tabular*}
\caption{\label{tab:1} The tenfold-way
symmetry classes are defined by the absence or presence time-reversal symmetry,
particle-hole antisymmetry, or chiral antisymmetry. The triple $\eta =
(\eta_{\cal T},\eta_{\cal P},\eta_{\cal C})$ (left column) indicates the
presence or absence of these symmetries as well as the square of the symmetry
operation if it is present. The second column gives the Cartan label for the
symmetry class; the third column lists the corresponding classifying group
$\mathfrak{K}_{\eta}[0]$ for zero-dimensional Hamiltonians. The rightmost
column lists the topological invariant $\nu$, where $\mathfrak{N}$ is the
number of eigenstates (Kramers pairs in class AII) with negative energy and
$\mathfrak{p}$ is the Pfaffian invariant as defined in
Eq.~(\ref{eq:definition_Pfaffian_invariant}).}
\end{table}

For physically relevant systems, the crystalline symmetry group $G$ commutes
with the time-reversal operation ${\cal T}$, which is what we assume
throughout. The algebraic relation between $G$ and the particle-hole
conjugation operation ${\cal P}$ or the chiral operation ${\cal C} = {\cal PT}$
is characterized by a one-dimensional representation $\Theta(g)$ of $G$, as
discussed in Sec.\ \ref{sec:symmetry_classification_orderparameter}. 

With a suitable choice of basis, the (projective) representation $U(g)$ for a
Bogoliubov-de Gennes Hamiltonian $H$ can be brought into a canonical form
analogous to Eq.~(\ref{eq:ugeneral}),
\begin{equation}
  U(g) = \mbox{diag}\,[r_1(g)\otimes \openone_{2 N_1},\ldots,r_n(g) \otimes \openone_{2 N_n}],
  \label{eq:UUgeneral}
\end{equation}
where the $r_{\alpha}(g)$ are irreducible representations of $G$. To keep the
notation simple, we make the choice $N_1 = N_2 = \ldots = N_n = N$. The
Bogoliubov-de Gennes Hamiltonian $H$ is then brought into a block-diagonal
form, 
\begin{equation}
  H = \mbox{diag}\,(\openone_{d_1} \otimes H_{1},\ldots,\openone_{d_n} \otimes H_{n}), \label{eq:HHgeneral}
\end{equation}
where $d_{\alpha}$ is the dimension of the irreducible representation
$r_{\alpha}$. The diagonal blocks $H_{\alpha}$ have dimension $2 N$ and are no
longer subject to additional unitary symmetries. However, because of the basis
change involved in representing $U$ in the canonical form (\ref{eq:UUgeneral}),
the action of the fundamental symmetry operations ${\cal T}$, ${\cal P}$,
and/or ${\cal C}$ on these blocks need not be the same as their action on the
original Hamiltonian $H$ and different blocks may be related to each other by
${\cal T}$, ${\cal P}$ and/or ${\cal C}$. Hence, additional considerations are
needed to determine which blocks are independent and to what tenfold-way class
they belong, which we now discuss.

{\it Time-reversal operation ${\cal T}$.---} If the irreducible representations
$r_{\alpha}$ and $r_{\alpha}^*$ are the same (up to a unitary transformation),
the time-reversal operation ${\cal T}$ acts within the diagonal block
corresponding to $r_{\alpha}$ and takes the form of an effective time-reversal
operation for $H_{\alpha}$, but with a square $W_{\cal T}(\alpha)$ that may
differ from $\eta_{\cal T}$. If the irreducible representations $r_{\alpha}$
and $r_{\alpha}^*$ are different, ${\cal T}$ interchanges the diagonal blocks
corresponding to $r_{\alpha}$ and $r_{\alpha}^*$. In most examples, which of
these cases applies can easily be found by inspection, although one may also
obtain the answer from the ``Wigner test'' by calculating the
quantity\cite{bradley1972,po2017,shiozaki2018} 
\begin{equation}
  W_{\cal T}(\alpha) = \eta_{\cal T}
  \left\langle \zz_{g,g} \chi_{\alpha}(g^2) \right\rangle_G,
  \label{eq:WignerT}
\end{equation}
where $\langle \ldots \rangle_G$ denotes an average over all $g \in G$,
$\zz_{g,g}$ is an element of the factor system, see Eq.\
(\ref{eq:factorsystem}), and $\chi_{\alpha}(g) = \mbox{tr}\, r_{\alpha}(g)$ is
the character of the irreducible representation. The quantity $W_{\cal
T}(\alpha)$ can take the values $W_{\cal T}(\alpha)= \pm 1$ or $0$,
corresponding to the two cases discussed above.

{\it Particle-hole conjugation ${\cal P}$.---} Similarly, if the irreducible
representations $r_{\alpha}$ and $\Theta r_{\alpha}^*$ are the same, ${\cal P}$
acts as an effective particle-hole conjugation operator for $H_{\alpha}$ with
square $W_{\cal P}(\alpha) = \pm 1$, whereas if $r_{\alpha}$ and $\Theta
r_{\alpha}^*$ are different irreducible representations of $G$, ${\cal P}$
interchanges the corresponding diagonal blocks. Again, one may find out which
of the three cases applies by calculating
\begin{equation}
  W_{\cal P}{(\alpha)} = \eta_{\cal P}
  \left\langle \Theta(g)^* \zz_{g,g} \chi_{\alpha}(g^2) \right\rangle_G,
  \label{eq:WignerP}
\end{equation} 
which takes the values $\pm 1$ or $0$, respectively, for the two cases described above. 

{\it Chiral operation ${\cal C}$.---} Finally, for the chiral operation ${\cal
C} = {\cal PT}$ one only needs to distinguish two cases: If $r_{\alpha}$ and
$\Theta r_{\alpha}$ are the same irreducible representation of $G$, ${\cal C}$
acts as an effective antisymmetry of $H_{\alpha}$, whereas ${\cal C}$
interchanges the diagonal blocks corresponding to $r_{\alpha}$ and $\Theta
r_{\alpha}$ if they are different. This defines the quantity $W_{\cal
C}{(\alpha)}$, which takes the values $0$ or $1$.

We denote the full classifying group for zero-dimensional Hamiltonians with the
additional symmetry group $G$ as $\mathfrak{K}_{\eta}[G,\Theta]$, where the
multiindex $\eta$ indicates the tenfold-way class, see Table \ref{tab:1}, and
$\Theta$ is the one-dimensional representation of $G$ that characterizes the
superconducting order parameter. The classifying group also depends on the
factor system of $G$, but we do not write this dependence explicitly. The
argument $\Theta$ is omitted for non-superconducting phases. The above
considerations then lead to the result
\begin{equation}
  \mathfrak{K}_{\eta}[G,\Theta] =
  \prod_{r_\alpha\, \text{irrep of}\, G} 
    \mathfrak{K}_{W(\alpha)}[0],
  \label{eq:KK}
\end{equation}
where the multiindex $W(\alpha) = (W_{\cal T}(\alpha),W_{\cal P}(\alpha),W_{\cal C}(\alpha))$ and the product is over the irreducible representations $r_{\alpha}$ of $G$, where only one representative appears in the product if multiple irreps
are linked by the fundamental symmetries ${\cal T}$, ${\cal
P}$, or ${\cal C}$. The classifying groups $\mathfrak{K}_{\eta}[0]$, which
apply to tenfold-way classes without additional crystalline
symmetries,\cite{ryu2010,kitaev2009} can be found in Table~\ref{tab:1}.  We
remark that the classification approach described here also works for
$d$-dimensional Hamiltonians $H(\vk)$, as long as $G$ contains only onsite
({\it i.e.} non-spatial, $\vk$-independent) crystalline symmetries.

{\it Examples.---} We illustrate this general procedure using the three
examples already discussed in the previous Section. The classification results
for these three examples as well as for other relevant point groups are
summarized in Tables~\ref{tab:classD_0d}--\ref{tab:class_0d_TRS_SpinRot}. 

\begin{table*}[!]
\begin{tabular*}{\textwidth}{c @{\extracolsep{\fill}} cc c c} \hline\hline
$G$  & $\Theta$  &  & $(r_{\alpha})_{W_{{\cal P}(\alpha)}}$ & $\mathfrak{K}_{\classD}[G,\Theta]$ \tabularnewline
\hline 
\hline 
$C_{1}$  & $A$  & $1$ & $(A)_{\eta_{{\cal P}}}$ & $\ZZ_{2}$ \tabularnewline
\hline 
$C_{i}$  & $A_{g}$  & $r_{+}$ & $(A_{g})_{\eta_{{\cal P}}},(A_{u})_{\eta_{{\cal P}}}$ & $\ZZ_{2}^{2}$ \tabularnewline
 & $A_{u}$  & $r_{-}$ & $(A_{g},A_{u})_{0}$ & $\ZZ$ \tabularnewline
\hline 
$C_{s}$  & $A'$  & $r_{+}$ & $(^{1}\bar{E},^{2}\bar{E})_{0}$ & $\ZZ$ \tabularnewline
 & $A''$  & $r_{-}$ & $(^{1}\bar{E})_{\eta_{{\cal P}}},(^{2}\bar{E})_{\eta_{{\cal P}}}$ & $\ZZ_{2}^{2}$ \tabularnewline
\hline 
$C_{2}$  & $A$  & $r_{+}$ & $(^{1}\bar{E},^{2}\bar{E})_{0}$ & $\ZZ$ \tabularnewline
 & $B$  & $r_{-}$ & $(^{1}\bar{E})_{\eta_{{\cal P}}},(^{2}\bar{E})_{\eta_{{\cal P}}}$ & $\ZZ_{2}^{2}$ \tabularnewline
\hline 
$C_{3}$  & $A_{1}$  & $r_{0}$ & $\bar{E}_{\eta_{{\cal P}}},(^{1}\bar{E},^{2}\bar{E})_{0}$ & $\ZZ\times\ZZ_{2}$ \tabularnewline
 & $^{2}E$  & $r_{2\pi/3}$ & $^{2}\bar{E}_{\eta_{{\cal P}}},(\bar{E},^{1}\bar{E})_{0}$ & $\ZZ\times\ZZ_{2}$ \tabularnewline
 & $^{1}E$  & $r_{4\pi/3}$ & $^{1}\bar{E}_{\eta_{{\cal P}}},(\bar{E},^{2}\bar{E})_{0}$ & $\ZZ\times\ZZ_{2}$ \tabularnewline
\hline 
$C_{4}$  & $A$  & $r_{0}$ & $(^{1}\bar{E}_{1},^{2}\bar{E}_{1})_{0},(^{1}\bar{E}_{2},^{2}\bar{E}_{2})_{0}$ & $\ZZ^{2}$ \tabularnewline
 & $B$  & $r_{\pi}$ & $(^{1}\bar{E}_{1},^{2}\bar{E}_{2})_{0},(^{1}\bar{E}_{2},^{2}\bar{E}_{1})_{0}$ & $\ZZ^{2}$ \tabularnewline
 & $^{2}E$  & $r_{\pi/2}$ & $(^{1}\bar{E}_{1})_{\eta_{{\cal P}}},(^{1}\bar{E}_{2})_{\eta_{{\cal P}}},(^{2}\bar{E}_{1},^{2}\bar{E}_{2})_{0}$ & $\ZZ\times\ZZ_{2}^{2}$ \tabularnewline
 & $^{1}E$  & $r_{3\pi/2}$ & $(^{2}\bar{E}_{1})_{\eta_{{\cal P}}},(^{2}\bar{E}_{2})_{\eta_{{\cal P}}},(^{1}\bar{E}_{1},^{1}\bar{E}_{2})_{0}$ & $\ZZ\times\ZZ_{2}^{2}$ \tabularnewline
\hline 
$C_{6}$  & $A$  & $r_{0}$ & $(^{1}\bar{E}_{1},^{2}\bar{E}_{1})_{0},(^{1}\bar{E}_{2},^{2}\bar{E}_{2})_{0},(^{1}\bar{E}_{3},^{2}\bar{E}_{3})_{0}$ & $\ZZ^{3}$ \tabularnewline
 & $B$  & $r_{\pi}$ & $(^{1}\bar{E}_{1})_{\eta_{{\cal P}}},({}^{2}\bar{E}_{1})_{\eta_{{\cal P}}},(^{1}\bar{E}_{3},^{2}\bar{E}_{2})_{0},(^{2}\bar{E}_{3},^{1}\bar{E}_{2})_{0}$ & $\ZZ^{2}\times\ZZ_{2}^{2}$ \tabularnewline
 & $^{2}E_{1}$  & $r_{4\pi/3}$ & $(^{1}\bar{E}_{1},^{2}\bar{E}_{3})_{0},(^{2}\bar{E}_{1},^{1}\bar{E}_{2})_{0},(^{2}\bar{E}_{2},^{1}\bar{E}_{3})_{0}$ & $\ZZ^{3}$ \tabularnewline
 & $^{2}E_{2}$  & $r_{\pi/3}$ & $(^{1}\bar{E}_{3})_{\eta_{{\cal P}}},({}^{2}\bar{E}_{2})_{\eta_{{\cal P}}},(^{2}\bar{E}_{1},^{2}\bar{E}_{3})_{0},(^{1}\bar{E}_{1},^{1}\bar{E}_{2})_{0}$ & $\ZZ^{2}\times\ZZ_{2}^{2}$ \tabularnewline
 & $^{1}E_{1}$  & $r_{2\pi/3}$ & $(^{1}\bar{E}_{1},^{2}\bar{E}_{2})_{0},(^{2}\bar{E}_{1},^{1}\bar{E}_{3})_{0},(^{2}\bar{E}_{3},^{1}\bar{E}_{2})_{0}$ & $\ZZ^{3}$ \tabularnewline
 & $^{1}E_{2}$  & $r_{-\pi/3}$ & $(^{1}\bar{E}_{2})_{\eta_{{\cal P}}},({}^{2}\bar{E}_{3})_{\eta_{{\cal P}}},(^{1}\bar{E}_{1},^{1}\bar{E}_{3})_{0},(^{2}\bar{E}_{1},^{2}\bar{E}_{2})_{0}$ & $\ZZ^{2}\times\ZZ_{2}^{2}$ \tabularnewline
\hline 
$C_{2v}$  & $A_{1}$  & $r_{++}$ & $(\bar{E})_{-\eta_{{\cal P}}}$ & 0\tabularnewline
 & $A_{2}$  & $r_{--}$ & $(\bar{E})_{\eta_{{\cal P}}}$ & $\ZZ_{2}$ \tabularnewline
 & $B_{1}$  & $r_{-+}$ & $(\bar{E})_{\eta_{{\cal P}}}$ & $\ZZ_{2}$ \tabularnewline
 & $B_{2}$  & $r_{+-}$ & $(\bar{E})_{\eta_{{\cal P}}}$ & $\ZZ_{2}$ \tabularnewline
\hline 
$C_{3v}$  & $A_{1}$  &  & $(^{1}\bar{E},^{2}\bar{E})_{0},(\bar{E}_{1})_{-\eta_{{\cal P}}}$ & $\ZZ$ \tabularnewline
 & $A_{2}$  &  & $(^{1}\bar{E})_{\eta_{{\cal P}}},(^{2}\bar{E})_{\eta_{{\cal P}}},(\bar{E}_{1})_{\eta_{{\cal P}}}$ & $\ZZ_{2}^{3}$ \tabularnewline
\hline 
$C_{4v}$  & $A_{1}$  &  & $(\bar{E}_{1})_{-\eta_{{\cal P}}},(\bar{E}_{2})_{-\eta_{{\cal P}}}$ & 0\tabularnewline
 & $A_{2}$  &  & $(\bar{E}_{1})_{\eta_{{\cal P}}},(\bar{E}_{2})_{\eta_{{\cal P}}}$ & $\ZZ_{2}^{2}$ \tabularnewline
 & $B_{1}$  &  & $(\bar{E}_{1},\bar{E}_{2})_{0}$ & $\ZZ$ \tabularnewline
 & $B_{2}$  &  & $(\bar{E}_{1},\bar{E}_{2})_{0}$ & $\ZZ$ \tabularnewline
\hline 
$C_{6v}$  & $A_{1}$  &  & $(\bar{E}_{1})_{-\eta_{{\cal P}}},(\bar{E}_{2})_{-\eta_{{\cal P}}},(\bar{E}_{3})_{-\eta_{{\cal P}}}$ & $0$ \tabularnewline
 & $A_{2}$  &  & $(\bar{E}_{1})_{\eta_{{\cal P}}},(\bar{E}_{2})_{\eta_{{\cal P}}},(\bar{E}_{3})_{\eta_{{\cal P}}}$ & $\ZZ_{2}^{3}$ \tabularnewline
 & $B_{1}$  &  & $(\bar{E}_{3})_{1},(\bar{E}_{1},\bar{E}_{2})_{0}$ & $\ZZ\times\ZZ_{2}$ \tabularnewline
 & $B_{2}$  &  & $(\bar{E}_{3})_{1},(\bar{E}_{1},\bar{E}_{2})_{0}$ & $\ZZ\times\ZZ_{2}$ \tabularnewline \hline \hline
\end{tabular*}
\caption{
    Classification of zero-dimensional BdG Hamiltonians with point group
	$G$ and one-dimensional representation $\Theta$ describing the symmetry
	of the superconducting order parameter in the absence of time-reversal
	symmetry and spin-rotation symmetry, corresponding to
	tenfold-way class D, $\eta_{\cal P} = 1$. 
	The second and third columns list the
	one-dimensional representation $\Theta$ using the standard
	crystallographic notation and the notation used in the examples in the
	main text, respectively. The fourth column lists the set $(
	r_\alpha)_{W_\mathcal{P} (\alpha)}$ of irreps $r_\alpha$ together with
	the result of the Wigner test $W_\mathcal{P}(\alpha)$. In case irreps
	$r_\alpha$ and $\Theta r_\alpha^*$ are paired by particle-hole
	antisymmetry ($W_\mathcal{P}(\alpha) = 0$), the paired representations
	are appear in brackets. From this information one can read off the
	classification, invariants, generators and representations using
	Table~\ref{tab:Wigner_cases_TP}. For convenience we list the result for
	the classifying group $\mathfrak{K}_\classD [G, \Theta]$ in
	the last column.}
\label{tab:classD_0d}
\end{table*}

\begin{table*}[!]
\begin{tabular*}{\textwidth}{c @{\extracolsep{\fill}} cc c c} \hline\hline
$G$  & $\Theta$  &  & $(r_{\alpha})_{W_{{\cal P}(\alpha)}}$ & $\mathfrak{K}_{\classC}[G,\Theta]$ \tabularnewline
\hline 
\hline
$C_{1}$  & $A$  & $1$ & $(A)_{\eta_{{\cal P}}}$ & 0\tabularnewline
\hline 
$C_{i}$  & $A_{g}$  & $r_{+}$ & $(A_{g})_{\eta_{{\cal P}}},(A_{u})_{\eta_{{\cal P}}}$ & 0\tabularnewline
 & $A_{u}$  & $r_{-}$ & $(A_{g},A_{u})_{0}$ & $\ZZ$ \tabularnewline
\hline 
$C_{s}$  & $A'$  & $r_{+}$ & $(A')_{\eta_{{\cal P}}},(A'')_{\eta_{{\cal P}}}$ & 0\tabularnewline
 & $A''$  & $r_{-}$ & $(A',A'')_{0}$ & $\ZZ$ \tabularnewline
\hline 
$C_{2}$  & $A$  & $r_{+}$ & $(A)_{\eta_{{\cal P}}},(B)_{\eta_{{\cal P}}}$ & 0\tabularnewline
 & $B$  & $r_{-}$ & $(A,B)_{0}$ & $\ZZ$ \tabularnewline
\hline 
$C_{3}$  & $A_{1}$  & $r_{0}$ & $(A_{1})_{\eta_{{\cal P}}},(^{1}E,^{2}E)_{0}$ & $\ZZ$ \tabularnewline
 & $^{2}E$  & $r_{2\pi/3}$ & $(^{1}E)_{\eta_{{\cal P}}},(A_{1},^{1}E)_{0}$ & $\ZZ$ \tabularnewline
 & $^{1}E$  & $r_{4\pi/3}$ & $(^{2}E)_{\eta_{{\cal P}}},(A_{1},^{2}E)_{0}$ & $\ZZ$ \tabularnewline
\hline 
$C_{4}$  & $A$  & $r_{0}$ & $A_{\eta_{\mathcal{P}}},B_{\eta_{\mathcal{P}}},(^{1}E,^{2}E)_{0}$ & $\ZZ$ \tabularnewline
 & $B$  & $r_{\pi}$ & $(^{1}E)_{\eta_{{\cal P}}},(^{2}E)_{\eta_{{\cal P}}},(A,B)_{0}$ & $\ZZ$ \tabularnewline
 & $^{2}E$  & $r_{\pi/2}$ & $(A,^{2}E)_{0},(B,^{1}E)_{0}$ & $\ZZ^{2}$ \tabularnewline
 & $^{1}E$  & $r_{3\pi/2}$ & $(A,^{1}E)_{0},(B,^{2}E)_{0}$ & $\ZZ^{2}$ \tabularnewline
\hline 
$C_{6}$  & $A$  & $r_{0}$ & $A_{\eta_{\mathcal{P}}},B_{\eta_{\mathcal{P}}},(^{1}E_{1},^{2}E_{1})_{0},(^{1}E_{2},^{2}E_{2})_{0}$ & $\ZZ^{2}$ \tabularnewline
 & $B$  & $r_{\pi}$ & $(A,B)_{0},(^{1}E_{1},^{2}E_{2})_{0},(^{1}E_{2},^{2}E_{1})_{0}$ & $\ZZ^{3}$ \tabularnewline
 & $^{2}E_{1}$  & $r_{4\pi/3}$ & $(^{1}E_{1})_{\eta_{\mathcal{P}}},(^{1}E_{2})_{\eta_{\mathcal{P}}},(A,^{2}E_{1})_{0},(B,^{2}E_{2})_{0}$ & $\ZZ^{2}$ \tabularnewline
 & $^{2}E_{2}$  & $r_{\pi/3}$ & $(A,^{2}E_{2})_{0},(B,^{2}E_{1})_{0},({}^{1}E_{1},{}^{1}E_{2})_{0}$ & $\ZZ^{3}$ \tabularnewline
 & $^{1}E_{1}$  & $r_{2\pi/3}$ & $(^{2}E_{1})_{\eta_{\mathcal{P}}},(^{2}E_{2})_{\eta_{\mathcal{P}}},(A,^{1}E_{1})_{0},(B,^{1}E_{2})_{0}$ & $\ZZ^{2}$ \tabularnewline
 & $^{1}E_{2}$  & $r_{-\pi/3}$ & $(A,^{1}E_{2})_{0},(B,^{1}E_{1})_{0},({}^{2}E_{1},{}^{2}E_{2})_{0}$ & $\ZZ^{3}$ \tabularnewline
\hline 
$C_{2v}$  & $A_{1}$  & $r_{++}$ & $(A_{1})_{\eta_{{\cal P}}},(A_{2})_{\eta_{{\cal P}}},(B_{1})_{\eta_{{\cal P}}},(A_{2})_{\eta_{{\cal P}}}$ & 0\tabularnewline
 & $A_{2}$  & $r_{--}$ & $(A_{1},A_{2})_{0},(B_{1},B_{2})_{0}$ & $\ZZ^{2}$ \tabularnewline
 & $B_{1}$  & $r_{-+}$ & $(A_{1},B_{1})_{0},(A_{2},B_{2})_{0}$ & $\ZZ^{2}$ \tabularnewline
 & $B_{2}$  & $r_{+-}$ & $(A_{1},B_{2})_{0},(B_{1},A_{2})_{0}$ & $\ZZ^{2}$ \tabularnewline
\hline 
$C_{3v}$  & $A_{1}$  &  & $(A_{1})_{\eta_{\mathcal{P}}},(A_{2})_{\eta_{\mathcal{P}}},(E)_{\eta_{\mathcal{P}}}$ & $\ZZ_{2}^{3}$ \tabularnewline
 & $A_{2}$  &  & $(A_{1},A_{2})_{0},(E)_{-\eta_{\mathcal{P}}}$ & $\ZZ$ \tabularnewline
\hline 
$C_{4v}$  & $A_{1}$  &  & $(A_{1})_{\eta_{{\cal P}}},(A_{2})_{\eta_{{\cal P}}},(B_{1})_{\eta_{{\cal P}}},(B_{2})_{\eta_{{\cal P}}},E_{\eta_{{\cal P}}}$ & 0\tabularnewline
 & $A_{2}$  &  & $(A_{1},A_{2})_{0},(B_{1},B_{2})_{0},E_{-\eta_{{\cal P}}}$ & $\ZZ^{2}\times\ZZ_{2}$ \tabularnewline
 & $B_{1}$  &  & $(A_{1},B_{1})_{0},(A_{2},B_{1})_{0},E_{\eta_{{\cal P}}}$ & $\ZZ^{2}$ \tabularnewline
 & $B_{2}$  &  & $(A_{1},B_{2})_{0},(B_{1},A_{2})_{0},E_{\eta_{{\cal P}}}$ & $\ZZ^{2}$ \tabularnewline
\hline 
$C_{6v}$  & $A_{1}$  &  & $(A_{1})_{\eta_{{\cal P}}},(A_{2})_{\eta_{{\cal P}}},(B_{1})_{\eta_{{\cal P}}},(B_{2})_{\eta_{{\cal P}}},(E_{1})_{\eta_{{\cal P}}},(E_{2})_{\eta_{{\cal P}}}$ & 0\tabularnewline
 & $A_{2}$  &  & $(A_{1},A_{2})_{0},(B_{1},B_{2})_{0},(E_{1})_{-\eta_{{\cal P}}},(E_{2})_{-\eta_{{\cal P}}}$ & $\ZZ^{2}\times\ZZ_{2}^{2}$ \tabularnewline
 & $B_{1}$  &  & $(A_{1},B_{1})_{0},(A_{2},B_{1})_{0},(E_{1},E_{2})_{0}$ & $\ZZ^{3}$ \tabularnewline
 & $B_{2}$  &  & $(A_{1},B_{1})_{0},(A_{2},B_{1})_{0},(E_{1},E_{2})_{0}$ & $\ZZ^{3}$ \tabularnewline
\hline \hline
\end{tabular*}
\caption{
	Same as Table~\ref{tab:classD_0d}, but in the presence of spin-rotation symmetry, {\it i.e.} for spinless representations of the crystalline symmetry group. BdG Hamiltonians with spin-rotation symmetry and without time-reversal symmetry correspond to tenfold-way class C, $\eta_{\cal P} = -1$.}
\label{tab:classC_0d}
\end{table*}

\begin{table*}[!]
\begin{tabular*}{\textwidth}{c @{\extracolsep{\fill}} cc c cc} \hline\hline
$G$  & $\Theta$  &  & $(r_{\alpha})_{W_{{\cal P}(\alpha)}}$ & $\mathfrak{K}_{\classDIII}[G,\Theta]$  & $\mathfrak{K}_{\classBDI}[G,\Theta]$ \tabularnewline
\hline 
\hline 
$C_{1}$  & $A$  & $1$ & $(A)_{(\eta_{{\cal T}},\eta_{{\cal P}},1)}$ & 0 & $\ZZ_{2}$ \tabularnewline
\hline 
$C_{i}$  & $A_{g}$  & $r_{+}$ & $(A_{g})_{(\eta_{{\cal T}},\eta_{{\cal P}},1)},(A_{u})_{(\eta_{{\cal T}},\eta_{{\cal P}},1)}$ & 0 & $\ZZ_{2}^{2}$ \tabularnewline
 & $A_{u}$  & $r_{-}$ & $(A_{g},A_{u})_{(\eta_{{\cal T}},0,0)}$ & $\ZZ$  & $\ZZ$ \tabularnewline
\hline 
$C_{s}$  & $A'$  & $r_{+}$ & $(^{1}\bar{E},^{2}\bar{E})_{(0,0,1)}$ & 0 & 0\tabularnewline
 & $A''$  & $r_{-}$ & $(^{1}\bar{E},^{2}\bar{E})_{(0,\eta_{{\cal P}},0)}$ & $\ZZ_{2}$  & $\ZZ_{2}$ \tabularnewline
\hline 
$C_{2}$  & $A$  & $r_{+}$ & $(^{1}\bar{E},^{2}\bar{E})_{(0,0,1)}$ & 0 & 0\tabularnewline
 & $B$  & $r_{-}$ & $(^{1}\bar{E},^{2}\bar{E})_{(0,\eta_{{\cal P}},0)}$ & $\ZZ_{2}$  & $\ZZ_{2}$ \tabularnewline
\hline 
$C_{3}$  & $A_{1}$  & $r_{0}$ & $\bar{E}_{(\eta_{{\cal T}},\eta_{{\cal P}},1)},(^{1}\bar{E},^{2}\bar{E})_{(0,0,1)}$ & 0 & $\ZZ_{2}$ \tabularnewline
\hline 
$C_{4}$  & $A$  & $r_{0}$ & $(^{1}\bar{E}_{1},^{2}\bar{E}_{1})_{(0,0,1)},(^{1}\bar{E}_{2},^{2}\bar{E}_{2})_{(0,0,1)}$ & 0 & 0\tabularnewline
 & $B$  & $r_{\pi}$ & $(^{1}\bar{E}_{1},{}^{2}\bar{E}_{1},^{2}\bar{E}_{2},^{1}\bar{E}_{2})_{(0,0,0)}$ & $\ZZ$  & $\ZZ$ \tabularnewline
\hline 
$C_{6}$  & $A$  & $r_{0}$ & $(^{1}\bar{E}_{1},^{2}\bar{E}_{1})_{(0,0,1)},(^{1}\bar{E}_{2},^{2}\bar{E}_{2})_{(0,0,1)},(^{1}\bar{E}_{3},^{2}\bar{E}_{3})_{(0,0,1)}$ & 0 & 0\tabularnewline
 & $B$  & $r_{\pi}$ & $(^{1}\bar{E}_{1},{}^{2}\bar{E}_{1})_{(0,\eta_{{\cal P}},0)},(^{1}\bar{E}_{2},^{2}\bar{E}_{2},{}^{2}\bar{E}_{3},^{1}\bar{E}_{3})_{(0,0,0)}$ & $\ZZ\times\ZZ_{2}$  & $\ZZ\times\ZZ_{2}$ \tabularnewline
\hline 
$C_{2v}$  & $A_{1}$  & $r_{++}$ & $(\bar{E})_{(-\eta_{{\cal T}},-\eta_{{\cal P}},1)}$ & 0 & 0\tabularnewline
 & $A_{2}$  & $r_{--}$ & $(\bar{E})_{(-\eta_{{\cal T}},\eta_{{\cal P}},1)}$ & $\ZZ_{2}$  & 0\tabularnewline
 & $B_{1}$  & $r_{-+}$ & $(\bar{E})_{(-\eta_{{\cal T}},\eta_{{\cal P}},1)}$ & $\ZZ_{2}$  & 0\tabularnewline
 & $B_{2}$  & $r_{+-}$ & $(\bar{E})_{(-\eta_{{\cal T}},\eta_{{\cal P}},1)}$ & $\ZZ_{2}$  & 0\tabularnewline
\hline 
$C_{3v}$  & $A_{1}$  &  & $(^{1}\bar{E},^{2}\bar{E})_{(0,0,1)},(\bar{E}_{1})_{(-\eta_{{\cal T}},-\eta_{{\cal P}},1)}$ & 0 & 0\tabularnewline
 & $A_{2}$  &  & $(^{1}\bar{E},{}^{2}\bar{E})_{(0,\eta_{{\cal P}},0)},(\bar{E}_{1})_{(-\eta_{{\cal T}},\eta_{{\cal P}},1)}$ & $\ZZ_{2}^{2}$  & $\ZZ_{2}$ \tabularnewline
\hline 
$C_{4v}$  & $A_{1}$  &  & $(\bar{E}_{1})_{(-\eta_{{\cal T}},-\eta_{{\cal P}},1)},(\bar{E}_{2})_{(-\eta_{{\cal T}},-\eta_{{\cal P}},1)}$ & 0 & 0\tabularnewline
 & $A_{2}$  &  & $(\bar{E}_{1})_{(-\eta_{{\cal T}},\eta_{{\cal P}},1)},(\bar{E}_{2})_{(-\eta_{{\cal T}},\eta_{{\cal P}},1)}$ & $\ZZ_{2}^{2}$  & 0\tabularnewline
 & $B_{1}$  &  & $(\bar{E}_{1},\bar{E}_{2})_{(-\eta_{{\cal T}},0,0)}$ & $\ZZ$  & $\ZZ$ \tabularnewline
 & $B_{2}$  &  & $(\bar{E}_{1},\bar{E}_{2})_{(-\eta_{{\cal T}},0,0)}$ & $\ZZ$  & $\ZZ$ \tabularnewline
\hline 
$C_{6v}$  & $A_{1}$  &  & $(\bar{E}_{1})_{(-\eta_{{\cal T}},-\eta_{{\cal P}},1)},(\bar{E}_{2})_{(-\eta_{{\cal T}},-\eta_{{\cal P}},1)},(\bar{E}_{3})_{(-\eta_{{\cal T}},-\eta_{{\cal P}},1)}$ & 0 & 0\tabularnewline
 & $A_{2}$  &  & $(\bar{E}_{1})_{(-\eta_{{\cal T}},\eta_{{\cal P}},1)},(\bar{E}_{2})_{(-\eta_{{\cal T}},\eta_{{\cal P}},1)},(\bar{E}_{3})_{(-\eta_{{\cal T}},\eta_{{\cal P}},1)}$ & $\ZZ_{2}^{3}$  & 0\tabularnewline
 & $B_{1}$  &  & $(\bar{E}_{3})_{(-\eta_{{\cal T}},\eta_{{\cal P}},1)},(\bar{E}_{1},\bar{E}_{2})_{(-\eta_{{\cal T}},0,0)}$ & $\ZZ\times\ZZ_{2}$  & $\ZZ$ \tabularnewline
 & $B_{2}$  &  & $(\bar{E}_{3})_{(-\eta_{{\cal T}},\eta_{{\cal P}},1)},(\bar{E}_{1},\bar{E}_{2})_{(-\eta_{{\cal T}},0,0)}$ & $\ZZ\times\ZZ_{2}$  & $\ZZ$ \tabularnewline \hline\hline
\end{tabular*}
\caption{
    Classification of zero-dimensional BdG Hamiltonians with point group
	$G$ and one-dimensional representation $\Theta$ describing the symmetry
	of the superconducting order parameter with time-reversal symmetry
	without spin-rotation symmetry, corresponding to tenfold-way
	classes DIII, $\eta = (-1,1,1)$. 
	We also include systems with emergent time-reversal
	symmetry, corresponding to class BDI, $\eta = (1,1,1)$.
	The second and third columns list the one-dimensional
	representation using the standard crystallographic notation and in the
	notation used in the examples in the main text, respectively. The
	fourth column lists the set $(r_\alpha)_{W_\mathcal{P} (\alpha)}$ of
	irreps $r_\alpha$, together with the result of the Wigner tests
	$W(\alpha)$. Irreps that are connected by application of the
	fundamental symmetry operations ${\cal T}$, ${\cal P}$, or ${\cal C}$
	are shown together, using brackets. If two of the three Wigner tests
	$W_\mathcal{T}(\alpha)$, $W_\mathcal{C}(\alpha)$, and
	$W_\mathcal{C}(\alpha)$ are zero, the irreps form a pair and the single
	symmetry operation that leave the irrep invariant is the one with the nonzero
	Wigner label $W(\alpha)$. If all three Wigner tests are zero,
	$W_{\mathcal{T}}(\alpha) = W_{\mathcal{P}}(\alpha) =
	W_{\mathcal{C}}(\alpha) = 0$, the irreps form a quartet. In that case
	the order of the four irreps between brackets is such, that the first
	two and last two irreps are interchanged by ${\cal T}$ and the first
	and third, and second and fourth irreps are interchanged by ${\cal P}$.
	From this information one can read off the classification, invariants,
	generators and representations using Table~\ref{tab:Wigner_cases_P}.
	For convenience we list the result for the classifying group
	$\mathfrak{K}_{\eta}[G, \Theta]$ in the last two columns.}
\label{tab:class_0d_TRS}
\end{table*}

\begin{table*}[!]
\begin{tabular*}{\textwidth}{c @{\extracolsep{\fill}} cc c cc} \hline\hline
$G$  & $\Theta$  &  & $(r_{\alpha})_{W_{{\cal P}(\alpha)}}$ & $\mathfrak{K}_{\classCI}[G,\Theta]$  & $\mathfrak{K}_{\classCII}[G,\Theta]$ \tabularnewline
\hline 
\hline 
$C_{1}$  & $A$  & $1$ & $(A)_{(\eta_{{\cal T}},\eta_{{\cal P}},1)}$ & 0 & 0\tabularnewline
\hline 
$C_{i}$  & $A_{g}$  & $r_{+}$ & $(A_{g})_{(\eta_{{\cal T}},\eta_{{\cal P}},1)},(A_{u})_{(\eta_{{\cal T}},\eta_{{\cal P}},1)}$ & 0 & 0\tabularnewline
 & $A_{u}$  & $r_{-}$ & $(A_{g},A_{u})_{(\eta_{{\cal T}},0,0)}$ & $\ZZ$  & $\ZZ$ \tabularnewline
\hline 
$C_{s}$  & $A'$  & $r_{+}$ & $(A')_{(\eta_{{\cal T}},\eta_{{\cal P}},1)},(A'')_{(\eta_{{\cal T}},\eta_{{\cal P}},1)}$ & 0 & 0\tabularnewline
 & $A''$  & $r_{-}$ & $(A',A'')_{(\eta_{{\cal T}},0,0)}$ & $\ZZ$  & $\ZZ$ \tabularnewline
\hline 
$C_{2}$  & $A$  & $r_{+}$ & $(A)_{(\eta_{{\cal T}},\eta_{{\cal P}},1)},(B)_{(\eta_{{\cal T}},\eta_{{\cal P}},1)}$ & 0 & 0\tabularnewline
 & $B$  & $r_{-}$ & $(A,B)_{(\eta_{{\cal T}},0,0)}$ & $\ZZ$  & $\ZZ$ \tabularnewline
\hline 
$C_{3}$  & $A_{1}$  & $r_{0}$ & $(A_{1})_{(\eta_{{\cal T}},\eta_{{\cal P}},1)},(^{1}E,^{2}E)_{(0,0,1)}$ & 0 & 0\tabularnewline
\hline 
$C_{4}$  & $A$  & $r_{0}$ & $A_{(\eta_{{\cal T}},\eta_{{\cal P}},1)},B_{(\eta_{{\cal T}},\eta_{{\cal P}},1)},(^{1}E,^{2}E)_{(0,0,1)}$ & 0 & 0\tabularnewline
 & $B$  & $r_{\pi}$ & $(^{1}E,^{2}E)_{(0,\eta_{{\cal P}},0)},(A,B)_{(\eta_{{\cal T}},0,0)}$ & $\ZZ$  & $\ZZ$ \tabularnewline
\hline 
$C_{6}$  & $A$  & $r_{0}$ & $A_{(\eta_{{\cal T}},\eta_{{\cal P}},1)},B_{(\eta_{{\cal T}},\eta_{{\cal P}},1)},(^{1}E_{1},^{2}E_{1})_{(0,0,1)},(^{1}E_{2},^{2}E_{2})_{(0,0,1)}$ & 0 & 0\tabularnewline
 & $B$  & $r_{\pi}$ & $(A,B)_{(\eta_{{\cal T}},0,0)},(^{1}E_{1},^{2}E_{1},{}^{2}E_{2},^{1}E_{2})_{(0,0,0)}$ & $\ZZ^{2}$  & $\ZZ^{2}$ \tabularnewline
\hline 
$C_{2v}$  & $A_{1}$  & $r_{++}$ & $(A_{1})_{(\eta_{{\cal T}},\eta_{{\cal P}},1)},(A_{2})_{(\eta_{{\cal T}},\eta_{{\cal P}},1)},(B_{1})_{(\eta_{{\cal T}},\eta_{{\cal P}},1)},(A_{2})_{(\eta_{{\cal T}},\eta_{{\cal P}},1)}$ & 0 & 0\tabularnewline
 & $A_{2}$  & $r_{--}$ & $(A_{1},A_{2})_{(\eta_{{\cal T}},0,0)},(B_{1},B_{2})_{(\eta_{{\cal T}},0,0)}$ & $\ZZ^{2}$  & $\ZZ^{2}$ \tabularnewline
 & $B_{1}$  & $r_{-+}$ & $(A_{1},B_{1})_{(\eta_{{\cal T}},0,0)},(A_{2},B_{2})_{(\eta_{{\cal T}},0,0)}$ & $\ZZ^{2}$  & $\ZZ^{2}$ \tabularnewline
 & $B_{2}$  & $r_{+-}$ & $(A_{1},B_{2})_{(\eta_{{\cal T}},0,0)},(B_{1},A_{2})_{(\eta_{{\cal T}},0,0)}$ & $\ZZ^{2}$  & $\ZZ^{2}$ \tabularnewline
\hline 
$C_{3v}$  & $A_{1}$  &  & $(A_{1})_{(\eta_{{\cal T}},\eta_{{\cal P}},1)},(A_{2})_{(\eta_{{\cal T}},\eta_{{\cal P}},1)},(E)_{(\eta_{{\cal T}},\eta_{{\cal P}},1)}$ & 0 & 0\tabularnewline
 & $A_{2}$  &  & $(A_{1},A_{2})_{(\eta_{{\cal T}},0,0)},(E)_{(\eta_{{\cal T}},-\eta_{{\cal P}},1)}$ & $\ZZ\times\ZZ_{2}$  & $\ZZ$ \tabularnewline
\hline 
$C_{4v}$  & $A_{1}$  &  & $(A_{1})_{(\eta_{{\cal T}},\eta_{{\cal P}},1)},(A_{2})_{(\eta_{{\cal T}},\eta_{{\cal P}},1)},(B_{1})_{(\eta_{{\cal T}},\eta_{{\cal P}},1)},(A_{2})_{(\eta_{{\cal T}},\eta_{{\cal P}},1)},E_{(\eta_{{\cal T}},\eta_{{\cal P}},1)}$ & 0 & 0\tabularnewline
 & $A_{2}$  &  & $(A_{1},A_{2})_{(\eta_{{\cal T}},0,0)},(B_{1},B_{2})_{(\eta_{{\cal T}},0,0)},E_{(\eta_{{\cal T}},-\eta_{{\cal P}},1)}$ & $\ZZ^{2}\times\ZZ_{2}$  & $\ZZ^{2}$ \tabularnewline
 & $B_{1}$  &  & $(A_{1},B_{1})_{(\eta_{{\cal T}},0,0)},(A_{2},B_{1})_{(\eta_{{\cal T}},0,0)},E_{(\eta_{{\cal T}},\eta_{{\cal P}},1)}$ & $\ZZ^{2}$  & $\ZZ^{2}$ \tabularnewline
 & $B_{2}$  &  & $(A_{1},B_{2})_{(\eta_{{\cal T}},0,0)},(B_{1},A_{2})_{(\eta_{{\cal T}},0,0)},E_{(\eta_{{\cal T}},\eta_{{\cal P}},1)}$ & $\ZZ^{2}$  & $\ZZ^{2}$ \tabularnewline
\hline 
$C_{6v}$  & $A_{1}$  &  & $(A_{1})_{(\eta_{{\cal T}},\eta_{{\cal P}},1)},(A_{2})_{(\eta_{{\cal T}},\eta_{{\cal P}},1)},(B_{1})_{(\eta_{{\cal T}},\eta_{{\cal P}},1)},(A_{2})_{(\eta_{{\cal T}},\eta_{{\cal P}},1)},(E_{1})_{(\eta_{{\cal T}},\eta_{{\cal P}},1)},(E_{2})_{(\eta_{{\cal T}},\eta_{{\cal P}},1)}$ & 0 & 0\tabularnewline
 & $A_{2}$  &  & $(A_{1},A_{2})_{(\eta_{{\cal T}},0,0)},(B_{1},B_{2})_{(\eta_{{\cal T}},0,0)},(E_{1})_{(\eta_{{\cal T}},-\eta_{{\cal P}},1)},(E_{2})_{(\eta_{{\cal T}},-\eta_{{\cal P}},1)}$ & $\ZZ^{2}\times\ZZ_{2}^{2}$  & $\ZZ^{2}$ \tabularnewline
 & $B_{1}$  &  & $(A_{1},B_{1})_{(\eta_{{\cal T}},0,0)},(A_{2},B_{1})_{(\eta_{{\cal T}},0,0)},(E_{1},E_{2})_{(\eta_{{\cal T}},0,0)}$ & $\ZZ^{3}$  & $\ZZ^{3}$ \tabularnewline
 & $B_{2}$  &  & $(A_{1},B_{1})_{(\eta_{{\cal T}},0,0)},(A_{2},B_{1})_{(\eta_{{\cal T}},0,0)},(E_{1},E_{2})_{(\eta_{{\cal T}},0,0)}$ & $\ZZ^{3}$  & $\ZZ^{3}$ \tabularnewline \hline\hline
\end{tabular*}
\caption{ 
	Same as Table~\ref{tab:class_0d_TRS}, but in the presence of spin-rotation symmetry, {\it i.e.} for spinless representations of the crystalline symmetry group. Time-reversal symmetric BdG Hamiltonians with spin-rotation symmetry correspond to tenfold-way class CI, $\eta = (1,-1,1)$. We also include systems with emergent time-reversal
	symmetry, corresponding to class CII, $\eta = (-1,-1,1)$.}
\label{tab:class_0d_TRS_SpinRot}
\end{table*}

We first consider a zero-dimensional superconductor with inversion symmetry.
The symmetry group $G = C_i = \{\id, {\cal I}\}$ and the two one-dimensional
representations $\Theta$ of $G$ that characterize the superconducting phase are the two irreducible representations $r_{\pm}$ of $G$, see Table~\ref{tab:rep_Ci}.
If the superconducting order is even under inversion, corresponding to $\Theta
= r_+$ (the ``$A_g$'' representation), Eqs.\ (\ref{eq:ugeneral}) and (\ref{eq:HBdGconstraint2}) give
$$
  U(\id) = \rho_0 \nootimes \tau_0,\ \ U({\cal I}) = \rho_3 \nootimes \tau_0,
$$%
where $\rho_3$ is the Pauli matrix in parity space. Correspondingly, the BdG
Hamiltonian $H = \mbox{diag}\,(H_+,H_-)$ is the diagonal sum of blocks with
even and odd parity under inversion. Since $U$ is real and does not involve the
particle-hole degree of freedom, the particle-hole conjugation operation ${\cal
P}$ commutes with $G$, so that it acts within each parity subblock. Its square
is $W_{\cal P} = \eta_{\cal P}$, {\it i.e.,} the same as for the full
Hamiltonian $H$. We conclude that
\begin{equation}
  \mathfrak{K}_{\eta}[C_i,A_g] = \mathfrak{K}_{\eta}[0]^2
\end{equation}
for all tenfold-way symmetry classes $\eta$. The groups $\mathfrak{K}_{\eta}[0]$ are listed in Table \ref{tab:1}. This conclusion is consistent with the Wigner test, which gives $W_{\cal P}(r_{\pm}) = 1$. If, on the other hand, $\Delta$ is odd under inversion, which corresponds to the one-dimensional representation $\Theta = r_-$ for the superconducting phase (the ``$A_u$ representation, see Table \ref{tab:rep_Ci}), the representation of $G$ for the full BdG Hamiltonian is 
$$
  U(\id) = \rho_0 \nootimes \tau_0,\ \ U({\cal I}) = \rho_3 \nootimes \tau_3,
$$%
see Eq.\ (\ref{eq:umm_inversion}). The block-diagonal structure of $H$
now involves one block with even-parity particle-like states and odd-parity
hole-like states and one block with odd-parity particle-like states and
even-parity hole-like states. The two blocks are interchanged by particle-hole
conjugation. One arrives at the same conclusion by observing that $r_+$ and
$\Theta r_+ = r_-$ are different irreducible representations of
$G$. For the classifying group $\mathfrak{K}_{\eta}[G,A_u]$ we thus find
\begin{equation}
  \mathfrak{K}_{(\eta_{\cal T},\eta_{\cal P},\eta_{\cal C})}[C_i,A_u] = \mathfrak{K}_{(\eta_{\cal T},0,0)}[0].
\end{equation}
Again, this conclusion is compatible with the Wigner test, which gives $W_{\cal P}(r_{\pm}) = 0$.

The second example deals with a spinful system with a fourfold rotation
symmetry, $G = C_4$. The one-dimensional representation $\Theta \equiv
r_{\theta}$ of $G$ that characterizes the superconducting phase is labeled by
an angle $\theta$ which can take the values $0$, $\pi/2$, $\pi$, or $3\pi/2$,
see Table \ref{tab:rep_C4}. The representation (\ref{eq:Ufourfold}) of the
rotation generator ${\cal R}_{\pi/2}$ implies that $H$ has a block-diagonal
structure in which particle-like states with angular momentum $j=\frac{1}{2}$,
$\frac{3}{2}$, $\frac{5}{2}$, or $\frac{7}{2}$ are combined with hole-like
states of angular momentum $-j+ 2 \theta/\pi \mod 4$. In the following
discussion, we use the angular momentum $j$ of the particle-like states to
label the blocks. If $\theta = 0$ (the ``$A$'' representation), particle-hole
conjugation ${\cal P}$ interchanges the blocks with $j=\frac{1}{2}$ and
$\frac{7}{2}$, as well as the blocks with $j=\frac{3}{2}$ and $\frac{5}{2}$. If
$\theta = \pi/2$ (the ``$^2\!E$'' representation), ${\cal P}$ interchanges the
blocks with $j=\frac{3}{2}$ and $\frac{7}{2}$, but acts within the blocks with
$j=\frac{1}{2}$ or $\frac{5}{2}$, squaring to one. If $\theta = \pi$ (the
``$A$'' representation), ${\cal P}$ interchanges the blocks with
$j=\frac{1}{2}$ and $\frac{3}{2}$, as well as the blocks with $j=\frac{5}{2}$
and $j=\frac{7}{2}$. Finally, if $\theta = 3\pi/2$ (the ``$^1\!E$''
representation), ${\cal P}$ interchanges the blocks with $j=\frac{1}{2}$ and
$\frac{5}{2}$, but acts within the blocks with $j=\frac{3}{2}$ and
$\frac{7}{2}$, again squaring to one. We conclude that
\begin{align}
  \mathfrak{K}_{\classD}[C_4,A] &=\, \mathfrak{K}_{\classD}[C_4,B] \nonumber \\ &=\,
\mathfrak{K}_{\classA}[0]^2, \nonumber \\ &=\, \ZZ^2
\end{align}
and
\begin{align}
  \mathfrak{K}_{\classD}[C_4,^{1,2}\!E] &=\, \mathfrak{K}_{\classA}[0] \times \mathfrak{K}_{\classD}[0]^2 \nonumber \\ &=\, \ZZ \times \ZZ_2^2.
\end{align}
The same conclusions can be obtained by performing the Wigner test.

\begin{table*}
\begin{tabular*}{\textwidth}{c @{\extracolsep{\fill}} ccccccc} \hline\hline
$W_{\mathcal{P}}(\alpha)$ & Cartan & $\mathfrak{K}_{(0,W_{P}(\alpha),0)}[0]$ & $\nu$ & $H_{\text{gen}}$ & $H_{\text{ref}}$ & $\mathcal{P}$ & $U$\tabularnewline
\hline 
1 & D & $\mathbb{Z}_{2}$ & $\mathfrak{p}_\alpha$ & $-\tau_{3}\otimes1_{d}$ & $\tau_{3}\otimes1_{d}$ & $\tau_{1}K$ & $r_{\alpha}\otimes\tau_{0}$\tabularnewline
-1 & C & $0$ & $-$ & - & $\tau_{3}\otimes1_{d}$ & $\tau_{2}K$ & $r_{\alpha}\otimes\tau_{0}$\tabularnewline
0 & A & $\mathbb{Z}$ & $\begin{cases}
\mathfrak{N}_\alpha\\
-\mathfrak{N}_{\Theta\alpha^{*}}
\end{cases}$ & $-\tau_{3}\otimes1_{d}$ & $\tau_{3}\otimes1_{d}$ & $\tau_{1,2}K$ & $\left(\begin{array}{cc}
r_{\alpha} & 0\\
0 & \Theta r_{\alpha}^{*}
\end{array}\right)_{\tau}$\tabularnewline \hline\hline
\end{tabular*}
\caption{Classifying groups $\mathfrak{K}_{W_{\cal P}(\alpha)}[0]$ and
topological invariants $\nu$ for the tenfold-way classes D and C, $\eta =
(0,\pm 1,0)$ (third and fourth columns). These depend on the irrep $r_\alpha$
through the outcome of the Wigner test $W_{\cal P}(\alpha)$ only, which is
listed in the first column. The second column lists the effective tenfold-way
class corresponding to $W_{\cal P}(\alpha)$. The fifth and sixth columns list a
generating Hamiltonian $H_{\rm gen}$ and a reference Hamiltonian
$H_\text{ref}$, respectively. The seventh and eighth columns give the
representations of particle-hole conjugation $\mathcal{P}$ and the
representation $U$ of the elements of the crystalline symmetry group $G$. The
dimension of the irreducible representation $r_\alpha$ is denoted $d_{\alpha}$.}
\label{tab:Wigner_cases_P}
\end{table*}

In the presence of time-reversal symmetry only the real one-dimensional
representations $A$ and $B$ are relevant (corresponding to $\theta = 0$, $\pi$,
respectively). In both cases the time-reversal operation ${\cal T}$
interchanges the blocks with angular momentum $j$ and angular momentum $-j \mod
4$. For the $A$ representation, ${\cal P}$ and ${\cal T}$ give the same pairing
of diagonal blocks of $H$ and the combined operation ${\cal C} = {\cal PT}$
leaves the diagonal blocks invariant. It follows that
\begin{align}
  \mathfrak{K}_{\classDIII}[C_4,A] &=\, \mathfrak{K}_{\classA}[0]^2 \nonumber \\ &=\, 0.
\end{align}
On the other hand, for the $B$ representation, ${\cal P}$, ${\cal T}$, and their
product ${\cal C} = {\cal PT}$ map all four diagonal blocks of $H$ to each
other and one has
\begin{align}
	\mathfrak{K}_{\classDIII}[C_4,B] &=\, \mathfrak{K}_{\classA}[0] \nonumber \\ &=\, \ZZ.
\end{align}

The third example is that of spinful particles with symmetry group $G =
C_{2v}$. We refer to Table \ref{tab:rep_C2v} for the irreducible
representations with nontrivial and trivial factor system. Since there is only
one (two-dimensional) irreducible representation $u$ --- the ``$\overline{E}$''
representation, see Table \ref{tab:rep_C2v} ---, it follows automatically that
$u^*$ and $r_{\sigma_x\sigma_y} u^*$ are the same representations of $G$.
Indeed, one easily verifies that $u^*$ and $r_{\sigma_x,\sigma_y} u^*$ are
equal to $u$ up to a unitary transformation. To find the square of the
effective time-reversal and particle-hole conjugation operations, we consider
the case of the one-dimensional irreducible representation $\Theta=r_{++}$ (the $A_1$ representation) in
detail and summarize results for the remaining three one-dimensional
representations $r_{+-}$, $r_{-+}$, and $r_{--}$ (the $B_2$, $B_1$, and $A_2$ representations, respectively).

For $\Theta = r_{++}$ (``$A_1$'' representation, see Table \ref{tab:rep_C2v}),
and choosing the representation of Table \ref{tab:rep_C2v} for the
representation of the generators ${\cal M}_x$ and ${\cal M}_y$ for the
normal-state block $h$, we find that their representation for the full BdG
Hamiltonian $H$ are $U({\cal M}_x) = i \sigma_1 \nootimes \tau_3$ and $U({\cal
M}_y) = i \sigma_2 \nootimes \tau_0$. To bring the representation matrices to the
canonical form (\ref{eq:UUgeneral}) we have to perform the basis transformation
$$
  H \to \begin{pmatrix} 1 & 0 \\ 0 & i \sigma_2 \end{pmatrix}
  H \begin{pmatrix} 1 & 0 \\ 0 & -i \sigma_2 \end{pmatrix}.
$$%
After this basis transformation, ${\cal M}_x$ and ${\cal M}_y$ are represented
as $i \sigma_1 \nootimes \tau_0$ and $i \sigma_2 \nootimes \tau_0$, respectively,
consistent with Eq.~(\ref{eq:UUgeneral}). The particle-hole conjugation
operation ${\cal P}$, which was represented by $\sigma_0 \nootimes \tau_1 K$
before the basis change, now reads $\sigma_2 \nootimes \tau_2 K$. Since the
$C_{2v}$ symmetry enforces that $H$ is of the form $\sigma_0 \nootimes
H_{\overline{E}}$, compare with Eq.\ (\ref{eq:HHgeneral}), the factor
$\sigma_2$ must be dropped from the representation of ${\cal P}$, since it does
not affect $H_{\overline{E}}$, and one arrives at the effective particle-hole
conjugation operator $\tau_2 K$: The effective particle-hole conjugation
operator squares to minus one. The time-reversal operation ${\cal T}$, which
was given by $\sigma_2 \nootimes \tau_0 K$ in the original basis, is unchanged by
the basis transformation. Again omitting the factor $\sigma_2$, one finds that
the effective time-reversal operation is $\tau_0 K$. It follows that
\begin{equation}
	\mathfrak{K}_{(\eta_{\mathcal{T}},1,\eta_{\mathcal{C}})}[C_{2v},A_1]  = 
    \mathfrak{K}_{(-\eta_{\mathcal{T}},-1,\eta_{\mathcal{C}})}[0].
\end{equation}
The same result is found if one performs the Wigner test, see
Eqs.~(\ref{eq:WignerT}) and (\ref{eq:WignerP}). 

For $\Theta = r_{+-}$ (the ``$B_2$'' representation) one finds that the
representation $u$ of Table \ref{tab:rep_C2v} for the normal-state block $h$
gives the representations $U({\cal M}_x) = i \sigma_1 \nootimes \tau_0$ and
$U({\cal M}_y) = i \sigma_2 \nootimes \tau_0$ for the full BdG Hamiltonian $H$,
which is already in the canonical form (\ref{eq:UUgeneral}). It then
immediately follows that the effective particle-hole conjugation operation is
$\tau_1 K$ and the effective time-reversal operation is $\tau_0 K$. For the
remaining two cases $r_{-+}$, and $r_{--}$ (the ``$B_1$'' and ``$A_2$''
representations), one may proceed as described above for the $A_1$
representation, by performing a suitable basis transformation, or use an
alternative normal-state representation with $u({\cal M}_x) = i \sigma_2$,
$u({\cal M}_y) = i \sigma_3$ for $B_1$ and $u({\cal M}_x) = i \sigma_3$,
$u({\cal M}_y) = i \sigma_1$ for $A_2$, to which the canonical form
(\ref{eq:UUgeneral}) applies without the need for a basis transformation.
Either way, we find that both the effective time-reversal operation and the
effective particle-hole conjugation operation square to unity, $W_{\cal T} =
-\eta_{\cal T} = 1$ and $W_{\cal P} = \eta_{\cal P} = 1$. As a result, we have
\begin{align}
	\mathfrak{K}_{(\eta_{\mathcal{T}},1,\eta_{\mathcal{C}})}[C_{2v},B_{1,2}] & = \mathfrak{K}_{(\eta_{\mathcal{T}},1,\eta_{\mathcal{C}})}[C_{2v},A_2] \nonumber \\ & = \mathfrak{K}_{(-\eta_{\mathcal{T}},1,\eta_{\mathcal{C}})}[0].
\end{align}

\begin{table*}
\begin{tabular*}{\textwidth}{c @{\extracolsep{\fill}} cclccccc} \hline\hline
$W(\alpha)$ & Cartan & $\mathfrak{K}_{W(\alpha)}[0]$ & $\nu$ & $H_{\text{gen}}$ & $H_{\text{ref}}$ & $\mathcal{T}$ & $\mathcal{P}$ & $U$\tabularnewline
\hline 
(0,0,0) & A & $\mathbb{Z}$ & $\begin{cases}
\mathfrak{N}_\alpha\\
\mathfrak{N}_{\alpha^{*}}\\
-\mathfrak{N}_{\Theta\alpha^{*}}\\
-\mathfrak{N}_{\Theta\alpha}
\end{cases}$ & $-\sigma_{0}\tau_{3}\otimes1_{d_\alpha}$ & $\sigma_{0}\tau_{3}\otimes1_{d_\alpha}$ & $\sigma_{1,2}K$ & $\tau_{1,2}K$ & $\left(\begin{array}{cc}
\left(\begin{array}{cc}
r & 0\\
0 & r_{\alpha}^{*}
\end{array}\right)_{\sigma} & 0\\
0 & \Theta\left(\begin{array}{cc}
r_{\alpha} & 0\\
0 & r_{\alpha}^{*}
\end{array}\right)_{\sigma}^{*}
\end{array}\right)_{\tau}$\tabularnewline
(0,0,1) & AIII & $0$ & - & - & $\sigma_{0}\tau_{3}\otimes1_{d_\alpha}$ & $\sigma_{1,2}K$ & $\sigma_{1,2}\tau_{1}K$ & $\left(\begin{array}{cc}
r_{\alpha} & 0\\
0 & r_{\alpha}^{*}
\end{array}\right)_{\sigma}\otimes\tau_{0}$\tabularnewline
(1,0,0) & AI & $\mathbb{Z}$ & $\begin{cases}
\mathfrak{N}_\alpha\\
-\mathfrak{N}_{\Theta\alpha^{*}}
\end{cases}$ & $-\tau_{3}\otimes1_{d_\alpha}$ & $\tau_{3}\otimes1_{d_\alpha}$ & $K$ & $\tau_{1,2}K$ & $\left(\begin{array}{cc}
r_{\alpha} & 0\\
0 & \Theta r_{\alpha}^{*}
\end{array}\right)_{\tau}$\tabularnewline
(1,1,1) & BDI & $\mathbb{Z}_{2}$ & $\mathfrak{p}_\alpha$ & $-\tau_{3}\otimes1_{d_\alpha}$ & $\tau_{3}\otimes1_{d_\alpha}$ & $K$ & $\tau_{1}K$ & $r_{\alpha}\otimes\tau_{0}$\tabularnewline
(0,1,0) & D & $\mathbb{Z}_{2}$ & $\begin{cases}
\mathfrak{p}_\alpha\\
\mathfrak{p}_{\alpha^{*}}
\end{cases}$ & $-\sigma_{0}\tau_{3}\otimes1_{d_\alpha}$ & $\sigma_{0}\tau_{3}\otimes1_{d_\alpha}$ & $\sigma_{1,2}K$ & $\tau_{1}K$ & $\left(\begin{array}{cc}
r_{\alpha} & 0\\
0 & r_{\alpha}^{*}
\end{array}\right)_{\sigma}\otimes\tau_{0}$\tabularnewline
(-1,1,1) & DIII & $0$ & - & - & $\sigma_{0}\tau_{3}\otimes1_{d_\alpha}$ & $\sigma_{2}K$ & $\tau_{1}K$ & $r_{\alpha}\otimes\sigma_{0}\tau_{0}$\tabularnewline
(-1,0,0) & AII & $\mathbb{Z}$ & $\begin{cases}
\mathfrak{N}_\alpha\\
-\mathfrak{N}_{\Theta\alpha^{*}}
\end{cases}$ & $-\sigma_{0}\tau_{3}\otimes1_{d_\alpha}$ & $\sigma_{0}\tau_{3}\otimes1_{d_\alpha}$ & $\sigma_{2}K$ & $\tau_{1,2}K$ & $\left(\begin{array}{cc}
r_{\alpha} & 0\\
0 & \Theta r_{\alpha}^{*}
\end{array}\right)_{\tau}\otimes\sigma_{0}$\tabularnewline
(-1,-1,1) & CII & $0$ & - & - & $\sigma_{0}\tau_{3}\otimes1_{d_\alpha}$ & $\sigma_{2}K$ & $\tau_{2}K$ & $r_{\alpha}\otimes\sigma_{0}\tau_{0}$\tabularnewline
(0,-1,0) & C & $0$ & - & - & $\sigma_{0}\tau_{3}\otimes1_{d_\alpha}$ & $\sigma_{1,2}K$ & $\tau_{2}K$ & $\left(\begin{array}{cc}
r_{\alpha} & 0\\
0 & r_{\alpha}^{*}
\end{array}\right)_{\sigma}\otimes\tau_{0}$\tabularnewline
(1,-1,1) & CI & $0$ & - & - & $\tau_{3}\otimes1_{d_\alpha}$ & $K$ & $\tau_{2}K$ & $r_{\alpha}\otimes\tau_{0}$\tabularnewline \hline\hline
\end{tabular*}
\caption{Classifying groups $\mathfrak{K}_{W_{\cal P}(\alpha)}[0]$ and
topological invariants $\nu$ for the tenfold-way classes with time-reversal
symmetry (third and fourth columns). These depend on the irrep $r_\alpha$ through
the outcome of the Wigner test $W_{\cal P}(\alpha)$ only, which is listed in
the first column. The second column lists the effective tenfold-way class
corresponding to $W_{\cal P}(\alpha)$. The fifth and sixth columns list a
generating Hamiltonian $H_{\rm gen}$ and a reference Hamiltonian
$H_\text{ref}$, respectively. The seventh, eighth, and ninth columns give the
representations of time-reversal ${\cal T}$, particle-hole conjugation
$\mathcal{P}$, and the representation $U$ of the elements of the crystalline
symmetry group $G$. The dimension of the irreducible representation $r_\alpha$ is
denoted $d_{\alpha}$.}
\label{tab:Wigner_cases_TP}
\end{table*}

{\it A remark on classifying groups.---} We close this section by making a few
remarks regarding the group structure of topological classification within
$K$-theory. Formally, the group structure within $K$-theory is given by the
Grothendieck construction,\cite{nakahara2003,turner2012} where the group
elements are represented by ordered pairs $(H_1,H_2)$ of hermitian
matrix-valued functions $H_{1,2}(\vk)$ of equal dimension. The two pairs $(H_1,H_2)$
and $(H_1^\prime,H_2^\prime)$ are topologically equivalent if $H_1\oplus
H_2^\prime$ is continuously deformable to $H_1^\prime\oplus H_2$. Loosely
speaking the ordered pair $(H_1,H_2)$ represents the ``difference'' of the two
Hamiltonians $H_1$ and $H_2$. 

For gapped phases with dimension $d > 0$, a meaningful concept of
``topologically nontrivial'' Hamiltonians can be obtained by defining a
reference atomic-limit Hamitonian $H_{\rm ref}(\vk)$ as a ``topologically
trivial phase''. (A precise definition of ``atomic-limit Hamiltonians'' will be
given in the next Section.) Such a strategy results in a unique and
well-defined topological classification that is independent of the choice of
the precise reference Hamiltonian $H_{\rm ref}(\vk)$ if one considers
Hamiltonians that differ by an atomic-limit Hamiltonian to be in the same
topological class. It is this classification principle that underlies the
classifying groups ${\cal K}_{\eta}[G,\Theta]$ used throughout the remainder of
this article for Hamiltonians of dimension $d \ge 1$. On other hand, for
zero-dimensional Hamiltonians, there is no natural choice for the trivial phase
and it is important to adhere to the notion that a topological classification
classifies pairs of Hamiltonians only. It is this stricter notion of
topological classification that is used for the definition of the classifying
groups $\mathfrak{K}_{\eta}[G,\Theta]$ of zero-dimensional Hamiltonians, which
play a key role in the construction of symmetry-based indicators in the next
Section.

{\it Generators and invariants for the classifying groups $\mathfrak{K}_\eta[G, \Theta]$.---}
The classifying groups $\mathfrak{K}_{\eta}[G,\Theta]$ of zero-dimensional
Hamiltonians with additional point group $G$ and with one-dimensional
representation $\Theta$ governing the pairing term $\Theta$ are determined by
the effective Cartan class of diagonal blocks corresponding to the irreducible
representation $\alpha$ or by pairs or quadruples of such blocks, as discussed
above. We tabulate the classification $\mathfrak{K}_\eta [0]$, the invariants
$\nu$, generators $H$ and representations $U(g)$ of all symmetry elements $g$,
$\mathcal{T}$ and $\mathcal{P}$ for all cases in
Tables~\ref{tab:Wigner_cases_P}, \ref{tab:Wigner_cases_TP}. The
zero-dimensional invariants can be given as the number $\mathfrak{N}_\alpha$ of negative energy eigenstates (Kramers pairs in case the block $H_\alpha$ is invariant under an antiunitary symmetry $U K$ with $(U K)^2 = -1$, {\it i. e.} in Cartan class AII) with representation $\alpha$ or the Pfaffian invariant $\mathfrak{p}_\alpha$ of
the block $H_\alpha$ belonging to the irreducible representation $r_\alpha$. In
case there are multiple blocks related by $\mathcal{T}$ or $\mathcal{P}$ we
present all equivalent invariants. 

%\MG{Suggestion to give the definition of the 0d invariants in equation form: } 
%In case the block $H_\alpha$ belonging to the irreducible representation $r_\alpha$ does not satisfy any antisymmetry, {\it i.e. }$H_{\alpha}$ is in Cartan class A, AI (AII), the zero dimensional invariant can be defined as 
%\begin{equation}
%\mathfrak{N}_\alpha
%\end{equation}
%the number of (Kramers pairs of) negative energy eigenstates with representation $\alpha$.  
%For a subblock $H_{\alpha}$
%invariant under an (effective) antiunitary antisymmetry with representation $U$
%with $( U K)^2 = 1$, {\it i.e. }$H_{\alpha}$ is in Cartan classes D, BDI, DIII, one can define the Pfaffian invariant 
%\begin{equation}
%(-1)^{\mathfrak{p}_\alpha} := \sign \mbox{Pf}\, (H_{\alpha} U) .
%\end{equation}
%Due to Kramers degeneracy in class DIII, the Pfaffian invariant of blocks with Cartan class DIII is always trivial. 
%In case there are multiple blocks related by $\mathcal{T}$ or $\mathcal{P}$ we
%present all equivalent invariants.

For the calculation the zero-dimensional invariants $\nu_\alpha =
\mathfrak{N}_\alpha$ or $\nu_{\alpha} =\mathfrak{p}_\alpha$ it may be helpful
to use the projector~\cite{dresselhaus2007} onto a subspace spanned by
irreducible representation $\alpha$ 
\begin{equation}
  P_\alpha = d_\alpha \langle \chi_\alpha^*( g ) U(g) \rangle_{G},
\label{eq:projector_onto_irrep}
\end{equation}
where $\langle \ldots \rangle_{G}$ denotes the average over all elements $g \in
G$ and $\chi_{\alpha}(g)$ is the character. Choosing a basis in which the
projector $P_{\alpha}$ is block diagonal --- which can be done for all irreps
$r_\alpha$ simultaneously, although this is not necessary for the calculation to
succeed ---, the projected Hamiltonian $P_{\alpha} H P_{\alpha}$ takes the form
$\text{diag}(H_\alpha, 0_{N - N_\alpha})$, and the topological invariant can be
computed as
\begin{equation}
  \nu_{\alpha}(H) \equiv \nu(H_\alpha). 
\end{equation}
The Pfaffian invariant $\mathfrak{p}_\alpha$ of a subblock $H_{\alpha}$
invariant under an (effective) antiunitary antisymmetry with representation $U$
with $( U K)^2 = 1$ is defined as 
\begin{equation}
(-1)^{\mathfrak{p}_\alpha} = \sign \mbox{Pf}\, (H_{\alpha} U) . 
\label{eq:definition_Pfaffian_invariant}
\end{equation}

\section{Symmetry-based indicators of band topology} \label{sec:SI}
Whereas a full topological classification of a gapped Hamiltonians in $d$
dimensions --- as described by the classifying group ${\cal
K}_{\eta}[G,\Theta]$ --- requires the analysis of matrix-valued {\it functions}
$H(\vk)$ with the momentum $\vk$ taken on the full Brillouin zone, partial
information on the topological phase can be already obtained by inspection of
the topological class of matrices $H(\vk_{\rm s})$ at a discrete set of
high-symmetry momenta $\vk_{\rm s}$. Such an approach has been developed by Po
{\it et al.} for non-superconducting insulators\cite{po2017} (see also Refs.\
\onlinecite{bradlyn2017,song2018b}.) We here present this approach in such a way
that it can immediately be generalized to Hamiltonians of Bogoliubov-de Gennes
type.

We consider a band structure defined by the Hamiltonian $H(\vk)$, with $\vk$ an
element of the Brillouin zone of a $d$-dimensional crystal with (discrete)
translation invariance and with symmorphic crystalline symmetry described by
the point group $G$. In addition, there may be nonspatial symmetries such as
time-reversal symmetry, particle-hole antisymmetry, or chiral antisymmetry,
which determine the tenfold-way symmetry class. These nonspatial symmetries are
characterized by the triple $\eta = (\eta_{\cal T},\eta_{\cal P},\eta_{\cal
C})$, as explained in the previous Section. For superconductors, {\it i.e.} for
Hamiltonians of BdG type, one further needs to specify a one-dimensional
representation $\Theta$ of $G$, which characterizes the superconducting phase,
see the discussion in Sec.\ \ref{sec:symmetry_classification_orderparameter}. 

\subsection{Construction of a reference set of high-symmetry momenta}\label{sec:4a}
%{\it Construction of a reference set of momenta $\XX$.---} 
To define a representative set of high-symmetry momenta, we consider the group $\tilde{G}$ of automorphisms of the Brillouin zone that are induced by elements of $G$ and by the operations ${\cal T}$ and ${\cal P}$, if present. (The element $g \in G$ induces an automorphism of the Brillouin zone by sending $\vk \to g \vk$, whereas ${\cal T}$ and ${\cal P}$ send $\vk$ to $-\vk$, in both cases identifying wavevectors that differ by a reciprocal lattice vector.) For each momentum $\vk$ we define the ``little group'' $\tilde{G}_{\vk}$ as the subgroup of elements $\tilde{g} \in \tilde{G}$ such that $\tilde{g} \vk = \vk$. Two momenta $\vk_1$ and $\vk_2$ are considered equivalent for classification purposes if there exists an element $\tilde{g} \in \tilde{G}$ and a continuous path between $\vk_1$ and $\tilde{g} \vk_2$ along which the little group $\tilde{G}_{\vk}$ does not change. (In particular, this implies that $\vk_1$ and $\vk_2$ are equivalent if $\vk_1 = \tilde{g} \vk_2$ and that equivalent momenta have the same little group.) A set of equivalent momenta is called of ``high-symmetry'' if it does not border to another set of equivalent momenta with a strictly larger little group. The representative set $\XX$ of high-symmetry momenta is then constructed by arbitrarily selecting one momentum from each set of equivalent high-symmetry momenta. 

\subsection{Definition of the groups ``$\BS$'' and ``$\SI$''}
To each high-symmetry momentum $\vkslist$ we may associate a subgroup
$G_{\vk_{\rm s}} \subset G$ of crystalline symmetry operations that leave
$\vk_{\rm s}$ invariant and a triple $\eta_{\vk_{\rm s}} = (\eta_{{\cal
T},\vk_{\rm s}},\eta_{{\cal P},\vk_{\rm s}},\eta_{\cal C})$ that indicate
whether $\vk_{\rm s}$ is invariant under the operations ${\cal T}$ and ${\cal
P}$, if present. The group $G_{\vk_{\rm s}}$, the symmetry indices
$\eta_{\vk_{\rm s}}$, and the one-dimensional representation $\Theta$ of $G$
(suitably restricted to $G_{\vk_{\rm s}}$) determine the symmetry of the
Hamiltonian $H(\vk_{\rm s})$ at the high-symmetry momentum $\vk_{\rm s}$. The
group of ``band labels'' $\BSz$ is defined as the combined set of topological
invariants that can be obtained from the topological information at these
high-symmetry momenta alone,~\cite{shiozaki2018}
\begin{align}
  \BSz_{\eta}[G,\Theta] = 
  \prod_{\vkslist} \mathfrak{K}_{\eta_{\vk_{\rm s}}}[G_{\vk_{\rm s}},\Theta].
	\label{eq:BS}
\end{align}
The classifying group $\mathfrak{K}_{\eta_{\vk}}[G_\vk,\Theta]$ describes
zero-dimensional topological phases protected by the onsite symmetry group
$G_\vk$ and can be calculated from Eq.~(\ref{eq:KK}).

Although this procedure associates a well-defined topological invariant with
each Hamiltonian $H(\vk)$, there are three reasons why $\BSz_{\eta}[G,\Theta]$
is different from the classifying group ${\cal K}_{\eta}[G,\Theta]$ of
topological phases with tenfold-way symmetries $\eta$, point group $G$, and a
superconducting phase with symmetry described by $\Theta$: (i) Not every
element in $\BSz$ represents the band labels of a gapped Hamiltonian $H(\vk)$,
(ii) there may exist $d$-dimensional Hamiltonians $H(\vk)$ without
topologically protected anomalous boundary states for which the band labels are
nontrivial nevertheless, and (iii) there may be Hamiltonians $H(\vk)$ with
topologically protected anomalous boundary states for which the band labels are
trivial. 

Problem (i) is addressed by the introduction of ``compatibility relations'',
constraints on the band labels, which follow from the fact that $H(\vk)$ is not
only gapped at the high-symmetry momenta $\vkslist$ --- which is what allows
the band labels to be defined in the first place ---, but also in the remainder
of the Brillouin zone.\cite{po2017,bradlyn2017,kruthoff2017} These constraints
appear, in the first place, because for a gapped Hamiltonian $H(\vk)$ any band
labels that can be defined for lower-symmetry momenta $\vk \notin \XX$ remain
well-defined and continuous if $\vk$ approaches a high-symmetry momentum
$\vkslist$. Since band labels are essentially zero-dimensional topological
invariants, see the discussion above, we call these compatibility relations of
``zero-dimensional'' (``0d'') type. Po {\it et al.} use these ``0d
compatibility relations'' to define the subgroup $\BS \subset \BSz$ of
``topological band labels'',
\begin{equation}
  \BS = \left. \BSz \right|_{\text{$0d$ compatibility relations}}.
  \label{eq:XBS}
\end{equation}
We use $B[H(\vk)]$ to denote the element of $\BS$ associated with the gapped
Hamiltonian $H(\vk)$. 

Whereas the group $\BSz$ of band labels in principle depends on the choice of
the set of high-symmetry momenta $\{ \vk_{\rm s} \}$, the group $\BS$ is
independent of this choice, as long as sufficiently many momenta included. To
see this, we note that the inclusion of additional momenta beyond those
appearing in the set $\{ \vk_{\rm s} \}$ of high-symmetry momenta of
Sec.~\ref{sec:4a} adds band labels to $\BSz$, but not to $\BS$, as band labels
at lower-symmetry-momenta are fully determined by the $0d$ compatibility
relations.

Additional constraints on topological band labels of gapped Hamiltonians ---
compatibility relations of ``one-dimensional'' (``$1d$'') or
``two-dimensional'' (``$2d$'') type --- may also appear because of the
existence of topological invariants on families of higher-dimensional subspaces
of the Brillouin zone,\cite{turner2012,song2018} as we discuss below in more
detail. Elements in $\BSz$ or $\BS$ that violate the compatibility constraints
describe ``representation-enforced'' gapless
phases.~\cite{po2017,song2018,ono2018} Depending on the dimension of the
lower-symmetry region in reciprocal space and the type of constraint imposed on
it, these representation-enforced gapless phases may have nodal points, nodal
lines, or nodal planes. 

Problem (ii) can be remedied by passing to the quotient group\cite{po2017}
\begin{equation}
  \SI = \BS/\AI,
\end{equation}
the group of ``symmetry-based indicators''. 
Here $\AI \subset \BS$ is the subgroup generated by the image under $B$ of all Hamiltonians $H(\vk)$ without boundary signature. 
A generating set of Hamiltonians without boundary states
consists of the ``atomic-limit'' Hamiltonians, $d$-dimensional Hamiltonians
that correspond to the arrangement of zero-dimensional Hamiltonians on a
suitably defined lattice. After dividing out $\AI$, a gapped
Hamiltonian $H(\vk)$ is
associated with a nontrivial element of $\SI$ only if it has topologically protected anomalous boundary states. Po {\it et al.} use the notation $X_{\rm BS}$ to refer to the group of symmetry-based indicators.

There is no general solution to address problem (iii), however. This is why
the group $\SI$ is said to contain ``indicators'' of the topology, not a full
classification.~\cite{po2017}

Although the topological invariants of zero-dimensional Hamiltonians are only
defined for pairs of Hamiltonians, the symmetry-based indicators are expressed
in terms of the band labels of a general Hamiltonian $H(\vk)$ with
$\vk$-independent symmetry representation, without comparing to a reference
Hamiltonian.

\subsection{Construction of the subgroup $\AI \subset \BS$} 
%{\it Construction of the subgroup $\AI \subset \BS$.---} 
Following Po {\it et al.}~\cite{po2017} and Bradlyn {\it et al.}~\cite{bradlyn2017}, the construction of the subgroup $\AI$ proceeds in
three steps: (1) One first selects a representative collection $W$ of
high-symmetry Wyckoff positions in the unit cell. Hereto, one defines the
site symmetry group $G_\vx$ for a position $\vx$ in the unit cell as the subgroup
of $G$ that leaves $\vx$ invariant, possibly up to lattice translations, and
arranges lattice positions with the same site symmetry group which are related by a
continuous path and/or by an element of $G$ into equivalence classes (``Wyckoff
positions''). A representative collection of positions $W$ in the unit cell is
then obtained by choosing a position from each equivalence class
that does not border on an equivalence class with a larger site symmetry group. (2)
For each position $\vx \in W$ the ``orbit'' of $\vx$ is defined as the set
$\{g \vx+\bm t|\, g \in G\}$ for $\bm t$ in Bravais lattice. The orbit defines
a $G$-symmetric Bravais lattice. We label the lattice sites within the unit
cell by $O_\vx=\{\vx_\sigma\}_{\sigma=1,2,\ldots}$, with the convention
$\vx_1\equiv\vx$.  (3) For each Wyckoff position $\vx$ and each pair
$(\alpha,\Theta)$ of an irreducible representation of the site symmetry group
$G_{\vx}$ and the associated one-dimensional representation $\Theta$ describing
the symmetry of the superconducting order parameter, we construct a {\it pair}
of atomic-limit Hamiltonians  $H_{\vx,{\rm ref}}^{(\alpha,\Theta)}$ and $H_{\vx,{\rm gen}}^{(\alpha,\Theta)}$ by placing the zero-dimensional reference and generating
Hamiltonians $H_{\rm ref}$ and $H_{\rm gen}$ of
$\mathfrak{K}_{\eta}[G_{\vx},\Theta]$ on the positions $\vx_\sigma$ of the
orbit of $\vx$. One verifies that invariance of the zero-dimensional
Hamiltonians $H_{\vx,{\rm ref}}^{(\alpha,\Theta)}$ and $H_{\vx,{\rm gen}}^{(\alpha,\Theta)}$
under the site symmetry group $G_{\vx}$ ensures that such a procedure yields 
well-defined $G$-symmetric, translationally invariant, Hamiltonians $H_{\vx,{\rm
ref}}^{(\alpha,\Theta)}(\vk)$ and $H_{\vx,{\rm gen}}^{(\alpha,\Theta)}(\vk)$ on a $d$-dimensional
lattice.  The subgroup $\AI \subset \BS$ is generated by the {\it differences}
$B[H_{\vx,{\rm gen}}^{(\alpha,\Theta)}(\vk)]-B[H_{\vx,{\rm ref}}^{(\alpha,\Theta)}(\vk)]$ of the
images of the Hamiltonians obtained in this manner.

The procedure of taking differences of images under $B$ is necessary for the
construction of $\AI$, because, although the reference Hamiltonians $H_{\rm
ref}$ in Tables~\ref{tab:Wigner_cases_P} and \ref{tab:Wigner_cases_TP} have
been chosen such that they map to the trivial element under $B$, the band
labels need not be trivial for atomic-limit Hamiltonian obtained by placing
copies of $H_{\rm ref}$ on nontrivial Wyckoff positions. The nontrivial band
labels for the reference Hamiltonian have their origin in the $\vk$-dependence
of the representation of the point group $G$, which is unavoidable if minimal
$0d$ superconductors are placed at nontrivial Wyckoff positions. Such a $\vk$
dependence of the representation is not compatible with our construction of the
topological band labels $B[H(\vk)]$, which assume a $\vk$-independent
representation of $G$. (The simplest example where this is the case is a
one-dimensional superconductor with inversion symmetry and odd-parity
superconducting order, see Sec.~\ref{sec:Ci_D}). 

One can resolve this issue either by adding multiple copies of the ``trivial''
$0d$ Hamiltonian $H_{\rm ref}$ at the same Wyckoff position, which eventually
allows one to construct a $\vk$-independent representation of $G$. We here
prefer to take an equivalent, but computationally more efficiently approach, in
which we remedy this situation by comparing atomic-limit Hamiltonians with the
same number of orbitals (per representation pair $(\alpha,\Theta)$) at each
Wyckoff position only. We construct a ``trivial'' reference atomic-limit
Hamiltonian for this case, and define the band labels of each other
atomic-limit Hamiltonian as the difference of the band labels with the
reference atomic-limit Hamiltonian. The trivial reference atomic-limit
Hamiltonian is chosen such that all Wyckoff positions and all representation
pairs $(\alpha,\Theta)$ are represented: It is the direct sum of $0d$ reference
Hamiltonians $H_{\rm ref}$ from Tables \ref{tab:Wigner_cases_P} and
\ref{tab:Wigner_cases_TP} for all pairs $(\alpha,\Theta)$, placed at all
Wyckoff positions. Generator atomic-limit Hamiltonians
$H^{(\alpha,\Theta)}_{\vx,{\rm gen}}(\vk)$ are obtained by replacing a $0d$
reference Hamiltonian in this direct sum by a generator from Tables
\ref{tab:Wigner_cases_P} and \ref{tab:Wigner_cases_TP} for the orbit $O_{\vx}$
and representation pair $(\alpha,\Theta)$ only. Since the map $B$ is additive
under the taking of direct sums, taking the difference of the band labels of
this generator Hamiltonian and the reference atomic-limit Hamiltonian is the
same as taking the difference $B[H_{\vx,{\rm
gen}}^{(\alpha,\Theta)}(\vk)]-B[H_{\vx,{\rm ref}}^{(\alpha,\Theta)}(\vk)]$ for
atomic-limit Hamiltonians involving a single Wyckoff position $\vx$ and
representation pair $(\alpha,\Theta)$ only, which justifies the procedure
outlined step (iii) above.

Each reference Hamiltonian or generator Hamiltonian is compactly expressed as\cite{bradley1972,po2017,bradlyn2017}
\begin{align}
	H_{\vx}^{(\alpha,\Theta)}(\vk)&=\mathbb{1}_{\vert O_\vx\vert\times\vert O_\vx\vert}\otimes H_\vx^{(\alpha,\Theta)},
\end{align}
where $|O_{\vx}|$ is the number of sites in the orbit of $\vx$. The representation of $g \in G$ is
\begin{align}
	[U^{\vx,\alpha}(g,\vk)]_{\sigma'j,\sigma i}&=\delta^\prime_{\vx_{\sigma^\prime},g\vx_\sigma}e^{-i\vk\cdot(g\vx_\sigma-\vx_{\sigma^\prime})}\\
	&\times\frac{z_{g,g_{\sigma}}}{z_{g_{\sigma'},g_{\sigma'}^{-1}gg_\sigma}}[u_\vx^\alpha(g_{\sigma'}^{-1}t_{\vx_{\sigma^\prime}-\vx_\sigma}gg_\sigma)]_{ji},\nonumber
\end{align}
where the indices $\sigma$ and $\sigma'$ label the positions $\vx_{\sigma}$ in
the orbit of $\vx$ and the indices $i$ and $j$ the degree of freedom of
$H_{\vx}^{(\alpha,\Theta)}$. Further, a choice $\{g_\sigma\}_{\sigma=1,\ldots,\vert
O_\vx\vert}$ from $G$ was made in such a way that $g_{\sigma=1}$ is the
identity and $g_\sigma\vx=\vx_\sigma$ for $\sigma=2,\ldots,\vert O_\vx\vert$.
Each group element is decomposed as $g=t_\vR g_\sigma h$ with $h\in G_\vx$,
$t_\vR\in T$ and $\vR=g\vx_\sigma-\vx_{\sigma^\prime}$. The Kronecker symbol
$\delta'_{\vx_1,\vx_2}$ is non-zero and equal to one only when
$\vx_1=\vx_2+\vR$ for some $\vR$ in Bravais lattice. The representation
$u_\alpha$ is given in Tables~\ref{tab:Wigner_cases_P} and
\ref{tab:Wigner_cases_TP}.
%, where $\alpha$ is a shorthand notation for the representation pair $(\alpha,\Theta)$.

In this construction it is important that time-reversal ${\cal T}$ and
particle-hole conjugation ${\cal P}$ are {\it local} operations. If ${\cal P}$
is not a local operation, an arrangement of disconnected onsite Hamiltonians
can acquire a spurious non-locality, in spite of the absence of inter-site
matrix elements. The suggestion of Ref.\ \onlinecite{skurativska2019} that no
atomic limit can be defined for Hamiltonians of Bogoliubov-de Gennes type can
be traced back to the use of a non-local particle-hole conjugation in that
reference.

\subsection{Compatibility relations}
In the examples that we consider in the next Sections, we find that it is
sufficient to use  compatibility relations for gapped hermitian matrices
$H(\vk)$ based on local-in-$\vk$ symmetries only. These compatibility relations
may be applied to the entire matrix $H(\vk)$ or to a diagonal block, if
$H(\vk)$ has a block structure that is preserved throughout a part of the
Brillouin zone. The compatibility relations involve topological invariants of
$H(\vk)$ defined using information about $H(\vk)$ away from the high-symmetry
points $\vk_{\rm s}$. Depending on the dimensionality of the subspace required
to define these topological invariants, we distinguish between compatibility
relations of $0d$ type (which are used to define the group $\BS$ of topological
band labels) and of $1d$ or $2d$ type.\cite{shiozaki2018} We now discuss the
construction of $0d$, $1d$, and $2d$ compatibility relations in detail.

{\it 0a.---} The number of negative eigenvalues of a hermitian matrix $H(\vk)$
with a gapped spectrum does not change if $\vk$ is changed continuously. This
gives a compatibility constraint if $\vk$ can be changed continuously between
high-symmetry points $\vk_{{\rm s},1}$ and $\vk_{{\rm s},2}$, while preserving
Hermiticity of (a subblock of) the Hamiltonian, and if the number of negative
eigenvalues can be related to the band labels at $\vk_{{\rm s},1}$ and
$\vk_{{\rm s},2}$.

{\it 0b.---} The sign of the Pfaffian of a gapped, antisymmetric matrix $U
H(\vk)$, with $U$ an appropriately chosen unitary operator, does not change if
$\vk$ is changed continuously. This results in a compatibility constraint for
topological band labels if $\vk$ can be changed continuously between
high-symmetry points $\vk_{{\rm s},1}$ and $\vk_{{\rm s},2}$, while preserving
the antisymmetry of (a subblock of) the matrix $U H(\vk)$, and if the sign of
the Pfaffian can be related to the band labels at $\vk_{{\rm s},1}$ and
$\vk_{{\rm s},2}$.

{\it 1.---} The topological invariant of a one-dimensional Hamiltonian $H(k)$,
obtained by restricting $H(\vk)$ to a one-dimensional closed contour in
reciprocal space, does not change if this contour is continuously deformed.
Such a topological invariant can be a winding number, if $H(k)$ has a chiral
antisymmetry, but it may also be a $\ZZ_2$ invariant (the first Stiefel-Whitney
number \cite{fang2015, ahn2018}), if $H(k)$ or a diagonal block of it satisfy
local-in-$\vk$ symmetries that place it effectively in tenfold-way classes AI
or BDI. This gives an additional compatibility relation
if the topological invariant can be related to the topological band labels of
$H(\vk_{\rm s})$ at high-symmetry points $\vk_{\rm s}$ on the contour and if
the contour can be deformed so that it can be made to pass through different
sets of high-symmetry momenta, while preserving local-in-$\vk$ symmetries of
$H(\vk)$. An example of such a $1d$ compatibility relation is given in the
example discussed in Sec.~\ref{sec:3d_CI_Ci} (tenfold-way class CI with
inversion symmetry).

{\it 2.---} The topological invariant of a gapped hermitian matrix
$H(k_1,k_2)$, which is defined on a two-dimensional plane cutting through the
Brillouin zone, does not change if the position of the plane is shifted
continuously. The topological invariant is a Chern number if the local-in-$\vk$
crystalline symmetries are such that $H(k_1,k_2)$ or a diagonal block of it are
effectively in tenfold-way classes A, D, or C, or it may be a $\ZZ_2$ invariant (the second Stiefel-Whitney number \cite{fang2015, ahn2018}),
if $H(k_1,k_2)$ is effectively in tenfold-way classes CI or AI. This yields an
additional compatibility relation if the topological invariant can be related
to the band labels of $H(\vk_{\rm s})$ at high-symmetry points $\vk_{\rm s}$ on
the plane and if the plane can be shifted continuously, such that it can be
made to pass through different sets of high-symmetry momenta while preserving
the hermiticity of (the relevant subblock of) $H(\vk)$. This ``2$d$
compatibility relation'' is used in Ref.~\onlinecite{po2017} to identify
representation-enforced nodal semimetals (see also
Ref.~\onlinecite{turner2012,song2018}). It also appears, {\it e.g.}, in the
example discussed in Sec.~\ref{sec:3d_D_Ci} (tenfold-way class D with
inversion symmetry).

Unlike the construction of the groups $\BS$ and $\AI$, we are not aware of a
method that allows one to implement the compatibility relations based on
higher-dimensional topological invariants in an algorithmic way. The
``bottleneck'' is the relation between the $n$-dimensional topological
invariant used to construct the compatibility relation and the topological band
labels of $H(\vk)$ at high-symmetry momenta $\vk_{\rm s}$, which requires
knowledge of the full classification of $n$-dimensional crystalline phases.
This is not a problem if the symmetry-based indicators are used to determine
the topological phase of a given BdG Hamiltonian with a gapped spectrum, but it
does affect the use of $\SI$ as a proxy for the full boundary classification
group ${\cal K}$.

Generalizing the definition of $\BS$, see Eq.~(\ref{eq:XBS}), we may define
the subgroup series
\begin{equation}
  \BS^{(2)} \subset \BS^{(1)} \subset \BS^{(0)} \equiv \BS,
\end{equation}
where $\BS^{(n)}$ is the subgroup of $\BSz$ obtained by imposing all
compatibility constraints involving topological invariants of dimension $\le
n$. Correspondingly, we may define the group $\SI^{(n)} \subset \SI$ of
symmetry-based indicators as
\begin{equation}
  \SI^{(2)} \subset \SI^{(1)} \subset \SI^{(0)} \equiv \SI,
\end{equation}
with
\begin{equation}
  \SI^{(n)} = \BS^{(n)}/\AI
\end{equation}
and $\SI^{(0)} = \SI$. For the examples we consider, we find that $\SI^{(d-1)}$
contains no gapless phases, implying that the compatibility relations based on
local-in-$\vk$ symmetries are sufficient for these cases. 

\subsection{Weak-pairing limit}

In the weak-pairing limit --- superconducting order parameter $\Delta$ much smaller than energy scales typical for the normal-state band structure --- the band labels $\mathfrak{N}_\alpha^{\vk_{\rm s}}$ and $\mathfrak{p}_\alpha^{\vk_{\rm s}}$ of the BdG Hamiltonian $H({\vk_{\rm s}})$ at the high-symmetry momentum $\vk_{\rm s}$ can be expressed in terms of zero-dimensional topological invariants of the normal-state Hamiltonians $h({\vk_{\rm s}})$ and $h(-{\vk_{\rm s}})$. For $\mathfrak{N}_\alpha^{\vk_{\rm s}}$ one has~\cite{ono2019b,shiozaki2019SI,ono2019c}
\begin{align}
	\label{eq:WPN_inv}
	\mathfrak{N}_\alpha^{\vk_{\rm s}} & = n_\alpha^{\vk_{\rm s}} |_\text{occ} + n_{\Theta \alpha^*}^{- {\vk_{\rm s}}} |_\text{unocc} \\ \nonumber
	& = n_\alpha^{\vk_{\rm s}} |_\text{occ} - n_{\Theta \alpha^*}^{- {\vk_{\rm s}}} |_\text{occ} + n_{\Theta \alpha^*}^{- {\vk_{\rm s}}} 
\end{align}
where $n^{\vk_{\rm s}}_\alpha$, $n^{\vk_{\rm s}}_\alpha|_\text{occ}$ and $n^{\vk_{\rm s}}_\alpha|_\text{unocc}$
are the total number, the number of occupied, and the number of unoccupied bands of the subblock $h_\alpha({\vk_{\rm s}})$ of the normal-state Hamiltonian, respectively.
Similarly, the Pfaffian invariant $\mathfrak{p}_\alpha^{\vk_{\rm s}}$ can be expressed as~\cite{shiozaki2019SI}
\begin{equation}
	\mathfrak{p}_\alpha^{\vk_{\rm s}} = n_\alpha^{\vk_{\rm s}}|_\text{occ} \mod 2.
	\label{eq:WPp_inv}
\end{equation}

\section{Examples: One dimension}\label{sec:examples}
We now discuss symmetry-based indicators for Bogoliubov-de Gennes-type
Hamiltonians for a selected set of point groups $G$ and tenfold-way classes. 

\subsection{Trivial point group $G = C_1$, class D}
Without crystalline symmetries, the Hamiltonian $H(k)$ is particle-hole
antisymmetric at both high-symmetry momenta $k = 0$ and $k = \pi$. In the
absence of time-reversal symmetry and spin-rotation symmetry (class D), this
gives a $\ZZ_2$-classification for $H(0)$ and $H(\pi)$, so that $\BSz =
\ZZ_2^2$. There are no compatibility relations, hence
\begin{equation}
  \BS = \ZZ_2^2.
\end{equation}
The corresponding topological
invariants are $(-1)^{\mathfrak{p}^{(0)}} = \sign \text{Pf}\, [H(0) \tau_1]$ and
$(-1)^{\mathfrak{p}^{(\pi)}} = \sign \text{Pf}\, [H(\pi) \tau_1]$, giving
\begin{equation}
  B[H(k)] = \{\mathfrak{p}^{(0)},\mathfrak{p}^{(\pi)}\}.
\end{equation}
Without crystalline symmetries there is only one Wyckoff position in a
one-dimensional crystal. Placing zero-dimensional Hamiltonians on the generic
Wyckoff position, one obtains two topologically different classes of
atomic-limit Hamiltonians,
$$
  \AI = \ZZ_2.
$$%
As a subgroup of $\BS$, $\AI$ corresponds to the pairs
$(\mathfrak{p}^{(0)},\mathfrak{p}^{(\pi)})$ with $\mathfrak{p}^{(0)} =
\mathfrak{p}^{(\pi)}$. The group $\SI = \BS/\AI = \ZZ_2$ is identical to the full
classification group 
\begin{equation}
  \SI_{\classD}[C_1] = {\cal K}_{\classD}[C_1] = \ZZ_2,
\end{equation}
the nontrivial element of which describes one-dimensional topological
superconductors (the ``Kitaev chain''). The symmetry-based indicator for the
topological superconductor phase is
\begin{equation}
  z_1 = \sum_{k_{\rm s}} \mathfrak{p}^{(k_{\rm s})} \mod 2.
  \label{eq:z1C1d1}
%  \mathfrak{p}(\pi) - \mathfrak{p}(0) \mod 2.
\end{equation}
%which in the weak-pairing limit can be written as
%\begin{equation}
%  z_1^\WP = \sum_{k_{\rm s}} n^{(k_{\rm s})}|_\text{occ} \mod 2.
%%  \mathfrak{p}(\pi) - \mathfrak{p}(0) \mod 2.
%\label{eq:1d_z1_WP}
%\end{equation}
 In the weak pairing limit, the indicator (\ref{eq:z1C1d1}) gives the parity of the number of Fermi level crossings between $k=0$ and $k=\pi$.~\cite{sato2010, fu2010} The same expression for the full BdG Hamiltonian was obtained in Ref.~\onlinecite{shiozaki2019SI}.

\subsection{Inversion symmetry $C_i$, class D}\label{sec:Ci_D}
{\it Representation $\Theta = A_g$.---} For a superconducting order parameter
that transforms according to the representation $A_g$, inversion may be
represented as $\rho_3 \nootimes \tau_0$, the Pauli matrices $\rho_3$ and
$\tau_0$ acting in parity and particle-hole space, respectively, see Sec.\
\ref{sec:symmetry_classification_orderparameter}. At the high symmetry momenta
$k_{\rm s} = (0), (\pi)$, the Hamiltonian $H(k_{\rm s}) = \mbox{diag}\,
(H_+(k_{\rm s}),H_-(k_{\rm s}))$ is the diagonal sum of blocks acting within
the even and odd parity subspaces, where the blocks correspond to the
irreducible representations $A_g$ and $A_u$ of $C_i$. The preceding discussion
of a one-dimensional Hamiltonian $H(k)$ without crystalline symmetries applies
to the two blocks separately. In particular, it follows that $\BSz=\ZZ_2^4$.

To find the compatibility constraints for gapped Hamiltonians, we note that the
combination of inversion and particle-hole conjugation gives an antiunitary
antisymmetry local in momentum space,
$$
  H(k) = -\rho_3 \tau_1 H(k)^* \rho_3 \tau_1.
$$%
As a result, $H(k) \rho_3 \tau_1$ is antisymmetric for all $k$, so that $\sign
\mbox{Pf}\,[H(k) \rho_3 \tau_1]$ is well-defined and $k$-independent if $H(k)$
is gapped. Considering that at the high-symmetry momenta $k=0$ and $k=\pi$ one
has $\sign \mbox{Pf}\,[H(k) \rho_3 \tau_1] = (-1)^{\mathfrak{p}_+^{(k)}
+\mathfrak{p}_-^{(k)}}$, one finds the compatibility constraint
\begin{equation}
  \mathfrak{p}_+^{(0)} + \mathfrak{p}_-^{(0)} = \mathfrak{p}_+^{(\pi)} + \mathfrak{p}_-^{(\pi)} \mod 2
  \label{eq:comp_Ci_D}
\end{equation}
for gapped Hamiltonians, where we use the subscripts $+$ and $-$ for the
representations $A_g$ and $A_u$, respectively. Using the compatibility relation
(\ref{eq:comp_Ci_D}) to eliminate $\mathfrak{p}_-^{(\pi)}$ as an independent
band label, we find
\begin{equation}
  \BS = \ZZ_2^3,\ \ B[H(k)] = \{\mathfrak{p}_+^{(0)},\mathfrak{p}_+^{(\pi)},\mathfrak{p}_-^{(0)}\},
\end{equation}

With inversion symmetry there are two inequivalent Wyckoff positions, labeled
$x = 0$ and $x=\frac{1}{2}$. A generating set of atomic-limit Hamiltonians is
obtained by placing zero-dimensional inversion-symmetric Hamiltonians at each
of the Wyckoff positions. Placing a generator of $\mathfrak{K}_D[0]$ with
topological invariant $\mathfrak{p}=1$ and irreducible representation $A_{g}$
or $A_{u}$ at position $x=0$ we obtain a $k$-independent Hamiltonian for which
inversion is represented as by $\tau_0$ or $-\tau_0$, respectively, and
particle-hole conjugation by $\tau_1 K$. Under the map $B$, these Hamiltonians
are mapped to $\{ 1,0,1\}$ and $\{ 0, 1,0\}$ for the $A_g$ and $A_u$
representations, respectively. Placing a generator of $\mathfrak{K}_D[0]$ with
topological invariant $\mathfrak{p}=1$ and irreducible representation $A_{g}$
or $A_{u}$ at the Wyckoff position $x=\frac{1}{2}$, one obtains a
$k$-independent Hamiltonian for which inversion is represented by $\tau_0 e^{i
k}$ or $- \tau_0 e^{i k}$ for $A_{g}$ and $A_{u}$, respectively. At $k=\pi$ the
irreducible representations are interchanged, so that under $B$ such a
Hamiltonian is mapped to $\{ 1,0,0\}$ and $\{ 0, 1,1\}$ for the irreducible
representations $A_g$ and $A_u$, respectively. The images of these four
generating atomic-limit Hamiltonians under $B$ span the whole group $\AI =
\BS$. The quotient group 
\begin{equation}
  \SI_{\classD}[C_i,A_g] = 0.
\end{equation}
The conclusion that $\SI$ contains no nontrivial gapped phases is consistent
with the triviality of the classifying group 
\begin{equation}
  {\cal K}_{\classD}[C_i,A_g] = 0
\end{equation}
for this symmetry class.~\cite{trifunovic2019}

{\it Representation $A_u$.---} For the representation $A_u$ the inversion
operation for the BdG Hamiltonian is represented as $\rho_3 \nootimes \tau_3$,
see Sec.~\ref{sec:symmetry_classification_orderparameter}. The Hamiltonian
$H(k) = \mbox{diag}\, (H_+(k),H_-(k))$ is block diagonal with respect to the
eigenvalues $\pm$ of $\rho_3 \tau_3$, corresponding to the fundamental
representations $A_{g,u}$ of $C_i$. At $k=0$ and $k=\pi$ the two blocks are
minus the particle-hole conjugate of each other. The topological invariants of
$H(0)$ and $H(\pi)$ are $\mathfrak{N}_+^{(0)}$ and $\mathfrak{N}_+^{(\pi)}$, where
$\mathfrak{N}_{\pm}^{(k)}$ is the number of positive eigenvalues of $H_{\pm}(k)$.
(Note that the topological invariants involve the numbers
$\mathfrak{N}_+^{(k)}$ for $k=0,\pi$ only; Particle-hole antisymmetry implies
that $\mathfrak{N}_+^{(k)} = -\mathfrak{N}_-^{(k)}$ at $k=0,\pi$.) Accordingly,
$\BSz=\ZZ^2$.

The combination of inversion and particle-hole conjugation gives an antiunitary
antisymmetry of $H(k)$ that is local in momentum space and squares to $-1$,
thus $H(k)$ belongs to class C with trivial classification in zero-dimension.
We conclude that there are no compatibility constraints for $\mathfrak{N}_+^{(0)}$
and $\mathfrak{N}_+^{(\pi)}$, so that
\begin{equation}
  \BS = \ZZ^2,\ \ B[H(k)] = \{\mathfrak{N}_+^{(0)},\mathfrak{N}_+^{(\pi)}\}.
\end{equation}

\begin{figure}
\includegraphics[width=0.95\columnwidth]{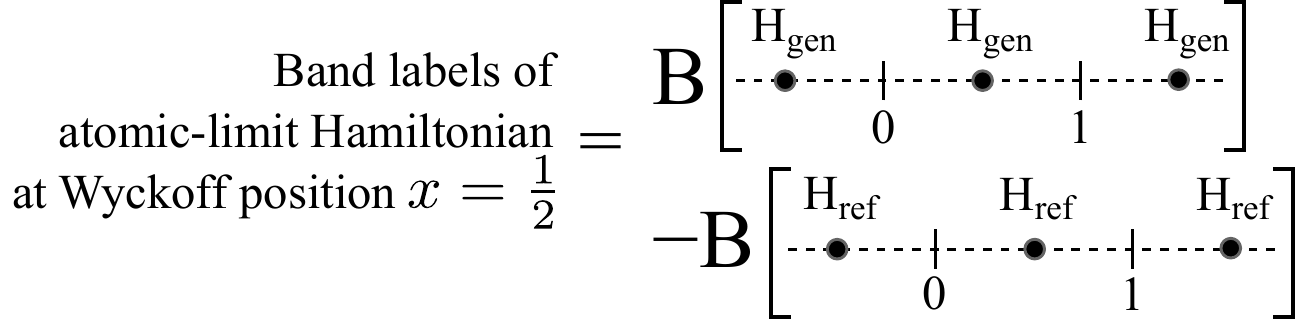}
\caption{Topological band labels associated with an atomic-limit superconductor are calculated as the difference of band labels of arrays with the $0d$ generator Hamiltonian $H_{\rm gen}$ and with the $0d$ reference Hamiltonian $H_{\rm ref}$ at Wyckoff position $x$. The figure shows this schematically for the Wyckoff position $x=\frac{1}{2}$. \label{fig:difference}}
\end{figure}

To construct the subgroup $\AI \subset \BS$ we place zero-dimensional generator
Hamiltonians at the Wyckoff positions $x=0$ or $x=\frac{1}{2}$, which gives the
$k$-independent Hamiltonian $H = \mbox{diag}\,(-1,1) = - \tau_3$, with representation
$\tau_1 K$ for particle-hole conjugation. For the Wyckoff position $x=0$
inversion is represented by $\tau_3$, so that the corresponding atomic-limit
Hamiltonian maps to the element $\{1,1\}$ of $\BS$. For the Wyckoff position
$x=\frac{1}{2}$ inversion is represented by $\tau_3 e^{i k}$. The $A_g$ and
$A_u$ blocks are interchanged at $k=\pi$ and the corresponding atomic-limit
Hamiltonian maps to $\{1,-1\}$, where the $-1$ originates from the difference
with an atomic-limit superconductor obtained by placing the reference
Hamiltonian at the same Wyckoff position, see Fig.~\ref{fig:difference} for a
schematic picture. The subgroup $\AI = \ZZ \times 2 \ZZ$ thus consists of all
elements $\{\mathfrak{N}_{+}^{(0)},\mathfrak{N}_{+}^{(\pi)}\}$ for which
$\mathfrak{N}_{+}^{(0)} - \mathfrak{N}_{+}^{(\pi)}$ is even. The quotient
$\SI=\BS/\AI = \ZZ_2$ contains the symmetry-based indicator
\begin{equation}
  z_1 = \sum_{k_{\rm s}} \mathfrak{N}_{+}^{k_{\rm s}}  \mod 2.
\end{equation}
%whose weak-pairing expression is
%\begin{align}
%  z_1^\WP & = \sum_{k_{\rm s}} n_{+}^{(k_{\rm s})}|_\text{occ} - n_{-}^{(k_{\rm s})}|_\text{occ}  \mod 2 \\ \nonumber
%  & = \sum_{k_{\rm s}} n^{(k_{\rm s})}|_\text{occ} \mod 2
%  \label{eq:1d_Ci_Au_z1_WP}
%\end{align}
%identical to Eq.~\eqref{eq:1d_z1_WP}.
The conclusion that 
\begin{equation}
  \SI_{\classD}[C_i,A_u] = \ZZ_2
\end{equation}
is consistent with the classifying group 
\begin{equation}
  {\cal K}_{\classD}[C_i,A_u] = \ZZ_2
\end{equation}
for this symmetry class,~\cite{trifunovic2019} which describes a topological
superconductor phase with a single zero-energy Majorana bound state at each
end. An example of a nontrivial Hamiltonian in this symmetry class is
$$
  H(k) = \mbox{diag}\,\left( \tau_3 (1 - m - \cos k)  + \tau_1 \sin k ,\tau_3 \right)
$$%
with $0<m<2$.

The weak-pairing limit~(\ref{eq:WPN_inv}) of the symmetry-based indicator $z_1$ agrees Refs.~\onlinecite{skurativska2019,shiozaki2019SI,ono2019c}. The symmetry-based indicator $z_1$ for the full BdG Hamiltonian agrees with the forms defined in Refs.~\onlinecite{shiozaki2019SI,ono2019c}. 

\section{Examples: Two dimensions}\label{sec:examples2}

\subsection{Two mirror symmetries $C_{2v}$, class D}\label{sec:C2v_d2}
The four high-symmetry momenta are $\vk_{\rm s} = (0,0)$, $(0,\pi)$, $(\pi,0)$, and
$(\pi,\pi)$. Each of these four momenta is invariant under the full group
$C_{2v}$ and under particle-hole conjugation, so that the Hamiltonian $H(\vk)$
has the symmetries corresponding to a zero-dimensional Hamiltonian of
tenfold-way class D and with symmetry group $C_{2v}$. As discussed in
Sec.~\ref{sec:0d}, such Hamiltonians are topologically trivial if the
superconducting phase transforms according to the $A_1$ representation of
$C_{2v}$, whereas there is a $\ZZ_2$ invariant $\mathfrak{p}$ for the
representations $A_2$, $B_1$, and $B_2$. Hence, we find that
$$
  \BSz = \left\{ \begin{array}{ll} 0 & \mbox{for $\Theta = A_1$}, \\
  \ZZ_2^4 & \mbox{for $\Theta = A_2$, $B_1$, or $B_2$.} \end{array} \right.
$$%
%For the representations $A_2$, $B_1$, and $B_2$ one has
%$$
%  B[H(\vk)] = \{ \mathfrak{p}(0,0),\mathfrak{p}(0,\pi),\mathfrak{p}(\pi,0),\mathfrak{p}(\pi,\pi) \}.
%$$%

{\it Representation $\Theta = A_1$.---} For the case $\Theta = A_1$ the group of symmetry-based indicators is
\begin{equation}
  \SI_{\classD}[C_{2v},A_1] = 0,
\end{equation}
consistent with the triviality
\begin{equation}
  {\cal K}_{\classD}[C_{2v},A_1] = 0
\end{equation}
of the full classifying group, see App.\ \ref{app:class_2d_C2v}.

{\it Representation $\Theta = B_2$.---} For this representation we have ${\cal
M}_x{\cal P}= -{\cal P}{\cal M}_x$, ${\cal M}_y{\cal P}={\cal P}{\cal M}_y$ and
${\cal M}_x^2={\cal M}_y^2=-1$. For $k_y=0$ one has the effective
``particle-hole antisymmetry'' ${\cal P}{\cal M}_x$ with $({\cal P}{\cal
M}_x)^2=1$, which is local in momentum space. The mirror operation $i {\cal
M}_y$ is also local in momentum space if $k_y = 0$, squares to one, and
commutes with the effective particle-hole symmetry, $({\cal P}{\cal
M}_x)(i{\cal M}_y)=(i{\cal M}_y)({\cal P}{\cal M}_x)$. We conclude that for
each $0 < k_x < \pi$ the Hamiltonian $H(k_x,0)$ satisfies an onsite ``inversion
symmetry'' with $\Theta = A_g$, so that it has a nontrivial topological
classification given by the classifying group
$\mathfrak{K}_{\mathrm{D}}(C_i,A_g)=\ZZ_2^2$. Upon taking the limits $k_x \to
0$, $\pi$, the classification of $H(k_x,0)$ for generic $k_x$ can be related to
the classification at the high-symmetry momenta $k_x = 0$, $\pi$, which is
given by the topological band labels $\mathfrak{p}^{(0,0)}$ and
$\mathfrak{p}^{(\pi,0)}$. Explicit calculation gives that only the diagonal
elements $(\mathfrak{p},\mathfrak{p})$ of
$\mathfrak{K}_{\mathrm{D}}(C_i,A_g)=\ZZ_2^2$ are allowed, with $\mathfrak{p} =
\mathfrak{p}^{(0,0)} = \mathfrak{p}^{(\pi,0)}$. It follows that one has the
compatibility constraint
\begin{equation}
  \mathfrak{p}^{(0,0)} = \mathfrak{p}^{(\pi,0)}.
  \label{eq:B2compatibility}
\end{equation}
In the same way one finds
\begin{equation}
 \mathfrak{p}^{(0,\pi)} = \mathfrak{p}^{(\pi,\pi)}.
  \label{eq:B2compatibility2}
\end{equation}
One may obtain the same compatibility relations using the explicit
representations ${\cal P} = \rho_0 \nootimes \tau_1 K$, $U({\cal M}_x) = i
\rho_1 \nootimes \tau_0$, $U({\cal M}_y)= i \rho_2 \nootimes \tau_0$ used
in Sec.~\ref{sec:0d}, so that $H(\vk)$ satisfies

\begin{align}
  H(k_x,k_y) &=\, \rho_1 \nootimes \tau_0 H(-k_x,k_y) \rho_1 \nootimes \tau_0 \nonumber \\ &=\,
  \rho_2 \nootimes \tau_0 H(k_x,-k_y) \rho_2 \nootimes \tau_0 \nonumber \\ &=\,
  - \rho_0 \nootimes \tau_1 H(-k_x,-k_y)^* \rho_0 \nootimes \tau_1. \nonumber
\end{align}
For $k_y = 0$ these constraints imply that $H(k_x,0) = -\rho_1 \nootimes
\tau_1 H(k_x,0)^* \rho_1 \nootimes \tau_1 = \rho_2 \nootimes \tau_0
H(k_x,0) \rho_2 \nootimes \tau_0$, so that we may write $H(k_x,0) = \rho_0 \nootimes h_0 + \rho_2 \nootimes h_2$ with $h_0 \tau_1$ and $h_2 \tau_1$
antisymmetric. Further, $h_0 \pm h_2$ is gapped for all $0 \le k_x \le \pi$, so
that the Pfaffian of $\tau_1(h_0 \pm h_2)$ is nonzero and cannot change sign
for $0 \le k_x \le \pi$. Since $h_2 = 0$ for $k_x = 0$, $\pi$, the
compatibility relations (\ref{eq:B2compatibility}) and
(\ref{eq:B2compatibility2}) follow immediately. It follows that
\begin{equation}
%  \BS_{\classD}[C_{2v},B_2] 
  \BS = \ZZ_2^2,\ \ B[H(\vk)] = \{ \mathfrak{p}(\pi,0),\mathfrak{p}(\pi,\pi) \}.
\end{equation}

To construct $\AI$, we place generators at one of the four Wyckoff positions.
The Wyckoff positions are $(x,y) = (0,0)$, $(0,\frac{1}{2})$,
$(\frac{1}{2},0)$, and $(\frac{1}{2},\frac{1}{2})$. From the onsite representations $U({\cal M}_x) = i \rho_1
\nootimes \tau_0$, $U({\cal M}_y) =
i \rho_2 \nootimes \tau_0$ we derive the $\vk$-dependent representations
\begin{align}
	U^{\vx}({\cal M}_x, \vk) &=\, i \rho_1 \nootimes \tau_0 e^{2 i k_x x}, \nonumber \\
	U^{\vx}({\cal M}_y, \vk) &=\, i \rho_2 \nootimes \tau_0 e^{2 i k_y y} \nonumber
\end{align}
for a zero-dimensional Hamiltonian placed at Wyckoff position $\vx=(x,y)$. The
$\vk$-dependent factors appear, because for the Wyckoff positions other than
$(0,0)$ the operations ${\cal M}_x$ and/or ${\cal M}_y$ correspond to an onsite
operation followed by a translation by a lattice vector. As these induced
representations differ from the onsite representations by a sign for half of
the elements of the symmetry group, which is a change that can be accommodated
by a basis transformation, the induced representation is the same at all
high-symmetry momenta. After verifying that placing the trivial reference
Hamiltonian $H_{\rm ref}$ at the same Wyckoff positions produces the trivial
element in $\BS$, we conclude that the subgroup $\AI \subset \BS$ consists of
the elements with
$\mathfrak{p}^{(0,0)}=\mathfrak{p}^{(0,\pi)}$. We conclude that
\begin{equation}
  \SI_{\classD}[C_{2v},B_2] = \ZZ_2,
\end{equation}
corresponding to the symmetry-based indicator
\begin{equation}
  z_{1;x} = 
  \sum_{\vk_{\rm s}|k_{{\rm s},x} = \pi} \mathfrak{p}^{\vk_{\rm s}} \mod 2.
\end{equation}
This symmetry-based indicator describes a weak phase of one-dimensional $C_{2v}$-symmetric topological superconductors in the $y$ direction, stacked in the $x$ direction, 
\begin{equation}
  {\cal K}_{\classD}[C_{2v},B_2] = \ZZ_2,
\end{equation}
see App.~\ref{app:class_2d_C2v}. An example of a generator Hamiltonian is
$$
  H_{(1;x)} = \rho_0 \nootimes \tau_3 (1 - m - \cos k_y)  
  + \rho_1 \nootimes \tau_2 \sin k_y,
$$%
with $0 < m < 2$. 

{\it Representation $\Theta = B_1$.---} The discussion for the $B_1$
representation is similar to the discussion above. One finds
\begin{align}
  \SI_{\classD}[C_{2v},B_1] =&\, \ZZ_2,\\
  {\cal K}_{\classD}[C_{2v},B_1] =&\, \ZZ_2.
\end{align}  
The only topologically nontrivial gapped phase is a weak phase of one-dimensional $C_{2v}$-symmetric topological superconductors in the $x$ direction, stacked in the $y$ direction, for which $z_{1;y} = \mathfrak{p}^{(0,0)} + \mathfrak{p}^{(\pi,0)} \mod 2$ is the associated symmetry-based indicator.

{\it Representation $\Theta = A_2$.---} The discussion of this representation
is easiest if we choose the representation $U({\cal M}_x) = i \rho_3
\nootimes \tau_0$, $U({\cal M}_y) = i \rho_1 \nootimes \tau_0$, see the
discussion in Sec.~\ref{sec:0d}. With this choice of representation, one finds
that $H(k_x,k_y) = h_0 \nootimes \rho_0$ at the high-symmetry momenta
$(k_x,k_y) = (0,0)$, $(0,\pi)$, $(\pi,0)$, and $(\pi,\pi)$. The matrix $\tau_1
h_0$ is antisymmetric and the sign $(-1)^{\mathfrak{p}}$ of its Pfaffian is
used as the topological invariant of $H(k_x,k_y)$. In contrast to the $B_1$ and
$B_2$ representations, there are no compatibility constraints for the $A_2$
representation, hence
\begin{align}
  & \BS =
%  & \BS_{\classD}[C_{2v},A_2] = 
  \ZZ_2^4,\ 
  \nonumber \\ & 
  B[H(\vk)] = \{ \mathfrak{p}(0,0),\mathfrak{p}(\pi,0),\mathfrak{p}(0,\pi),\mathfrak{p}(\pi,\pi) \}.
\end{align}
The subgroup $\AI$ for the $A_2$ representation consists of the elements with
$\mathfrak{p}^{(0,0)}=\mathfrak{p}^{(0,\pi)}=\mathfrak{p}^{(\pi,0)}=\mathfrak{p}^{(\pi,\pi)}$,
so that 
\begin{equation}
  \SI_{\classD}[C_{2v},A_2] = \ZZ_2^3.
\end{equation}
The classifying group ${\cal K}$ for the $A_2$ representation is 

\begin{equation}
  {\cal K}_{\classD}[C_{2v},A_2] = \ZZ_2^4,
\end{equation}
see App.~\ref{app:class_2d_C2v}. Two factors $\ZZ_2$ correspond to weak phases
stacked in the $x$ and $y$ directions, with labels ``$1;x$'' and ``$1;y$'',
symmetry-based indicators
\begin{align}
  z_{1;j} =&\, \sum_{\vk_{\rm s}|k_{{\rm s},j} = \pi} \mathfrak{p}^{\vk_{\rm s}} \mod 2, \ \ j=x,y,
%  z_{1;x} =&\, \mathfrak{p}^{(0,0)} + \mathfrak{p}^{(0,\pi)} \mod 2,\nonumber \\
%  z_{1;y} =&\, \mathfrak{p}^{(0,0)} + \mathfrak{p}^{(\pi,0)} \mod 2,
\end{align}
The corresponding generating Hamiltonians are
$$
  H_{(1;x)} = \rho_0 \nootimes \tau_3 (1 - m - \cos k_y)  
  + \rho_3 \nootimes \tau_1 \sin k_y,
$$%
and
$$
  H_{(1;y)} = \rho_0 \nootimes \tau_3 (1 - m - \cos k_x) 
  + \rho_1 \nootimes \tau_1 \sin k_x,
$$%
with $0 < m < 2$, respectively. The remaining two factors $\ZZ_2$ in the
classifying group ${\cal K}$ and the single remaining factor $\ZZ_2$ in $\SI$
correspond to strong second-order phases. These have generator Hamiltonians
\begin{align*}
  H_{(2,\pm)}' = & \rho_0 \nootimes \tau_3 (2 - m - \cos k_x - \cos k_y)   \\
  & \pm \rho_1 \nootimes \tau_1 \sin k_x
  + \rho_3 \nootimes \tau_1 \sin k_y,
\end{align*}
with $0 < m < 2$, but only a single associated symmetry-based indicator
\begin{equation}
  z_{2} = \sum_{\vk_{\rm s}} \mathfrak{p}^{\vk_{\rm s}} \mod 2.
%   \mathfrak{p}^{(0,0)} + \mathfrak{p}^{(0,\pi)} + \mathfrak{p}^{(\pi,0)} + \mathfrak{p}^{(\pi,\pi)} \mod 2.
\end{equation}
As discussed in App.~\ref{app:class_2d_C2v}, the generator Hamiltonians
$H_{(2;\pm)}$ describe a second-order phase with a single Majorana corner state
at each mirror-symmetric corner, whereas the direct sum $H_{(2;+)} \oplus
H_{(2;-)}$, which is mapped to the trivial element in $\SI$, has pairs of even
and odd parity Majorana corner states at the corners bisected by one mirror
axis, but no corner states at the corners bisected by the other mirror axis. 

The results are summarized in Table~\ref{tab:ex_C2v_A2_bandlabels}. Following
the notation of Ref.~\onlinecite{trifunovic2019}, we list a full
boundary-signature-resolved subgroup series for each singly-generated group
${\cal K}_i$ contributing to the full classification group ${\cal K} =
\prod_{i} {\cal K}_i$ that admits topological phases with a higher-order
boundary signature. The subgroup ${\cal K}_i' \subset {\cal K}_i$ is the subset
of all topological phases with boundary signature of order larger than one.
Weak topological phases in two dimensions, which are essentially stacks of
one-dimensional phases, do not have a higher-order boundary signature, so that
we do not give a subgroup sequence for factors ${\cal K}_i$ representing a weak
phase.

Table~\ref{tab:ex_C2v_A2_bandlabels} also lists the topological band labels for
the four Hamiltonians generating the classifying group ${\cal K}$, as well as
their image in the quotient group $\SI$. Here we use the symbol
$\gen_j\!^{(n)}$ to denote the $j$th generator of $\SI$. The superscript $n$
indicates its order, {\it i.e.}, $n \gen_j\!^{(n)} = 0$.

\begin{table}
\begin{tabular*}{\columnwidth}{c @{\extracolsep{\fill}} c | cccc | c} \hline\hline
 &   & \multicolumn{4}{c|}{BS} & $\SI$\tabularnewline
$\mathcal{K}_{i}'\subseteq\mathcal{K}_{i}$ & Phase & $\mathfrak{p}^{(0,0)}$ & $\mathfrak{p}^{(\pi,0)}$ & $\mathfrak{p}^{(0,\pi)}$ & $\mathfrak{p}^{(\pi,\pi)}$ & $\mathbb{Z}_{2}^{3}$\tabularnewline
\hline 
 & $\vec{x}=(0,0)$ & 1 & 1 & 1 & 1 & $\idn$\tabularnewline
 & $\vec{x}=(\frac{1}{2},0)$ & 1 & 1 & 1 & 1 & $\idn$\tabularnewline
 & $\vec{x}=(0,\frac{1}{2})$ & 1 & 1 & 1 & 1 & $\idn$\tabularnewline
 & $\vec{x}=(\frac{1}{2},\frac{1}{2})$ & 1 & 1 & 1 & 1 & $\idn$\tabularnewline
\hline 
$\mathbb{Z}_{2}$ & $(1;x)$  %\includegraphics[scale=0.5]{figs/C2v/2d_1x.pdf} 
& 1 & 1 & 0 & 0 & $\gen_{1;x}^{(2)}$\tabularnewline
$\mathbb{Z}_{2}$ & $(1;y)$  %\includegraphics[scale=0.5]{figs/C2v/2d_1y.pdf} 
& 1 & 0 & 1 & 0 & $\gen_{1;y}^{(2)}$\tabularnewline
$\mathbb{Z}_{2}\subseteq\mathbb{Z}_{2}$ & $(2;+)'$  %\includegraphics[scale=0.5]{figs/C2v/2d_2+.pdf} 
& 1 & 0 & 0 & 0 & $\gen_{2}^{(2)}$\tabularnewline
$\mathbb{Z}_{2}\subseteq\mathbb{Z}_{2}$ & $(2;-)'$  %\includegraphics[scale=0.5]{figs/C2v/2d_2-.pdf} 
& 1 & 0 & 0 & 0 & $\gen_{2}^{(2)}$\tabularnewline \hline\hline
\end{tabular*}\bigskip

\begin{tabular}{cccc}
 \includegraphics[scale=0.7]{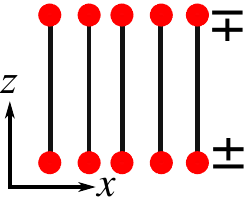} 
& \includegraphics[scale=0.7]{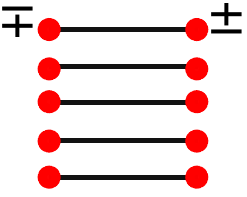}
& \includegraphics[scale=0.7]{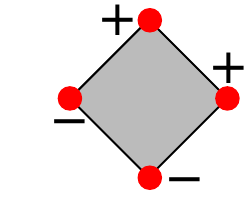}
& \includegraphics[scale=0.7]{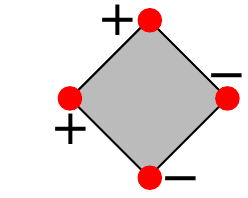} \\
$(1;x)$ & $(1;y)$ & $(2;+)'$ & $(2;-)'$ \\
$z_{1;x} = 1$ & $z_{1;y} = 1$ & $z_2 = 1$ & $z_2 = 1$
\end{tabular}

\caption{Band labels and symmetry-based indicators for atomic-limit
	Hamiltonians obtained by placing 0d generators at Wyckoff position
	$\vx$ (upper four rows) and for the generators $H_{(1;x)}$,
	$H_{(1;y)}$, $H_{(2,+)}'$, $H_{(2,-)}'$ of the weak and second-order
	phases for the symmetry group $C_{2v}$ and representation $\Theta =
	A_2$ in two dimensions, class D. The weak phases $(1;x)$ and $(1;y)$
	can be constructed from one dimensional topological superconductors
	with Majorana bound states, as schematically indicated below (red
	dots with indicated mirror parity $\pm$), stacked in the $x$ or $y$
	direction, respectively. A possible boundary signature of the
	generators of second-order phases consists of Majorana bound states at
	mirror symmetric corners with indicated mirror parity. The first column
	contains the boundary-signature-resolved subgroup sequence for each of
	the factor groups contributing to the full classification group ${\cal
	K} = \prod_{i} {\cal K}_i$.
\label{tab:ex_C2v_A2_bandlabels}}
\end{table}

\subsection{Fourfold rotation symmetry $C_4$, class D}
\label{sec:2d_C4_D}
{\it General considerations.---} There are three non-equivalent high-symmetry
momenta, $\XX= \{ (0,0), (0, \pi), (\pi, \pi)\}$. All three non-equivalent high
symmetry momenta are invariant under particle-hole conjugation. In addition,
the momenta $(0,0)$ and $(\pi, \pi)$ are invariant under fourfold rotation,
while $(0, \pi)$ is invariant under twofold rotation only.

On a square lattice we can choose the set $W$ of representative Wyckoff
positions as $W = \{(0,0), (\frac{1}{2}, \frac{1}{2}), (\frac{1}{2}, 0)\}$. The
two Wyckoff positions $\vx = (0,0)$ and $\vx =  (\frac{1}{2}, \frac{1}{2})$ are
invariant under fourfold rotations; the Wyckoff position $\vx = (\frac{1}{2},
0)$ is invariant under twofold rotations only. This Wyckoff position has a
nontrivial orbit consisting of the positions $(\frac{1}{2}, 0)$ and $(0,
\frac{1}{2})$, generated by fourfold rotation. The induced representations
$U^{\vx,\alpha} (g, \vk)$ of orbitals located at $\vx$ can be written in terms of the
onsite ({\it i.e.}, zero-dimensional) representations $U^\alpha(g)$ using
\begin{align}
\label{eq:ex_C4_indrep}
  U^{(0,0), \alpha} ({\cal R}_{\pi/2}, \vk) & = U^\alpha({\cal R}_{\pi/2}),
  \nonumber \\ 
  U^{(\frac{1}{2},\frac{1}{2}), \alpha} ({\cal R}_{\pi / 2}, \vk) & = U^\alpha({\cal R}_{\pi / 2}) e^{i k_x}, \\ \nonumber
%  U^{(\frac{1}{2},\frac{1}{2}), \alpha} ({\cal R}_{\pi}, \vk) & = U^\alpha({\cal R}_{\pi}) e^{i k_x + i k_y} \\ \nonumber
  U^{(\frac{1}{2},0), \alpha} ({\cal R}_{\pi / 2}, \vk) & = U^\alpha({\cal R}_{\pi / 2}) \otimes \begin{pmatrix}
  0 & 1 \\
  e^{i k_y} & 0
  \end{pmatrix},
%  \\ \nonumber 
%U^{(\frac{1}{2},0), \alpha} ({\cal R}_{\pi}, \vk) & = U^\alpha({\cal R}_{\pi}) \otimes \begin{pmatrix}
%e^{i k_x} & 0 \\
%0 & e^{i k_y}
%\end{pmatrix}
\end{align} 
where ${\cal R}_{\pi/2}$ is a rotation by $\pi/2$ and the matrices for the
Wyckoff position $(\frac{1}{2},0)$ act in the space of orbitals contained in
its orbit.  As in Sec.\ \ref{sec:symmetry_classification_orderparameter} we use
the half-integer angular momentum $j$ to denote the irreps $\alpha$ of the
rotation symmetry. The angular momentum $j$ is defined modulo $4$ for fourfold
rotation and modulo $2$ for twofold rotation.

The computation of the band labels can be performed with the help of the
projector $P_j(\vk_{\rm s })$ of Eq.~(\ref{eq:projector_onto_irrep}), which
projects the Hamiltonian $H(\vk_{\rm s})$ at the high-symmetry momentum
$\vk_{\rm s}$ onto its diagonal block with irreducible representation
$\alpha = j$. In addition to using Eq.\ (\ref{eq:projector_onto_irrep}) to
project on the angular momentum $j$ subspace, a unitary basis transformation
$V_j$ must be implemented to ensure that $V_j P_j(\vk_{\rm s}) V_j^{\dagger}$
is block diagonal. (The unitary matrix $V_j$ depends on $\vk_{\rm s}$, but we
suppress this dependence to keep the notation simple.) In the present case, all
irreps are one-dimensional and the characters $\chi_{j}$ of
Eq.~(\ref{eq:projector_onto_irrep}) are given by 
$$
  \chi_j({\cal R}_{\pi/4}^n) = r_j^n({\cal R}_{\pi/4}) = e^{i n j \pi / 2}.
$$%
We may then choose the unitary matrix $V \equiv V_j$ independent of $j$ by
requiring that $V U(g,\vk_{\rm s}) V^{\dagger}$ be diagonal as in
Eq.~\eqref{eq:ug_diag_C4}, with $U(g) = \mbox{diag}_{\alpha}
[U^{\alpha}(g,\vk_{\rm s})]$ and $U^{\alpha}(g,\vk_{\rm s})$ given in
Eq.~(\ref{eq:ex_C4_indrep}). Indeed, one verifies that with this choice $V
P_j(\vk_{\rm s}) V^{\dagger}$ is a diagonal matrix with unit entries for each
band $n$ with $(V U({\cal R}_{\pi/4}) V^\dagger)_{nn} = r_j({\cal R}_{\pi/4})$
and zeroes otherwise. The transformed Hamiltonian $VH(\vk_{\rm s})V^\dagger$ is
block diagonal and all zero-dimensional invariants can be computed from the
block $V P_j(\vk_{\rm s}) H(\vk_{\rm s}) P_j(\vk_{\rm s}) V^\dagger$. In case
the block is characterized by a Pfaffian invariant $\mathfrak{p}_j(\vk_{\rm
s})$, the transformed representation of the corresponding antiunitary
antisymmetry $V U({\cal P})V^T$ acts within the block such that
$(-1)^\mathfrak{p}_j(\vk_{\rm s}) = \sign \text{Pf}(V P_j(\vk_{\rm s}) 
H(\vk_{\rm s}) U({\cal P}) V^T V P_j(\vk_{\rm s}) V^\dagger)$ is well defined. 

\textit{Representations $\Theta = A$ and $\Theta = B$.---} These two
representations have the same algebraic structure, so that it is sufficient to
discuss the case $\Theta = A$ only. (To see this, one considers $i {\cal
R}_{\pi/2}$ as the generator of $C_{4}$ for the $B$ representation and verifies
that $i {\cal R}_{\pi/2}$ commutes with particle-hole conjugation ${\cal P}$.)

The Hamiltonian has fourfold rotation symmetry at the high-symmetry momenta
$(0,0)$ and $(\pi,\pi)$. As discussed in Sec.~\ref{sec:0d}, for
zero-dimensional BdG Hamiltonians with additional symmetry group $C_4$ and
representation $\Theta = A$, particle-hole conjugation pairs the eigensectors
corresponding to the irreducible representations $j=\frac{1}{2}$ and
$j=\frac{7}{2}$ (mod $4$), as well as $j=\frac{3}{2}$ and $j=\frac{5}{2}$ (mod $4$). The
classifying group $\mathfrak{K}_{\classD}[C_4, A] = \ZZ^2$ and the topological
invariants are $\mathfrak{N}_{\frac{1}{2}}$ and $\mathfrak{N}_{\frac{5}{2}}$.
At the high-symmetry momentum $(0,\pi)$ there is twofold rotation symmetry
only. In this case, particle-hole conjugation pairs the angular momenta
$j=\frac{1}{2}$ and $j=\frac{3}{2}$ (mod $2$) and one has the classification
group $\mathfrak{K}_{\classD}[C_2, A] = \ZZ$ and topological invariant
$\mathfrak{N}_{\frac{1}{2}}$. A general momentum $\vk$ is invariant under
${\cal R}_\pi \mathcal{P}$ with $({\cal R}_\pi \mathcal{P})^2 = -1$ such that
the corresponding zero-dimensional Hamiltonians are in class C with trivial
classification. Hence, there are no compatibility relations in this symmetry
class.  We conclude that
\begin{align}
%	& \BS_{\classD}[C_4,A] = \ZZ^5 .
  & \BS = \ZZ^5, \\
  & B[H(\vk)] = \{ \mathfrak{N}_{\frac{1}{2}}^{(0,0)},\mathfrak{N}_{\frac{5}{2}}^{(0,0)},\mathfrak{N}_{\frac{1}{2}}^{(\pi,0)},\mathfrak{N}_{\frac{1}{2}}^{(\pi,\pi)},\mathfrak{N}_{\frac{5}{2}}^{(\pi,\pi)}\}. \nonumber
\end{align}

The reference and generating Hamiltonians for all Wyckoff positions and
(paired) irreducible representations labeled by the angular momentum $j$, as
well as the onsite representations of the symmetry group, can be taken directly
from Table \ref{tab:Wigner_cases_P}. In particular, we note that
\begin{align}
%  H_{j}^\text{gen} &=\, -\tau_3, \nonumber \\
%  H_{j}^\text{ref} &=\, \tau_3, \nonumber \\
  U^{j,4-j}({\cal R}_{\pi/2}) &=\, \mbox{diag}\, (e^{i j \pi/2}, e^{i (4-j) \pi/2})_{\tau},
%  U({\cal P}) &=\, \tau_1, \nonumber
\end{align}
for $j=\frac{1}{2}$ and $\frac{5}{2}$, where we recall that particle-hole
conjugation pairs $j=\frac{1}{2}$ with $j=\frac{7}{2}$ and $j=\frac{3}{2}$ with
$j=\frac{5}{2}$.
{}From the induced representation (\ref{eq:ex_C4_indrep}) one can then directly
compute the band labels of the reference Hamiltonian as well as all generators, see Table \ref{tab:SM_ex_C4_A_bandlabels} in the supplementary material for details.
Upon computing the quotient $\BS/\AI$, one finds that the group of symmetry
indicators is
\begin{equation}
	\SI_{\classD}[C_4,A] = \SI_{\classD}[C_4,B] = \ZZ_2 \times \ZZ_8 .
	\label{eq:SI_2d_AB_D}
\end{equation}

To interpret the symmetry-based indicators for this representation, we note
that this symmetry class has a classifying group~\cite{teo2013, benalcazar2014}
\begin{equation}
  {\cal K}_{\classD}[C_4,A] =
  {\cal K}_{\classD}[C_4,B] =
  \ZZ \times \ZZ_2^2.
  \label{eq:K_2d_AB_D}
\end{equation}
The factor $\ZZ$ in the above group corresponds to a Chern superconductor
phase, for which the generating Hamiltonian is
$$
	H_{(2)}(\vk) = \tau_3 (2 - m - \cos k_x - \cos k_y) + \tau_1 \sin k_x + \tau_2 \sin k_y
$$%
with $0 < m < 2$ and with (standard) representations $U(\mathcal{P}) = \tau_1$,
$U({\cal R}_{\pi /2}) = e^{-i \pi \tau_3 / 4}$. Chern superconductors with even Chern numbers $\mbox{Ch}$ can be deformed such that they have a BdG Hamiltonian with zero pairing-potential $\Delta$ and normal part $h$ corresponding to a Chern insulator with Chern number $\mbox{Ch}/2$.

Next, one of the factors $\ZZ_2$ of the group~(\ref{eq:K_2d_AB_D}) correspond to a
weak phase, consisting of two ``layers'' of $C_2$-symmetric Kitaev chains
related by a fourfold rotation. (Note that ${\cal K}_{\classD}[C_2, A] = \ZZ_2$
in one dimension, see Refs.~\onlinecite{shiozaki2014,trifunovic2019}.) The
generator Hamiltonian is
\begin{align}
  H_{(1;x,y)}(\vk) =&\, [\tau_3 (1 - m - \cos k_x) + \tau_1 \sin k_x ]
  \nonumber \\ &\, \mbox{}
   \oplus_\mu  [\tau_3 (1 - m - \cos k_y) - \tau_2 \sin k_y ]
  \label{eq:HweakC4}
\end{align}
with $0 < m < 2$, where $\oplus_\mu$ denotes that the direct sum acts in
``layer space'' with Pauli matrices $\mu_\alpha$. The representations are
$U(\mathcal{P}) = \mu_0 \nootimes \tau_1$, $U({\cal R}_{\pi /2}) = \mu_1
\nootimes e^{i \pi \tau_3 / 4}$. The second factor $\ZZ_2$ corresponds to a
second-order phase with four Majorana corner states. The generating Hamiltonian
is the direct sum of a twofold symmetric second-order topological
superconductor~\cite{geier2018} and a copy related by a fourfold rotation,
\begin{align}
  H_{(2)}'(\vk) =&\, \left[\rho_0 \nootimes \tau_3 (2 - m - \cos k_x - \cos k_y)  
  \right. \nonumber \\ &\, \left. \mbox{} 
  + \rho_3 \nootimes \tau_1 \sin k_x + \rho_0 \nootimes \tau_2 \sin k_y \right] 
  \nonumber \\ & \mbox{}
  \oplus_\mu
  \left[ \rho_0 \nootimes \tau_3 (2 - m - \cos k_x - \cos k_y) 
  \right. \nonumber \\ &\, \left. \mbox{} 
  + \rho_0 \nootimes \tau_1 \sin k_x + \rho_3 \nootimes \tau_2 \sin k_y
  \right]
\end{align}
with $0 < m < 2$, and representations $U(\mathcal{P}) = \mu_0 \nootimes \rho_0 \nootimes \tau_1 $, $U({\cal R}_{\pi /2}) = \mu_1 \nootimes \rho_0 \nootimes e^{i \pi \tau_3 / 4}$.

\begin{figure}
\begin{tabular}{ccc}
  $\ZZ_2$ & $0 \subseteq \ZZ$ & $\ZZ_2 \subseteq \ZZ_2$ \\
  $z_{1;x,y}=1$ & $z_2=1$ & $z_2=4$ \\ 
  \includegraphics[scale=0.8]{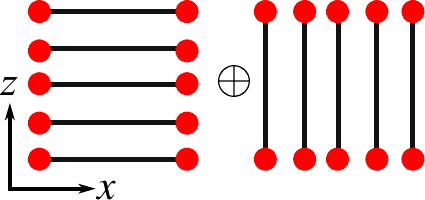} &
  \includegraphics[scale=0.8]{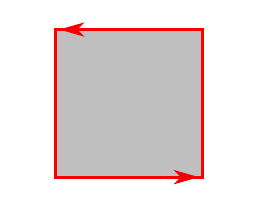} &
  \includegraphics[scale=0.8]{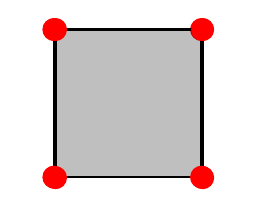}
\end{tabular}
\caption{Topological phases of a two-dimensional superconductor in tenfold-way class D and with additional $C_4$ symmetry with one-dimensional representation $\Theta=A,\,B$. For each boundary signature, the boundary subgroup sequence (top row) and the nonzero symmetry-based indicators for a generator of that phase are given (middle row). \label{fig:2d_AB_D}}
\end{figure}

One verifies that the weak phase generates the factor $\ZZ_2$ of $\SI$ and has
the symmetry-based indicator
\begin{equation}
  z_{1;x,y} = \mathfrak{N}_{\frac{1}{2}}^{(\pi,0)} + 
   \mathfrak{N}_{\frac{1}{2}}^{(\pi,\pi)} + \mathfrak{N}_{\frac{5}{2}}^{(\pi,\pi)} \mod 2. \label{eq:z1_2d_C4_A}
\end{equation}
The factor $\ZZ_8$, with the symmetry-based indicator 
\begin{align}
  z_2 =&\, - \mathfrak{N}_{\frac{1}{2}}^{(0,0)} 
        + 3  \mathfrak{N}_{\frac{5}{2}}^{(0,0)} 
        - 2 \mathfrak{N}_{\frac{1}{2}}^{(\pi,0)} \nonumber \\ &\, \mbox{}
        + 3 \mathfrak{N}_{\frac{1}{2}}^{(\pi,\pi)}
        - \mathfrak{N}_{\frac{5}{2}}^{(\pi,\pi)} \mod 8, \label{eq:z2_2d_C4_A}
\end{align}
is generated by the band labels of the Chern superconductor. The element ``$z_2
= 4$'' of the factor $\ZZ_8$ of $\SI$ is ambiguous, as it corresponds to
the second-order phase or the Chern superconductor with four Majorana modes, see
Fig.~\ref{fig:2d_AB_D}. These classification results and symmetry-based indicators agree with the results from Refs.~\onlinecite{teo2013, benalcazar2014}.

\textit{Representations $\Theta={}^1E$ and $\Theta={}^2E$.---} Here we consider
the case $\Theta={}^1E$ only and note that the case $\Theta={}^2E$ is
analogous. In Sec.~\ref{sec:0d} we found that the particle-hole conjugation
${\cal P}$ with $\Theta={}^1E$ relates the eigensectors with angular momentum
$j=\frac{3}{2}$ and $j=\frac{7}{2}$, while ${\cal P}$ acts within the
$j=\frac{1}{2}$ and $j=\frac{5}{2}$ eigensectors. The eigensectors $j =
\frac{1}{2}$ and $j=\frac{3}{2}$ of twofold rotation symmetry are left
invariant by particle-hole conjugation. From Table~\ref{tab:classD_0d} we
conclude that at each of the high-symmetry momenta $(0,0)$ and $(\pi,\pi)$ with
fourfold rotation symmetry we obtain a $\mathfrak{K}_{\classD}[C_4, {}^1E] =
\ZZ \times \ZZ_2^2$ classification of the band labels, with invariants
$\mathfrak{N}_{\frac{3}{2}}$, $\mathfrak{p}_{\frac{1}{2}}$, and
$\mathfrak{p}_{\frac{5}{2}}$, whereas for the momentum $(0,\pi)$ with twofold
rotation symmetry the classifying group is $\mathfrak{K}_{\classD}[C_2, B] =
\ZZ_2^2$, with invariants $\mathfrak{p}_{\frac{1}{2}}$ and
$\mathfrak{p}_{\frac{3}{2}}$. It follows that $\BSz=\ZZ^2\times\ZZ_2^6$.

\begin{figure}
\begin{tabular}{c}
  $0 \subseteq \ZZ$ \\
  $z_2 = 1$ \\
  \includegraphics[scale=1]{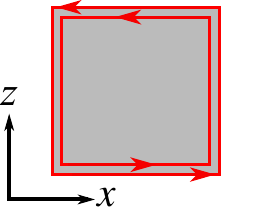}
\end{tabular}
\caption{The only topological phase with nontrivial boundary signature for a two-dimensional superconductor in tenfold-way class D and with additional $C_4$ symmetry with one-dimensional representation $\Theta = \Theta={}^{1,2}E$ is a Chern superconductor with an even number of chiral Majorana boundary modes. The boundary subgroup sequence for this phase is given in the top row; the symmetry-based indicator for a generator of the phase is given in the middle row. \label{fig:2d_12E_D}}
\end{figure}

The combination of twofold rotation and particle-hole conjugation provides
an antiunitary antisymmetry operation that is local in momentum space and
squares to one. For gapped Hamiltonians, the quantity
$(-1)^{\mathfrak{p}({\cal R}_\pi {\cal P})}= \mbox{sign}\, \text{Pf}[H(\vk) U({\cal
R}_\pi {\cal P})]$ is well-defined and constant throughout the Brillouin
zone, from which one derives the ($0d$) compatibility constraints
\begin{align}
\label{eq:ex_C4_E_compability}
  & \mathfrak{p}_{\frac{1}{2}}^{(0,\pi)} + \mathfrak{p}_{\frac{3}{2}}^{(0,\pi)} 
  \\ & =  \mathfrak{p}_{\frac{1}{2}}^{(0,0)} + \mathfrak{p}_{\frac{5}{2}}^{(0,0)} + \mathfrak{N}_{\frac{3}{2}}^{(0,0)}  \nonumber \\ &=
  \mathfrak{p}_{\frac{1}{2}}^{(\pi,\pi)} + \mathfrak{p}_{\frac{5}{2}}^{(\pi,\pi)}  + \mathfrak{N}_{\frac{3}{2}}^{(\pi,\pi )} \mod 2. \nonumber
\end{align}
We therefore conclude that
\begin{align}
%	\BS_{\classD}[C_4,{}^{1}\!E] 
  & \BS = \ZZ^2\times\ZZ_2^4, \\
  & B[H(\vk)] = \{ \mathfrak{p}_{\frac{1}{2}}^{(0,0)}, \mathfrak{p}_{\frac{5}{2}}^{(0,0)}, \mathfrak{N}_{\frac{3}{2}}^{(0,0)}, \mathfrak{p}_{\frac{3}{2}}^{(\pi,0)}, \mathfrak{p}_{\frac{5}{2}}^{(\pi,\pi)}, \mathfrak{N}_{\frac{3}{2}}^{(\pi,\pi)} \}. \nonumber
\end{align}

As before, we take the onsite reference Hamiltonian and generating Hamiltonians
as well as the onsite representations from Table~\ref{tab:Wigner_cases_P}. For
the angular momenta $j=\frac{3}{2}$ and $j=\frac{7}{2}$, which are paired by
particle-hole conjugation, one has the representation
\begin{equation}
  U^{\frac{3}{2},\frac{7}{2}}({\cal R}_{\pi/2}) = \mbox{diag}\, (e^{3 \pi i/4},e^{7 \pi i/4})_{\tau}.
\end{equation}
With the help of the induced representation Eq.~(\ref{eq:ex_C4_indrep}) the band
labels of the atomic-limit Hamiltonians then easily follow, see
Table~\ref{tab:SM_ex_C4_E_bandlabels} in the Supplementary Material. By taking
the quotient group, one arrives at
\begin{equation}
	\SI_{\classD}[C_4,^1\!\!E] =
	\SI_{\classD}[C_4,^2\!\!E] = \ZZ_4 .
	\label{eq:SI_2d_12E_D}
\end{equation}
The group is generated by a Chern superconductor with Chern number 2, 
\begin{align}
H_{(2)}(\vk) =&\, \rho_0 \nootimes \tau_3 (2 - m - \cos k_x - \cos k_y)
  \nonumber \\ &\, 
  \mbox{} + \rho_1 \nootimes \tau_2 \sin k_x + \rho_1 \nootimes \tau_1 \sin k_y, 
  \label{eq:H2_2d_C4_1E}
\end{align}
with $0 < m < 2$, and representations 
\begin{equation*}
U(\mathcal{P}) = \rho_0 \nootimes \tau_1, \quad U({\cal R}_{\pi/2}) = \tau_3 e^{- i \pi /4} \oplus_\rho \tau_0 e^{i \pi /4}.
\end{equation*}
The symmetry-based indicator is
\begin{align}
  z_2 =&\, -\mathfrak{N}_{\frac{3}{2}}^{(0,0)} - \mathfrak{N}_{\frac{3}{2}}^{(\pi,\pi)} + 
  2 \mathfrak{p}_{\frac{5}{2}}^{(0,0)} 
  \nonumber \\ &\, \mbox{}
  + 2 \mathfrak{p}_{\frac{3}{2}}^{(\pi,0)}
  + 2 \mathfrak{p}_{\frac{5}{2}}^{(\pi,\pi)} \mod 4.
  \label{eq:z2_2d_C4_1E}
\end{align}
A Hamiltonian with Chern number $1$ is not compatible with the constrains given
by the algebraic relations between the representations in this symmetry class.

For comparison, we note that the classifying group is 
\begin{equation}
  {\cal K}_{\classD}[C_4,^1\!\!E] = 
  {\cal K}_{\classD}[C_4,^2\!\!E] =
  \ZZ,
  \label{eq:K_2d_12E_D}
\end{equation}
which is generated by the even-Chern-number superconductors, see
Fig.~\ref{fig:2d_12E_D}. There are no weak or second-order phases in this
symmetry class. This observation is compatible with the absence of weak or
second-order phases for a twofold rotation symmetry with $B$ pairing
symmetry.~\cite{shiozaki2014,geier2018,trifunovic2019} 
We note that the Chern superconductor with even Chern number $\mbox{Ch}$ is topologically 
equivalent to a BdG Hamiltonian~(\ref{eq:HBdG}) with zero pairing-potential,
corresponding to a non-superconducting Chern insulator with the Chern number $\mbox{Ch}/2$.

\section{Examples: Three dimensions}\label{sec:examples3}

\subsection{Inversion symmetry, class D} \label{sec:3d_D_Ci}
There are eight high-symmetry momenta $\vk_{\rm s}$ with $k_{{\rm s},x}$,
$k_{{\rm s},y}$, $k_{{\rm s},z} \in \{ 0, \pi \}$. With the classifying group
$\mathfrak{K}_\eta[C_i, A_g] = \mathfrak{K}_\eta[0]^2 = \ZZ_2^2$ and
$\mathfrak{K}_\eta[C_i, A_u] = \mathfrak{K}_{(0, 0, 0)}[0] = \ZZ$ the group of
the band labels is
\begin{equation}
\BSz = \begin{cases}
  \ZZ_2^{16} & \text{for } \Theta = A_g, \\
  \ZZ^8 & \text{for } \Theta = A_u .
\end{cases}
\end{equation}
The band labels are given by the sets of topological invariants
$\{\mathfrak{p}_{\pm}(\vk_{\rm s}) \}$ or $\{ \mathfrak{N}_+(\vk_{\rm s}) \}$
for all high-symmetry momenta $\vk_{\rm s}$ for $\Theta = A_g$ and $\Theta = A_u$, respectively.

In an inversion-symmetric cubic lattice the set of representative Wyckoff
positions $W$ consists of elements $(x,y,z)$ with $x,y,z \in \{0, \frac{1}{2}
\}$. The induced represenstation $U^{\vx,\alpha} ({\cal I}, \vk)$ can be written in
terms of the onsite representations $U^{\alpha} ({\cal I})$ as
\begin{equation}
U^{\vx,\alpha} ({\cal I}, \vk) = U^\alpha ({\cal I}) e^{2 i \vk \cdot \vx}.
\end{equation}

{\it Representation $\Theta = A_g$.---}
The Hamiltonian $H(\vk)$ satisfies the antiunitary antisymmetry ${\cal IP}$
with $({\cal IP})^2 = 1$, which is local in momentum space. As a result, the
quantity $(-1)^{\mathfrak{p}({\cal IP})} = \mbox{sign}\, \text{Pf}[H(\vk) U({\cal
IP})]$ is well-defined and constant throughout the Brillouin zone,
which gives the compatibility relations
\begin{align}
  & \mathfrak{p}_+^{\vk_{\rm s}} + \mathfrak{p}_-^{\vk_{\rm s}} =
  \mathfrak{p}_+^{\bf 0} + \mathfrak{p}_-^{\bf 0} \mod 2
  \label{eq:ex_Ci_3d_D_compability_0d}
\end{align}
for the high-symmetry momenta $\vk_{\rm s}$. It follows that

\begin{align}
	\label{eq:BS_3d_D_Ag}
	& \BS=\ZZ_2^9\\
	& B[H(\vk)] = \left\{ \mathfrak{p}_+^{(0,0,0)}, \mathfrak{p}_-^{(0,0,0)}, \mathfrak{p}_{+}^{(\pi,0,0)}, \mathfrak{p}_{+}^{(0,\pi,0)}, \mathfrak{p}_{+}^{(\pi,\pi,0)}, \right. \nonumber \\
	& \hphantom{B[H(\vk)]=}\ \ \ \left. \mathfrak{p}_{+}^{(0,0,\pi)}, \mathfrak{p}_{+}^{(\pi,0,\pi)}, \mathfrak{p}_{+}^{(0,\pi,\pi)}, \mathfrak{p}_{+}^{(\pi,\pi,\pi)}  \right\}. \nonumber 
\end{align}

Further, for a gapped Hamiltonian $H(\vk)$ a Chern number $\mbox{Ch}_i$ can be
defined on planes with fixed $k_i$ with $i=x$, $y$, or $z$. For the
representation $\Theta = A_g$ the combination of particle-hole and inversion
symmetry allows for even Chern numbers $\mbox{Ch}_i$ only. At high-symmetry
planes with $k_i = 0$ or $\pi$, the Chern number $\text{Ch}_i$ is related to
the band labels as
\begin{align}
  \text{Ch}_i =&\, 2 \left( \sum_{\vkslist|_{k_{{\rm s}, i} = 0}} \mathfrak{p}_+^{\vk_{\rm s}} \right) \mod 4 \nonumber \\ =&\,
  2 \left( \sum_{\vkslist|_{k_{{\rm s}, i} = \pi}} \mathfrak{p}_+^{\vk_{\rm s}} \right) \mod 4,
  \label{eq:Ch_Ci_3d_D}
\end{align}
where we used Eq.~\eqref{eq:ex_Ci_3d_D_compability_0d} to arbitrarily select
the invariant $\mathfrak{p}_+^{\vk_{\rm s}}$ in the even inversion parity
subspace. Imposing this $2d$ compatibility relation further reduces the group
of band labels to $\BS^{(2)} = \ZZ_2^8$, the independent band
labels~(\ref{eq:BS_3d_D_Ag}) with the exception of
$\mathfrak{p}_{+}^{(\pi,\pi,\pi)}$.

\begin{figure}
\begin{tabular}{ccc}
  $0 \subseteq \ZZ$ & $0 \subseteq \ZZ$ & $0 \subseteq \ZZ$ \\
  $z_{2;z}=1$ & $z_{2;y}=1$ & $z_{2;x}=1$ \\
  \includegraphics[scale=1]{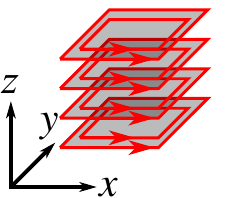} &
  \includegraphics[scale=1]{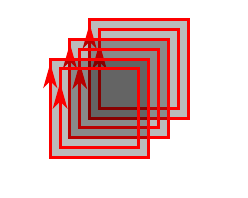} &
  \includegraphics[scale=1]{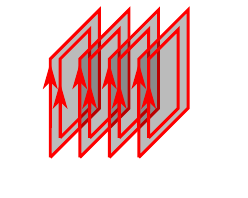} 
\end{tabular}
\caption{Topological phases of a three-dimensional superconductor in tenfold-way class D and with additional $C_i$ symmetry with one-dimensional representation $\Theta=A_g$. For each boundary signature, the boundary subgroup sequence (top row) and the nonzero symmetry-based indicators for a generator of that phase are given (middle row). \label{fig:3d_Ag_D}}
\end{figure}

The band labels of atomic insulators, which span the subgroup $\AI \subset
\BS$, are listed in the supplementary material. Computing the quotient, we find
that
\begin{equation}
	\label{eq:SI_3d_Ag_D}
  \SI_{\classD}[C_i,A_g] = \ZZ_2^{4},\ \ \
  \SI_{\classD}^{(2)}[C_i,A_g] = \ZZ_2^{3},
\end{equation} 
where the four generators of $\SI$ are related to the symmetry-based indicators
\begin{align}
	z_{2;i} =&\, \sum_{\vkslist|k_{{\rm s},i} = \pi} \mathfrak{p}_+^{\vk_{\rm s}} \mod 2,\ \ i=x,y,z, \nonumber \\
        z_3 =&\, \sum_{\vkslist}\mathfrak{p}_+^{\vk_{\rm s}} \mod 2.
	\label{eq:2d_D_Au}
\end{align}
(The three generators of $\SI^{(2)}$ have symmetry-based indicators $z_{2;i}$,
$i=x,y,z$.) The $\ZZ_2$ indicator $z_3$ corresponds to a
representation-enforced nodal superconductor with different Chern numbers at
parallel planes in the Brillouin zone, see Eq.~(\ref{eq:Ch_Ci_3d_D}), and a
nodal point at a generic position in the Brillouin zone. The three $\ZZ_2$
indicators $z_{2;i}$, $i=x,y,z$, correspond to weak Chern superconductors with
even Chern number in stacking planes, see Fig.~\ref{fig:3d_Ag_D}.  Generator
Hamiltonians are
\begin{align}
 H_{(2;l)}(\vk) =&\, \rho_0 \nootimes \tau_3 (2 - m - \cos k_i - \cos k_j) 
  \nonumber \\ &\, + \rho_1 \nootimes \tau_1 \sin k_i  + \rho_1 \nootimes \tau_2 \sin k_j ,
\end{align}
with $(k_i, k_j) = (k_y, k_z)$, $(k_z, k_x)$, or $(k_x, k_y)$ for $l=x$, $y$,
or $z$, respectively. For all three cases, the representations are
\begin{equation}
  U(\mathcal{P}) = \rho_0 \nootimes \tau_1, \quad U({\cal I}) = \rho_3 \nootimes \tau_0.
\end{equation}

For comparison, we note that this symmetry class has a classifying group 
\begin{equation}
  {\cal K}_{\classD}[C_i,A_g] = \ZZ^3,
  \label{eq:K_3d_Ag_D}
\end{equation}
which contains the weak phases with even Chern numbers in the three stacking
planes. Generators for these weak phases are shown schematically in
Fig.~\ref{fig:3d_Ag_D}. Since each factor $\ZZ$ in ${\cal K}$ describes a weak
topological phase, which is obtained by stacking two-dimensional
superconductors, only second-order boundary signatures are allowed in
principle, which is why the subgroup sequences listed in Fig.~\ref{fig:3d_Ag_D}
contains one subgroup only. The three weak Chern superconductors with even
Chern numbers $\mbox{Ch}$ per layer
can be continuously deformed to normal-state weak Chern insulators with vanishing superconducting correlations and Chern number $\mbox{Ch}/2$ per layer.

For comparison, we note that Ref.~\onlinecite{ono2019c} found no symmetry-based
indicators due to the absence of Pfaffian band labels in their construction.

\begin{figure}
\begin{tabular}{ccc}
  $\ZZ_2$ & $\ZZ_2$ & $\ZZ_2$ \\
  $z_{1;y,z} = 1$ & $z_{1;x,z} = 1$ & $z_{1;x,y} = 1$ \\
  \includegraphics[scale=1]{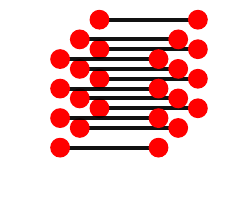} &
  \includegraphics[scale=1]{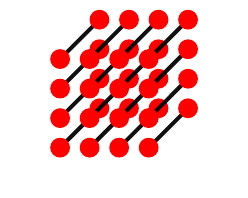} &
  \includegraphics[scale=1]{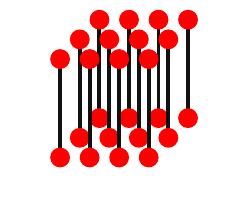} \\
  $0 \subseteq \ZZ$ & $0 \subseteq \ZZ$ & $0 \subseteq \ZZ$ \\
  $z_{2;z}=1$ & $z_{2;y}=1$ & $z_{2;x}=1$ \\  
  \includegraphics[scale=1]{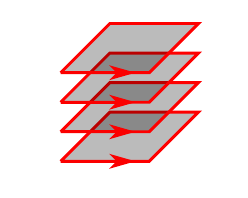} &
  \includegraphics[scale=1]{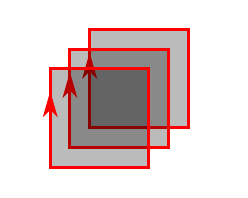} &
  \includegraphics[scale=1]{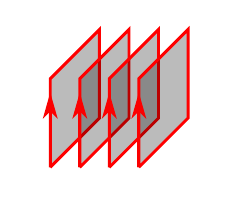} \\
  $\ZZ_2 \subseteq \ZZ_2$ & $\ZZ_2 \subseteq \ZZ_2$ & $\ZZ_2 \subseteq \ZZ_2$ \\
  $z_{2;z}=2$ & $z_{2;y}=2$ & $z_{2;x}=2$ \\
  \includegraphics[scale=1]{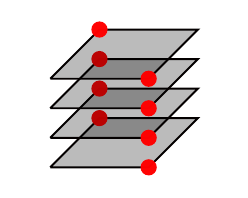} &
  \includegraphics[scale=1]{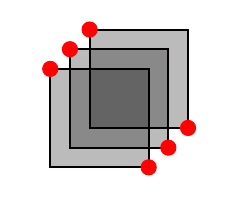} &
  \includegraphics[scale=1]{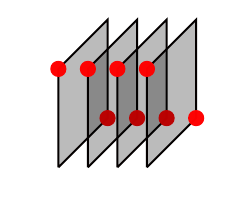} \\
  \multicolumn{2}{c}{$\ZZ_2 \subseteq \ZZ_4 \subseteq \ZZ_4$} & \\
  $z_3 = 2,6$ & $z_3 = 4$ & \\
  \includegraphics[scale=1]{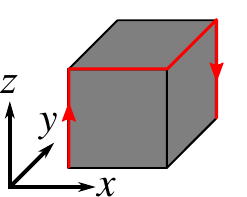} & 
  \includegraphics[scale=1]{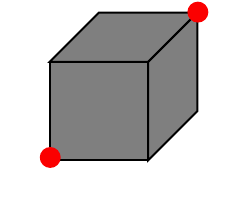} & \\
\end{tabular}
\caption{Topological phases of a three-dimensional superconductor in tenfold-way class D, with additional $C_i$ symmetry and one-dimensional representation $\Theta=A_u$. For each boundary signature, the subgroup sequence and the symmetry-based indicators of the generators of that phase are given. The third-order topological superconductor with $z_3 = 4$ can be constructed as the direct sum of two second-order topological superconductors with $z_3 = 2$.
	\label{fig:3d_Au_D}}
\end{figure}

{\it Representation $\Theta = A_u$.---} For the case of the $A_u$
representation, there are no compatibility relations of $0d$ type, so that
\begin{equation}
  \BS_{\classD}[C_i,A_u]=\BSz_{\classD}[C_i,A_u]=\ZZ^8.
\end{equation}
There is a compatibility relation of $2d$ type, however, which follows from the
existence of Chern numbers $\mbox{Ch}_i$ along planes at constant $k_i$,
$i=x,y,z$,~\cite{turner2012}
\begin{align}
  \mbox{Ch}_i =&\, 
  \sum_{\vkslist|_{k_{{\rm s},i} = 0}} \mathfrak{N}_+^{\vk_{\rm s}} \nonumber
  \\ =&\,
  \sum_{\vkslist|_{k_{{\rm s},i} = \pi}} \mathfrak{N}_+^{\vk_{\rm s}}   \mod 2,
  \label{eq:2dcomp_D}
\end{align}
with $i=x,y,z$. Computing the quotient $\SI=\BS/\AI$, we find that the group of
symmetry-based indicators is
\begin{equation}
  \SI_{\classD}[C_i,A_u] = \ZZ_2^3 \times \ZZ_4^3 \times \ZZ_8.
  \label{eq:SI_3d_D_Au}
\end{equation}
Here the three factors $\ZZ_2$ with symmetry-based indicators
\begin{align}
  z_{1;i,j} =&\, \sum_{\vk_{\rm s}|k_{{\rm s},i} = k_{{\rm s},j} = \pi}
  \mathfrak{N}_{+}^{\vk_{\rm s}}\mod2, 
%  z_{1;y,z} =&\, \mathfrak{N}_{+}^{(0,0,0)} - \mathfrak{N}_{+}^{(\pi,0,0)} \mod 2,
%  \nonumber \\
%  z_{1;x,z} =&\, \mathfrak{N}_{+}^{(0,0,0)} - \mathfrak{N}_{+}^{(0,\pi,0)} \mod 2, \\
%  z_{1;x,y} =&\, \mathfrak{N}_{+}^{(0,0,0)} - \mathfrak{N}_{+}^{(0,0,\pi)} \mod 2, \nonumber
\end{align}
for $(i,j) = (x,y)$, $(z,x)$, and $(y,z)$ correspond to weak phases consisting of stacks of one-dimensional topological
superconductors, see Fig.~\ref{fig:3d_Au_D}. Generator Hamiltonians are
\begin{equation}
  H_{(1;x,y)}(\vk) = \tau_3 (1 - m - \cos k_z) + \tau_1 \sin k_z,\ \
  \mbox{cycl.},
\end{equation}
%with $(l,m) = (y,z), (x,z), (x,y)$ for $i=x,y,z$, respectively, and 
with the representations
\begin{equation}
  U(\mathcal{P}) = \tau_1, \quad U({\cal I}) = \tau_3.
  \label{eq:D_Au_repr}
\end{equation}
The three factors $\ZZ_4$, which have indicators
\begin{align}
\label{eq:z2_3d_Ci_Au}
  z_{2;l} =&\, - \sum_{\vkslist|_{k_{{\rm s},l} = \pi}} \mathfrak{N}_+^{\vk_{\rm s}} (-1)^{(k_{{\rm s},x}+k_{{\rm s},y}+k_{{\rm s},z})/\pi} \mod 4,
\end{align}
for $l=x,y,z$, correspond to weak Chern superconductors, see
Fig.~\ref{fig:3d_Au_D}. Generator Hamiltonians are
\begin{align}
  H_{(2,x)}(\vk) =&\, \tau_3 (2 - m - \cos k_y - \cos k_z) \nonumber \\
  & \mbox{} + \tau_1 \sin k_y + \tau_2 \sin k_z, \ \mbox{cycl.}
  \label{eq:Z4_generator}
\end{align}
%with $(k_i, k_j) = (k_x, k_y), (k_x, k_z), (k_y, k_z)$ for $l=z$, $y$, and $x$,respectively, and 
with $0 < m < 2$, and the same representations as above. Similar to previous examples, we find that the weak even-Chern-number superconductors are continuously deformable to weak normal-state Chern insulators. The symmetry-based
indicators $z_{2,l} = 2$ are ambiguous as they may also
correspond to a weak second-order topological superconductor,
\begin{align}
H_{(2,x)}'(\vk) =&\, \rho_0 \nootimes \tau_3 (2 - m - \cos k_y - \cos k_z) 
  \nonumber \\ &\, \mbox{} + \rho_3 \nootimes \tau_1 \sin k_y + \rho_0 \nootimes \tau_2 \sin k_z,\ \mbox{cycl.}
\end{align}
with $0 < m < 2$, and the representations
\begin{equation}
  U(\mathcal{P}) = \rho_0 \nootimes \tau_1, \quad U({\cal I}) = \rho_0 \nootimes \tau_3.
  \label{eq:D_Au_repr2}
\end{equation}
Finally, the factor $\ZZ_8$ with indicator 
\begin{equation}
  z_3 = \sum_{\vkslist} \mathfrak{N}_+^{\vk_{\rm s}} (-1)^{(k_{{\rm s},x}+k_{{\rm s},y}+k_{{\rm s},z})/\pi} \mod 8
  \label{eq:SI_Z8}
\end{equation}
is generated by a representation-enforced nodal superconductor with different
Chern number at parallel planes. The direct sum of two representation
enforced-nodal superconductors may produce a strong second-order topological
superconductor with chiral Majorana hinge states and generator Hamiltonian
\begin{align}
  H_{(3)}'(\vk) =&\, \rho_3 \nootimes \tau_3 (3 - m - \cos k_x - \cos k_y - \cos k_z)  \nonumber \\
&\, \mbox{} 
  + \rho_0 \nootimes \tau_1 \sin k_x + \rho_0 \nootimes \tau_2 \sin k_y
%  \\ \nonumber &\, \mbox{} 
  + \rho_2 \nootimes \tau_3 \sin k_z
\end{align}
with $0 < m < 2$, and with representations
\begin{equation}
U(\mathcal{P}) = \rho_0 \nootimes \tau_1, \quad U({\cal I}) = \rho_3 \nootimes \tau_3.
\end{equation}
The direct sum of two strong second-order topological superconductors in this
symmetry class, corresponding to $z_3 = 4$, generates a third-order topological
superconductor.~\cite{khalaf2018b,trifunovic2019} This identifies $z_3=2$ and
$z_3=6$ as indicators of strong second-order phases, whereas $z_3 = 4$
indicates a third-order phase, see Fig.~\ref{fig:3d_Au_D}.

For comparison, we note that the classifying group for this case is 
\begin{equation}
  {\cal K}_{\classD}[C_i,A_u] = \ZZ_2^6 \times \ZZ_4 \times \ZZ^3,
  \label{eq:K_3d_Au_D}
\end{equation} where
three factors $\ZZ_2$ correspond to weak phases consisting of stacks of
one-dimensional topological superconductors, the other three factors $\ZZ_2$
correspond to stacks of second-order two-dimensional topological
superconductors, the three factors $\ZZ$ correspond to weak phases consisting
of stacks of two-dimensional Chern superconductors, whereas the factor $\ZZ_4$
consists of (strong) second- and third-order topological superconductors, as
described above. Following the notation of Ref.~\onlinecite{trifunovic2019},
Fig~\ref{fig:3d_Au_D} lists the appropriate boundary-resolved
subgroup sequence ${\cal K}_i'' \subset {\cal K}_i' \subset {\cal K}$ for each
of the factors ${\cal K}_i$ contained in the full classification group ${\cal
K} = \prod_i {\cal K}_i$, where ${\cal K}_i'$ and ${\cal K}_i''$ are the
subgroups of ${\cal K}$ containing topological phases with boundary signature
of order larger than one or two, respectively. Weak phases have a shorter
subgroup sequence, because they do not admit boundary states of order larger
than one or two for stacks of one-dimensional topological phases or
two-dimensional topological phases, respectively.

Symmetry-based indicators corresponding to the class considered in this example
were previously considered in Refs.~\onlinecite{ono2019c,ono2019b,skurativska2019}. The explicit expression for $z_{3}$ in Ref.\ \onlinecite{skurativska2019} in the weak pairing limit differs from our Eqs.\ (\ref{eq:z2_3d_Ci_Au}) and (\ref{eq:SI_Z8}) by the absence of the sign factors in these references. (Reference \onlinecite{ono2019c} contains no explicit expression for the symmetry-based indicators; Reference \onlinecite{ono2019b} defines a $\ZZ_2$ invariant only, for which the sign factors are not important.) This difference affects the assignment of a strong (higher-order) topological index $z_3$ to weak phases, such as those described by the Hamiltonian $H_{(2,x)}(\vk)$ of Eq.\ (\ref{eq:Z4_generator}). The assignment of a strong index to these weak phases is ambiguous, as the presence of gapless surfaces in the weak phases makes it impossible to uniquely associate a boundary signature with a nonzero value of the strong indicator in a weak phase. We here follow the convention of Refs.\ \onlinecite{turner2012} and \onlinecite{trifunovic2017}, according to which weak phases are constructed as stack of layers containing the unit cell center, {\it i.e.} with momentum independent representations. 

\subsection{Inversion symmetry, class DIII}
{\it Representation $\Theta = A_g$. ---} Bogoliubov-de Gennes-type Hamiltonians
in tenfold-way class DIII with inversion symmetry and with a superconducting
order parameter transforming according to the $A_g$ representation have a
trivial classifying group $\mathfrak{K}_\classDIII[C_i, A_g] =
\mathfrak{K}_\classDIII[0]^2 = 0$ at the high-symmetry momenta, so that no
topological band labels can be defined. Also, the boundary classifying group of
topological phases ${\cal K} = 0$ is trivial in this symmetry class in three
dimensions.~\cite{trifunovic2019}

{\it Representation $\Theta = A_u$. ---}
For the $A_u$ representation we have the classifying group of inversion
symmetric momenta and Wyckoff positions $\mathfrak{K}_\classDIII[C_i, A_u] =
\mathfrak{K}_\classAII[0] = \ZZ$.  The results for class DIII can be
constructed from the class D results by taking the direct sum
\begin{equation}
  H_\text{DIII}(\vk) = H(\vk) \oplus_{\sigma} H(-\vk)^*,
  \label{eq:HDIIIdef}
\end{equation}
where $H(\vk)$ is a class-D Hamiltonian satisfying particle-hole antisymmetry
and inversion symmetry with representation $\Theta = A_u$. Time-reversal is
represented as ${\cal T} = \sigma_2 K$; Particle-hole conjugation and inversion
are diagonal with respect to the $\sigma$ degree of freedom, where it is
important that the representation for inversion be real. With the
relation~(\ref{eq:HDIIIdef}) the construction of symmetry-based indicators is
the same as in class D, with the exception of the 2d compabitility
relation~(\ref{eq:2dcomp_D}), which does not apply to class DIII since no Chern
numbers can be defined at two-dimensional planes in the Brillouin zone.

\begin{figure}
\begin{tabular}{ccc}
  $\ZZ_2$ & $\ZZ_2$ & $\ZZ_2$ \\
  $z_{1;y,z} = 1$ & $z_{1;x,z} = 1$ & $z_{1;x,y} = 1$ \\
  \includegraphics[scale=1]{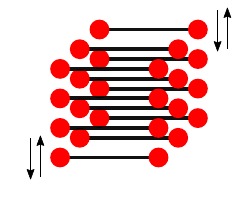} &
  \includegraphics[scale=1]{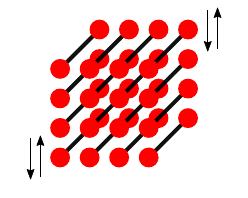} &
  \includegraphics[scale=1]{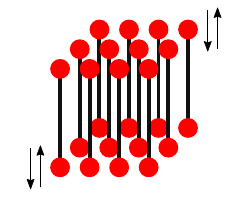} \vspace{0.4cm}\\
  $\ZZ_2 \subseteq \ZZ_4$ & $\ZZ_2 \subseteq \ZZ_4$ & $\ZZ_2 \subseteq \ZZ_4$ \\
  $z_{2;z}=1$ & $z_{2;y}=1$ & $z_{2;x}=1$ \\  
  \includegraphics[scale=1]{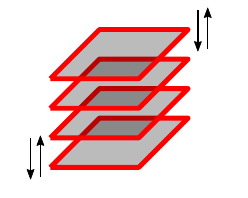} &
  \includegraphics[scale=1]{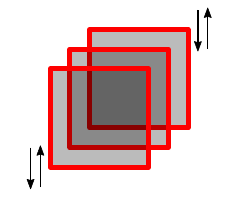} &
  \includegraphics[scale=1]{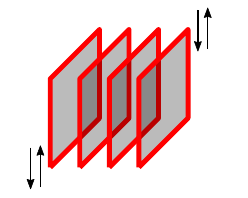} \vspace{-0.2cm}\\
  $z_{2;z}=2$ & $z_{2;y}=2$ & $z_{2;x}=2$ \\
  \includegraphics[scale=1]{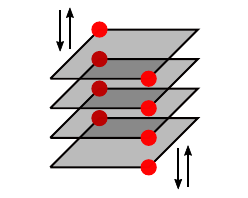} &
  \includegraphics[scale=1]{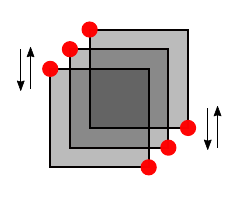} &
  \includegraphics[scale=1]{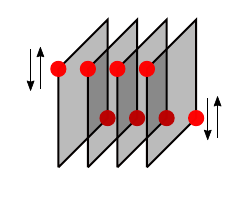} \vspace{0.4cm} \\
  $0 \subseteq 0 \subseteq \ZZ$ & \multicolumn{2}{c}{$\ZZ_2 \subseteq \ZZ_4 \subseteq \ZZ_4$}\\
  $z_3 = 1$ & $z_3 = 2,6$ & $z_3 = 4$ \\
  \includegraphics[scale=1]{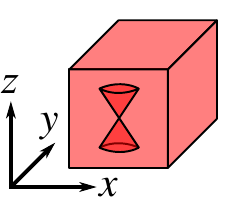} &
  \includegraphics[scale=1]{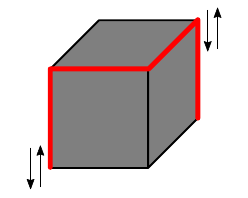} &
  \includegraphics[scale=1]{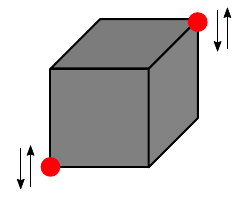} \\
\end{tabular}
\caption{Topological phases of a three-dimensional superconductor in tenfold-way class DIII, with additional $C_i$ symmetry and one-dimensional representation $\Theta=A_u$. For each boundary signature, the subgroup sequence and the symmetry-based indicators of the generators of that phase are given. The two-dimensional second-order topological superconductor with $z_{2;l} = 2$ can be constructed as the direct sum of two two-dimensional first-order topological superconductors with $z_{2;l} = 1$, $l = x,y,z$. Similarly, the three dimensional third-order topological superconductor with $z_3 = 4$ can be constructed as the direct sum of two three dimensional second-order topological superconductors with $z_3 = 2$. \label{fig:3d_Au_DIII}}
\end{figure}

The boundary classifying group for class DIII is 
\begin{equation}
  {\cal K}_{\classDIII}[C_i,A_u] = \ZZ_2^3 \times \ZZ_4^4 \times \ZZ.
  \label{eq:KDIIICiAu}
\end{equation}
With the exception of the one factor $\ZZ$, the interpretation of these factors
and their relation to the topological band labels is the same as for class D.
In particular, the construction~(\ref{eq:HDIIIdef}) takes an
inversion-symmetric Kitaev chain to its time-reversed couterpart and each Chern
superconductor with odd Chern number to a two-dimensional topological
superconductor with helical Majorana edge states. Both the even-Chern number
superconductor and the two-dimensional second-order topological superconductor
in class D map to the two-dimensional second-order topological superconductor
in class DIII. Upon adding weak odd-parity superconducting correlation to an
inversion-symmetric quantum spin Hall insulator or three-dimensional strong
topological insulator in class AII, the edge states gap and create a
second-order topological superconductor with a Kramers pair of Majorana corner
states or helical Majorana hinge states, respectively. Correspondingly, the
inversion-symmetry-protected second-order topological insulator in class AII
turns into an odd-parity third-order topological superconductor. For the
remaining factor $\ZZ$, we observe that the construction~(\ref{eq:HDIIIdef})
maps the representation-induced nodal superconductor in class D, which has
difference in Chern numbers for parallel planes cutting through the Brillouin
zone, to the generator of the three-dimensional (gapped) time-reversal
symmetric topological superconductors, which has indicator $z_3 = 1$.  The
classification results together with the subgroup sequences are illustrated in
Figure~\ref{fig:3d_Au_DIII}.

The generator of the first-order time-reversal symmetric topological
superconductor is
\begin{align*}
H_{(3)}^\text{DIII}(\vk) =&\, \sigma_2 \nootimes \tau_2 (3 - m - \cos k_x - \cos k_y - \cos k_z)  \\ \nonumber
& \mbox{} + \sigma_1 \nootimes \tau_0 \sin k_x + \sigma_2 \nootimes \tau_3  \sin k_y \\ \nonumber
& \mbox{} + \sigma_3 \nootimes \tau_0 \sin k_z,
\end{align*}
with $0 < m < 2$ where we used the representations
$$
U(\mathcal{P}) = \sigma_0 \nootimes \tau_1, \ U({\cal I}) = \sigma_2 \nootimes \tau_2, \ U({\cal T}) = \sigma_2 \nootimes \tau_0.
$$%
Direct summation of this phase generates the free ($\ZZ$) group of first-order topological superconducting phases. Direct summation of the two generators as $H_{(3)}^\text{DIII} \otimes \mu_3$, with inversion
represented as $U({\cal I}) \otimes \mu_3$, constructs the generator of the
factor $\ZZ_4$ in Eq.~(\ref{eq:KDIIICiAu}), a second-order TSC that becomes a third-order topological superconductor upon taking the direct sum with itself.
Consequently, all ``even'' symmetry-based indicators belonging to the factor
$\ZZ_8$ are ambiguous: The elements $z_3=2$ or $6$ and $z_3 = 4$ may correspond
to two or four copies of the strong three dimensional first-order TSC or to the
strong second or third-order topological superconductor, respectively. 

The group of symmetry indicators $\SI = \ZZ_2^3 \times \ZZ_4^3 \times \ZZ_8$
that we obtain agrees with results from Refs.~\onlinecite{shiozaki2019SI,
ono2019c}. Reference \onlinecite{ono2019b} finds a subgroup $\ZZ_2^3 \otimes
\ZZ_4 \subseteq \SI$ consisting of the ``even'' elements only. The explicit
expressions for the symmetry-based indicators differ from those in Refs.\
\onlinecite{ono2019b,ono2019c,skurativska2019} by the presence of the sign
factors, see the discussion at the end of Sec.~\ref{sec:3d_D_Ci}.\footnote{For
symmetry class DIII, the gapless surface states of the weak phases can be
removed by breaking translation symmetry. If inversion symmetry is maintained,
a strong higher-order phase can result, which has helical Majorana modes along
hinges. Such breaking of translation symmetry requires a doubling of the unit
cell. Our convention that Hamiltonians such as Eq.~(\ref{eq:Z4_generator}) have
no strong index requires that the inversion center is shifted by half a unit
cell upon doubling the unit cell, see, {\it e.g.}, the discussion in
Ref.~\onlinecite{trifunovic2017}.} The expression for the symmetry-based
indicator for strong phases $z_3$ in Eq.~\eqref{eq:SI_Z8} agrees with the
corresponding expression in Ref.~\onlinecite{shiozaki2019SI}.  However, in
Ref.~\onlinecite{shiozaki2019SI} the symmetry-based indicators for the weak
phases are defined on planes or lines with $k_l = k_m = 0$, while our
symmetry-based indicators $z_{1;l,m}$ and $z_{2;l}$ are defined on planes or
lines with $k_l = k_m = \pi$. This definition ensures that no spurious weak
indices are assigned to other weak phases or to strong phases.  In the
weak-pairing limit, the criterion $z_3 = 1 \mod 2$ for the three-dimensional
first-order topological superconductor agrees with the well-known condition
that there must be an odd number of Fermi level crossings between high-symmetry
momenta, see Refs.~\onlinecite{sato2010, fu2010}. 

Very recently, the possibility of hybrid higher-order topology was discussed
in the literature.~\cite{bultinck2019} We note that for such ``hybrid'' of
first- and second-order topology, it is sufficient to consider inversion
symmetric superconductor as the current example shows: A ``hybrid'' phase can be
identified as a direct sum of a first-order phase (last row, first column in
Fig.~\ref{fig:3d_Au_DIII}) and one of the weak phases from the third row of
Fig.~\ref{fig:3d_Au_DIII}.

\subsection{Inversion symmetry, class C} \label{sec:3d_C_Ci}

{\it General considerations.}
With spin-rotation symmetry, the Hamiltonian is of the form $H(\vk) = H_C(\vk)
\otimes \sigma_0$ where $H_C(\vk)$ satisfies an effective particle-hole
antisymmetry $\mathcal{P} = \tau_2 K$ squaring to $-1$. This is the tenfold-way
class C.

{\it Representation $\Theta = A_g$. ---}
With the $A_g$ representation for the transformation behavior of the
superconducting order parameter, the classifying group of the Hamiltonian at
each of the high-symmetry momenta is trivial, $\mathfrak{K}_\classC[C_i, A_g]
= \mathfrak{K}_\classC[0]^2 = 0 $, so that no topological band labels can be
defined. The classifying group is 
\begin{equation}
  {\cal K}_{\classC}[C_i,A_g] = \ZZ^3 \times \ZZ_2,
\end{equation}
see Ref.~\onlinecite{trifunovic2019}. It contains a factor $\ZZ^3$
corresponding to weak Chern superconductors in all three stacking directions
and a factor $\ZZ_2$ corresponding to a strong second-order TSC with chiral
hinge states. None of those phases can be detected by symmetry-based
indicators. The boundary signatures of all topological phases in this class are pairs of chiral Majorana modes. In the limit of vanishing superconducting correlations, they can be adiabatically deformed to corresponding normal-state topological insulators with chiral fermionic modes on their boundaries.
Our results for the $\Theta = A_g$ representation are consistent with the results of Ref.~\onlinecite{ono2019c}.

{\it Representation $\Theta = A_u$. ---}
The discussion in class C with the one-dimensional representation $\Theta =
A_u$ of inversion symmetry is analogous to the discussion in class D. Each of
the high-symmetry momenta and each of the Wyckoff positions comes with the
classification group $\mathfrak{K}_\classC[C_i, A_u] = \mathfrak{K}_\classA[0]
= \ZZ$. Compatibility relations of $0d$ type follow from the observation that
the product ${\cal I} {\cal P}$ acts as an effective particle-hole symmetry
that is local in reciprocal space and squares to one. As a result, one may
define a Pfaffian invariant $\mathfrak{p}({\cal IP})$ for $H(\vk)$ at generic
$\vk$. Relating $\mathfrak{p}({\cal IP})$ to the integer invariants
$\mathfrak{N}_+^{\vk_{\rm s}}$ at the eight high-symmetry momenta $\vk_{\rm s}$,
we obtain the seven compatibility relations
\begin{equation}
  \mathfrak{N}_+^{\vk_{\rm s}} = \mathfrak{N}_{+}^{(\pi,\pi,\pi)} \mod 2
  \label{eq:Ch_Ci_3d_C1}
\end{equation}
for $\vk_{\rm s} \neq (\pi,\pi,\pi)$. Defining the integer band labels 
\begin{equation}
  \mathfrak{n}_+^{\vk_{\rm s}}=(\mathfrak{N}_+^{\vk_{\rm s}}- \mathfrak{N}_{+}^{(\pi,\pi,\pi)})/2,
\end{equation}
we arrive at 
\begin{align}
%  \BS_{\classC}[C_i,A_u]&=\ZZ^8,\\
  & \BS = \ZZ^8, \label{eq:BS_3d_C_Au}\\ \nonumber 
 &	B[H(\vk)]= \left\{ \mathfrak{n}_+^{(0,0,0)} ,  \mathfrak{n}_{+}^{(\pi,0,0)} , \mathfrak{n}_{+}^{(0,\pi,0)} , \mathfrak{n}_{+}^{(\pi,\pi,0)} ,\right. \\
	& \hphantom{B[H(\vk)]=}\ \ \ \left.\mathfrak{n}_{+}^{(0,0,\pi)} , \mathfrak{n}_{+}^{(\pi,0,\pi)} , \mathfrak{n}_{+}^{(0,\pi,\pi)} , \mathfrak{N}_{+}^{(\pi,\pi,\pi)}\right\}.\nonumber
\end{align}
The group $\SI$ of symmetry-based indicators is
\begin{equation}
	\SI_{\classC}[C_i,A_u] = \ZZ_2^3 \times \ZZ_4.
  \label{eq:SI_C}
\end{equation}

\begin{figure}
\begin{tabular}{cccc}
  $0 \subseteq \ZZ_2 \subseteq \ZZ_2$ & $0 \subseteq \ZZ$ & $0 \subseteq \ZZ$ & $0 \subseteq \ZZ$ \\
  $z_3 = 2$ & $z_{2;z}=1$ & $z_{2;y}=1$ & $z_{2;x}=1$ \\  
  \includegraphics[scale=0.8]{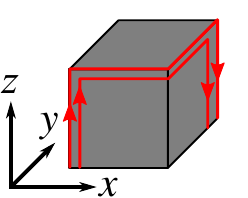} & 
  \includegraphics[scale=0.8]{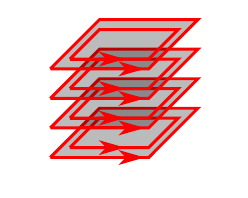} &
  \includegraphics[scale=0.8]{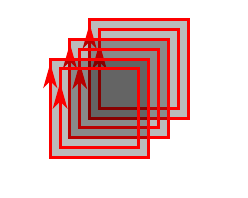} &
  \includegraphics[scale=0.8]{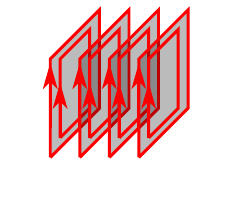} \\
\end{tabular}
\caption{Topological phases of a three-dimensional superconductor in
	tenfold-way class C, with additional $C_i$ symmetry and one-dimensional
	representation $\Theta=A_u$. For each boundary signature, the subgroup
	sequence and the symmetry-based indicators of the generators of that
	phase are given. \label{fig:3d_Au_C}}
\end{figure}

An additional compatibility relation of $2d$ type follows by noting that or a
gapped Hamiltonian a Chern number can be defined for on planes with fixed $k_l$
with $l=x$, $y$, or $z$ and that this Chern number is always even. At
high-symmetry planes with $k_l = 0$ or $\pi$ the Chern number $\mbox{Ch}_{l}$
is related to the topological band labels as 
\begin{align}
  \frac{1}{2} \text{Ch}_l =&\, \sum_{\vkslist |_{k_{{\rm s},l} = 0}} \mathfrak{n}_+^{\vk_{\rm s}} \mod 2 \nonumber \\ =&\,
  \sum_{\vkslist \neq (\pi,\pi,\pi)|_{k_{{\rm s},l} = \pi}} \mathfrak{n}_+^{\vk_{\rm s}} \mod 2.
  \label{eq:Ch_Ci_3d_C2}
\end{align}
One verifies that the generator of the factor $\ZZ_4$ in Eq.~(\ref{eq:SI_C}),
which has symmetry-based indicator
\begin{equation}
  z_3 = \sum_{\vk_{\rm s} \neq (\pi,\pi,\pi)} \mathfrak{n}_+^{\vk_{\rm s}} 
  (-1)^{(k_{{\rm s},x}+k_{{\rm s},y}+k_{{\rm s},z})/\pi} \mod 4,
\end{equation}
is a representation-enforced nodal superconductor, which violates the
compatibility relation~(\ref{eq:Ch_Ci_3d_C2}). The three factors $\ZZ_2$ with
indicators
\begin{equation}
  z_{2;i} = 
  \sum_{\vk_{\rm s} \neq (\pi,\pi,\pi)|k_{{\rm s},i = \pi}} \mathfrak{n}_+^{\vk_{\rm s}} \mod 2,\ \ i=x,y,z,
\end{equation}
correspond to weak Chern superconductor phases, see Fig.~\ref{fig:3d_Au_C}.

For comparison, we note that the boundary classification group
is~\cite{trifunovic2019}
\begin{equation}
  {\cal K}_{\classC}[C_i,A_u] = \ZZ^3 \times \ZZ_2.
  \label{eq:K_3d_Au_C}
\end{equation}
Here the three factors $\ZZ$ correspond to weak Chern superconductors with
generator Hamiltonians
\begin{align*}
   H_{(2,x)}^\text{C}(\vk) =&\, \mu_0 \nootimes \tau_3 (2 - m - \cos k_y - \cos k_z) 
  \\ &\, \mbox{} + \mu_1 \nootimes \tau_1 \sin k_y
  + \mu_1 \nootimes \tau_2 \sin k_z,\ \ \mbox{cycl.}
\end{align*}
with $0 < m < 2$.
%with $(k_i, k_j) = (k_x, k_y), (k_x, k_z), (k_y, k_z)$ for $l = z,y,x$, respectively. 
The corresponding representations are
$$
U(\mathcal{P}) = \mu_3 \tau_2 , \quad U({\cal I}) = \mu_0 \tau_3  .
$$%
The Hamiltonians $H_{(2,l)}^{\text{C}}$ have Chern number $\mbox{Ch}_l = 2$ and
generate the elements ``$1$'' in the three factors $\ZZ_2$ of $\SI$, see
Eq.~(\ref{eq:SI_C}). The remaining factor $\ZZ_2$ of $\SI$ is
generated by a second-order topological superconductor with chiral hinge states
\begin{align}
  H_{(3)}^{\prime \text{C}}(\vk) =&\, \mu_0 \nootimes \rho_3 \nootimes \tau_3 (3 - m - \cos k_x - \cos k_y - \cos k_z) \\ \nonumber
  & \mbox{} + \mu_1 \nootimes \rho_0 \nootimes \tau_1 \sin k_x + \mu_1 \nootimes \rho_0 \nootimes \tau_2 \sin k_y + \mu_1 \nootimes \rho_1 \nootimes \tau_3 \sin k_z
\end{align}
with $0 < m < 2$, where we used the representations
$$
U(\mathcal{P}) =  \mu_3 \rho_0 \tau_2, \quad U({\cal I}) = \mu_0 \rho_3 \tau_3 .
$$%
The above phase has symmetry-based indicator $z_3 = 2$, see
Fig.~\ref{fig:3d_Au_C}. As for the $A_g$ representation, the boundary signatures of all topological phases in this class are pairs of chiral Majorana modes. They can be adiabatically deformed to corresponding normal-state topological insulators with chiral fermionic boundary modes.

Our result (\ref{eq:SI_C}) for the group of symmetry-based indicators is more constrained than the corresponding result from Ref.~\onlinecite{ono2019c}, which finds $\SI = \ZZ_2^3 \times \ZZ_4^3 \times \ZZ_8$. The origin of the difference is that we include the $0d$ compatibility constraint arising from the conservation of the Pfaffian invariant.

\subsection{Inversion symmetry, class CI}
\label{sec:3d_CI_Ci}
{\it Representation $\Theta = A_g$. ---} Here the discussion is the same as for
class C. Both the group $\SI$ of symmetry-based indicators and the classifying
group ${\cal K} = 0$ are trivial.~\cite{trifunovic2019}

{\it Representation $\Theta = A_u$. ---} In the presence of spin-rotation
symmetry, time-reversal symmetry can be represented as $\mathcal{T} = K$. Class
CI has the same topological band labels as class C, and the generators of
atomic limits in classes C and CI have the same band labels, too. The 0d
compatibility relations~(\ref{eq:Ch_Ci_3d_C1}) continue to be valid. Thus, it
follows that the group $\SI$ of symmetry-based indicators is again given by
Eq.~(\ref{eq:SI_C}).

\begin{figure}
\begin{tabular}{c}
  $0 \subseteq \ZZ$ \\
  $z_3 = 2$ \\  
  \includegraphics[scale=1]{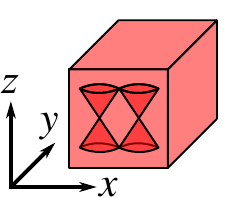} \\
\end{tabular}
\caption{In tenfold-way class CI with additional $C_i$ symmetry and one-dimensional
	representation $\Theta=A_u$, there is a single topological phase with first order boundary signature consisting of pairs of Majorana Dirac cones. The subgroup sequence and the symmetry-based indicator of the generator of this phase is given. \label{fig:3d_Au_CI}}
\end{figure}

The higher-dimensional compatibility relations differ from those in class C. On
the one hand, there is no Chern number for two-dimensional cuts of the
Brillouin zone. On the other hand, $H(\vk)$ is subject to the chiral
antisymmetry ${\cal P T}$ and a local-in-$\vk$ antiunitary symmetry ${\cal I
T}$ for generic $\vk$, which effectively places it in tenfold-way class BDI.
For a one-parameter family of gapped Hamiltonian that are (effectively) in
class BDI for each $\vk$, there exists a $\ZZ_2$ topological
invariant, the first Stiefel-Whitney number \cite{fang2015,
ahn2018}.~\cite{teo2010} Considering this $\ZZ_2$ invariant along
one-dimensional cuts through the Brillouin zone and using its relation to the
band labels at high-symmetry momenta, one finds four compatibility relations of
$1d$ type (see App.~\ref{app:c}),
\begin{equation}
  \mathfrak{n}_{+}^{\vk_{\rm s}} -
  \mathfrak{n}_{+}^{\vk_{\rm s} +  \pi \ve_l} =
  \mathfrak{n}_{+}^{(\pi, \pi, \pi) - \pi \ve_l} \mod 2,
  \label{eq:Ch_Ci_3d_CI}
\end{equation}
for $k_{{\rm s},l} = 0$, $\vk_{{\rm s}} +  \pi \ve_l \neq (\pi,\pi,\pi)$ where $\ve_l$ is the unit vector
in the $l$-direction, $l=x$, $y$, $z$. These 1d compatibility relations impose
the conditions $z_{2;l} = 0$ and $z_3 = 0 \mod 2$ for gapped phases. Phases
violating these conditions, corresponding to the elements ``$1$'' in the three
factors $\ZZ_2$ and the elements ``$1$'' and ``$3$'' in the factor $\ZZ_4$ are
representation-enforced nodal-line superconductors. The group of symmetry-based
indicators for phases satifying all compatibility relations is, hence,
\begin{equation}
  \SI^{(2)}_{\classCI}[C_i,A_u] = \ZZ_2. \label{eq:CI_3d_Ci_SI2}
\end{equation}

For comparison, we note that the boundary classification group is 
\begin{equation}
  {\cal K}_{\classCI}[C_i,A_u] = \ZZ.
\end{equation}
It is generated by a strong first-order phase in three dimensions, which has
the generator
\begin{align}
  H_{(3)}^{\text{CI}}(\vk) =& \mu_0 \nootimes \rho_0 \nootimes \tau_3 (3 - m - \cos k_x - \cos k_y - \cos k_z)\\ \nonumber
  &+ \mu_2 \nootimes \rho_1 \nootimes \tau_1 \sin k_x + \mu_0 \nootimes \rho_2 \nootimes \tau_1 \sin k_y + \mu_2 \nootimes \rho_3 \nootimes \tau_1 \sin k_z 
\end{align}
with $0 < m < 2$. Here we used the representations
$$
U(\mathcal{P}) =  \mu_0 \rho_0 \tau_2, \ U({\cal I}) = \mu_0 \rho_0 \tau_3, \ U({\cal T}) = \mu_0 \rho_0 \tau_0.
$$%
It has band labels $\mathfrak{n}_+^{\bf 0}=2$, $\mathfrak{N}_+^{(\pi,\pi,\pi)}
= 0$, and $\mathfrak{n}_+^{\vk_{\rm s}}= 0$ for $\vk_{\rm s} \neq (0,0,0)$,
$(\pi,\pi,\pi)$. Its band labels generate the element ``$2$'' in the factor
$\ZZ_4$ of $\SI$ and the element ``$1$'' in the factor $\ZZ_2$ of $\SI^{(2)}$,
see Eq.\ (\ref{eq:CI_3d_Ci_SI2}).

Similar to the previous section, our group (\ref{eq:SI_C}) of symmetry-based indicators is more constrained than the corresponding result from Ref.~\onlinecite{ono2019c}, as we include the compatibility relation due to the conservation of the Pfaffian invariant. As discussed above, one finds an even smaller group of symmetry-based indicators if $1d$ compatibility relations based on the Stiefel-Whitney number are included as well.

\section{Conclusion} \label{sec:conclusion}
Symmetry-based indicators have proven to be a pragmatic substitute for a full
classification of topological insulator phases using a complete set of
topological
invariants.~\cite{kruthoff2017,po2017,bradlyn2017,song2018,song2018b} Their main
advantage is that symmetry-based indicators are easier to calculate than other
types of topological invariants, since they require local information in
reciprocal space only. On the other hand, a nonzero value of a symmetry-based
indicators is sufficient to establish that a phase has nontrivial topology and
may, in addition, contain information of the type of anomalous boundary
states.~\cite{khalaf2018,zhang2019} 

In this work, we extend the concept of symmetry-based indicators to
Hamiltonians of Bogoliubov-de Gennes (BdG) type, which appear in the mean-field
theory superconducting phases. Hamiltonians of BdG type are antisymmetric with
respect to particle-hole conjugation. Additionally, for Hamiltonians of BdG
type, a crystalline symmetry class is defined by the presence or absence of
time-reversal and spin-rotation symmetry, by the point group $G$, and by a
one-dimensional representation $\Theta$ of $G$ that describes how the
superconducting order parameter $\Delta$ transforms under the crystalline
symmetry.~\cite{sigrist1991} 

Like the symmetry-based indicators for non-superconducting insulating
phases,~\cite{po2017,bradlyn2017,song2018,song2018b} the symmetry-based
indicators for BdG Hamiltonians can be constructed in a fully algorithmic
manner. Input for our construction are ``band labels'', which are the complete
set of ``zero-dimensional'' topological invariants of the BdG Hamiltonian
$H(\vk_{\rm s})$ evaluated at high-symmetry points $\vk_{\rm s}$ in Brillouin
zone. Such band labels were first considered by Shiozaki, Sato, and Gomi 
in Ref.~\onlinecite{shiozaki2018}.
Unlike previous works on the symmetry-based indicators for topological 
superconductors,~\cite{ono2018,ono2019b,ono2019c,skurativska2019} the
set of zero-dimensional topological invariants considered by us (and by
Ref.\ \onlinecite{shiozaki2018}) is provably
complete --- it is not possible to resolve more boundary signatures through the
prism of zero-dimensional bulk topological invariants. In particular, the set
of topological invariants considered in this work includes not only integer
topological invariants of the type used in the construction of symmetry-based
indicators for non-superconducting Hamiltonians --- counting the number of
occupied bands corresponding to a certain symmetry representation of the
crystalline symmetry ---, but also $\ZZ_2$ topological invariants constructed
using Pfaffians, which do not appear in
Refs.~\onlinecite{ono2018,ono2019b,ono2019c,skurativska2019}. Pfaffians play a
role not only as useful topological band labels, but they also give rise to
additional compatibility relations,\cite{shiozaki2018} 
even in cases in which there are
symmetry-based indicators of integer type only. The latter point is well
illustrated by the example of tenfold-way classes C and CI with inversion
symmetry and $\Theta = A_u$, discussed in Secs.~\ref{sec:3d_C_Ci} and
\ref{sec:3d_CI_Ci}.
 
In the weak-pairing limit (superconducting gap $\Delta$ small in comparison to characteristic energy scales of the normal-state band structure), the band labels of the BdG Hamiltonian can be expressed in terms of conventional integer band labels of the normal-state Hamiltonian, provided the symmetry of the superconducting order parameter is known.\cite{ono2019c,shiozaki2019SI} This applies both to the integer invariant and to the $\ZZ_2$ Pfaffian invariant. This means that in the weak-pairing the symmetry-based indicators constructed here can be calculated using the vast amount of known band structure data in the normal state.

The main difference between zero-dimensional band labels and symmetry-based
indicators~\cite{po2017} is that the latter are designed to ``see'' only the
topological phases with non-trivial boundary signatures. Topologically
non-trivial superconductors without gapless boundary states are deformable to
atomic-limit phases. In this work we combine the complete set of
zero-dimensional band labels~\cite{shiozaki2018} and the definition of
atomic-limit superconductors as an ``array'' of
zero-dimensional superconductors,~\cite{trifunovic2019} to arrive at an 
extension of symmetry-based
indicators~\cite{po2017} to Hamiltonians of Bogoliubov-de Gennes type. 
Our definition of atomic-limit superconductors agrees with the definitions of Refs.\ \onlinecite{ono2019c,shiozaki2019SI}. Since Pfaffian invariants do not appear in Refs.\ \onlinecite{skurativska2019,ono2019b,ono2019c}, the symmetry-based indicators we obtain may be expected to be consistent with these references once all topological band labels and all compatibility relations associated with Pfaffian invariants are omitted from our construction. (For inversion-symmetric superconductors, however, our concrete expressions for the symmetry-based indicators differ from those of Refs.\ \onlinecite{skurativska2019,ono2019b,ono2019c}. This is discussed in detail in Sec.\ \ref{sec:examples3}.) Reference~\onlinecite{shiozaki2019SI} contains results closely related to and consistent with ours.

If the symmetry-based indicators are used as a substitute for a full
classification, ideally, one wants symmetry-based indicators to detect gapped
phases only. The algorithmic construction defined in this work partially meets
this goal: It only guarantees that no indicator corresponds to gapless
topological superconductor phases with its gapless points on high-symmetry
lines in Brillouin zone that connect the high-symmetry points. Nevertheless,
for the examples considered in this work, we were able to explicitly relate
{\it all} gapless phases, including gapless phases with gapless points occurring
on high-symmetry planes and in the bulk of the Brillouin zone, to
symmetry-based indicators by invoking additional compatibility relations that
involve winding numbers, Chern numbers, as well as first and second Stiefel-Whitney numbers. These compatibility relations are
defined on one-dimensional and two-dimensional cuts through the Brillouin zone,
respectively, involve local-in-$\vk$ properties of $H(\vk)$ only,
and generalize the compatibility relations that make use of the
continuity of zero-dimensional invariants in the Brillouin
zone.~\cite{kruthoff2017,po2017,bradlyn2017} Formally, the inclusion of higher
(up to $n$)-dimensional compatibility relations allows one to define a smaller
group $\SI^{(n)} \subset \SI$ of symmetry-based indicators. The relevant
(smallest) groups $\SI^{(d-1)}$ are listed in Table~\ref{tab:comparison} for
the examples considered. For the examples we considered, we find that all
phases indexed by $\SI^{(d-1)}$ are gapped. It would be interesting to find
out, whether this feature holds in general, {\it i.e.}, whether the inclusion
of these two ``higher dimensional'' compatibility relations involving local-in-$\vk$ symmetries of $H(\vk)$ only is sufficient to identify all symmetry-based
indicators that correspond to enforced gapless phases. 

\begin{table}
\begin{tabular*}{\columnwidth}{c @{\extracolsep{\fill}} ccc ccc} \hline\hline
$d$ & $G$ & \!\!Cartan\!\! & $\Theta$ & 
%$\BSz$ & $\BS$ & 
$\SI$ & $\SI^{(d-1)}$ & ${\cal K}$\tabularnewline
\hline 
1 & $C_{1}$ & D & $A$ &
%$\ZZ_{2}^{2}$ & $\ZZ_{2}^{2}$ &
$\ZZ_{2}$ & $\ZZ_{2}$ & $\ZZ_{2}$\tabularnewline
 &  & DIII & $A$ & 
%0 & 0 & 
0 & 0 & $\ZZ_{2}$\tabularnewline
 &  & C & $A$ & 
%0 & 0 & 
0 & 0 & $0$\tabularnewline
 &  & CI & $A$ & 
%0 & 0 & 
0 & 0 & $0$\tabularnewline
\hline 
1 & $C_{i}$ & D & $A_{g}$ & 
%$\ZZ_{2}^{4}$ & $\ZZ_{2}^{3}$ & 
0 & 0 & 0\tabularnewline
 &  &  & $A_{u}$ & 
%$\ZZ^{2}$ & $\ZZ^{2}$ & 
$\ZZ_{2}$ & $\ZZ_{2}$ & $\ZZ_{2}$\tabularnewline
\hline 
2 & $C_{1}$ & D & $A$ & 
%$\ZZ_{2}^{8}$ & $\ZZ_{2}^{8}$ & 
$\ZZ_{2}^{3}$ & $\ZZ_{2}^{3}$ & $\ZZ_{2}^{2}\times\ZZ$\tabularnewline
 &  & DIII & $A$ & 
%0 & 0 & 
0 & 0 & $\ZZ_{2}^{3}$\tabularnewline
 &  & C & $A$ & 
%0 & 0 & 
0 & 0 & $\ZZ$\tabularnewline
 &  & CI & $A$ & 
%0 & 0 & 
0 & 0 & $0$\tabularnewline
\hline 
2 & $C_{i}$ & D & $A_g$ & 
%$\ZZ_{2}^{8}$ & $\ZZ_{2}^{5}$ & 
$\ZZ_{2}^{}$ & $\ZZ_{2}^{}$ & $\ZZ$\tabularnewline
 &  &  & $A_u$ & 
%$\ZZ^{4}$ & $\ZZ^{4}$ & 
$\ZZ_{2}^{2}\times\ZZ_{4}$ & $\ZZ_{2}^{2}\times\ZZ_{4}$ & $\ZZ_{2}^{3}\times\ZZ$\tabularnewline
\hline 
2 & $C_{s}$ & D & $A'$ & 
%$\ZZ^{4}$ & $\ZZ^{2}$ & 
$\ZZ_{2}$ & $\ZZ_{2}$ & $\ZZ_{2}$\tabularnewline
 &  &  & $A''$ & 
%$\ZZ_{2}^{8}$ & $\ZZ_{2}^{6}$ & 
$\ZZ_{2}^{3}$ & $\ZZ_{2}^{3}$ & $\ZZ_{2}^{3}$\tabularnewline
\hline 
2 & $C_{2}$ & D & $A$ & 
%$\ZZ^{4}$ & $\ZZ^{4}$ & 
$\ZZ_{2}^{2}\times\ZZ_{4}$ & $\ZZ_{2}^{2}\times\ZZ_{4}$ & $\ZZ_{2}^{3}\times\ZZ$\tabularnewline
 &  &  & $B$ & 
%$\ZZ_{2}^{8}$ & $\ZZ_{2}^{5}$ & 
$\ZZ_{2}^{}$ & $\ZZ_{2}^{}$ & $\ZZ$\tabularnewline
\hline 
2 & $C_{2v}$ & D & $A_{1}$ &
%0 & 0 & 
0 & 0 & 0\tabularnewline
 &  &  & $A_{2}$ & 
%$\ZZ_{2}^{4}$ & $\ZZ_{2}^{4}$ & 
$\ZZ_{2}^{3}$ & $\ZZ_{2}^{3}$ & $\ZZ_{2}^{4}$\tabularnewline
 &  &  & $B_{1}$ & 
%$\ZZ_{2}^{4}$ & $\ZZ_{2}^{2}$ & 
$\ZZ_{2}$ & $\ZZ_{2}$ & $\ZZ_{2}$\tabularnewline
 &  &  & $B_{2}$ & 
%$\ZZ_{2}^{4}$ & $\ZZ_{2}^{2}$ & 
$\ZZ_{2}$ & $\ZZ_{2}$ & $\ZZ_{2}$\tabularnewline
\hline 
2 & $C_{4}$ & D & $A,B$ & 
%$\ZZ^{5}$ & $\ZZ^{5}$ & 
$\ZZ_{2}\times\ZZ_{8}$ & $\ZZ_{2}\times\ZZ_{8}$ & $\ZZ_{2}^{2}\times\ZZ$\tabularnewline
 &  &  & $^{1,2}E$ & 
%$\ZZ^{2}\times\ZZ_{2}^{6}$ & $\ZZ^{2}\times\ZZ_{2}^{4}$ & 
$\ZZ_{4}$ & $\ZZ_{4}$ & $\ZZ$\tabularnewline
\hline 
3 & $C_{1}$ & D & $A$ & 
%$\ZZ_{2}^{8}$ & $\ZZ_{2}^{8}$ & 
$\ZZ_{2}^{7}$ & $\ZZ_{2}^{6}$ & $\ZZ_{2}^{3}\times\ZZ^{3}$\tabularnewline
 &  & DIII & $A$ & 
%0 & 0 & 
0 & 0 & $\ZZ_{2}^{6}\times\ZZ$\tabularnewline
 &  & C & $A$ & 
%0 & 0 & 
0 & 0 & $\ZZ^{3}$\tabularnewline
 &  & CI & $A$ & 
%0 & 0 & 
0 & 0 & $\ZZ$\tabularnewline
\hline 
3 & $C_{i}$ & D & $A_{g}$ & 
%$\ZZ_{2}^{16}$ & $\ZZ_{2}^{9}$ & 
$\ZZ_{2}^{4}$ & $\ZZ_{2}^{3}$ & $\ZZ^{3}$\tabularnewline
 &  &  & $A_{u}$ & 
%$\ZZ^{8}$ & $\ZZ^{8}$ & 
$\ZZ_{2}^{3}\times\ZZ_{4}^{3}\times\ZZ_{8}$ & $\ZZ_{2}^{3}\times\ZZ_{4}^{4}$ & $\ZZ_{2}^{6}\times\ZZ_{4}\times\ZZ^{3}$\tabularnewline
 &  & DIII & $A_{g}$ & 
%0 & 0 & 
0 & 0 & 0\tabularnewline
 &  &  & $A_{u}$ & 
%$\ZZ^{8}$ & $\ZZ^{8}$ & 
$\ZZ_{2}^{3}\times\ZZ_{4}^{3}\times\ZZ_{8}$ & $\ZZ_{2}^{3}\times\ZZ_{4}^{3}\times\ZZ_{8}$ & $\ZZ_{2}^{3}\times\ZZ_{4}^4\times\ZZ$\tabularnewline
 &  & C & $A_{g}$ & 
%0 & 0 & 
0 & 0 & $\ZZ^{3}\times\ZZ_{2}$\tabularnewline
 &  &  & $A_{u}$ & 
%$\ZZ^{8}$ & $\ZZ^{8}$ & 
$\ZZ_{2}^{3}\times\ZZ_{4}$ & $\ZZ_{2}^{4}$ & $\ZZ^{3}\times\ZZ_{2}$\tabularnewline
 &  & CI & $A_{g}$ & 
%0 & 0 & 
0 & 0 & 0\tabularnewline
 &  &  & $A_{u}$ & 
%$\ZZ^{8}$ & $\ZZ^{8}$ & 
$\ZZ_{2}^{3}\times\ZZ_{4}$ & $\ZZ_{2}$ & $\ZZ$\tabularnewline
\hline 
3 & $C_{s}$ & D & $A'$ & 
%$\ZZ^{8}$ & $\ZZ^{2}$ & 
$\ZZ_{2}$ & $\ZZ_{2}$ & $\ZZ_{2}\times\ZZ$\tabularnewline
 &  &  & $A''$ & 
%$\ZZ_{2}^{16}$ & $\ZZ_{2}^{12}$ & 
$\ZZ_{2}^{9}$ & $\ZZ_{2}^{9}$ & $\ZZ_{2}^{7}\times\ZZ^{2}$\tabularnewline
\hline 
3 & $C_{2}$ & D & $A$ & 
%$\ZZ^{8}$ & $\ZZ^{4}$ & 
$\ZZ_{2}^{2}\times\ZZ_{4}$ & $\ZZ_{2}^{2}\times\ZZ_{4}$ & $\ZZ_{2}^{3}\times\ZZ$\tabularnewline
 &  &  & $B$ & 
%$\ZZ_{2}^{16}$ & $\ZZ_{2}^{10}$ & 
$\ZZ_{2}^{6}$ & $\ZZ_{2}^{5}$ & $\ZZ_{2}^{4}\times\ZZ$\tabularnewline
\hline 
3 & $C_{2v}$ & D & $A_{1}$ & 
%0 & 0 & 
0 & 0 & 0\tabularnewline
 &  &  & $A_{2}$ & 
%$\ZZ_{2}^{8}$ & $\ZZ_{2}^{8}$ & 
$\ZZ_{2}^{7}$ & $\ZZ_{2}^{7}$ & $\ZZ_{2}^{5}\times\ZZ^{4}$\tabularnewline
 &  &  & $B_{1}$ & 
%$\ZZ_{2}^{8}$ & $\ZZ_{2}^{4}$ & 
$\ZZ_{2}^{3}$ & $\ZZ_{2}^{3}$ & $\ZZ_{2}^{3}\times\ZZ^{2}$\tabularnewline
 &  &  & $B_{2}$ & 
%$\ZZ_{2}^{8}$ & $\ZZ_{2}^{4}$ & 
$\ZZ_{2}^{3}$ & $\ZZ_{2}^{3}$ & $\ZZ_{2}^{3}\times\ZZ^{2}$\tabularnewline 
\hline 
3 & $C_{4}$ & D & $A,B$ & 
%$\ZZ^{10}$ & $\ZZ^{5}$ & 
$\ZZ_{2}\times\ZZ_{8}$ & $\ZZ_{2}\times\ZZ_{8}$ & $\ZZ_{2}^{2}\times\ZZ$\tabularnewline
 &  &  & $^{1,2}E$ & 
%$\ZZ_{2}^{12}\times\ZZ^{4}$ & $\ZZ_{2}^{8}\times\ZZ^{2}$ & 
$\ZZ_{2}^{4}\times\ZZ_{4}$ & $\ZZ_{2}^{3}\times\ZZ_{4}$ & $\ZZ_{2}^{3}\times\ZZ$\tabularnewline
  \hline\hline
\end{tabular*}
\caption{\label{tab:comparison} The group of symmetry-based indicators $\SI$ obtained by including compatiblity relations of $0d$ type only, the group $\SI^{(d-1)}$ obtained by including compatibility relations of $n$-dimensional type, with $n < d$, and the full boundary classification group ${\cal K}$ for the combinations of tenfold-way classes, dimension  $d$, and point group $G$ considered in this work.}
\end{table}

To see to what extent symmetry-based indicators offer a faithful representation
of all (crystalline) topological phases we compared the symmetry-based
indicators with the complete classification information for selected examples.
To this end, we used the classifying group ${\cal K}$, which classifies all
topological phases with protected boundary states. (This excludes atomic-limit
phases with nontrivial topology from the topological classification, which is
consistent with the fact that symmetry-based indicators of atomic-limit phases
are {\it defined} to be zero.\cite{po2017,bradlyn2017}) A summary of this
comparison is shown in Table \ref{tab:comparison}. Only for a small number of
the examples we consider --- such as tenfold-way class D in three dimensions
with symmetry groups $C_{s}$ or $C_{2v}$ and representations $\Theta = A'$ and
$\Theta = B_2$, respectively ---, entire classes of topological phases are
missed by the symmetry-based indicators, whereas for some crystalline symmetry
classes the full classifying group ${\cal K}$ and the group $\SI$ of
symmetry-based indicators are identical. In most cases, all generators of
topological phases are detectable by symmetry-based indicators, although there
may be ambiguities preventing a unique identification of the precise nature of
the topological phase. Although these examples clearly show that symmetry-based
indicators are not equivalent to a complete classification, it remains an
interesting observation that zero-dimensional invariants alone perform so well
at this task.

\acknowledgements
We thank Andrei~Bernevig, Titus~Neupert, Seishiro~Ono, Felix~von~Oppen, Ken~Shiozaki and Haruki~Watanabe for
stimulating discussions. Reference \onlinecite{shiozaki2019SI} and \onlinecite{ono2019c} appeared when this manuscript was close to completion. We acknowledge support by project A03 of the CRC-TR 183 (MG and PWB) and from the FNS/SNF Ambizione Grant PZ00P2\_179962 (LT). 

\appendix

\section{More examples in three dimensions}
\label{app:b}
We discuss the classifying group $\mathcal{K}_\eta[G,\Theta]$, band labels, compatibility relations and
symmetry-based indicators of the topological phases with nontrivial boundary
signatures for a selection of point groups not covered in the main text.

\subsection{The trivial point group $C_1$, classes D, DIII, C, CI} 
In the absence of crystalline symmetries (other than translation), for classes
DIII, C and CI there are no topological band labels, as the classification of
the inversion symmetric momenta $\mathfrak{K}_\eta[C_1,A]$ is trivial for
those classes. The classifying groups are ${\cal K}_{\classDIII}[C_1,A] =
\ZZ_2^6 \times \ZZ$, with three factors $\ZZ_2$ for weak phases corresponding
to stacks of one-dimensional time-reversal symmetric topological
superconductors, three factors $\ZZ_2$ for stacks of two-dimensional
time-reversal symmetric topological superconductors, and one factor $\ZZ$ for a
three-dimensional strong first-order superconductor phase , ${\cal
K}_{\classC}[C_1,A] = \ZZ^3$, the three factors $\ZZ$ describing weak phases
corresponding to stacks of two-dimensional Chern superconductors with even Chern number, which can be adiabatically deformed to normal-state Chern insulators, and ${\cal
K}_{\classCI}[C_1,A] = \ZZ$, corresponding to a three-dimensional strong phase. None of these phases can be detected using symmetry-based indicators.
The symmetry-based indicators for tenfold-way class D are nontrivial, as we
discuss below.

\begin{figure}
\begin{tabular}{ccc}
  $\mathbb{Z}_{2}$ & $\mathbb{Z}_{2}$ & $\mathbb{Z}_{2}$ \\
  $z_{1;y,z}=1$ & $z_{1;x,z}=1$ & $z_{1;x,y}=1$ \\
  \includegraphics[scale=1]{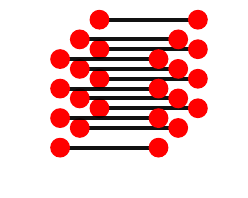} &
  \includegraphics[scale=1]{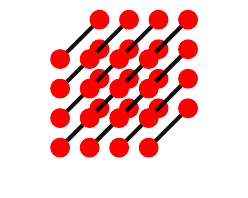} &
  \includegraphics[scale=1]{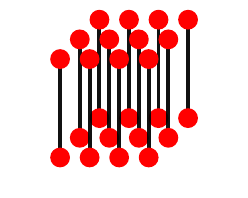} \\
  $0 \subseteq \ZZ$ & $0 \subseteq \ZZ$ & $0 \subseteq \ZZ$ \\ 
  $z_{2;z}=1$ & $z_{2;y}=1$ & $z_{2;x}=1$ \\ 
  \includegraphics[scale=1]{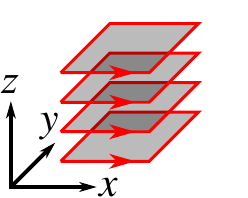} &
  \includegraphics[scale=1]{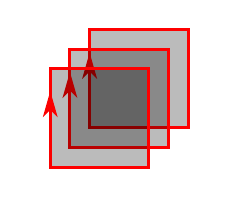} &
  \includegraphics[scale=1]{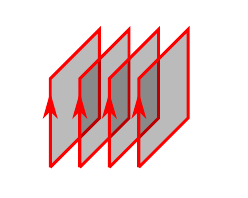} \\
\end{tabular}
\caption[]{
Topological phases of a three-dimensional superconductor in tenfold-way class D and with translation symmetry only (point group $C_1$). For each boundary signature, the boundary subgroup sequence (top row) and the nonzero symmetry-based indicators for a generator of that phase are given (middle row).
\label{tab:ex_3d_C1_D}}
\end{figure}

{\it Classifying group.} The boundary classifying group for tenfold-way class D is
$${\cal K}_{\classD}[C_1,A] = \mathbb{Z}_{2}^{3}\times \mathbb{Z}^{3}.$$ The
factor $\mathbb{Z}_{2}^{3}$ corresponds to Kitaev chains stacked in the
$y$ and $z$, $x$ and $z$ or $x$ and $y$ directions (labels $(1;y,z)$,
$(1;x,z)$, and $(1;x,y)$) and the factor $\mathbb{Z}^{3}$ corresponds to Chern
superconductors stacked in the $z$, $y$, or $x$ direction (labels $(2;z)$,
$(2;y)$, and $(2;x)$). The even-Chern-number superconductors can be deformed to Chern insulators with vanishing superconducting order parameter. The topological phases generate a $\mathbb{Z}_{2}^{6}$
subgroup of $\SI_\classD[C_1,A] =\mathbb{Z}_{2}^{7}$. The boundary signatures
of the topological phases together with their symmetry-based indicators are
shown in Fig.~\ref{tab:ex_3d_C1_D}.

{\it Band labels.} There are $\BSz \simeq \ZZ_2^8$ topological band labels
given by the Pfaffian invariant ${\frak p}^{\vk_{\rm s}}$ at all high-symmetry
momenta $\vk_{\rm s}$. 

{\it Compatibility relations.} 
There are no $0d$ compatibility relations. Hence $\BSz \simeq \BS$ and 
\[
  B[H(\vk)] = \{{\frak p}^{\vk_{\rm s}}\} .
\]
There is a $2d$ compability relation required by the conservation of the Chern number between parallel planes,
\[
%\text{Ch}_{i}(0)=
  \sum_{\vkslist| k_{{\rm s},i} = 0} \mathfrak{p}^{\vec{k}_{\rm s}} =
  \sum_{\vkslist| k_{{\rm s},i} = \pi} \mathfrak{p}^{\vec{k}_{\rm s}} \mod 2,\ \ i=x,y,z.
%=\text{Ch}_{i}(\pi)\mod2
\]
%with $\text{Ch}_{i}(k_{i}')$ the Chern number in a plane with $k_{i}=k_{i}'$
%and $X_{i}[k_{i}']$ the set of all inversion symmetric momenta with
%$k_{i}=k_{i}'$. 
The violation of this compatibility relation signals a gapless phase with nodal points. 

{\it Symmetry-based indicators.} The group of symmetry-based indicators 
$$\SI_\classD[C_1,A] = \BS / \AI \simeq \ZZ_2^7$$ 
contains a factor $\ZZ_2^6$ corresponding to stacks of one and two dimensional topological superconductors,
\begin{align*}
  z_{1;i,j} =&\, \sum_{\vk_{\rm s}|k_{{\rm s},i} = k_{{\rm s},j} = \pi}
  \mathfrak{p}^{\vk_{\rm s}} \mod 2, \nonumber \\
%  z_{1;y,z} =&\, \mathfrak{p}^{(0,0,0)} + \mathfrak{p}^{(\pi,0,0)} \mod 2, \nonumber \\
%  z_{1;x,z} =&\, \mathfrak{p}^{(0,0,0)} + \mathfrak{p}^{(0,\pi,0)} \mod 2, \nonumber \\
%  z_{1;x,y} =&\, \mathfrak{p}^{(0,0,0)} + \mathfrak{p}^{(0,0,\pi)} \mod 2, \nonumber \\
  z_{2;j} =&\, \sum_{\vk_{\rm s}|k_{{\rm s},j}= \pi} \mathfrak{p}^{\vk_{\rm s}} \mod 2,
%
%  z_{2;x} =&\, \sum_{\vk_{\rm s}|k_{{\rm s},x}= 0} \mathfrak{p}^{(\vk_{\rm s})} \mod 2\nonumber \\
%  z_{2;y} =&\, \sum_{\vk_{\rm s}|k_{{\rm s},y}= 0} \mathfrak{p}^{(\vk_{\rm s})} \mod 2 \nonumber \\
%  z_{2;z} =&\, \sum_{\vk_{\rm s}|k_{{\rm s},z}= 0} \mathfrak{p}^{(\vk_{\rm s})} \mod 2 \nonumber \\
%  z_{2;z} =&\,
% \mathfrak{p}^{(0,0,0)} + \mathfrak{p}^{(\pi,0,0)} + \mathfrak{p}^{(0,\pi,0)} + \mathfrak{p}^{(\pi,\pi,0)} \mod 2, \nonumber \\
%  z_{2;y} =&\, \mathfrak{p}^{(0,0,0)} + \mathfrak{p}^{(\pi,0,0)} + \mathfrak{p}^{(0,0,\pi)} + \mathfrak{p}^{(\pi,0,\pi)} \mod 2, \nonumber \\
%  z_{2;x} =&\, \mathfrak{p}^{(0,0,0)} + \mathfrak{p}^{(0,0,\pi)} + \mathfrak{p}^{(0,\pi,0)} + \mathfrak{p}^{(0,\pi,\pi)} \mod 2, \nonumber \\
%  z_{3} =&\, \sum_{\vk_{\rm s}} \mathfrak{p}^{(\vk_{\rm s})} \mod 2.
\end{align*}
with $i \neq j = x$, $y$, $z$.
The remaining factor $\ZZ_2$ corresponds to a nodal superconductor detected by the violation of the 2d compatibility relation with symmetry-based indicator
\[
	z_{3} = \, \sum_{\vk_{\rm s}} \mathfrak{p}^{\vk_{\rm s}} \mod 2.
\]

%In \onlinecite{ono2019c}, no band labels or symmetry-based indicators are defined for classes D, DIII, C, CI.

\subsection{Point group $C_s$, class D}

\subsubsection{Representation $\Theta = A'$}

\begin{figure}
\begin{tabular}{cc}
$\mathbb{Z}_{2}$ & $0 \subseteq \ZZ$ \\
$z_{1;y,z}=1$ & $-$ \\
  \includegraphics[scale=1]{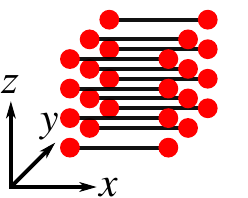} &
  \includegraphics[scale=1]{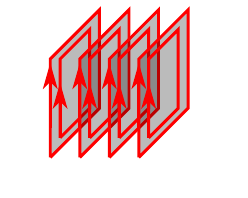} 
\end{tabular}
\caption[]{
Topological phases of a three-dimensional superconductor in tenfold-way class D with point group $C_s$ and representation $\Theta = A'$. 
\label{tab:ex_3d_Cs_Ap_D}}
\end{figure}

{\it Classifying group:} 
$${\cal K}_\classD[C_s,A']=\mathbb{Z}_{2}\times\mathbb{Z}.$$%
The factor $\mathbb{Z}_{2}$ corresponds to a stack of mirror symmetric Kitaev chains in the $y$ and $z$ directions (label $(1;y,z)$) and the factor $\mathbb{Z}$ corresponds to a stack of two-dimensional Chern superconductors with even Chern number within the mirror plane (label $(2;x)$). The latter phase can not be detected by symmetry-based indicators. It can be adiabatically deformed to a Chern insulator that is robust to the introduction of odd-parity superconducting correlations. The boundary signatures of the topological phases together with their symmetry-based indicators are displayed in Fig.~\ref{tab:ex_3d_Cs_Ap_D}. 

{\it Band labels.} There are $\BSz = \ZZ^8$ topological band labels given by the invariants ${\frak N}_+^{\vk_{\rm s}}$ at all high-symmetry momenta $\vk_{\rm s}$. 

{\it Compatibility relations.} Within a mirror plane $k_{x}=0,\pi$ the invariant $\mathfrak{N}_{+}^{\vec{k}}$ is preserved for a gapped phase, which leads to the $0d$ compatibility relations
\begin{minipage}{\linewidth}
\begin{align*}
  \mathfrak{N}_{+}^{(0,0,0)}=&\, \mathfrak{N}_{+}^{(0,\pi,0)}=\mathfrak{N}_{+}^{(0,0,\pi)}=\mathfrak{N}_{+}^{(0,\pi,\pi)}, \\ \mathfrak{N}_{+}^{(\pi,0,0)}=&\, \mathfrak{N}_{+}^{(\pi,\pi,0)}=\mathfrak{N}_{+}^{(\pi,0,\pi)}=\mathfrak{N}_{+}^{(\pi,\pi,\pi)},
\end{align*}
\end{minipage}
which identifies a factor $\mathbb{Z}^{6}$ of the group of topological band labels $\BSz$ as representation enforced gapless nodal-line superconductors such that we have $\BS \simeq \ZZ^2$ and
\[
B[H(\vk)] = \{ {\frak N}_+^{(0,\pi,\pi)}, {\frak N}_+^{(\pi,\pi,\pi)}\}.
\]

{\it Symmetry-based indicators.} The group of symmetry-based indicators 
\begin{equation}
  \SI_\classD[C_s,A'] \simeq \ZZ_2
  \label{eq:SIDCsA}
\end{equation}
contains a single factor corresponding to the stack of Kitaev chains with indicator
\[
	z_{1;y,z} = \mathfrak{N}_{+}^{(\pi,\pi,\pi)} + \mathfrak{N}_{+}^{(0,\pi,\pi)} \mod 2 .
\] 
Our result (\ref{eq:SIDCsA}) for the group of symmetry-based indicators agrees with the corresponding result from \onlinecite{ono2019c}.

\subsubsection{Representation $\Theta = A''$}

\begin{figure}
\begin{tabular}{ccc}
$\mathbb{Z}_{2}$ & $\mathbb{Z}_{2}$ & $0 \subseteq \mathbb{Z}_{2} \subseteq \mathbb{Z}_{2}$\\
$z_{1,\pm;x,z}=1$ & $z_{1,\pm;x,y}=1$ & $z_3=1$\\
  \includegraphics[scale=1]{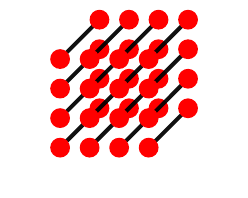} &
  \includegraphics[scale=1]{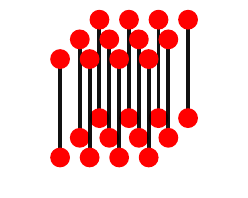} & 
  \includegraphics[scale=1]{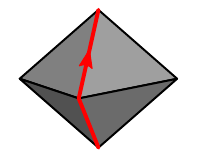} \\
$0 \subseteq \ZZ$ & $\mathbb{Z}_{2} \subseteq \mathbb{Z}_{2}$ & $\mathbb{Z}_{2} \subseteq \mathbb{Z}_{2}$ \\
$z_{2,\pm;x}=1$ & $z_{2;y}=1$ & $z_{2;z}=1$ \\
  \includegraphics[scale=1]{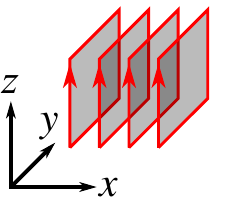} &
  \includegraphics[scale=1]{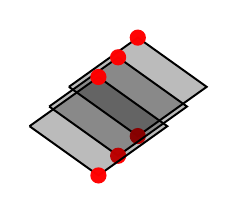} &
  \includegraphics[scale=1]{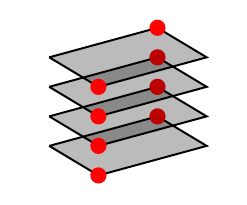}
\end{tabular}
\caption[]{
Topological phases of a three-dimensional superconductor in tenfold-way class D with point group $C_s$ and representation $\Theta = A''$.  
\label{tab:ex_3d_Cs_App_D}}
\end{figure}

{\it Classifying group:} 
$$\mathcal{K}_\classD[C_s,A'']=\mathbb{Z}_{2}^{7}\times \mathbb{Z}^{2}.$$%
A factor $\mathbb{Z}_{2}^{4}$ corresponds to one-dimensional superconductors in the $y$ or $z$ directions, stacked in the $x$ and $z$ or $x$ and $y$ directions, respectively, and with ${\cal M}_x$ parities $\pm$ (labels $(1,\pm;x,z)$ and $(1,\pm;x,y)$). The factor $\ZZ^2$ corresponds to two-dimensional Chern superconductors in the $yz$ plane with even or odd ${\cal M}_x$ parity, stacked in the $x$ direction (labels $(2,\pm;x)$). Systems with an even Chern number in either mirror plane can be adiabatically deformed to Chern insulator with vanishing superconducting order parameter. A factor $\mathbb{Z}_{2}^{2}$ corresponds to two-dimensional second-order topological superconductors stacked in the $z$ or $y$ direction (labels $(2;z)$ and $(2;y)$). The remaining factor $\mathbb{Z}_{2}$ corresponds to a three dimensional second-order topological superconductor (label $(3)$). The boundary signatures of the topological phases together with their symmetry-based indicators are displayed in Fig.~\ref{tab:ex_3d_Cs_App_D}.

{\it Band labels.} There are $\BSz \simeq \ZZ_2^{16}$ topological band labels given by the Pfaffian invariants ${\frak p}_\pm^{\vk_{\rm s}}$ in both even and odd mirror parity subspaces at all high-symmetry momenta $\vk_{\rm s}$. 

{\it Compatibility relations.} From the fact that $\mathcal{PM}$ is a local antisymmetry along lines in reciprocal space one deduces the compatibility relations
% a gapped system preserves the quantity $(-1)^{\mathfrak{p}^{\mathcal{MP}}}=\text{Pf}(U(\mathcal{MP})H(\vec{k})$ along the high-symmetry lines $(k_{x},k_y',k_z')$ 
\begin{align*}
\sum_{s=\pm}\mathfrak{p}_{s}^{(0,k_y',k_z')} =\sum_{s=\pm}\mathfrak{p}_{s}^{(\pi,k_y',k_z')},\ \
  \mbox{for $k_y'$, $k_z' = 0$, $\pi$}.
\end{align*}
The compatibility relations identify a factor $\mathbb{Z}_{2}^{4}$ of $\BSz$ as representation enforced gapless superconductors with nodal points on high-symmetry lines in the Brillouin zone such that we have $\BS \simeq \ZZ_2^{12}$ and
\begin{align*}
B[H(\vk)] = \{ & {\frak p}_-^{(0,0,0)}, {\frak p}_-^{(0,0,\pi)}, {\frak p}_-^{(0,\pi,0)}, {\frak p}_-^{(0,\pi,\pi)} \\
& {\frak p}_+^{(\pi,0,0)}, {\frak p}_+^{(\pi,0,\pi)}, {\frak p}_+^{(\pi,\pi,0)}, {\frak p}_+^{(\pi,\pi,\pi)} \\
& {\frak p}_-^{(\pi,0,0)}, {\frak p}_-^{(\pi,0,\pi)}, {\frak p}_-^{(\pi,\pi,0)}, {\frak p}_-^{(\pi,\pi,\pi)}\}.
\end{align*}

{\it Symmetry-based indicators.} The group of symmetry-based indicators is 
\begin{equation}
  \SI_\classD[C_s,A''] \simeq \ZZ_2^9,
  \label{eq:SI_DCsAA}
\end{equation}
where each factor corresponds to a different topological phase. For the stacks of Kitaev chains we have
\begin{align*}
z_{1,\pm;x,z} & =\mathfrak{p}_{\pm}^{(\pi,\pi,\pi)}+\mathfrak{p}_{\pm}^{(\pi,0,\pi)} \mod 2 , \\
z_{1,\pm;x,y} & =\mathfrak{p}_{\pm}^{(\pi,\pi,\pi)}+\mathfrak{p}_{\pm}^{(\pi,\pi,0)} \mod 2 ,
\end{align*}
for the Chern superconductors within the mirror plane with even or odd mirror parity 
\begin{align*}
z_{2,\pm;x} & =\sum_{\vec{k}_{s}|_{k_{{\rm s}, x}=\pi}}\mathfrak{p}_{\pm}^{\vec{k}_{\rm s}} \mod2,
\end{align*}
for two-dimensional second-order topological superconductors stacked in the $l = z$ or $y$ direction,
\begin{align*}
z_{2;l}=\sum_{\vec{k}_{s}|_{k_{{\rm s}, l}=\pi}}\mathfrak{p}_{+}^{\vec{k}_{\rm s}}=\sum_{\vec{k}_{\rm s}|_{k_{{\rm s}, l}=\pi}}\mathfrak{p}_{-}^{\vec{k}_{\rm s}} \mod 2,
\end{align*}
where the equality follows from the compatibility constraint, and for the three dimensional second-order topological superconductor we have 
\begin{align*}
z_{3}=\sum_{\vec{k}_{\rm s}}\mathfrak{p}_{+}^{\vec{k}_{\rm s}}=\sum_{\vec{k}_{\rm s}}\mathfrak{p}_{-}^{\vec{k}_{\rm s}} \mod 2 .
\end{align*}
The group (\ref{eq:SI_DCsAA}) of symmetry-based indicators differs from the result obtained in Ref.\ \onlinecite{ono2019c}, where no symmetry-based indicators are found due to the absence of Pfaffian band labels.

\subsection{Point group $C_2$, class D}

\subsubsection{Representation $\Theta = A$}

\begin{figure}
\begin{tabular}{cccc}
  $\mathbb{Z}_{2}$ & $\mathbb{Z}_{2}$ & $0\subseteq\mathbb{Z}$ & $\mathbb{Z}_{2}\subseteq\mathbb{Z}_{2}$ \\
  $z_{1;y,z}=1$ & $z_{1;x,z}=1$ & $z_{2,z}=1$ & $z_{2;z}=2$ \\
  \includegraphics[scale=0.8]{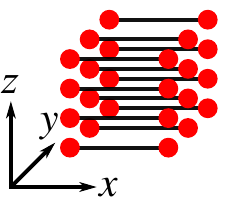} &
  \includegraphics[scale=0.8]{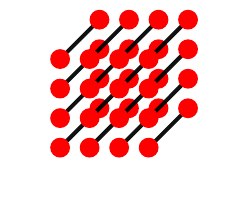} &
  \includegraphics[scale=0.8]{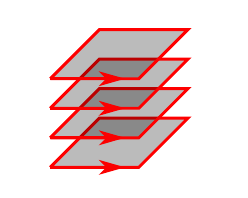} &
  \includegraphics[scale=0.8]{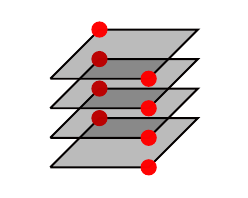} \\
\end{tabular}
\caption[]{Topological phases of a three-dimensional superconductor in tenfold-way class D with point group $C_2$ and representation $\Theta = A$.  
\label{tab:ex_3d_C2_A_D}}
\end{figure}

{\it Classifying group:} 
$${\cal K}_\classD[C_2,A] = \mathbb{Z}_{2}^{3}\times\mathbb{Z}.$$%
A factor $\mathbb{Z}_{2}^{2}$ corresponds to Kitaev chains perpendicular to the rotation axis ($z$) and stacked in the in the $x$ and $z$ or $y$ and $z$ directions (labels $(1;y,z)$ and $(1;x,z)$). The remaining factors $\mathbb{Z}_{2}$ and $\mathbb{Z}$ correspond to two-dimensional topological superconductors stacked perpendicular to the rotation axis (labels $(2,z)$' and $(2,z)$, respectively), where the factor $\mathbb{Z}_{2}$ describes a two-dimensional second-order phase and the factor $\mathbb{Z}$ a Chern superconductor. Even-Chern number superconductors can be adiabatically deformed to normal-state Chern insulators. The weak second-order phase and the stack of Chern superconductors with Chern number two have identical band labels. The boundary signatures of the topological phases together with their symmetry-based indicators are displayed in Fig.~\ref{tab:ex_3d_C2_A_D}.

{\it Band labels.} There are $\BSz \simeq \ZZ^8$ topological band labels given by the invariants ${\frak N}_+^{\vk_{\rm s}}$ at all high-symmetry momenta $\vk_{\rm s}$. 

{\it Compatibility relations.} For gapped phases, the invariant $\mathfrak{N}_{+}(\vec{k})$ is preserved along high-symmetry lines in reciprocal space parallel to the rotation axis, which gives the compatibility relation
\[
\mathfrak{N}_{+}^{(k_{x}',k_{y}',0)}=\mathfrak{N}_{+}^{(k_{x}',k_{y}',\pi)}, \ \
  \mbox{for $k_{x}'$, $k_{y}' = 0$, $\pi$}.
\]
This compatibility relation identifies a factor $\mathbb{Z}^{4}$ of $\BSz$ as representation-enforced gapless superconductors with nodal points such that we have $\BS \simeq \ZZ^4$ and
\[
B[H(\vk)] = \{ {\frak N}_+^{(0,0,\pi)}, {\frak N}_+^{(\pi,0,\pi)}, {\frak N}_+^{(0,\pi,\pi)}, {\frak N}_+^{(\pi,\pi,\pi)}\}.
\]

{\it Symmetry-based indicators.} The group of symmetry-based indicators is 
\begin{equation}
  \SI_\classD[C_2,A] \simeq \ZZ_2^2 \times \ZZ_4.
  \label{eq:SI_DC2A}
\end{equation}
The factor $\ZZ_2^2$ correspond to stacks of Kitaev chains with indicators
\begin{align*}
z_{1;y,z} & =\mathfrak{N}_{+}^{(0,\pi,\pi)} + \mathfrak{N}_{+}^{(\pi,\pi,\pi)} \mod 2, \\
z_{1;x,z} & =\mathfrak{N}_{+}^{(\pi,0,\pi)} + \mathfrak{N}_{+}^{(\pi,\pi,\pi)} \mod 2
\end{align*}
and the factor $\ZZ_4$ corresponds to stacks of two dimensional topological superconductors in the $z$ direction with indicator
\[
z_{2;z}=\sum_{\vec{k}_{s}|_{k_{{\rm s}, z}=\pi}}\mathfrak{N}_{+}^{\vec{k}_{s}} 
  (-1)^{(k_{{\rm s},x} + k_{{\rm s},y})/\pi} \mod 4 . 
\]
The value $z_{2;z} = 2$ is ambiguous, since it may either correspond to a stack
of first-order two-dimensional superconductors with Chern number two or to a
stack of second-order two-dimensional superconductors, see
Fig.~\ref{tab:ex_3d_C2_A_D}.
Our result (\ref{eq:SI_DC2A}) for the group of symmetry-based indicators agrees with the corresponding result from \onlinecite{ono2019c}.

\subsubsection{Representation $\Theta = B$}

\begin{figure}
\begin{tabular}{cccc}
  $\mathbb{Z}_{2}$ & $\mathbb{Z}_{2}\subseteq\mathbb{Z}_{2}$ & $\mathbb{Z}_{2}\subseteq\mathbb{Z}_{2}$ & $0\subseteq\mathbb{Z}$ \\
  $z_{1,\pm;x,y}=1$ & $z_{2;x}=1$ & $z_{2,y}=1$ & $z_{2;z}=1$ \\
  \includegraphics[scale=0.8]{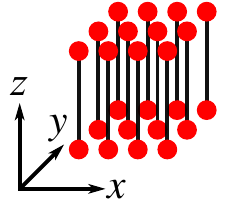} &
  \includegraphics[scale=0.8]{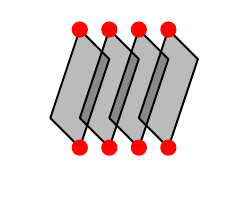} &
  \includegraphics[scale=0.8]{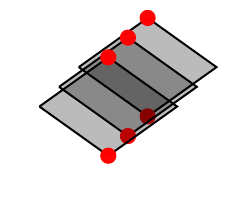} & 
  \includegraphics[scale=0.8]{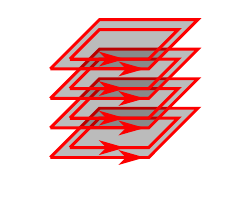} \\
\end{tabular}
\caption[]{Topological phases of a three-dimensional superconductor in tenfold-way class D with point group $C_2$ and representation $\Theta = B$.
\label{tab:ex_3d_C2_B_D}}
\end{figure}

{\it Classifying group:} 
$$\mathcal{K}_\classD[C_2,B] = \mathbb{Z}_{2}^{4}\times\mathbb{Z}.$$%
A factor $\mathbb{Z}_{2}^{2}$ corresponds to one-dimensional superconductors parallel to the rotation ($z$) axis, stacked in the $x$ and $y$ directions, and with rotation parities $\pm$ (labels $(1,\pm;x,y)$). A factor $\mathbb{Z}_{2}^{2}$ corresponds to two-dimensional second-order topological superconductors stacked in the $x$ or $y$ directions (labels $(2;x)$' and $(2;y)$'). The remaining factor $\mathbb{Z}$ describes a two-dimensional even-Chern-number superconductor stacked in the $z$ direction with label $(2;z)$. They can be adiabatically deformed to normal-state Chern insulators. The boundary signatures of the topological phases together with their symmetry-based indicators are displayed in Fig.~\ref{tab:ex_3d_C2_B_D}.

{\it Band labels.} There are $\BSz \simeq \ZZ_2^{16}$ topological band labels given by the Pfaffian invariants ${\frak p}_\pm^{\vk_{\rm s}}$ in both even and odd rotation parity subspaces at all high-symmetry momenta $\vk_{\rm s}$. 

{\it Compatibility relations.} From the fact that $\mathcal{R}_{\pi}\mathcal{P}$ is a local symmetry in reciprocal space in high-symmetry planes, we derive the compatibility relation
\begin{align*}
\sum_{s=\pm}\mathfrak{p}_{s}^{(0,0,k_{z})} & = \sum_{s=\pm}\mathfrak{p}_{s}^{(\pi,0,k_{z})} \\ 
& = \sum_{s=\pm}\mathfrak{p}_{s}^{(0,\pi,k_{z})} \\
& = \sum_{s=\pm}\mathfrak{p}_{s}^{(\pi,\pi,k_{z})}\ \ \mbox{for $k_z = 0$, $\pi$}.
\end{align*}
The compatibility relations identify a factor $\mathbb{Z}_{2}^{6}$ of $\BSz$ as representation-enforced gapless nodal-line superconductors such that $\BS \simeq \ZZ_2^{10}$ and 
\begin{align*}
B[H(\vk)] = \{ & {\frak p}_-^{(0,0,0)}, {\frak p}_-^{(\pi,0,0)}, {\frak p}_-^{(0,\pi,0)}, {\frak p}_+^{(\pi,\pi,0)}, {\frak p}_-^{(\pi,\pi,0)} \\
& {\frak p}_-^{(0,0,\pi)}, {\frak p}_-^{(\pi,0,\pi)}, {\frak p}_-^{(0,\pi,\pi)}, {\frak p}_+^{(\pi,\pi,\pi)}, {\frak p}_-^{(\pi,\pi,\pi)}\}.
\end{align*}
A second compatibility relation for gapped phases, identical to Eq.\ (\ref{eq:Ch_Ci_3d_D}) in the main text, follows by considering Chern numbers at two-dimensional cuts through the Brillouin zone. 
%The remaining factor $\mathbb{Z}_{2}$ of $\SI$ accounts for phases that violate this compatibility relation. These are representation-enforced superconductor with nodal points at generic positions in the Brillouin zone.

{\it Symmetry-based indicators.} The group of symmetry-based indicators is 
\begin{equation}
  \SI_\classD[C_2,B] \simeq \ZZ_2^6,
  \label{eq:SI_DC2B}
\end{equation}
out of which a factor $\ZZ_2^5$ corresponds to gapped topological phases. For phases corresponding to stacks of Kitaev chains we have
\[
z_{1,\pm;x,y}=\mathfrak{p}_{\pm}^{(\pi,\pi,\pi)}+\mathfrak{p}_{\pm}^{(\pi,\pi,0)}\mod2
\]
and for phases corresponding to stacks of two dimensional superconductors in the $l=x,y,z$ direction we have
\[
z_{(2;l)}=\sum_{\vec{k}_{\rm s}|_{k_{{\rm s}, l}=\pi}}\mathfrak{p}_{+}^{\vec{k}_{\rm s}}=\sum_{\vec{k}_{\rm s}|_{k_{{\rm s}, l}=\pi}}\mathfrak{p}_{-}^{\vec{k}_{\rm s}}\mod 2,
\]
where the equality follows from the 0d compatibility relation. The remaining factor $\ZZ_2$ corresponds to a symmetry-based indicator for a nodal superconductor with different Chern number parity detected by the band labels,
\[
z_{3}=\sum_{\vec{k}_{\rm s}}\mathfrak{p}_{+}^{\vec{k}_{\rm s}}=\sum_{\vec{k}_{\rm s}}\mathfrak{p}_{-}^{\vec{k}_{\rm s}} \mod 2 .
\]
The group (\ref{eq:SI_DC2B}) of symmetry-based indicators differs from the result obtained in Ref.\ \onlinecite{ono2019c}, where no symmetry-based indicators are found due to the absence of Pfaffian band labels.

\subsection{Point group $C_4$, class D}

\subsubsection{Representation $\Theta = A$ or $\Theta = B$}

\begin{figure}
\begin{tabular}{ccc}
  $\mathbb{Z}_{2}$ & $0\subseteq\mathbb{Z}$ & $\mathbb{Z}_{2}\subseteq\mathbb{Z}_{2}$  \\
  $z_{1;x,y,z}=1$ & $z_{2;z}=1$ & $z_{2,z}=4$ \\
  \includegraphics[scale=0.8]{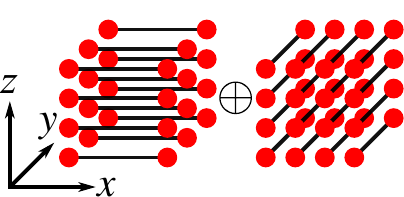} &
  \includegraphics[scale=0.8]{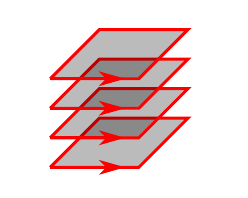} &
  \includegraphics[scale=0.8]{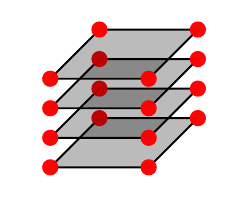}   
\end{tabular}
\caption[]{Topological phases of a three-dimensional superconductor in tenfold-way class D with point group $C_4$ and representation $\Theta = A$ or $\Theta = B$.
\label{tab:ex_3d_C4_A_D}}
\end{figure}

{\it Classifying group:} 
$${\cal K}_\classD[C_4,A]={\cal K}_\classD[C_4,B]=\mathbb{Z}_{2}^{2}\times \mathbb{Z}.$$%
A factor $\mathbb{Z}_{2}$ corresponds to Kitaev chains aligned perpendicular to the rotation axis and to each other and stacked in all three spatial directions (label $(1;x,y,z)$). The remaining factor $\mathbb{Z}_{2}\times \mathbb{Z}$ corresponds to topological superconductors stacked perpendicular to the rotation axis (label $(2,z)$), where the factor $\mathbb{Z}_{2}$ describes a second-order phase and the factor $\mathbb{Z}$ a Chern superconductor. Even-Chern number superconductors can be adiabatically deformed to Chern insulators with vanishing superconducting correlations. The boundary signatures of the topological phases together with their symmetry-based indicators are displayed in Fig.~\ref{tab:ex_3d_C4_A_D}.

{\it Band labels.} There are $\BSz \simeq \ZZ^{10}$ topological band labels given by the invariants ${\frak N}_j^{\vk_{\rm s}}$ with $j=\frac{1}{2}, \frac{5}{2}$ at high-symmetry momenta $\vk_{\rm s}$ with fourfold rotation symmetry and $j = \frac{1}{2}$ at high-symmetry momenta $\vk_{\rm s}$ with twofold rotation symmetry.

{\it Compatibility relations.} For high-symmetry lines parallel to the rotation axis the invariant $\mathfrak{N}_{j}^{\vec{k}}$ is preserved for gapped phases, which gives the compatibility relation
\[
\mathfrak{N}_{j}^{(k_{x}',k_{y}',0)}=\mathfrak{N}_{j}^{(k_{x}',k_{y}',\pi)},\ \
  \mbox{for $k_{x}'$, $k_{y}' = 0$, $\pi$}.
\]
This identifies a factor $\mathbb{Z}^{5}$ of $\BSz$ as representation-enforced gapless superconductors with nodal points on high-symmetry lines in the Brillouin zone. We find that $\BS \simeq \ZZ^5$ and choose the independent band labels as
\[
B[H(\vk)] = \{ \mathfrak{N}_{\frac{1}{2}}^{(0,0,\pi)},\mathfrak{N}_{\frac{5}{2}}^{(0,0,\pi)},\mathfrak{N}_{\frac{1}{2}}^{(\pi,0,\pi)},\mathfrak{N}_{\frac{1}{2}}^{(\pi,\pi,\pi)},\mathfrak{N}_{\frac{5}{2}}^{(\pi,\pi,\pi)}\}.
\]

{\it Symmetry-based indicators.} The group of symmetry-based indicators is 
\begin{equation}
  \SI_\classD[C_4,A] = \SI_\classD[C_4,B] \simeq \ZZ_2 \times \ZZ_8,
  \label{eq:SI_DC4A}
\end{equation} 
isomorphic to the result of the two dimensional
plane discussed in Sec.~\ref{sec:2d_C4_D}. The symmetry-based indicators for
the stack of Kitaev chains $z_{1;x,y,z}$, two dimensional second-order and
Chern superconductors $z_{2;z}$ are identical to the symmetry-based indicators
of the two dimensional plane, Eqs.~(\ref{eq:z1_2d_C4_A}) and
(\ref{eq:z2_2d_C4_A}), respectively. As in the two-dimensional case, the weak
second-order phase and the weak phase with Chern number $4$ have identical band
labels. In three dimensions, the symmetry-based indicators are expressed as 
\begin{align*}
  z_{1;x,y,z} =&\, \mathfrak{N}_{\frac{1}{2}}^{(0,0,\pi)} + \mathfrak{N}_{\frac{5}{2}}^{(0,0,\pi)} +
  \mathfrak{N}_{\frac{1}{2}}^{(\pi,0,\pi)} \mod 2, \\
  z_{2;z} =&\, - \mathfrak{N}_{\frac{1}{2}}^{(0,0,\pi)} 
        + 3  \mathfrak{N}_{\frac{5}{2}}^{(0,0,\pi)} 
        - 2 \mathfrak{N}_{\frac{1}{2}}^{(\pi,0,\pi)} \nonumber \\ &\, \mbox{}
        + 3 \mathfrak{N}_{\frac{1}{2}}^{(\pi,\pi,\pi)}
        - \mathfrak{N}_{\frac{5}{2}}^{(\pi,\pi,\pi)} \mod 8.
  \label{eq:z2_2d_C4_1E}
\end{align*}
The group of symmetry-based indicators (\ref{eq:SI_DC4A}) is identical with the result from \onlinecite{ono2019c}.

\subsubsection{Representation $\Theta = {}^1 E$ or $\Theta = {}^2 E$}

\begin{figure}
\begin{tabular}{ccc}
  $\mathbb{Z}_{2}$  & $0\subseteq\mathbb{Z}$ & $\mathbb{Z}_{2}\subseteq\mathbb{Z}_{2}$ \\
  $z_{1,\{ \frac{1}{2}, \frac{5}{2} \};x,y}=1$ & $z_{2;z}=1$ & $z_{2;x,y}=1$ \\  
  \includegraphics[scale=0.8]{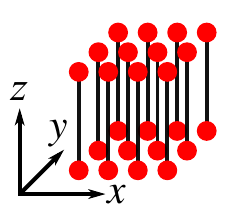} &
  \includegraphics[scale=0.8]{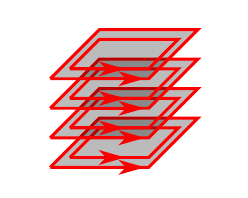} &
  \includegraphics[scale=0.8]{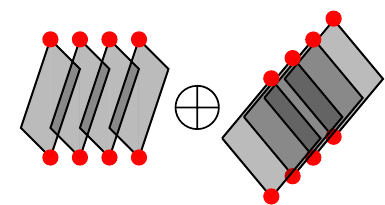} \\
\end{tabular}
\caption[]{Topological phases of a three-dimensional superconductor in tenfold-way class D with point group $C_4$ and representation $\Theta = {}^1 E$ or $\Theta = {}^2 E$.
\label{tab:ex_3d_C4_E_D}}
\end{figure}

{\it Classifying group}:
$${\cal K}_\classD[C_4,{}^2 E] = {\cal K}_\classD[C_4,{}^1 E] =\mathbb{Z}_{2}^{3}\times\mathbb{Z}.$$% 
A factor $\mathbb{Z}_{2}^{2}$ corresponds to one-dimensional topological superconductors parallel to the rotation axis $(z)$ in the angular momentum subspaces $j=\frac{1}{2},\frac{5}{2}$ (labels $(1,j;x,y)$). A factor $\mathbb{Z}$ corresponds to Chern superconductors with even Chern number stacked perpendicular to the rotation axis (label $(2,z)$). These phases can be adiabatically deformed to Chern insulators with vanishing superconducting correlations. The remaining factor $\mathbb{Z}_{2}$ corresponds to a pair of two-dimensional second-order topological superconductors parallel to the rotation axis, exchanged by fourfold rotation and stacked in both $x$ and $y$ directions (label $(2;x,y)$). The boundary signatures of the topological phases together with their symmetry-based indicators are displayed in Fig.~\ref{tab:ex_3d_C4_E_D}.

{\it Band labels.} There are $\BSz \simeq \ZZ_2^{12} \times \ZZ^4$ topological band labels given by the invariants ${\frak N}_{{3}/{2}}^{\vk_{\rm s}}$, ${\frak p}_{{1}/{2}}^{\vk_{\rm s}}$, ${\frak p}_{{5}/{2}}^{\vk_{\rm s}}$ at fourfold rotation-symmetric high-symmetry momenta $\vk_{\rm s} = (0,0,0)$, $(0,0,\pi)$, $(\pi,\pi,0)$, and $(\pi,\pi,\pi)$, and ${\frak p}_{{1}/{2}}^{\vk_{\rm s}}$, ${\frak p}_{{3}/{2}}^{\vk_{\rm s}}$ at twofold rotation symmetric high-symmetry momenta $\vk_{\rm s} = (0,\pi,0)$ and $(0,\pi,\pi)$.

{\it Compatibility relations.} From the fact that ${\cal R}_{\pi}{\cal P}$ is a local symmetry for high-symmetry planes in the Brillouin zone one derives the compatibility relation \eqref{eq:ex_C4_E_compability} of the main text. From the fact that $\mathfrak{N}_{j}$ is conserved along the fourfold symmetric lines in the Brillouin zone parallel to the rotation axis  for $j=\frac{3}{2},\frac{7}{2}$ one derives the further compatibility relations
\begin{align*}
\mathfrak{N}_{\frac{3}{2}}^{(k_{x}',k_{y}',0)}=&\, \mathfrak{N}_{\frac{3}{2}}^{(k_{x}',k_{y}',\pi)},
\end{align*}
for $(k_{x}',k_{y}') = (0,0)$, $(\pi,\pi)$.
%The invariant $\mathfrak{p}_{j}$ for $j=\frac{1}{2},\frac{5}{2}$
%at the fourfold symmetric momenta is affected by this constraint as
%the pair $H_{gen}=-\tau_{3}$, $H_{ref}=\tau_{3}$ that generates
%the invariant $\mathfrak{p}_{j}$ at the high-symmetry momenta $k_{z}=0$
%is trivial at a general momentum $k_{z}$, $\mathfrak{N}_{j}(H)-\mathfrak{N}_{j}%(H_{ref})=0$.
The above compatibility relations associate factors $\mathbb{Z}_{2}^{4}$ and $\mathbb{Z}^{2}$ in $\BSz$ with as representation-enforced gapless superconductors with nodal-line and nodal points, respectively. Hence we have $\BS \simeq \ZZ_2^8 \times \ZZ^2$ with
\begin{align*}
 B[H(\vk)] =&\, \{ \mathfrak{p}_{\frac{1}{2}}^{(0,0,0)}, \mathfrak{p}_{\frac{5}{2}}^{(0,0,0)}, \mathfrak{p}_{\frac{3}{2}}^{(\pi,0,0)}, \mathfrak{p}_{\frac{5}{2}}^{(\pi,\pi,0)}, \mathfrak{p}_{\frac{1}{2}}^{(0,0,\pi)},\\
&\, \ \  \mathfrak{p}_{\frac{5}{2}}^{(0,0,\pi)}, \mathfrak{N}_{\frac{3}{2}}^{(0,0,\pi)}, \mathfrak{p}_{\frac{3}{2}}^{(\pi,0,\pi)}, \mathfrak{p}_{\frac{5}{2}}^{(\pi,\pi,\pi)},  \mathfrak{N}_{\frac{3}{2}}^{(\pi,\pi,\pi)}
 \} .
\end{align*}
A third compatibility relation for gapped phases follows by considering Chern
numbers at two-dimensional cuts through the Brillouin zone.
\begin{align*}
& -\mathfrak{N}_{\frac{3}{2}}^{(0,0,0)} - \mathfrak{N}_{\frac{3}{2}}^{(\pi,\pi,0)} \\ & + 2 \mathfrak{p}_{\frac{5}{2}}^{(0,0,0)} + 2 \mathfrak{p}_{\frac{3}{2}}^{(\pi,0,0)} 
  + 2 \mathfrak{p}_{\frac{5}{2}}^{(\pi,\pi,0)} = \\
& -\mathfrak{N}_{\frac{3}{2}}^{(0,0,\pi)} - \mathfrak{N}_{\frac{3}{2}}^{(\pi,\pi,\pi)} \\ & + 2 \mathfrak{p}_{\frac{5}{2}}^{(0,0,\pi)} + 2 \mathfrak{p}_{\frac{3}{2}}^{(\pi,0,\pi)} 
  + 2 \mathfrak{p}_{\frac{5}{2}}^{(\pi,\pi,\pi)} \mod 4 .
\end{align*}

{\it Symmetry-based indicators.} The group of symmetry-based indicators is 
\begin{equation}
  \SI_\classD[C_4, {}^1 E] = \SI_\classD[C_4, {}^2 E] \simeq \ZZ_2^4 \times \ZZ_4,
\end{equation}
out of which two factors $\ZZ_2$ correspond to
stacks of Kitaev chains with indicator
\[
z_{1,j;x,y}=\mathfrak{p}_{j}^{(\pi,\pi,0)}+\mathfrak{p}_{j}^{(\pi,\pi,\pi)} \mod 2
\]
for $j=\frac{1}{2}, \frac{5}{2}$, the factor $\ZZ_4$ corresponds to a stack of
Chern superconductors perpendicular to the rotation axis $z_{2;z}$ with 
indicator given by 
\begin{align*}
  z_{2; z} =&\, -\mathfrak{N}_{\frac{3}{2}}^{(0,0,\pi)} - \mathfrak{N}_{\frac{3}{2}}^{(\pi,\pi,\pi)} + 
  2 \mathfrak{p}_{\frac{5}{2}}^{(0,0,\pi)} 
  \nonumber \\ &\, \mbox{}
  + 2 \mathfrak{p}_{\frac{3}{2}}^{(\pi,0,\pi)}
  + 2 \mathfrak{p}_{\frac{5}{2}}^{(\pi,\pi,\pi)} \mod 4,
\end{align*}
and a factor $\ZZ_2$ corresponds to a stack of second-order Chern
superconductors parallel to the rotation axis with symmetry-based indicator
\[
z_{2;x,y}=\mathfrak{p}_{\frac{3}{2}}^{(\pi,0,0)}+\mathfrak{p}_{\frac{3}{2}}^{(\pi,0,\pi)} \mod 2 .
\] 

The remaining factor $\ZZ_2$ corresponds to a nodal superconductor the Chern number of which changes by $4n+2$ between parallel planes, with the difference detected by the band labels as 
\begin{align*}
z_{3} = \sum_{k_z = 0,\pi} &   2 \mathfrak{p}_{\frac{5}{2}}^{(0,0,k_z)} 
  + 2 \mathfrak{p}_{\frac{3}{2}}^{(\pi,0,k_z)}
  + 2 \mathfrak{p}_{\frac{5}{2}}^{(\pi,\pi,k_z)} \mod 4.
\end{align*}
Note that a change of the Chern number by an odd number is prohibited by the zero dimensional compatibility constraint along rotation lines.

Reference \onlinecite{ono2019c} finds the group $\SI = \ZZ_2$. Upon omitting band labels corresponding to Pfaffians in our discussion, this single factor $\ZZ_2$ is identified as the parity of $z_{2; z}$, the Chern number of a stack of two dimensional topological superconductors in parallel to the rotation axis. 

\subsection{Point group $C_{2v}$, class D}

The classification and symmetry labels are trivial for representation $\Theta =
A_1$. The $B_1$ representation is analogous to the $B_2$ representation by
exchanging the labels $x$, $y$.

For the $A_1$ representation, our results agree with those of Ref.\ \onlinecite{ono2019c}. For the other representations $A_2, B_1, B_2$, the band labels are defined exclusively in terms of Pfaffians. Correspondingly, in Ref.\ \onlinecite{ono2019c}, no symmetry-based indicators are found due to the absence of Pfaffian band labels in their construction.

\subsubsection{Representation $\Theta = A_2$}

\begin{figure}
\begin{tabular}{cccc}
  $\mathbb{Z}_{2}$ & $\mathbb{Z}_{2}$ & $\mathbb{Z}_{2}$ & \\
  $z_{1;y,z}=1$ & $z_{1;x,z}=1$ & $z_{1;x,y}=1$ & \\
  \includegraphics[scale=0.8]{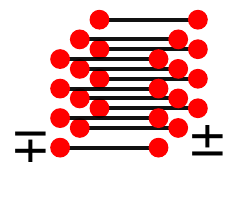} &
  \includegraphics[scale=0.8]{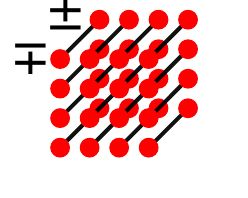} &
  \includegraphics[scale=0.8]{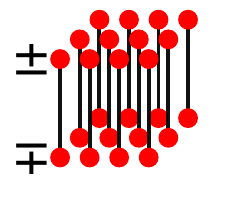} & \\
  $\mathbb{Z}_{2}\subseteq\mathbb{Z}_{2}$ & $\mathbb{Z}_{2}\subseteq\mathbb{Z}_{2}$ & $0\subseteq\mathbb{Z}$ & $0\subseteq\mathbb{Z}$ \\
  $z_{2;z}=1$ & $z_{2;z}=1$ & $z_{2;y}=1$ & $z_{2;x}=1$ \\
  \includegraphics[scale=0.8]{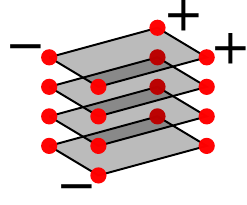} &
  \includegraphics[scale=0.8]{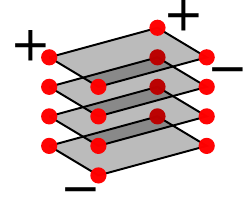} &
  \includegraphics[scale=0.8]{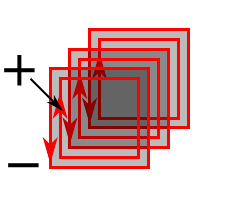} &
  \includegraphics[scale=0.8]{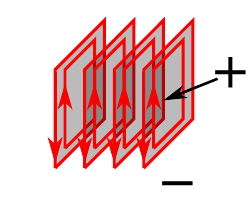} \\
  $0\subseteq\mathbb{Z}\subseteq\mathbb{Z}$ & $0\subseteq\mathbb{Z}\subseteq\mathbb{Z}$ & & \\
  $z_3=1$ & $z_3=1$ & & \\
  \includegraphics[scale=0.8]{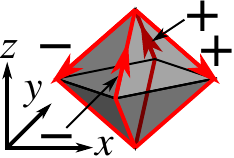} & 
  \includegraphics[scale=0.8]{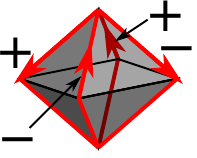} & & \\
\end{tabular}
\caption{Topological phases of a three-dimensional superconductor in
	tenfold-way class D with point group $C_{2v}$ and representation
	$\Theta = A_2$. The parity of the gapless states under corresponding
	mirror symmetry are denoted by $\pm$.
\label{tab:ex_3d_C2v_A2_D}}
\end{figure}

{\it Classifying group:} 
$$\mathcal{K}_\classD[C_{2v},A_2] = \ZZ_2^5 \times \ZZ^4,$$%
see App.\ \ref{app:class_3d_C2v} for a derivation and for the definition of the labels. The boundary signatures of the topological phases together with their symmetry-based indicators are displayed in Fig.~\ref{tab:ex_3d_C2v_A2_D}.

{\it Band labels.} There are $\BSz \simeq \ZZ_2^8$ topological band labels given by the Pfaffian invariant ${\frak p}^{\vk_{\rm s}}$ at all high-symmetry momenta $\vk_{\rm s}$. 

{\it Compatibility relations.}
There are no compatibility relations restricting the symmetry-based indicators for gapped phases. Thus $\BS = \BSz$ and
\[
B[H(\vk)] = \{{\frak p}^{\vk_{\rm s}}\} .
\]

{\it Symmetry-based indicators.} The group of symmetry-based indicators is 
$$\SI_\classD[C_{2v},A_2] \simeq \ZZ_2^7 .$$
The symmetry-based indicators of the stacks of one dimensional topological superconductors with stacking directions $(i,j) = (y,z),(x,z),(x,y)$ are
\[
z_{1;i,j} = \, \sum_{\vk_{\rm s}|_{k_{{\rm s},i} = k_{{\rm s},j} = \pi}}
  \mathfrak{p}^{\vk_{\rm s}} \mod 2, \nonumber \\
\]
and for the stacks of two dimensional topological superconductors with stacking directions $l=x,y,z$ they are
\[
z_{2,l}=\sum_{\vec{k}_{s}|_{k_{{\rm s}, l}=\pi}}\mathfrak{p}^{\vec{k}_{\rm s}}\mod2
\]
and for the three dimensional second-order topological superconductors the symmetry-based indicator is
\[
z_{3}=\sum_{\vec{k}_{\rm s}}\mathfrak{p}^{\vec{k}_{\rm s}}\mod2 .
\]

\subsubsection{Representation $\Theta = B_2$}
\begin{figure}
\begin{tabular}{ccc}
  $\mathbb{Z}_{2}$ & $\mathbb{Z}_{2}$ & \\
  $z_{1;x,z}=1$ & $z_{1;x,y}=1$ & \\
  \includegraphics[scale=1]{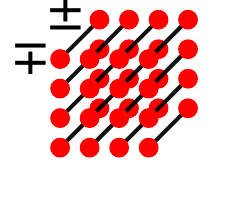} &
  \includegraphics[scale=1]{figs/C2v/3d_A2_1xy} & \\
  $\mathbb{Z}_{2}\subseteq\mathbb{Z}_{2}$ & $0\subseteq\mathbb{Z}$ & $0\subseteq\mathbb{Z}\subseteq\mathbb{Z}$ \\
  $ - $ & $z_{2;x}=1$ & $ - $ \\
  \includegraphics[scale=1]{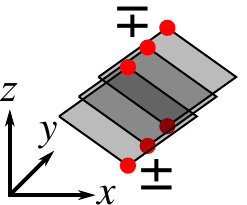} &
  \includegraphics[scale=1]{figs/C2v/3d_A2_2x} &
  \includegraphics[scale=1]{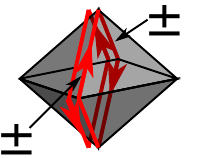} 
\end{tabular}
\caption[]{Topological phases of a three-dimensional superconductor in
	tenfold-way class D with point group $C_{2v}$ and representation
	$\Theta = B_2$. The parity of the gapless states under corresponding
        mirror symmetry are denoted by $\pm$.
\label{tab:ex_3d_C2v_B2_D}}
\end{figure}

{\it Classifying group:} 
$$\mathcal{K}_\classD[C_{2v},B_2] =\mathbb{Z}_{2}^{3}\times\mathbb{Z}^{2},$$%
see App.\ \ref{app:class_3d_C2v} for a derivation and for the definition of the
labels. The boundary signatures of the topological phases together with their
symmetry-based indicators are displayed in Fig.~\ref{tab:ex_3d_C2v_B2_D}.

{\it Band labels.} There are $\BSz \simeq \ZZ_2^8$ topological band labels
given by the Pfaffian invariant ${\frak p}^{\vk_{\rm s}}$ at all high-symmetry
momenta $\vk_{\rm s}$. 

{\it Compatibility relations.} From the fact that ${\cal P M}_x$ acts locally
on high-symmetry lines in reciprocal space we derive the compatibility relation 
\begin{align*} 
  \mathfrak{p}^{(0,k_y',k_z')} = \mathfrak{p}^{(\pi,k_y',k_z')},\ \ \mbox{for $k_{y}'$, $k_{z}' = 0$, $\pi$}.
\end{align*} 
The compatibility relations identify a factor $\mathbb{Z}_{2}^{4}$ of $\BSz$ as
representation-enforced gapless superconductors with nodal points such that
$\BS \simeq \ZZ_2^4$ and 
\[
B[H(\vk)] = \{{\frak p}^{\vk_{\rm s}|_{k_x = \pi}}\} .
\]

{\it Symmetry-based indicators.} The group of symmetry-based indicators is 
$$\SI_\classD[C_{2v},B_2]\simeq \ZZ_2^3 .$$  
The symmetry-based indicators of the stacks of one
dimensional topological superconductors with stacking directions $(i,j) =
(x,z),(x,y)$ are
\[
z_{1;i,j} = \, \sum_{\vk_{\rm s}|k_{{\rm s},i} = k_{{\rm s},j} = \pi}
  \mathfrak{p}^{\vk_{\rm s}} \mod 2, \nonumber \\
\]
and for the stacks of two dimensional topological superconductors with stacking
direction $x$ it is
\[
z_{2,x}=\sum_{\vec{k}_{\rm s}|_{k_{{\rm s}, x}=\pi}}\mathfrak{p}^{\vec{k}_{\rm s}} \mod 2 .
\]
The two dimensional second-order topological superconductor stacked in the $y$
direction and the three dimensional second-order topological superconductor can
not be detected from symmetry-based indicators.

\section{Boundary classification groups ${\cal K}$}
\label{app:class}
For tenfold-way classes in $d$ dimensions the classification groups ${\cal K}$
follow from the ``periodic table of topological
phases''.~\cite{kitaev2009,schnyder2009} They are the direct product of groups
classifying strong phases and groups classifying weak phases obtained by
stacking topological phases in dimension $d-n$, $n=1,\ldots,d-1$. 

In the presence of an additional order-two crystalline symmetry, such as
inversion, mirror, or twofold rotation, the classifying group ${\cal K}$ can be
obtained using the known classifications of strong phases given in
Ref.~\onlinecite{shiozaki2014,trifunovic2019} (see also
Refs.~\onlinecite{chiu2013,morimoto2013,trifunovic2017}), again accounting for
weak phases by taking direct products of the appropriate classification groups
in lower dimensions.~\cite{shiozaki2014} It is important to point out that the
classification group ${\cal K}$ we consider in this article classifies
topological phases with nontrivial boundary signature only --- see the
discussion at the end of Sec.\ \ref{sec:0d}. These ``boundary classification
groups'' ${\cal K}$ are obtained from the groups $K$ classifying the bulk band
structure by dividing out the classifying group $K^{(d)}$ of atomic-limit
phases.\cite{trifunovic2019}

Classifications of strong tenfold-way phases with other additional crystalline
symmetry groups can be found in the
literature.~\cite{cornfeld2019,shiozaki2019classification} We here give classification
results for the examples discussed in the main text and in App.~\ref{app:b}.
These classifications are derived from the enumeration of the possible
anomalous boundary states compatible with the symmetries defining the
topological class. Boundary states that can be removed by a change of lattice
termination are removed from the classification, because
they are not a consequence of the topology of the bulk band structure. (Such
boundary states are called ``extrinsic'' in
Refs.~\onlinecite{geier2018,trifunovic2019}.)

Below we compute the classifying groups ${\cal K}_\eta[G\vert G_{\cal O},
\Theta, d]$ of $d$ dimensional topological phases with point group $G$, where
its normal subgroup $G_{\cal O}$ is assumed to act as onsite symmetry. For each
factor of ${\cal K}_\eta[G\vert G_{\cal O}, \Theta, d]$ we present an
interpretation in terms of higher order topological phases or stacks of lower
dimensional topological (weak) phases and generators from which the band labels
can be computed.  We not only consider the dimensions $d$ corresponding to the
examples used in the main text, but also lower dimensions, since the
classification results for lower dimensions $d$ provides information on weak
and higher-order phases.

\subsection{One dimension, class D, $C_{2v}\vert C_s$}
\label{app:class_1d_C2v/Cs}
We consider a one dimensional system extended in the $x$ direction, such that
it lies within the mirror plane of ${\cal M}_y$. Hence ${\cal M}_y$ is an
onsite symmetry, wheres ${\cal M}_x$ acts non-locally. For spinful fermions there is a single irreducible
representation $\alpha = \bar{E}$ with dimension $d_{\bar E} = 2$. 

{\it Representation $\Theta = A_1$.---} For the representation $\Theta = A_1$,
${\cal M}_x$ and ${\cal M}_y$ commute with particle-hole conjugation ${\cal
P}$. The parallel mirror plane ${\cal M}_y$ with ${\cal M}_y^2 = -1$ forbids a
topological superconducting phase, as it effectively turns the one-dimensional
system into class A$^{{\cal P}^-{\cal M}}$ with trivial classification, see the
notation of Ref.~\onlinecite{trifunovic2019}. We conclude that

\begin{equation}
  {\cal K}_{\classD}[C_{2v}\vert C_s,A_1,d=1] = 0.
\end{equation}

{\it Representation $\Theta = A_2$.---} In this case ${\cal M}_x$ and ${\cal
M}_y$ anticommute with ${\cal P}$. The parallel mirror plane allows to block
diagonalize the system into two blocks according to ${\cal M}_y$-parity $\pm$.
Particle-hole conjugation acts within the blocks, whereas the perpendicular
mirror plane ${\cal M}_x$ interchanges the two blocks. The system is thus
completely specified by a single block in tenfold-way class D, which allows a
Kitaev-chain topological phase. A generator of the nontrivial phase is
$$
  H_{(1)}(k_{x})=\rho_{0}\tau_{3} (1 - m - \cos k_x)+ \rho_{1}\tau_{1} \sin k_{x},
$$%
with $0 < m < 2$ and representations $\mathcal{P} = \rho_0 \tau_1 K$, $U({\cal R}_\pi)
=i\rho_{2}\tau_{0}$, $U({\cal M}_x) =i\rho_{3}\tau_{0}$, $U({\cal M}_y)
=i\rho_{1}\tau_{0}$. 

{\it Representation $\Theta = B_1$.---} In this case ${\cal M}_x$ commutes with
${\cal P}$, whereas ${\cal M}_y$ anticommutes with ${\cal P}$. The discussion
can be mapped to that of the case $\Theta = A_2$ by noting that ${\cal
R}_{\pi}$ anticommutes with ${\cal P}$, so that the arguments put forward for
$\Theta = A_1$ can be applied to the case $\Theta = B_2$ by exchanging the
roles of ${\cal M}_x$ and ${\cal R}_{\pi}$. It follows that
\begin{equation}
  {\cal K}_{\classD}[C_{2v}\vert C_s,B_1,d=1] = \ZZ_2.
\end{equation}

{\it Representation $\Theta = B_2$.---} The parallel mirror plane ${\cal M}_y$
with ${\cal M}_y^2 = -1$ forbids a topological superconducting phase, as it
effectively turns the one-dimensional system into class A$^{{\cal P}^+{\cal
M}}$ with trivial classification,~\cite{trifunovic2019} so that

\begin{equation}
  {\cal K}_{\classD}[C_{2v}\vert C_s,B_2,d=1] = 0.
\end{equation}

\subsection{One dimension, class D, $C_{2v}\vert C_{2v}$}
\label{app:class_1d_C2v/C2v}

This case applies to a one-dimensional system along the $z$ direction, such
that it lies in the intersection of the mirror planes ${\cal M}_x$, ${\cal
M}_y$ and the full crystalline symmetry group $C_{2v}$ acts onsite. As shown in
Table \ref{tab:classD_0d} and discussed in Sec.\ \ref{sec:0d}, for spinful
fermions there is a single irreducible representation $\alpha = \bar{E}$ with
dimension $d_{\bar E} = 2$, which effectively changes the tenfold-way
symmetries from those of class D to those of class C for the case $\Theta =
A_1$, but leaves them unchanged for $\Theta = A_2$, $B_1$, and $B_2$. It
follows for $d=1$ that
\begin{align}
  {\cal K}_{\classD}[C_{2v}\vert C_{2v}, A_1,d=1] =&\, 0 \\
  {\cal K}_{\classD}[C_{2v}\vert C_{2v}, \Theta,d=1] =&\, \ZZ_2,\ \
  \Theta = A_2, B_1, B_2.
\end{align}
A generator for the nontrivial phases is 
$$
  H_{(1)}(k_{z})=\rho_{0}\tau_{3} (1 - m - \cos k_z)+ \rho_{1}\tau_{1} \sin k_{z},
$$%
with $0 < m < 2$ and accordingly chosen representations.

\subsection{Two dimensions, class D, $C_{2v}\vert C_s$}
\label{app:class_2d_C2v/Cs}

This case applies to a two-dimensional system in the $xz$ plane, such that
${\cal M}_y$ acts as an onsite symmetry. A system with boundaries parallel to
the coordinate axes has two ${\cal M}_x$-symmetric boundaries and two
boundaries that are mapped to each other by ${\cal M}_x$.

{\it Representation $\Theta = A_1$.---} From an analysis of the reflection
symmetric boundary, we find that there is no first-order topological
superconducting phase. (The nontrivial first-order phase in class D has a
chiral edge mode, which is not compatible with a mirror symmetry.) There are
also no second-order or weak phases in this representation,
\begin{equation}
  {\cal K}_{\classD}[C_{2v}\vert C_s,A_1,d=2] = 0.
\end{equation}
This follows from the triviality of the classifying groups
$\mathcal{K}_{\classD}[C_{2v}\vert C_s, A_1,d]$ and
$\mathcal{K}_{\classD}[C_{2v}\vert C_{2v}, A_1,d]$ for $d=1$, into which an
eventual higher-order phase can be deformed or which could be stacked to
form a weak phase, see the discussions in Apps.~\ref{app:class_1d_C2v/Cs} and
\ref{app:class_1d_C2v/C2v}.

{\it Representation $\Theta = A_2$.---} A ${\cal M}_x$-symmetric boundary
allows counter-propagating chiral Majorana modes in opposite ${\cal
M}_y$-parity subspaces. They are the first-order boundary signature of a pair
of Chern superconductors related to each other by ${\cal M}_x$. Superconductors with even Chern number in a mirror plane can be adiabatically deformed to Chern insulators with vanishing superconducting correlations. From the
nontriviality of the classifying groups $\mathcal{K}_{\classD}[C_{2v}\vert C_s,
A_2,d] = \mathcal{K}_{\classD}[C_{2v}\vert C_{2v}, A_2,d] = \ZZ_2$ for $d=1$ we
conclude that two types of weak phases are possible, obtained by stacking
one-dimensional superconductors in the $x$ and $z$ direction. There is no
second-order phase, despite the fact that pairs of even and odd-${\cal
M}_y$-parity Majorana zero-energy bound states are allowed at corners. Such
states are protected against local perturbations at corners, but they can be
removed by a change of termination along the crystal edges. (Formally, this
follows because the group $\mathcal{K}_{\classD}[C_{2v}/C_{2v}, A_2]$
classifying ${\cal M}_x$-symmetric corner states is ``separable'' in the
language of Ref.~\onlinecite{trifunovic2019}.) We thus conclude that

\begin{equation}
  {\cal K}_{\classD}[C_{2v}\vert C_s,A_2,d=2] = \ZZ_2^2 \times \ZZ.
  \label{eq:class_2d_C2v/Cs_A2}
\end{equation}

The same result can be obtained by arguing that the onsite symmetry ${\cal
M}_y$ commutes with ${\cal P}$, so that the Hamiltonian can be written as the
diagonal sum of ${\cal M}_y$-even and ${\cal M}_y$-odd blocks that each satisfy
particle-hole symmetry. The mirror symmetry ${\cal M}_x$ interchanges the two
blocks. Hence, the topological classification is the same as of a
two-dimensional Hamiltonian in tenfold-way class D without crystalline
symmetries, which also gives Eq.~(\ref{eq:class_2d_C2v/Cs_A2}).

{\it Representation $\Theta = B_1$.---} This case can be mapped to $\Theta = A_2$ by interchanging the roles of ${\cal M}_x$ and ${\cal R}_{\pi}$. One thus finds
\begin{equation}
  {\cal K}_{\classD}[C_{2v}\vert C_s,B_1,d=2] = \ZZ_2^2 \times \ZZ.\\
\end{equation}

{\it Representation $\Theta = B_2$.---} In this case no first-order boundary
signatures are allowed on a ${\cal M}_x$-symmetric boundary, because they are
incompatible with the ${\cal M}_x$ symmetry. To check for the existence of
second-order or weak phases, we note that one-dimensional superconductors in
the $z$ direction ({\it i.e.}, perpendicular to the mirror plane of ${\cal
M}_x$) have classifying group $\mathcal{K}_{\classD}[C_{2v}\vert C_{s},
B_2,d=1]$, which is trivial, see App.~\ref{app:class_1d_C2v/Cs}. Also,
one-dimensional superconductors in the $z$ direction, {\it i.e.}, in the
intersection of both mirror planes, have classifying group
$\mathcal{K}_{\classD}[C_{2v}\vert C_{2v}, B_2, d=1] = \mathbb{Z}_2$, as shown
in App.~\ref{app:class_1d_C2v/C2v}. This has two consequences: (i) A
second-order boundary signature corresponding to a pair of Majorana bound
states at the ${\cal M}_x$-symmetric corner is possible.  The in-plane mirror
symmetry ${\cal M}_y$ with ${\cal M}_y {\cal P} = {\cal P} {\cal M}_y $ forbids
one-dimensional topological phases within the mirror ${\cal M}_y$ plane that do
not satisfy ${\cal M}_x$ mirror symmetry. Hence the boundary signature can not
be removed by a decoration of the crystal boundary and there is a second-order
topological superconducting phase. A generator for this phase is
\begin{align*}
  H_{2}'(\vk)=&\,\mu_{0}\rho_0\tau_{3} (\cos k_x + \cos k_z + m)
  \nonumber \\ &\, \mbox{}
  + \mu_{1}\rho_0\tau_{1} \sin k_{x} + \mu_2 \rho_2 \tau_1 \sin k_z ,
\end{align*}
with $0 < m < 2$ and representations $\mathcal{P} = \mu_0 \rho_0 \tau_1 K$, $U({\cal R}_\pi)
=i\mu_{0}\rho_3\tau_{3}$, $U({\cal M}_x) =i\mu_{3}\rho_3\tau_{0}$, $U({\cal
M}_y) =i\mu_{3}\rho_0\tau_{3}$. Alternatively, the parallel mirror plane allows
to block diagonalize the system into two blocks according to ${\cal
M}_y$-parity $\pm$. Particle-hole conjugation interchanges the two blocks, and
each block belongs to class A$^{ {\cal P}^+{\cal M}}$ with
$0\subseteq\ZZ_2\subseteq\ZZ_2$ bulk subgroup sequence for
$d=2$.~\cite{trifunovic2019} (ii) There is a weak phase corresponding to
one-dimensional topological superconductors in the $z$ direction, which are
stacked in the $x$ direction.  We thus conclude that
\begin{equation}
  {\cal K}_{\classD}[C_{2v}\vert C_s,B_2,d=2] = \ZZ_2^2.
\end{equation}

\subsection{Two dimensions, class D, $C_{2v}$}
\label{app:class_2d_C2v}
We now derive the boundary classification for the example discussed in
Sec.~\ref{sec:C2v_d2} of the main text. The two mirror symmetries forbid a
phase with a nonzero Chern number. Hence, there are no first-order phases for
all representations of the pairing term $\Theta$. 

\begin{figure}
\includegraphics[width=\columnwidth]{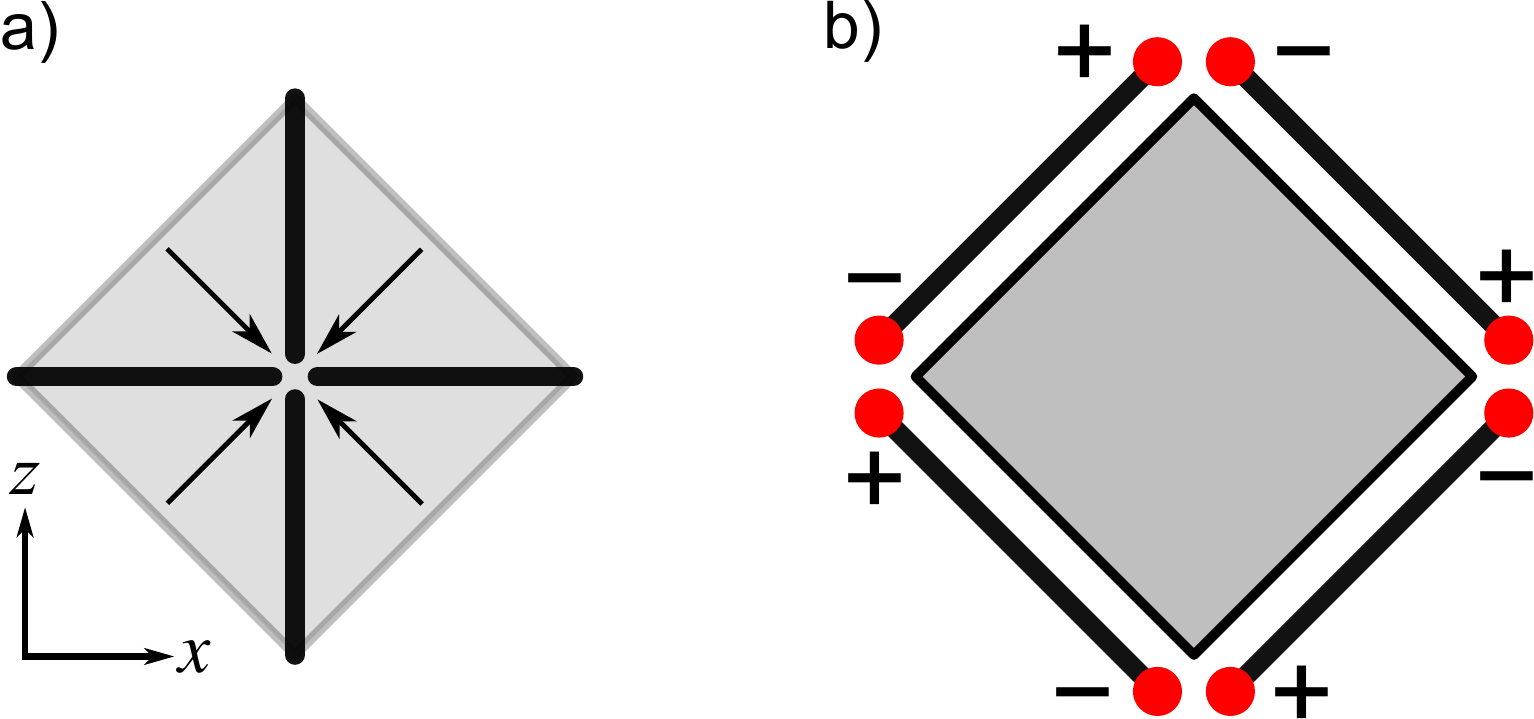}
\caption{a) A $C_{2v}$-symmetric two-dimensional crystal with two perpendicular mirror symmetries with gapped edges may be deformed into a four one-dimensional chains in a cross-like arrangement. Corner states of the two-dimensional crystal are in one-to-one correspondence with end states of the one-dimensional structure. b) A $C_{2v}$-symmetric crystal may be decorated symmetrically with Kitaev chains, resulting in the appearance of extrinsic pairs of corner states at all four mirror-symmetric corners. \label{fig:decoration_C2v_A2}}
\end{figure}

{\it Representation $\Theta = A_1$.---} For this representation, both ${\cal
M}_x$ and ${\cal M}_y$ with ${\cal M}_x^2={\cal M}_y^2 = -1$ commute with
${\cal P}$, ruling out the existence of zero-energy Majorana bound states at
mirror-symmetric corners. There are no weak phases as the classifying group
$\mathcal{K}_{\classD}[C_{2v}\vert C_s, A_1,d=1] = 0$ is trivial. We
conclude that
\begin{equation}
  {\cal K}_{\classD}[C_{2v},A_1,d=2] = 0.
\end{equation}
{\it Representation $\Theta = A_2$.---} In this case, both ${\cal M}_x$ and
${\cal M}_y$ anticommute with ${\cal P}$. We first investigate the possibility
of second-order phases with non-degenerate zero-energy Majorana corner states
at mirror-symmetric corners of a $C_{2v}$-symmetric crystal. Since the edges of
such a crystal are gapped, it may be deformed into a ``cross'' shape consisting
of four one-dimensional chains, see Fig.\ \ref{fig:decoration_C2v_A2}a. Each
chain has a parallel mirror plane, which effectively acts as a local symmetry.
Perpendicular mirror planes connect chains on opposite sides of the cross, but
no longer act inside a chain. The parallel mirror planes allow the Hamiltonian
to be block-diagonalized into blocks with odd and even mirror parity. Each of
these blocks is in tenfold-way class D, thus in principle allowing for the
existence of up to two zero-energy Majorana end states, which turn into corner
states if the system is deformed back to a $C_{2v}$-symmetric two-dimensional
crystal. The presence of the perpendicular mirror planes and the condition that
the center of the cross be gapped restrict the possible configurations of
corner states: Opposite corners must have Majorana states of opposite mirror
parities, since the two mirror operations anticommute, and the total number of
Majorana modes must be a multiple of four. This leaves a $\ZZ_2^3$ extrinsic
classification of allowed Majorana corner modes. To obtain the classification
of intrinsic second-order phases, we must divide out configurations of corner
states that differ by a change of termination. Hereto we note that the four
edges of a $C_{2v}$-symmetric crystal allow a ``decoration'' with Kitaev
chains, which yield opposite-parity pairs of Majorana states at all four
corners of the crystal, see Fig.\ \ref{fig:decoration_C2v_A2}b, so that a
$\ZZ_2^2$ classifying group of second-order phases remains. The generators of
the two distinct $\ZZ_2$ second-order phases can be given as configurations
with single zero-energy Majorana bound states at all corners of a $C_{2v}$
symmetric sample with mirror parities as indicated in
Table~\ref{tab:ex_C2v_A2_bandlabels}.

\begin{figure}
\includegraphics[width=1\columnwidth]{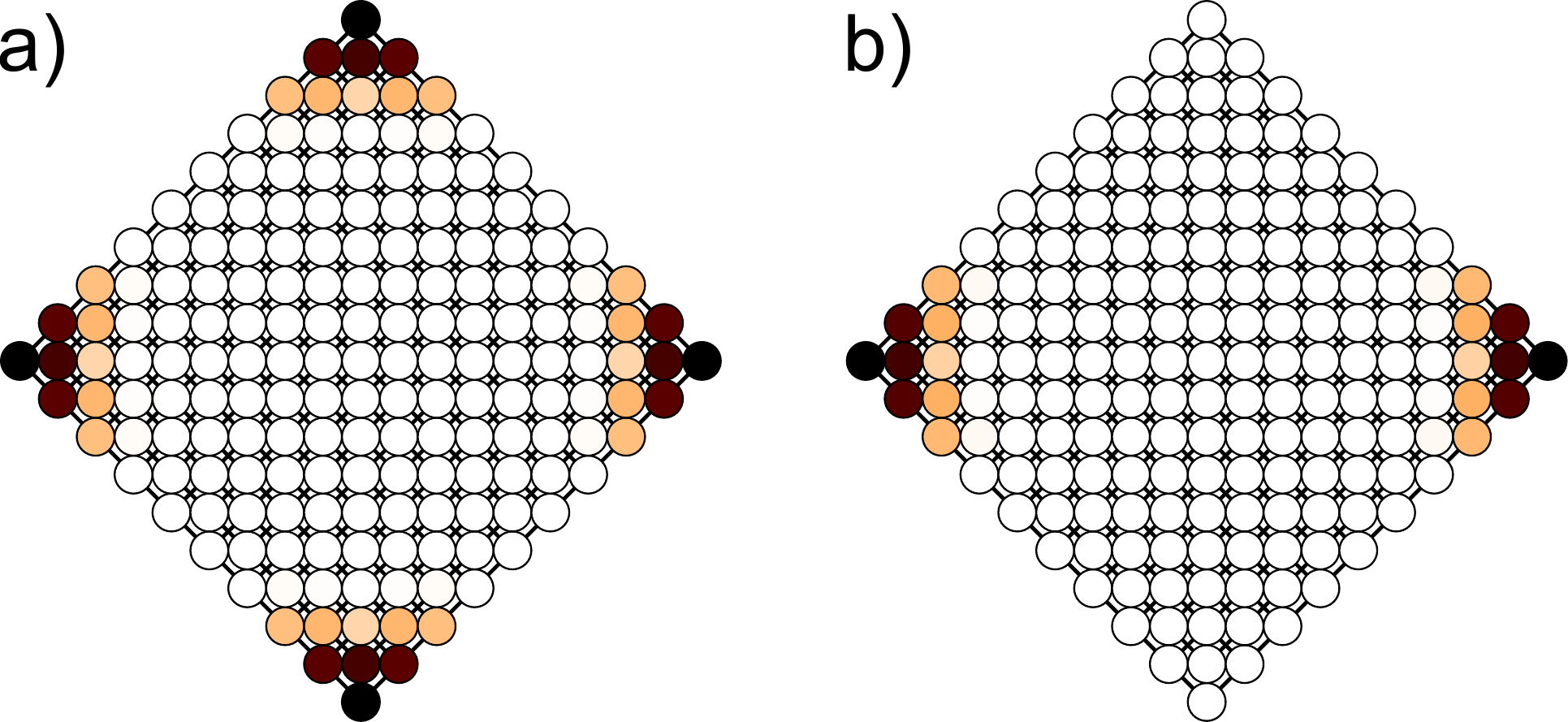}
\caption{a) Support of Majorana bound states for the Hamiltonians $H'_{2,\pm}$ of Eq.\ (\ref{eq:class_2d_C2v_A2_HOTSC12}). A symmetry-preserving next-nearest neighbor term $m_1 \rho_2 \tau_1 \sin k_x \sin k_y$ with $m_1 = 0.4$ is added to remove a spurious gapless edge mode. b) Support of Majorana bound states for $H'_{2,+} \oplus H'_{2,-}$, after adding an additional weak symmetry-preserving hybridization term $m_2 \rho_0 \tau_1 \mu_2$ with strength $m_2 = 0.1$. The corner modes may appear at the other pair of corners if a different hybridization term is chosen.
\label{fig:WF_D_C2v_A2_m04}}
\end{figure}

Furthermore there are two weak phases as the classifying group
$\mathcal{K}_{\classD}[C_{2v}\vert C_s, A_2, d=1] =\ZZ_2$ (see
App.~\ref{app:class_1d_C2v/Cs}) allows stacking of one dimensional topological
superconductors both in $x$ and $y$ directions. Combining everything, we have
the classifying group 
\begin{align}
  \mathcal{K}_{\classD}[C_{2v}, A_2,d=2] = &\, \ZZ_2^4.
\end{align}
Generators of the second-order phases are
\begin{align}
\label{eq:class_2d_C2v_A2_HOTSC12}
  H'_{2,\pm}(\vk) =&\,\rho_{0}\tau_{3} (2 - m - \cos k_{x} - \cos k_{y})
  \nonumber \\ &\, \mbox{}  
  \pm \rho_{1}\tau_{1} \sin k_{x} + \rho_{3}\tau_{1} \sin k_{y}
\end{align}
with $0 < m < 2$ and representations $\mathcal{P}= \tau_1 K$, $U({\cal R}_\pi)
=i\rho_{2}\tau_{0}$, $U({\cal M}_x) =i\rho_{3}\tau_{0}$, $U({\cal M}_y)
=i\rho_{1}\tau_{0}$. To verify that these are indeed second-order phases with
the desired properties, one may either count the number crystalline symmetry
breaking mass terms,\cite{trifunovic2019} or simply verify that these model
Hamiltonians have the correct corner-state structure. Results of such a
calculation, using the {\it kwant} software,\cite{groth2014} are shown in Fig.\
\ref{fig:WF_D_C2v_A2_m04}a. As shown in Fig.~\ref{fig:WF_D_C2v_A2_m04}b,
taking the direct sum of and hybridizing the two generating Hamiltonians $H_{2,\pm}'$ yields a
pair of Majorana bound states with opposite mirror parity at a single pair
corners within a single mirror plane only.

{\it Representations $\Theta=B_{1,2}$.---}
The discussions for the cases $\Theta =B_1$ and $\Theta = B_2$ are analogous,
as they are related by a $\pi / 2$ rotation of the system. In the following
we focus on the $B_2$ case, for which ${\cal M}_x$ anticommutes with ${\cal P}$
and ${\cal M}_y$ commutes with ${\cal P}$. As before, we first consider the
possibility of second-order phases with zero-energy Majorana corner states.
Only corners bisected by the mirror plane ${\cal M}_x$ can host Majorana bound
states, as corners bisected by ${\cal M}_y$ are effectively in class A. Hence,
the system can be deformed to a one dimensional system within the mirror plane
${\cal M}_x$. This one-dimensional system has classifying group is
$\mathcal{K}_{\classD}[C_{2v}\vert C_s, B_1,d=1] = \ZZ_2$, as discussed in
App.~\ref{app:class_1d_C2v/Cs}. However, the corresponding configuration of
boundary states, a pair of Majorana bound states in both mirror parity sectors
at both corners bisected by the mirror ${\cal M}_x$ plane, can be removed by a
decoration with Kitaev chains on each surface of a symmetric sample. We thus
conclude that there are no intrinsic second-order phases for $\Theta =B_1$ or
$\Theta = B_2$ 

Weak phases can be obtained by stacking one-dimensional chains in the $y$
direction. Since ${\cal M}_x$ is an onsite symmetry for such chains, their
classification is given by the one-dimensional classifying group
$\mathcal{K}_{\classD}[C_{2v}\vert C_s, B_1,d=1] = \ZZ_2$, see
App.~\ref{app:class_1d_C2v/Cs}. (Note that for this stacking direction ${\cal
M}_x$ and ${\cal M}_y$ are exchanged in comparison to the discussion in
App.~\ref{app:class_1d_C2v/Cs}.) Superconductors in the perpendicular stacking
direction $x$ are trivial since $\mathcal{K}_{\classD}[C_{2v}\vert C_s, B_2,d=1] = 0$,
as shown in Sec.~\ref{app:class_1d_C2v/Cs}. We conclude that
\begin{equation}
  {\cal K}_{\classD}[C_{2v},B_{1,2},d=2] = \ZZ_2.
\end{equation}

\subsection{Two dimensions, class D, $C_4$}
\label{app:class_2d_C4}

\begin{figure}
\includegraphics[width=0.4\columnwidth]{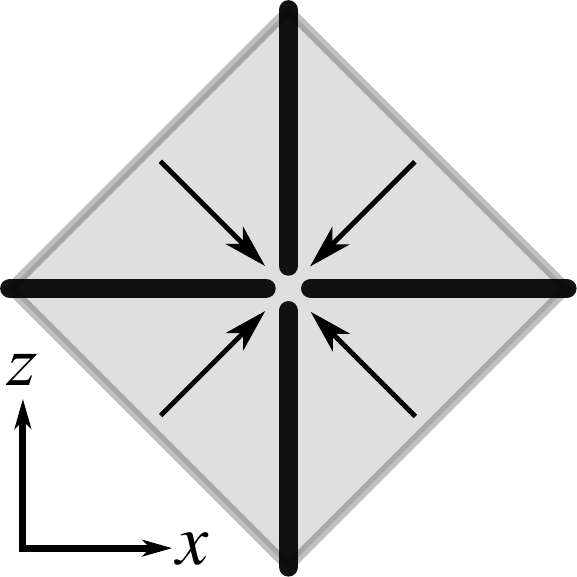}
\caption{\label{fig:C4_d2} A $C_{4}$-symmetric two-dimensional crystal may be deformed into a four one-dimensional chains in a cross-like arrangement. Corner states of the two-dimensional crystal are in one-to-one correspondence with end states of the one-dimensional structure.}
\end{figure}

{\it Representations $\Theta = A,B$.} The representations $\Theta = A$ and
$\Theta = B$ have the same algebraic structure, so that we may limit the
discussion to the case $\Theta = A$. This case allows a strong first-order
phase with a single chiral Majorana mode, corresponding to a $\ZZ$ topological
classification. To check for higher-order phases with four Majorana bound
states, we smoothly deform the two-dimensional system to four one-dimensional
chains arranged in a $C_4$-symmetric cross-like shape, as in Fig.\
\ref{fig:C4_d2}. Each chain may harbor a zero-energy Majorana bound state at
its end. The end states at the four chains are related to each other by $C_4$
symmetry. Since classification of zero-energy Majorana modes protected by $C_4$
is trivial for A and B superconducting pairing, see discussion in
Sec.~\ref{sec:0d}, the four Majorana states that would appear at the center of
the cross can gap out, so that one obtains a true second-order phase with a
$\ZZ_2$ classification. Finally, a weak phase may be obtained by stacking
one-dimensional $C_2$-symmetric $x$ lines (with representation $\Theta = A$) in
the $y$ direction and superimposing the same stack, rotated by $\pi/2$. The
one-dimensional superconductors that are the building blocks of this phase have
zero-energy Majorana states at their ends and a $\ZZ_2$ classification. We
conclude that
\begin{align}
  {\cal K}_{\classD}[C_4,A,d=2] =&\,
  {\cal K}_{\classD}[C_4,B,d=2] \nonumber \\ =&\, \ZZ_2^2 \times \ZZ.
\end{align}

{\it Representations $\Theta ={}^{1,2}E$.---} The representations $\Theta
={}^1E$ and $\Theta={}^2E$ allow a strong first-order phase with an even number
of chiral Majorana modes. One verifies that a single Majorana mode is not
allowed for this representation, since there is no one-dimensional
representation of ${\cal R}_{\pi}$ and ${\cal P}$ meeting the conditions that
(i) $i {\cal R}_{\pi}$ squares to one and (ii) $i {\cal R}_{\pi}$ anticommutes
with ${\cal P}$. An even number of Majorana modes is allowed, since there are
two-dimensional representations meeting these requirements. One also verifies
that the constructions of a second-order phase and a weak phase used for the
representations $\Theta = A,B$ discussed above do not work for the case $\Theta
={}^{1,2}E$. For the second-order phase, the reason is that at the center
cross, where $C_4$ is a local symmetry, one now ends up with states at all four
allowed angular momenta $j=\frac{1}{2}$, $\frac{3}{2}$, $\frac{5}{2}$, and $\frac{7}{2}$, which can not gap out
because particle-hole conjugation acts within two of the angular momentum
sectors, see the discussion in Sec.~\ref{sec:0d}. For the weak phases, this
follows because the underlying $C_2$-symmetric one-dimensional building blocks
have the representation $\Theta = B$, which does not allow for a topological
phase. We thus conclude that
\begin{equation}
  {\cal K}_{\classD}[C_4,{}^{1,2}E,d=2] = \ZZ.
\end{equation}

\subsection{Three dimensions, class D, $C_{2v}$}\label{app:class_3d_C2v}
\begin{figure}
\includegraphics[width=0.8\columnwidth]{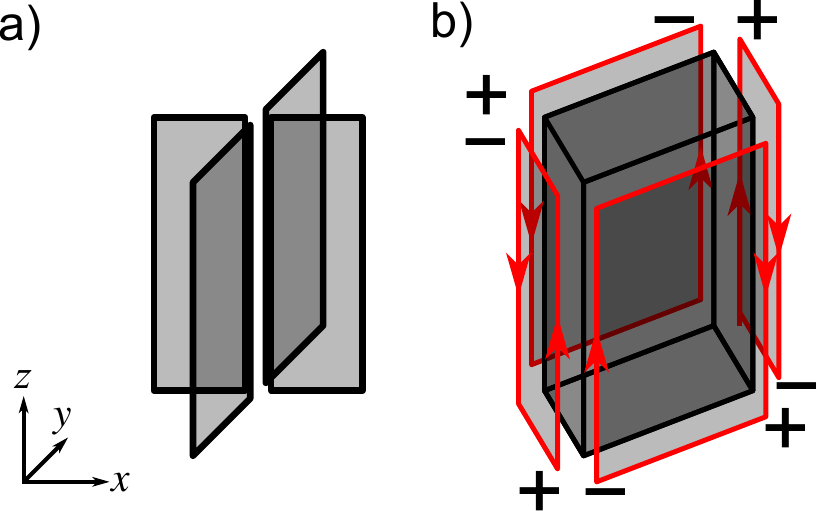}
\caption{\label{fig:decoration_C2v_d3} a) A $C_{2v}$-symmetric three-dimensional crystal with two perpendicular mirror symmetries may be deformed into a four two-dimensional planes in a cross-like arrangement. Hinge states of the two-dimensional crystal are in one-to-one correspondence with edge states of the two-dimensional structure. b) A $C_{2v}$-symmetric crystal may be decorated symmetrically with quantum Hall planes, resulting in the appearance of extrinsic configurations of chiral hinge modes at all four mirror-symmetric hinges.}
\end{figure}

In class D, there are no three dimensional strong first-order topological
phases without crystalline symmetries, so that no first-order phase is
possible.  To determine whether a strong phase is possible, with protected
gapless Majorana modes along mirror-symmetric hinges, we deform the
three-dimensional crystal into a ``cross'' of four two-dimensional planes, as
shown schematically in Fig.~\ref{fig:decoration_C2v_d3}a and determine below
for each representation $\Theta$, whether a phase with chiral hinge states is
possible and whether it is intrinsic or extrinsic ({\it i.e.}, whether it can
be removed with a decoration of the surface). Intrinisic second-order boundary
signatures are possible in representations $\Theta = A_2, B_1, B_2$ as we show
below.  To determine whether a third-order TSC is possible, we note that a
protected corner state can exist only for a corner on the intersection of the
two mirror planes. We find that pairs of corner states are allowed for the
representations $\Theta = A_2$, $B_1$, and $B_2$, but not for $\Theta = A_1$.
However, the corner state for $\Theta = A_2$, $B_1$, or $B_2$ is extrinsic, as
they can be removed by the decoration of a pair of hinges in a mirror plane
with one-dimensional topological superconductors.

{\it Representation $\Theta = A_1$.---} For the representation $A_1$ there are
no intrinsic second-order boundary signatures: Although the edges of the planes
in the deformed structure of Fig.~\ref{fig:decoration_C2v_d3}a not allow for
pairs of co-propagating chiral Majorana modes, one for each mirror parity, such
configurations of hinge modes can be removed by decorating the four
symmetry-related crystal faces by two-dimensional quantum Hall phases, as shown
schematically in Fig.~\ref{fig:decoration_C2v_d3}b. Further, all
lower-dimensional building blocks that can be used for stacking are trivial,
which rules out the existence of weak phases. We conclude that
\begin{equation}
 \mathcal{K}_{\classD}[C_{2v}, A_1, d=3] = 0.
\end{equation}

{\it Representation $\Theta = A_2$.---} We note that for each of the planes in
the deformed structure of Fig.\ \ref{fig:decoration_C2v_d3}a one of the mirror
symmetries acts as an onsite symmetry, whereas the other mirror symmetry maps
planes on opposite sides of the central ``cross'' onto each other. For
representation $A_2$, each of the planes in the deformed structure of
Fig.~\ref{fig:decoration_C2v_d3}a allows for chiral Majorana modes in both
parity sectors of the onsite mirror symmetry. These Majorana modes turn into
hinge modes upon deforming back to the full three-dimensional structure. The
perpendicular mirror symmetry imposes the condition that opposite hinges have
Majorana modes with the same propagation direction, but with opposite mirror
parity. The condition that the center of the cross be gapped imposes the
requirement that the net number of chiral modes, weighed with propagation
direction, is zero. Hence, there is an extrinsic $\ZZ^3$ classification of
$C_{2v}$-compatible hinge modes. To obtain the intrinsic second-order phases,
configurations of hinge modes that differ by a change of surface termination
must be divided out. Noting that the four surfaces of a $C_{2v}$-symmetric
crystal admits a ``decoration'' with quantum Hall phases, which gives a pair of
co-propagating opposite-parity hinge modes at each mirror-symmetric hinge, see
Fig.~\ref{fig:decoration_C2v_d3}b, we find that a classifying group $\ZZ^2$ of
intrinsic second-order phases remains. Generator Hamiltonians for these
second-order phases are
\begin{align}
  H_{(3,\pm)}' =&\, \rho_0 \tau_3 (3 - m - \sum_{i=x,y,z} \cos k_i)   
  \nonumber \\ &\, \mbox{}
  \pm \rho_1 \tau_1 \sin k_x + \rho_3 \tau_1 \sin k_y + \rho_0 \tau_2 \sin k_z
  \nonumber
\end{align}
with $0 < m < 2$ and representations $U({\cal M}_x) = i \rho_3 \tau_0,$ $U({\cal M}_y) = i \rho_1 \tau_0$, $\mathcal{P}=\rho_0 \tau_1 K$. 
%The plane perpendicular to both mirror axis does not allow a second-order boundary signature. Possible third-order boundary signatures from this plane can be removed with a decoration with Kitaev chains.

The weak phases in this symmetry class can be constructed as stacks of one or
two-dimensional topological phases. We find the following possibilities:
\begin{itemize}
\item Stack of two-dimensional $C_{2v}$-symmetric $xy$ planes in the $z$ direction with second-order topology. Such planes have a $\ZZ_2^2$ classification, see App.\  \ref{app:class_2d_C2v}. These phases are labeled ``$(2;z)$'' in Fig.~\ref{tab:ex_3d_C2v_A2_D}.
\item Stack of two-dimensional ``$C_{2v}\vert C_{s}$''-symmetric $yz$ or $xz$ planes in the $x$ or $y$ direction, respectively. These planes have a $\ZZ$ classification, corresponding to a first-order two-dimensional topological superconductor with counterpropagating Majorana modes in opposite in-plane mirror eigensectors, as shown in App.\ \ref{app:class_2d_C2v/Cs}. These phases are labeled ``$(2;x)$'' and ``$(2;y)$'' in Fig.~\ref{tab:ex_3d_C2v_A2_D}. Superconductors with even Chern number in a mirror plane can be adiabatically deformed to Chern insulators with vanishing superconducting correlations.
\item Stacks of one-dimensional ``$C_{2v}\vert C_{2v}$''-symmetric $z$ lines in the $x$ and $y$ directions. These have a $\ZZ_2$ classification, corresponding to a one-dimensional topological superconductor with a pair of Majorana bound states at each end as shown in App.\ \ref{app:class_1d_C2v/Cs}. These phases are labeled ``$(1;x;y)$'' in Fig.~\ref{tab:ex_3d_C2v_A2_D}.
\item Stacks of one-dimensional ``$C_{2v}\vert C_s$''-symmetric $y$ or $x$
	lines in the $x$ and $z$ and in the $y$ and $z$ directions,
	respectively. These, too, have a $\ZZ_2$ classification, corresponding
	to a one-dimensional topological superconductor with a pair of Majorana
	bound states at each end as shown in App.\ \ref{app:class_1d_C2v/Cs}.
	These phases are labeled ``$(1;x;z)$'' and ``$(1;y;z)^\prime$ in
	Fig.~\ref{tab:ex_3d_C2v_A2_D}.
\end{itemize}
The complete classifying group is hence 
\begin{equation}
  \mathcal{K}_{\classD}[C_{2v}, A_2,d=3] = \ZZ_2^5 \times \ZZ^4.
\end{equation}

{\it Representations $\Theta = B_{1,2}$.---} The representations $B_1$ and
$B_2$ are related by a rotation. For concreteness, we discuss the $B_2$ case in
the following. In this phase, only one pair of mirror-symmetric hinges allows
for intrinsic chiral Majorana modes, whereas any hinge modes appearing at the
other pair of mirror-symmetric hinges can always be removed by adding a
suitable decoration to the four mirror-related surfaces, as shown schematically
in Fig.~\ref{fig:decoration_C2v_d3}b. As a result, for the discussion of
second-order phases, the three-dimensional crystal may be deformed to a
two-dimensional one with $C_{2v}\vert C_s$ symmetry. As shown in
App.~\ref{app:class_2d_C2v/Cs}, there exists a strong two-dimensional phase
with counterpropagating Majorana edge modes with opposite mirror parity. This
phase corresponds to a second-order phase of the three-dimensional crystal. One
verifies that this second-order phase is intrinsic, since a surface decoration
with quantum Hall planes is forbidden by the perpendicular mirror symmetry. A
generator for the strong second-order phase is

\begin{align}
H_{(3)}'(\vk) =&\, \mu_0 \rho_0 \tau_3 (3 - m - \cos k_x - \cos k_y - \cos k_z)
  \nonumber \\ &\, \mbox{} + \mu_3 \rho_0 \tau_1 \sin k_y + \mu_0 \rho_0 \tau_2 \sin k_z + \mu_2 \rho_2 \tau_1 \sin k_x \nonumber
\end{align}
with $0 < m < 2$ and the representations $U({\cal M}_x) = i \mu_3 \rho_0 \tau_0 $, $U({\cal M}_y) = i \mu_2 \rho_0 \tau_0$,
$\mathcal{P}= \mu_0 \rho_0 \tau_1 K$. Second order topological superconductors with a pairs of counterpropagating Majorana modes within in a mirror plane can be adiabatically deformed to normal-state second-order topological insulators with chiral hinge mode.

The weak phases in this symmetry class can be constructed as stacks of one- or two-dimensional topological phases:
\begin{itemize}
\item A stack of two-dimensional $xy$ planes with $C_{2v}$ symmetry in the $z$
	direction is not possible, since there are no strong phases with
	$C_{2v}$ symmetry and $\Theta = B_2$ in two dimensions, see
	App.~\ref{app:class_2d_C2v}.
\item A stack of two-dimensional $xz$ planes with ``$C_{2v}\vert C_s$'' symmetry in the
	$y$ direction. These planes have a $\ZZ_2$ classification,
	corresponding to a second-order topological superconducting phase with
	a pair of Majorana bound states at a corner bisected by the mirror
	${\cal M}_x$ plane, as shown in App.~\ref{app:class_2d_C2v/Cs}. These
	phases are labeled ``$(2;y)$'' in Fig.~\ref{tab:ex_3d_C2v_B2_D}.
\item A stack of two-dimensional $yz$ planes with ``$C_{2v}\vert C_s$'' symmetry in
	the $x$ directions. These planes have a $\ZZ$ classification,
	corresponding to a first-order topological superconducting phase with
	counterpropagating Majorana modes in opposite in-plane mirror
	eigensectors, as shown in Sec.~\ref{app:class_2d_C2v/Cs}. These
	phases are labeled ``$(2;x)$'' in Fig.~\ref{tab:ex_3d_C2v_B2_D}. Superconductors with even Chern number in a mirror plane can be adiabatically deformed to normal-state Chern insulators.
\item A stack of one-dimensional $z$ lines with ``$C_{2v}\vert C_{2v}$'' symmetry in the
	$x$ and $y$ directions. These lines have a $\ZZ_2$ classification,
	corresponding to a one-dimensional topological superconductor with a
	pair of Majorana bound states at each end, as shown in
	Sec.~\ref{app:class_1d_C2v/C2v}. These phases are labeled ``$(1;x,y)$''
	in Fig.~\ref{tab:ex_3d_C2v_B2_D}.
\item A stack of one-dimensional $x$ lines with ``$C_{2v}\vert C_s$'' symmetry in the
	$y$ and $z$ directions does not give a weak phase, since there are no
	appropriate strong phases in one dimension, see App.~\ref{app:class_1d_C2v/Cs}.
\item A stack of one-dimensional $y$ lines with ``$C_{2v}\vert C_s$'' symmetry
	in the $x$ and $z$ directions. These lines have a $\ZZ_2$
	classification, corresponding to a one dimensional topological
	superconducting phase with a pair of Majorana bound states at each end,
	as shown in App.\ \ref{app:class_1d_C2v/Cs}. These phases are labeled
	``$(1;x, z)$'' in Fig.~\ref{tab:ex_3d_C2v_B2_D}.
\end{itemize}
The complete classifying group is hence 
\begin{equation}
  \mathcal{K}_{\classD}[C_{2v}, B_{1,2},d=3] = \ZZ_2^3 \times \ZZ^2.
\end{equation}

%The mirror ${\cal M}_y$ plane has classifying group $\mathcal{K}_{\classD}[C_{2v}/C_s, B_1, d=2] = \ZZ \times \ZZ_2$ corresponding to a first and second-order TSC that can be stacked in the $y$ direction. Furthermore, one dimensional TSC pointing in the $z$ direction with classifying group $\mathcal{K}_{\classD}[C_{2v}/C_{2v}, B_1, d=1] = \ZZ_2$ and in the $x$ direction with classifying group $\mathcal{K}_{\classD}[C_{2v}/C_{s}, B_1, d=1] = \ZZ_2$ can be stacked. 
%The total classifying group is $\mathcal{K}_{\classD}[C_{2v}, \{B_1, B_2\}, d=3] = \ZZ_2^3 \times \ZZ^2$.

\subsection{Three dimensions, class D, $C_4$}
Three dimensional, fourfold rotation-symmetric topological superconductors can
not have a first-order boundary signatures, as there are no first-order
topological phases in three dimensions for class D. A strong second-order
boundary signature is also forbidden as i) a rotation-symmetric chiral Majorana
mode in a plane perpendicular to the rotation axis can be shrunk to a point and
ii) planes parallel to the rotation axis satisfy a $(x,z) \to (-x,z)$ symmetry,
which forbids a chiral mode.  

In the following we determine whether a third-order boundary signature may
exist at the rotation axis.

{\it Representations $\Theta = A$ and $\Theta = B$.---}
For $\Theta = A$ or $\Theta = B$ the irreducible representations of fourfold rotation are exchanged under particle-hole conjugation as shown in Table~\ref{tab:classD_0d}. Each pair is effectively in class A, forbidding a one dimensional topological phase with boundary signature along rotation axis. Hence there are no third-order boundary signatures in these representations. Accounting for weak phases stacked in the $z$ direction (parallel to the rotation axis), we thus find that
\begin{align}
  {\cal K}_{\classD}[C_4,A,d=3] =&\,
  {\cal K}_{\classD}[C_4,B,d=3] \nonumber \\ =&\, \ZZ_2^2 \times \ZZ,
\end{align}
where we used the classification results for two-dimensional phases obtained in
App.~\ref{app:class_2d_C4}.

{\it Representations $\Theta = {}^1 E$ and $\Theta = {}^2 E$.---}
As in the main text, the discussion for the two representations $\Theta = {}^1
E$ or $\Theta = {}^2 E$ are analogous. We focus on the $\Theta = {}^1 E$
representation in the following.
As shown in Table~\ref{tab:classD_0d}, the $j=\frac{1}{2}$ and $j=\frac{5}{2}$
rotation eigenspaces are invariant under particle-hole conjugation while the
$j=\frac{3}{2}, \frac{7}{2}$ eigenspaces are exchanged in this symmetry class.
The $j=\frac{1}{2}$ and $j=\frac{5}{2}$ eigenspaces belong to class D allowing
stable zero-dimensional gapless Majorana states. However, below we show that a
three-dimensional superconductor with a single Majorana fermion on the fourfold
rotation axis is an extrinsic (non-anomalous) third-order phase.

\begin{figure}
\includegraphics[width=0.8\columnwidth]{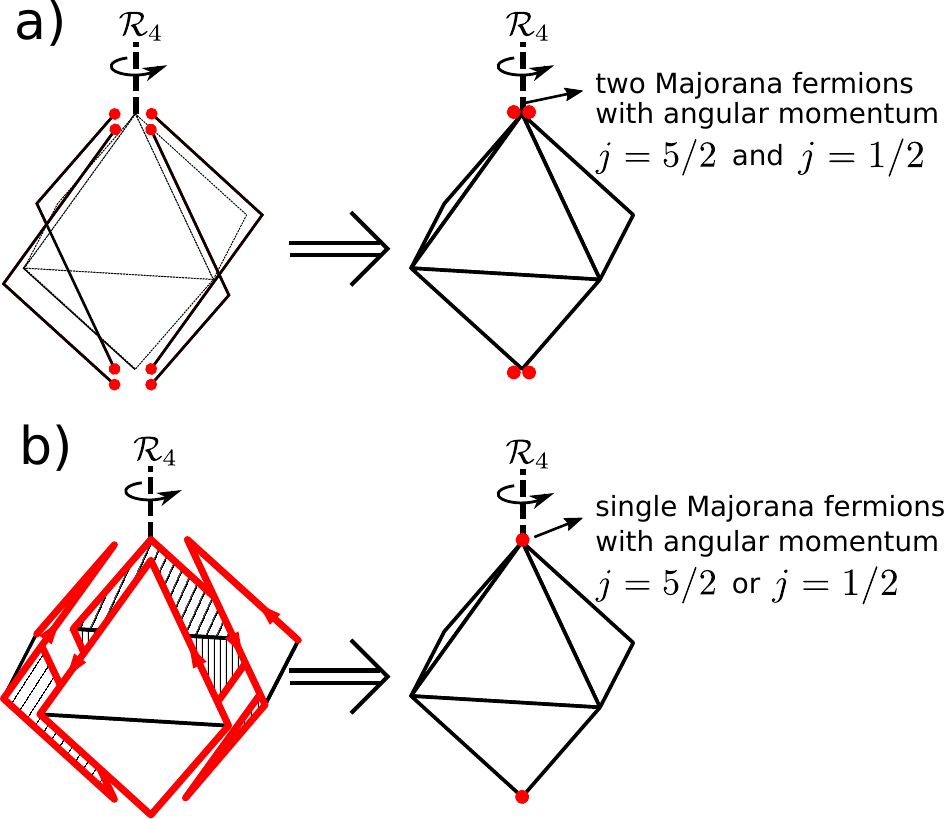}
\caption{a) Decoration consisting of four copies of one-dimensional Kitaev
	chain. This decoration results in two Majorana fermions with angular
momentum $j=\frac{1}{2}$ and $j=\frac{5}{2}$. b) Decoration consisting of four copies of
two-dimensional $p$-wave superconductors, resulting in single Majorana fermion
with angular momentum either $j=\frac{1}{2}$ or $j=\frac{5}{2}$. 
\label{fig:decoC4}}
\end{figure}

We consider two kinds of decorations,~\cite{trifunovic2019} see
Fig.~\ref{fig:decoC4}. The first kind consists of four copies of
one-dimensional Kitaev chains placed on the crystal surface and beginning and
ending on the rotation axis. The four Kitaev chains are related to each other
by a fourfold rotation, which results in four Majorana end states, all with
different angular momentum $j$. Majorana end states with $j=\frac{3}{2}$ and $j=\frac{7}{2}$
are not stable and can be gapped out, while Majorana stats with $j=\frac{1}{2}$ and
$j=\frac{5}{2}$ remain. The second type of decoration consists of covering the crystal
surface by four copies of a two-dimensional topological superconductor, such
that the four two-dimensional superconductors are mapped onto each other by
$C_4$ symmetry. Majorana modes running along the surface can be gapped out,
Fig.~\ref{fig:decoC4}b. Since $\Theta = {}^1 E$ or $\Theta = {}^2 E$
representations correspond to an odd-parity superconductor, the resulting
$C_4$-symmetric spinful superconductor needs to have a vortex at the rotation
axis.~\cite{trifunovic2019} We therefore conclude that this type of decoration
gives rise to a single Majorana bound state at the rotation axis, which needs
to have angular momentum either $j=\frac{1}{2}$ or $j=\frac{5}{2}$. The combination of the two
types of decorations described here can account for all configurations of
Majorana corner states compatible with the $C_4$ rotation symmetry. We conclude
that there are no intrinsic third-order anomalous boundary states in this
class.

We conclude that the only possible topological phases in three dimensions are weak phases:
\begin{itemize}
\item A stack of $xy$ planes with $C_4$ symmetry in the $z$ direction allows for a Chern superconductor with even Chern number, see App.\ \ref{app:class_2d_C4}. This phase is labeled ``$(2;z)$'' in Fig.~\ref{tab:ex_3d_C4_E_D}.
\item A stack of $xz$ and $yz$ planes related by fourfold rotation symmetry stacked in both $x$ and $y$ direction. As the rotation axis lies in the intersection of the two planes, their classifying group is $\mathcal{K}_{(0,1,0)}[C_{s}, A'',d=2] = \ZZ_2$ allowing a second-order topological superconductor. The configuration has pairs of Majorana bound states at the rotation axis. This phase is labeled ``$(2;x,y)$'' in Fig.~\ref{tab:ex_3d_C4_E_D}.
\item There is no stack of one dimensional superconductors pointing in the $x$ and $y$ direction, related by fourfold rotation symmetry, as the corresponding classifying group $\mathcal{K}_{(0,1,0)}[C_{s}, A'', d=1] = 0$ is trivial.
\item A stack of one dimensional $z$ lines within the rotation axis in the $x$ and $y$ direction. These lines have a classifying group $\mathcal{K}_{(0,1,0)}[C_{4}, {}^1 E, d=1] = \ZZ_2^2$ corresponding to one dimensional TSC with Majorana bound states in the class D rotation subspaces $j=\frac{1}{2}, \frac{5}{2}$. The phases are labeled ``$(1,j;x, y)$'' in Fig.~\ref{tab:ex_3d_C4_E_D}.
\end{itemize}
The complete classifying group is hence 
\begin{equation}
  \mathcal{K}_{\classD}[C_{4}, {}^{1,2} E, d = 3] = \ZZ_2^3 \times \ZZ .
\end{equation}

\section{$1d$ compatibility relation for class CI with inversion symmetry} 
\label{app:c}

We here discuss how to obtain the compatibility relation (\ref{eq:Ch_Ci_3d_CI})
of Sec.~\ref{sec:3d_CI_Ci}. For definiteness, we choose the representations
\begin{equation}
  {\cal T} = \rho_0 \tau_0 K,\ \ {\cal P} = \rho_0 \tau_2 K,\ \ {\cal I} = \rho_0 \tau_3.
\end{equation}
The following combinations of symmetries are local in $\vk$:
\begin{equation}
  {\cal IT} = \rho_0 \tau_3 K,\ \ {\cal I P} = \rho_0 \tau_1 K,
\end{equation}
as well as their product, which is a chiral antisymmetry. To construct the
compatibility relation (\ref{eq:Ch_Ci_3d_CI}) we consider the one-parameter
family of one-dimensinal Hamiltonians $H_t(k)$, defined by restricting $H(\vk)$
to the line $k_{x} = 0$, $k_{y} = t$, $k_{z} = k$. For $t=0$ and $t=\pi$, the
one-dimensional Hamiltonian $H_{t}(k)$ is a one-dimensional Hamiltonian in
class CI with inversion symmetry and the $A_u$ representation. The topological
invariants at high-symmetry momenta for this one-dimensional Hamiltonian
coincide with the corresponding band labels defined for the full
three-dimensional Hamiltonian $H(\vk)$ and satisfy the same $0d$ compatibility
relations, see Secs.~\ref{sec:3d_C_Ci} and \ref{sec:3d_CI_Ci}. For generic $0
< t < \pi$, $H_t(k)$ is a one-dimensional Hamiltonian that has the local-in-$k$
symmetries ${\cal IT}$ and ${\cal IP}$ only. Such a one-parameter family has a
strong $\ZZ_2$ invariant,~\cite{teo2010} which we call $\mathfrak{S}$. We can
construct a compatibility relation by relating $\mathfrak{S}$ to the
topological band labels of $H_t(k)$ at $t=0$ and $t=\pi$. We find that this can
be accomplished without knowledge of a general expression for the strong invariant
$\mathfrak{S}$.

Before we can construct such a compatibility relation, it is necessary to
obtain the full classification of the inversion-symmetric Hamiltonians $H_t(k)$
at $t=0$ and $t=\pi$, including topologically nontrivial atomic-limit states.
We recall that the topological band labels for this case are
$\{\mathfrak{n}_+^{(0,t,0)}, \mathfrak{N}_+^{(0,t,\pi)}\}$, where
\begin{equation}
  2 \mathfrak{n}_+^{(0,t,0)} = \mathfrak{N}_+^{(0,t,0)} - \mathfrak{N}_+^{(0,t,\pi)}\ \ t=0,\pi,
\end{equation}
see the discussion in Sec.~\ref{sec:3d_C_Ci}.  It follows that the group of
topological band labels is $\ZZ \times 2\ZZ$, where we use $2 \ZZ$ to indicate
the integers spanned by the label $\mathfrak{n}_+^{(0,t,\pi)}$. Interpreting
the topological band labels in terms of a topological classification, we note
that the factor $\ZZ$ corresponds to weak phases, whereas the factor $2 \ZZ$
describes strong one-dimensional atomic-limit phases.~\cite{trifunovic2019} We
use $ \left. K_{t} \right|_{t = 0,\pi} \simeq 2 \ZZ$, to denote the strong
phases of the model at $t=0$ and $t=\pi$ and note that the topological band
label $2\mathfrak{n}_+^{(0,t,0)}$ can be used as the corresponding topological
invariant.

To obtain the compatibility relation, we have to determine, what values of
$\mathfrak{n}_+^{(0,t,0)}$ are compatible with a given value of
$\mathfrak{S}$. Such a problem requires understanding the homomorphism
\begin{align*}
  \left. K_{t} \right|_{t = 0,\pi} \rightarrow &\ \left. K_{t} \right|_{0 < t < \pi},
\end{align*}
where $\left. K_{t} \right|_{0 < t < \pi} \simeq \ZZ_2$ is the group
classifying strong phases of $H_t(k)$ for generic $t$. To construct this
homomorphism, we notice that
\begin{equation}
  H(k) =  \rho_0 \tau_3 (1 - m - \cos k) + \rho_2 \tau_1 \sin k  \label{eq:example}
\end{equation}
with $0 < m < 2$ is an example of a Hamiltonian that satisfies ${\cal T}$, ${\cal P}$, and
${\cal I}$ symmetries and that is nontrivial if only the local-in-$k$
symmetries ${\cal IT}$ and ${\cal IP}$ are kept, {\it i.e.}, it has invariant
$\mathfrak{S} = 1$. (Although no general expression for $\mathfrak{S}$ is
available, $\mathfrak{S}$ can be calculated as the parity of a winding number
if $H(k)$ is a $4 \times 4$ matrix.) It follows that the mapping $\left.
K_{t} \right|_{t = 0,\pi} \rightarrow \ \left. K_{t} \right|_{0 < t < \pi}$
must be surjective. One verifies that the Hamiltonian~(\ref{eq:example}) has
topological band label $\mathfrak{n}_+^{(0)} = 1$. Since the trivial phase
has $\mathfrak{n}_+^{(0)} = 0$ and since $\mathfrak{S}$ has a $\ZZ_2$ group
structure, it directly follows that
\begin{align}
  \mathfrak{S} =&\, \mathfrak{n}_+^{(0,0,0)} \mod 2 \nonumber \\
  =&\, \mathfrak{n}_+^{(0,\pi,0)} \mod 2. \label{eq:compatibility}
\end{align}
This is one of the compatibility relations of Eq.~(\ref{eq:Ch_Ci_3d_CI}). The
other relations follow in analogous manner by considering other appropriately
chosen families of one-dimensional Hamiltonians $H_t(k)$.

\bibliography{refs}
\clearpage

\widetext

\section*{Supplementary material}
\setcounter{table}{0}
\renewcommand{\thetable}{S-\Roman{table}}

\begin{table*}[h]
\begin{tabular}{cc|ccccc|c}
 &  & \multicolumn{5}{c|}{BL} & $\SI$\tabularnewline
$\mathcal{K}_{i}'\subseteq\mathcal{K}_{i}$ & Phase & $\mathfrak{N}_{\frac{1}{2}}^{(0,0)}$ & $\mathfrak{N}_{\frac{5}{2}}^{(0,0)}$ & $\mathfrak{N}_{\frac{1}{2}}^{(\pi,0)}$ & $\mathfrak{N}_{\frac{1}{2}}^{(\pi,\pi)}$ & $\mathfrak{N}_{\frac{5}{2}}^{(\pi,\pi)}$ & $\mathbb{Z}_{2}\otimes\mathbb{Z}_{8}$\tabularnewline
\hline 
- & $\vec{x}=(0,0),\ j=\frac{1}{2}$ & 1 & 0 & 1 & 1 & 0 & $\idn$\tabularnewline
- & $\vec{x}=(0,0),\ j=\frac{5}{2}$ & 0 & 1 & 1 & 0 & 0 & $\idn$\tabularnewline
- & $\vec{x}=(\frac{1}{2},\frac{1}{2}),\ j=\frac{1}{2}$ & 1 & 0 & -1 & 0 & 0 & $\idn$\tabularnewline
- & $\vec{x}=(\frac{1}{2},\frac{1}{2}),\ j=\frac{5}{2}$ & 0 & 1 & -1 & 1 & 0 & $\idn$\tabularnewline
- & $\vec{x}=(\frac{1}{2},0),\ j=\frac{1}{2}$ & 1 & 1 & 0 & -1 & -1 & $\idn$\tabularnewline
\hline 
$\mathbb{Z}_{2}$ & $(1;x,y)$ %\includegraphics[scale=0.5]{figs/C4/2d_1xy.pdf} 
& 1 & 1 & 1 & 0 & 0 & $\gen_{1;x,y}^{(2)}$\tabularnewline
$0\subseteq\mathbb{Z}$ & $(2)$ %\includegraphics[scale=0.5]{figs/C4/2d_2.pdf} 
& 0 & 0 & 1 & 1 & 0 & $\gen_{2}^{(8)}$\tabularnewline
$\mathbb{Z}_{2}\subseteq\mathbb{Z}_{2}$ & $(2)$' % \includegraphics[scale=0.5]{figs/C4/2d_2p.pdf} 
& 2 & 2 & 0 & 0 & 0 & $4\gen_{2}^{(8)}$\tabularnewline
\end{tabular}
\caption{Band labels and symmetry-based indicators for atomic-limit Hamiltonians obtained by placing 0d generators for representation $j$ at Wyckoff position $\vx$ for the symmetry group $C_{4}$ in two dimensions, class D with representation $\Theta = A$ or $\Theta = B$ (upper five rows) and band labels for the generators of the weak, Chern and second-order phases (lower three rows). 
\label{tab:SM_ex_C4_A_bandlabels}}
\end{table*}

\begin{table*}[h]
\begin{tabular}{cc|cccccccc|c}
 &  & \multicolumn{8}{c|}{BL} & $\SI$\tabularnewline
$\mathcal{K}_{i}'\subseteq\mathcal{K}_{i}$ & Phase & $\mathfrak{p}_{\frac{1}{2}}^{(0,0)}$ & $\mathfrak{p}_{\frac{5}{2}}^{(0,0)}$ & $\mathfrak{N}_{\frac{3}{2}}^{(0,0)}$ & $\mathfrak{p}_{\frac{1}{2}}^{(\pi,0)}$ & $\mathfrak{p}_{\frac{3}{2}}^{(\pi,0)}$ & $\mathfrak{p}_{\frac{1}{2}}^{(\pi,\pi)}$ & $\mathfrak{p}_{\frac{5}{2}}^{(\pi,\pi)}$ & $\mathfrak{N}_{\frac{3}{2}}^{(\pi,\pi)}$ & $\mathbb{Z}_{4}$\tabularnewline
\hline 
- & $\vec{x}=(0,0),\ j=\frac{1}{2}$ & 1 & 0 & 0 & 1 & 0 & 1 & 0 & 0 & $\idn$\tabularnewline
- & $\vec{x}=(0,0),\ j=\frac{5}{2}$ & 0 & 1 & 0 & 1 & 0 & 0 & 1 & 0 & $\idn$\tabularnewline
- & $\vec{x}=(0,0),\ j=\frac{3}{2}$ & 0 & 0 & 1 & 0 & 1 & 0 & 0 & 1 & $\idn$\tabularnewline
- & $\vec{x}=(\frac{1}{2},\frac{1}{2}),\ j=\frac{1}{2}$ & 1 & 0 & 0 & 0 & 1 & 0 & 1 & 0 & $\idn$\tabularnewline
- & $\vec{x}=(\frac{1}{2},\frac{1}{2}),\ j=\frac{5}{2}$ & 0 & 1 & 0 & 0 & 1 & 1 & 0 & 0 & $\idn$\tabularnewline
- & $\vec{x}=(\frac{1}{2},\frac{1}{2}),\ j=\frac{3}{2}$ & 0 & 0 & 1 & 1 & 0 & 0 & 0 & -1 & $\idn$\tabularnewline
- & $\vec{x}=(\frac{1}{2},0),\ j=\frac{1}{2}$ & 1 & 1 & 0 & 1 & 1 & 0 & 0 & 0 & $\idn$\tabularnewline
- & $\vec{x}=(\frac{1}{2},0),\ j=\frac{3}{2}$ & 0 & 0 & 0 & 1 & 1 & 1 & 1 & 0 & $\idn$\tabularnewline
\hline 
$0\subseteq\mathbb{Z}$ & $(2) \quad $ %\includegraphics[scale=0.5]{figs/C4/2d_2_Ch2.pdf} 
& 0 & 0 & 0 & 1 & 1 & 1 & 0 & 1 & $\gen_{2}^{(4)}$\tabularnewline
\end{tabular}
\caption{Band labels and symmetry-based indicators for atomic-limit Hamiltonians obtained by placing 0d generators for representation $j$ at Wyckoff position $\vx$ for the symmetry group $C_{4}$ in two dimensions, class D  with representation $\Theta = {}^1 E$ or $\Theta = {}^2 E$ (upper eight rows) and band labels for the generator of the Chern phase (lowest row). 
\label{tab:SM_ex_C4_E_bandlabels}}
\end{table*}

%
%\section{Examples: Three dimensions}

\begin{table*}[h]
\begin{tabular}{cc|cccccccc|c}
 &  & \multicolumn{8}{c|}{BL} & $\SI$\tabularnewline
$\mathcal{K}_{i}''\subseteq\mathcal{K}_{i}'\subseteq\mathcal{K}_{i}$ & Phase & $\mathfrak{p}^{(0,0,0)}$ & $\mathfrak{p}^{(\pi,0,0)}$ & $\mathfrak{p}^{(0,\pi,0)}$ & $\mathfrak{p}^{(\pi,\pi,0)}$ & $\mathfrak{p}^{(0,0,\pi)}$ & $\mathfrak{p}^{(\pi,0,\pi)}$ & $\mathfrak{p}^{(0,\pi,\pi)}$ & $\mathfrak{p}^{(\pi,\pi,\pi)}$ & $\mathbb{Z}_{2}^{7}$\tabularnewline
\hline 
- & $\vec{x}=(0,0,0)$ & 1 & 1 & 1 & 1 & 1 & 1 & 1 & 1 & $\idn$\tabularnewline
\hline 
$\mathbb{Z}_{2}$ & $(1;y,z)$ & 1 & 0 & 1 & 0 & 1 & 0 & 1 & 0 & $\gen_{1;y,z}^{(2)}$\tabularnewline
$\mathbb{Z}_{2}$ & $(1;x,z)$ & 1 & 1 & 0 & 0 & 1 & 1 & 0 & 0 & $\gen_{1;x,z}^{(2)}$\tabularnewline
$\mathbb{Z}_{2}$ & $(1;x,y)$ & 1 & 1 & 1 & 1 & 0 & 0 & 0 & 0 & $\gen_{1;x,y}^{(2)}$\tabularnewline
$0\subseteq\mathbb{Z}$ & $(2;z)$ & 1 & 0 & 0 & 0 & 1 & 0 & 0 & 0 & $\gen_{2;z}^{(2)}$\tabularnewline
$0\subseteq\mathbb{Z}$ & $(2;y)$ & 1 & 0 & 1 & 0 & 0 & 0 & 0 & 0 & $\gen_{2;y}^{(2)}$\tabularnewline
$0\subseteq\mathbb{Z}$ & $(2;x)$ & 1 & 1 & 0 & 0 & 0 & 0 & 0 & 0 & $\gen_{2;x}^{(2)}$\tabularnewline
\end{tabular}
\caption[]{Topological band labels of three-dimensional superconductors in tenfold-way class D with translation symmetry only (point group $C_1$). 
\label{tab:SM_ex_3d_C1_D}}
\end{table*}

\begin{table*}
\begin{tabular}{cc|cccccccc|c}
 &  & \multicolumn{8}{c|}{BS} & $\text{SI}$\tabularnewline
$\mathcal{K}_{i}''\subseteq\mathcal{K}_{i}'\subseteq\mathcal{K}_{i}$ & Phase & $\mathfrak{p}_{(+,-)}^{(0,0,0)}$ & $\mathfrak{p}_{(+,-)}^{(\pi,0,0)}$ & $\mathfrak{p}_{(+,-)}^{(0,\pi,0)}$ & $\mathfrak{p}_{(+,-)}^{(\pi,\pi,0)}$ & $\mathfrak{p}_{(+,-)}^{(0,0,\pi)}$ & $\mathfrak{p}_{(+,-)}^{(\pi,0,\pi)}$ & $\mathfrak{p}_{(+,-)}^{(0,\pi,\pi)}$ & $\mathfrak{p}_{(+,-)}^{(\pi,\pi,\pi)}$ & $\mathbb{Z}_{2}^{4}$\tabularnewline
\hline 
- & $\vec{x}=(0,0,0),\ \alpha=\begin{cases}
+\\
-
\end{cases}$ & $\begin{array}{c}
(1,0)\\
(0,1)
\end{array}$ & $\begin{array}{c}
(1,0)\\
(0,1)
\end{array}$ & $\begin{array}{c}
(1,0)\\
(0,1)
\end{array}$ & $\begin{array}{c}
(1,0)\\
(0,1)
\end{array}$ & $\begin{array}{c}
(1,0)\\
(0,1)
\end{array}$ & $\begin{array}{c}
(1,0)\\
(0,1)
\end{array}$ & $\begin{array}{c}
(1,0)\\
(0,1)
\end{array}$ & $\begin{array}{c}
(1,0)\\
(0,1)
\end{array}$ & $\begin{array}{c}
\idn\\
\idn
\end{array}$\tabularnewline
- & $\vec{x}=(\frac{1}{2},0,0),\ \alpha=\begin{cases}
+\\
-
\end{cases}$ & $\begin{array}{c}
(1,0)\\
(0,1)
\end{array}$ & $\begin{array}{c}
(0,1)\\
(1,0)
\end{array}$ & $\begin{array}{c}
(1,0)\\
(0,1)
\end{array}$ & $\begin{array}{c}
(0,1)\\
(1,0)
\end{array}$ & $\begin{array}{c}
(1,0)\\
(0,1)
\end{array}$ & $\begin{array}{c}
(0,1)\\
(1,0)
\end{array}$ & $\begin{array}{c}
(1,0)\\
(0,1)
\end{array}$ & $\begin{array}{c}
(0,1)\\
(1,0)
\end{array}$ & $\begin{array}{c}
\idn\\
\idn
\end{array}$\tabularnewline
- & $\vec{x}=(0,\frac{1}{2},0),\ \alpha=\begin{cases}
+\\
-
\end{cases}$ & $\begin{array}{c}
(1,0)\\
(0,1)
\end{array}$ & $\begin{array}{c}
(1,0)\\
(0,1)
\end{array}$ & $\begin{array}{c}
(0,1)\\
(1,0)
\end{array}$ & $\begin{array}{c}
(0,1)\\
(1,0)
\end{array}$ & $\begin{array}{c}
(1,0)\\
(0,1)
\end{array}$ & $\begin{array}{c}
(1,0)\\
(0,1)
\end{array}$ & $\begin{array}{c}
(0,1)\\
(1,0)
\end{array}$ & $\begin{array}{c}
(0,1)\\
(1,0)
\end{array}$ & $\begin{array}{c}
\idn\\
\idn
\end{array}$\tabularnewline
- & $\vec{x}=(\frac{1}{2},\frac{1}{2},0),\ \alpha=\begin{cases}
+\\
-
\end{cases}$ & $\begin{array}{c}
(1,0)\\
(0,1)
\end{array}$ & $\begin{array}{c}
(0,1)\\
(1,0)
\end{array}$ & $\begin{array}{c}
(0,1)\\
(1,0)
\end{array}$ & $\begin{array}{c}
(1,0)\\
(0,1)
\end{array}$ & $\begin{array}{c}
(1,0)\\
(0,1)
\end{array}$ & $\begin{array}{c}
(0,1)\\
(1,0)
\end{array}$ & $\begin{array}{c}
(0,1)\\
(1,0)
\end{array}$ & $\begin{array}{c}
(1,0)\\
(0,1)
\end{array}$ & $\begin{array}{c}
\idn\\
\idn
\end{array}$\tabularnewline
- & $\vec{x}=(0,0,\frac{1}{2}),\ \alpha=\begin{cases}
+\\
-
\end{cases}$ & $\begin{array}{c}
(1,0)\\
(0,1)
\end{array}$ & $\begin{array}{c}
(1,0)\\
(0,1)
\end{array}$ & $\begin{array}{c}
(1,0)\\
(0,1)
\end{array}$ & $\begin{array}{c}
(1,0)\\
(0,1)
\end{array}$ & $\begin{array}{c}
(0,1)\\
(1,0)
\end{array}$ & $\begin{array}{c}
(0,1)\\
(1,0)
\end{array}$ & $\begin{array}{c}
(0,1)\\
(1,0)
\end{array}$ & $\begin{array}{c}
(0,1)\\
(1,0)
\end{array}$ & $\begin{array}{c}
\idn\\
\idn
\end{array}$\tabularnewline
- & $\vec{x}=(\frac{1}{2},0,\frac{1}{2}),\ \alpha=\begin{cases}
+\\
-
\end{cases}$ & $\begin{array}{c}
(1,0)\\
(0,1)
\end{array}$ & $\begin{array}{c}
(0,1)\\
(1,0)
\end{array}$ & $\begin{array}{c}
(1,0)\\
(0,1)
\end{array}$ & $\begin{array}{c}
(0,1)\\
(1,0)
\end{array}$ & $\begin{array}{c}
(0,1)\\
(1,0)
\end{array}$ & $\begin{array}{c}
(1,0)\\
(0,1)
\end{array}$ & $\begin{array}{c}
(0,1)\\
(1,0)
\end{array}$ & $\begin{array}{c}
(1,0)\\
(0,1)
\end{array}$ & $\begin{array}{c}
\idn\\
\idn
\end{array}$\tabularnewline
- & $\vec{x}=(0,\frac{1}{2},\frac{1}{2}),\ \alpha=\begin{cases}
+\\
-
\end{cases}$ & $\begin{array}{c}
(1,0)\\
(0,1)
\end{array}$ & $\begin{array}{c}
(1,0)\\
(0,1)
\end{array}$ & $\begin{array}{c}
(0,1)\\
(1,0)
\end{array}$ & $\begin{array}{c}
(0,1)\\
(1,0)
\end{array}$ & $\begin{array}{c}
(0,1)\\
(1,0)
\end{array}$ & $\begin{array}{c}
(0,1)\\
(1,0)
\end{array}$ & $\begin{array}{c}
(1,0)\\
(0,1)
\end{array}$ & $\begin{array}{c}
(1,0)\\
(0,1)
\end{array}$ & $\begin{array}{c}
\idn\\
\idn
\end{array}$\tabularnewline
- & $\vec{x}=(\frac{1}{2},\frac{1}{2},\frac{1}{2}),\ \alpha=\begin{cases}
+\\
-
\end{cases}$ & $\begin{array}{c}
(1,0)\\
(0,1)
\end{array}$ & $\begin{array}{c}
(0,1)\\
(1,0)
\end{array}$ & $\begin{array}{c}
(0,1)\\
(1,0)
\end{array}$ & $\begin{array}{c}
(1,0)\\
(0,1)
\end{array}$ & $\begin{array}{c}
(0,1)\\
(1,0)
\end{array}$ & $\begin{array}{c}
(1,0)\\
(0,1)
\end{array}$ & $\begin{array}{c}
(1,0)\\
(0,1)
\end{array}$ & $\begin{array}{c}
(0,1)\\
(1,0)
\end{array}$ & $\begin{array}{c}
\idn\\
\idn
\end{array}$\tabularnewline
\hline 
$0\subseteq\mathbb{Z}$ & $(2;z) \quad$ %\includegraphics[scale=0.5]{figs/Ci/3d_Ag_2z_Ch2}
& $(1,1)$ & $(0,0)$ & $(0,0)$ & $(0,0)$ & $(1,1)$ & $(0,0)$ & $(0,0)$ & $(0,0)$ & $\gen_{2;z}^{(2)}$\tabularnewline
$0\subseteq\mathbb{Z}$ & $(2;y) \quad$ %\includegraphics[scale=0.5]{figs/Ci/3d_Ag_2y_Ch2}
& $(1,1)$ & $(0,0)$ & $(1,1)$ & $(0,0)$ & $(0,0)$ & $(0,0)$ & $(0,0)$ & $(0,0)$ & $\gen_{2;y}^{(2)}$\tabularnewline
$0\subseteq\mathbb{Z}$ & $(2;x) \quad$ %\includegraphics[scale=0.5]{figs/Ci/3d_Ag_2x_Ch2}
& $(1,1)$ & $(1,1)$ & $(0,0)$ & $(0,0)$ & $(0,0)$ & $(0,0)$ & $(0,0)$ & $(0,0)$ & $\gen_{2;x}^{(2)}$\tabularnewline
\end{tabular}
\caption{Topological band labels of three-dimensional superconductors in tenfold-way class D with point group $C_{i}$ and representation $\Theta=A_g$. 
\label{tab:SM_ex_Ci_D_Ag_bandlabels}}
\end{table*}

\begin{table*}
\begin{tabular}{cc|cccccccc|c}
 & & \multicolumn{8}{c|}{BS} & $\text{SI}$\tabularnewline
$\mathcal{K}_{i}''\subseteq\mathcal{K}_{i}'\subseteq\mathcal{K}_{i}$ & Phase & $\mathfrak{N}_{+}^{(0,0,0)}$ & $\mathfrak{N}_{+}^{(\pi,0,0)}$ & $\mathfrak{N}_{+}^{(0,\pi,0)}$ & $\mathfrak{N}_{+}^{(\pi,\pi,0)}$ & $\mathfrak{N}_{+}^{(0,0,\pi)}$ & $\mathfrak{N}_{+}^{(\pi,0,\pi)}$ & $\mathfrak{N}_{+}^{(0,\pi,\pi)}$ & $\mathfrak{N}_{+}^{(\pi,\pi,\pi)}$ & $\mathbb{Z}_{2}^{3}\otimes\mathbb{Z}_{4}^{3}\otimes\mathbb{Z}_{8}$\tabularnewline
\hline 
- & $\vec{x}=(0,0,0)$ & 1 & 1 & 1 & 1 & 1 & 1 & 1 & 1 & $\idn$\tabularnewline
- & $\vec{x}=(\frac{1}{2},0,0)$ & 1 & -1 & 1 & -1 & 1 & -1 & 1 & -1 & $\idn$\tabularnewline
- & $\vec{x}=(0,\frac{1}{2},0)$ & 1 & 1 & -1 & -1 & 1 & 1 & -1 & -1 & $\idn$\tabularnewline
- & $\vec{x}=(\frac{1}{2},\frac{1}{2},0)$ & 1 & -1 & -1 & 1 & 1 & -1 & -1 & 1 & $\idn$\tabularnewline
- & $\vec{x}=(0,0,\frac{1}{2})$ & 1 & 1 & 1 & 1 & -1 & -1 & -1 & -1 & $\idn$\tabularnewline
- & $\vec{x}=(\frac{1}{2},0,\frac{1}{2})$ & 1 & -1 & 1 & -1 & -1 & 1 & -1 & 1 & $\idn$\tabularnewline
- & $\vec{x}=(0,\frac{1}{2},\frac{1}{2})$ & 1 & 1 & -1 & -1 & -1 & -1 & 1 & 1 & $\idn$\tabularnewline
- & $\vec{x}=(\frac{1}{2},\frac{1}{2},\frac{1}{2})$ & 1 & -1 & -1 & 1 & -1 & 1 & 1 & -1 & $\idn$\tabularnewline
\hline 
$\mathbb{Z}_{2}$ & $(1;y,z)$  %\includegraphics[scale=0.5]{figs/Ci/3d_1yz} 
& 1 & 0 & 1 & 0 & 1 & 0 & 1 & 0 &$\gen_{1;y,z}^{(2)}$\tabularnewline
$\mathbb{Z}_{2}$ & $(1;x,z)$  %\includegraphics[scale=0.5]{figs/Ci/3d_1xz}  
& 1 & 1 & 0 & 0 & 1 & 1 & 0 & 0 & $\gen_{1;x,z}^{(2)}$\tabularnewline
$\mathbb{Z}_{2}$ & $(1;x,y)$  %\includegraphics[scale=0.5]{figs/Ci/3d_1xy}  
& 1 & 1 & 1 & 1 & 0 & 0 & 0 & 0 & $\gen_{1;x,y}^{(2)}$\tabularnewline
$0\subseteq\mathbb{Z}$ & $(2;z)$  %\includegraphics[scale=0.5]{figs/Ci/3d_Au_2z}  
& 1 & 0 & 0 & 0 & 1 & 0 & 0 & 0 & $\gen_{2;z}^{(4)}$\tabularnewline
$0\subseteq\mathbb{Z}$ & $(2;y)$  %\includegraphics[scale=0.5]{figs/Ci/3d_Au_2y}  
& 1 & 0 & 1 & 0 & 0 & 0 & 0 & 0 & $\gen_{2;y}^{(4)}$\tabularnewline
$0\subseteq\mathbb{Z}$ & $(2;x)$  %\includegraphics[scale=0.5]{figs/Ci/3d_Au_2x}  
& 1 & 1 & 0 & 0 & 0 & 0 & 0 & 0 & $\gen_{2;x}^{(4)}$\tabularnewline
$\mathbb{Z}_{2}\subseteq\mathbb{Z}_{2}$ & $(2;z)'$  %\includegraphics[scale=0.5]{figs/Ci/3d_Au_2pz} 
& 2 & 0 & 0 & 0 & 2 & 0 & 0 & 0 & $2\gen_{2;z}^{(4)}$\tabularnewline
$\mathbb{Z}_{2}\subseteq\mathbb{Z}_{2}$ & $(2;y)'$  %\includegraphics[scale=0.5]{figs/Ci/3d_Au_2py}
& 2 & 0 & 2 & 0 & 0 & 0 & 0 & 0 & $2\gen_{2;y}^{(4)}$\tabularnewline
$\mathbb{Z}_{2}\subseteq\mathbb{Z}_{2}$ & $(2;x)'$  %\includegraphics[scale=0.5]{figs/Ci/3d_Au_2px}
& 2 & 2 & 0 & 0 & 0 & 0 & 0 & 0 & $2\gen_{2;x}^{(4)}$\tabularnewline
$\mathbb{Z}_{2}\subseteq\mathbb{Z}_{4}\subseteq\mathbb{Z}_{4}$ & $(3)'$  %\includegraphics[scale=0.5]{figs/Ci/3d_3p} 
& 2 & 0 & 0 & 0 & 0 & 0 & 0 & 0 & $2\gen_{3}^{(8)}$\tabularnewline
\end{tabular}
\caption{Topological band labels of three-dimensional superconductors in tenfold-way class D with point group $C_{i}$ and representation $\Theta=A_u$. 
\label{tab:ex_Ci_D_Au_bandlabels}}
\end{table*}

\begin{table*}
\begin{tabular}{cc|cccccccc|c}
 & & \multicolumn{8}{c|}{BS} & $\text{SI}$\tabularnewline
$\mathcal{K}_{i}''\subseteq\mathcal{K}_{i}'\subseteq\mathcal{K}_{i}$ & Phase & $\mathfrak{N}_{+}^{(0,0,0)}$ & $\mathfrak{N}_{+}^{(\pi,0,0)}$ & $\mathfrak{N}_{+}^{(0,\pi,0)}$ & $\mathfrak{N}_{+}^{(\pi,\pi,0)}$ & $\mathfrak{N}_{+}^{(0,0,\pi)}$ & $\mathfrak{N}_{+}^{(\pi,0,\pi)}$ & $\mathfrak{N}_{+}^{(0,\pi,\pi)}$ & $\mathfrak{N}_{+}^{(\pi,\pi,\pi)}$ & $\mathbb{Z}_{2}^{3}\otimes\mathbb{Z}_{4}^{3}\otimes\mathbb{Z}_{8}$\tabularnewline
\hline 
- & $\vec{x}=(0,0,0)$ & 1 & 1 & 1 & 1 & 1 & 1 & 1 & 1 & $\idn$\tabularnewline
- & $\vec{x}=(\frac{1}{2},0,0)$ & 1 & -1 & 1 & -1 & 1 & -1 & 1 & -1 & $\idn$\tabularnewline
- & $\vec{x}=(0,\frac{1}{2},0)$ & 1 & 1 & -1 & -1 & 1 & 1 & -1 & -1 & $\idn$\tabularnewline
- & $\vec{x}=(\frac{1}{2},\frac{1}{2},0)$ & 1 & -1 & -1 & 1 & 1 & -1 & -1 & 1 & $\idn$\tabularnewline
- & $\vec{x}=(0,0,\frac{1}{2})$ & 1 & 1 & 1 & 1 & -1 & -1 & -1 & -1 & $\idn$\tabularnewline
- & $\vec{x}=(\frac{1}{2},0,\frac{1}{2})$ & 1 & -1 & 1 & -1 & -1 & 1 & -1 & 1 & $\idn$\tabularnewline
- & $\vec{x}=(0,\frac{1}{2},\frac{1}{2})$ & 1 & 1 & -1 & -1 & -1 & -1 & 1 & 1 & $\idn$\tabularnewline
- & $\vec{x}=(\frac{1}{2},\frac{1}{2},\frac{1}{2})$ & 1 & -1 & -1 & 1 & -1 & 1 & 1 & -1 & $\idn$\tabularnewline
\hline 
$\mathbb{Z}_{2}$ & $(1;y,z)$  %\includegraphics[scale=0.5]{figs/Ci/3d_1yz} 
& 1 & 0 & 1 & 0 & 1 & 0 & 1 & 0 &$\gen_{1;y,z}^{(2)}$\tabularnewline
$\mathbb{Z}_{2}$ & $(1;x,z)$  %\includegraphics[scale=0.5]{figs/Ci/3d_1xz}  
& 1 & 1 & 0 & 0 & 1 & 1 & 0 & 0 & $\gen_{1;x,z}^{(2)}$\tabularnewline
$\mathbb{Z}_{2}$ & $(1;x,y)$  %\includegraphics[scale=0.5]{figs/Ci/3d_1xy}  
& 1 & 1 & 1 & 1 & 0 & 0 & 0 & 0 & $\gen_{1;x,y}^{(2)}$\tabularnewline
$\mathbb{Z}_{2}\subseteq\mathbb{Z}_{4}$ & $(2;z)$  %\includegraphics[scale=0.5]{figs/Ci/3d_Au_2z}  
& 1 & 0 & 0 & 0 & 1 & 0 & 0 & 0 & $\gen_{2;z}^{(4)}$\tabularnewline
$\mathbb{Z}_{2}\subseteq\mathbb{Z}_{4}$ & $(2;y)$  %\includegraphics[scale=0.5]{figs/Ci/3d_Au_2y}  
& 1 & 0 & 1 & 0 & 0 & 0 & 0 & 0 & $\gen_{2;y}^{(4)}$\tabularnewline
$\mathbb{Z}_{2}\subseteq\mathbb{Z}_{4}$ & $(2;x)$  %\includegraphics[scale=0.5]{figs/Ci/3d_Au_2x}  
& 1 & 1 & 0 & 0 & 0 & 0 & 0 & 0 & $\gen_{2;x}^{(4)}$\tabularnewline
$0 \subseteq 0\subseteq\mathbb{Z}$ & $(3)$  %\includegraphics[scale=0.5]{figs/Ci/3d_3p} 
& 1 & 0 & 0 & 0 & 0 & 0 & 0 & 0 & $\gen_{3}^{(8)}$\tabularnewline
$\mathbb{Z}_{2}\subseteq\mathbb{Z}_{4}\subseteq\mathbb{Z}_{4}$ & $(3)'$  %\includegraphics[scale=0.5]{figs/Ci/3d_3p} 
& 2 & 0 & 0 & 0 & 0 & 0 & 0 & 0 & $2\gen_{3}^{(8)}$\tabularnewline
\end{tabular}
\caption{Topological band labels of three-dimensional superconductors in tenfold-way class DIII with point group $C_{i}$ and representation $\Theta=A_u$. 
\label{tab:ex_Ci_DIII_Au_bandlabels}}
\end{table*}

\begin{table*}
\begin{tabular}{cc|cccccccc|c}
 & & \multicolumn{8}{c|}{BS} & $\text{SI}$\tabularnewline
$\mathcal{K}_{i}''\subseteq\mathcal{K}_{i}'\subseteq\mathcal{K}_{i}$ & Phase & $\mathfrak{N}_{+}^{(0,0,0)}$ & $\mathfrak{N}_{+}^{(\pi,0,0)}$ & $\mathfrak{N}_{+}^{(0,\pi,0)}$ & $\mathfrak{N}_{+}^{(\pi,\pi,0)}$ & $\mathfrak{N}_{+}^{(0,0,\pi)}$ & $\mathfrak{N}_{+}^{(\pi,0,\pi)}$ & $\mathfrak{N}_{+}^{(0,\pi,\pi)}$ & $\mathfrak{N}_{+}^{(\pi,\pi,\pi)}$ & $\mathbb{Z}_{2}^{3}\otimes\mathbb{Z}_{4}$\tabularnewline
\hline 
- & $\vec{x}=(0,0,0)$ & 1 & 1 & 1 & 1 & 1 & 1 & 1 & 1 & $\idn$\tabularnewline
- & $\vec{x}=(\frac{1}{2},0,0)$ & 1 & -1 & 1 & -1 & 1 & -1 & 1 & -1 & $\idn$\tabularnewline
- & $\vec{x}=(0,\frac{1}{2},0)$ & 1 & 1 & -1 & -1 & 1 & 1 & -1 & -1 & $\idn$\tabularnewline
- & $\vec{x}=(\frac{1}{2},\frac{1}{2},0)$ & 1 & -1 & -1 & 1 & 1 & -1 & -1 & 1 & $\idn$\tabularnewline
- & $\vec{x}=(0,0,\frac{1}{2})$ & 1 & 1 & 1 & 1 & -1 & -1 & -1 & -1 & $\idn$\tabularnewline
- & $\vec{x}=(\frac{1}{2},0,\frac{1}{2})$ & 1 & -1 & 1 & -1 & -1 & 1 & -1 & 1 & $\idn$\tabularnewline
- & $\vec{x}=(0,\frac{1}{2},\frac{1}{2})$ & 1 & 1 & -1 & -1 & -1 & -1 & 1 & 1 & $\idn$\tabularnewline
- & $\vec{x}=(\frac{1}{2},\frac{1}{2},\frac{1}{2})$ & 1 & -1 & -1 & 1 & -1 & 1 & 1 & -1 & $\idn$\tabularnewline
\hline 
$0\subseteq\mathbb{Z}$ & $(2;z)$  %\includegraphics[scale=0.5]{figs/Ci/3d_Au_2z}  
& 2 & 0 & 0 & 0 & 2 & 0 & 0 & 0 & $\gen_{2;z}^{(2)}$\tabularnewline
$0\subseteq\mathbb{Z}$ & $(2;y)$  %\includegraphics[scale=0.5]{figs/Ci/3d_Au_2y}  
& 2 & 0 & 2 & 0 & 0 & 0 & 0 & 0 & $\gen_{2;y}^{(2)}$\tabularnewline
$0\subseteq\mathbb{Z}$ & $(2;x)$  %\includegraphics[scale=0.5]{figs/Ci/3d_Au_2x}  
& 2 & 2 & 0 & 0 & 0 & 0 & 0 & 0 & $\gen_{2;x}^{(2)}$\tabularnewline
$0\subseteq\mathbb{Z}_{2}\subseteq\mathbb{Z}_{2}$ & $(3)'$  %\includegraphics[scale=0.5]{figs/Ci/3d_3p} 
& 4 & 0 & 0 & 0 & 0 & 0 & 0 & 0 & $2\gen_{3}^{(4)}$\tabularnewline
\end{tabular}
\caption{Topological band labels of three-dimensional superconductors in tenfold-way class C with point group $C_{i}$ and representation $\Theta=A_u$. 
\label{tab:ex_Ci_C_Au_bandlabels}}
\end{table*}

\begin{table*}
\begin{tabular}{cc|cccccccc|c}
 & & \multicolumn{8}{c|}{BS} & $\text{SI}$\tabularnewline
$\mathcal{K}_{i}''\subseteq\mathcal{K}_{i}'\subseteq\mathcal{K}_{i}$ & Phase & $\mathfrak{N}_{+}^{(0,0,0)}$ & $\mathfrak{N}_{+}^{(\pi,0,0)}$ & $\mathfrak{N}_{+}^{(0,\pi,0)}$ & $\mathfrak{N}_{+}^{(\pi,\pi,0)}$ & $\mathfrak{N}_{+}^{(0,0,\pi)}$ & $\mathfrak{N}_{+}^{(\pi,0,\pi)}$ & $\mathfrak{N}_{+}^{(0,\pi,\pi)}$ & $\mathfrak{N}_{+}^{(\pi,\pi,\pi)}$ & $\mathbb{Z}_{2}^{3}\otimes\mathbb{Z}_{4}$\tabularnewline
\hline 
- & $\vec{x}=(0,0,0)$ & 1 & 1 & 1 & 1 & 1 & 1 & 1 & 1 & $\idn$\tabularnewline
- & $\vec{x}=(\frac{1}{2},0,0)$ & 1 & -1 & 1 & -1 & 1 & -1 & 1 & -1 & $\idn$\tabularnewline
- & $\vec{x}=(0,\frac{1}{2},0)$ & 1 & 1 & -1 & -1 & 1 & 1 & -1 & -1 & $\idn$\tabularnewline
- & $\vec{x}=(\frac{1}{2},\frac{1}{2},0)$ & 1 & -1 & -1 & 1 & 1 & -1 & -1 & 1 & $\idn$\tabularnewline
- & $\vec{x}=(0,0,\frac{1}{2})$ & 1 & 1 & 1 & 1 & -1 & -1 & -1 & -1 & $\idn$\tabularnewline
- & $\vec{x}=(\frac{1}{2},0,\frac{1}{2})$ & 1 & -1 & 1 & -1 & -1 & 1 & -1 & 1 & $\idn$\tabularnewline
- & $\vec{x}=(0,\frac{1}{2},\frac{1}{2})$ & 1 & 1 & -1 & -1 & -1 & -1 & 1 & 1 & $\idn$\tabularnewline
- & $\vec{x}=(\frac{1}{2},\frac{1}{2},\frac{1}{2})$ & 1 & -1 & -1 & 1 & -1 & 1 & 1 & -1 & $\idn$\tabularnewline
\hline 
$0\subseteq 0\subseteq\mathbb{Z}$ & $(3)$  %\includegraphics[scale=0.5]{figs/Ci/3d_3p} 
& 4 & 0 & 0 & 0 & 0 & 0 & 0 & 0 & $2\gen_{3}^{(4)}$\tabularnewline
\end{tabular}
\caption{Topological band labels of three-dimensional superconductors in tenfold-way class CI with point group $C_{i}$ and representation $\Theta=A_u$. 
\label{tab:ex_Ci_CI_Au_bandlabels}}
\end{table*}

\begin{table*}
\begin{tabular}{cc|cccccccc|c}
 &  & \multicolumn{8}{c|}{BL} & $\SI$\tabularnewline
$\mathcal{K}_{i}''\subseteq\mathcal{K}_{i}'\subseteq\mathcal{K}_{i}$ & Phase & $\mathfrak{N}_{+}^{(0,0,0)}$ & $\mathfrak{N}_{+}^{(\pi,0,0)}$ & $\mathfrak{N}_{+}^{(0,\pi,0)}$ & $\mathfrak{N}_{+}^{(\pi,\pi,0)}$ & $\mathfrak{N}_{+}^{(0,0,\pi)}$ & $\mathfrak{N}_{+}^{(\pi,0,\pi)}$ & $\mathfrak{N}_{+}^{(0,\pi,\pi)}$ & $\mathfrak{N}_{+}^{(\pi,\pi,\pi)}$ & $\mathbb{Z}_{2}$\tabularnewline
\hline 
- & $\vec{x}=(0,0,0)$ & 1 & 1 & 1 & 1 & 1 & 1 & 1 & 1 & $\idn$\tabularnewline
- & $\vec{x}=(\frac{1}{2},0,0)$ & 1 & -1 & 1 & -1 & 1 & -1 & 1 & -1 & $\idn$\tabularnewline
\hline 
$\mathbb{Z}_{2}$ & $(1;y,z)$ & 1 & 0 & 1 & 0 & 1 & 0 & 1 & 0 & $\gen_{1;y,z}^{(2)}$\tabularnewline
$0\subseteq\mathbb{Z}$ & $(2;x)$ & 0 & 0 & 0 & 0 & 0 & 0 & 0 & 0 & $\idn$\tabularnewline
\end{tabular}
\caption[]{
Topological band labels of three-dimensional superconductors in tenfold-way class D with point group $C_{s}$ and representation $\Theta=A'$. 
\label{tab:SM_ex_3d_Cs_Ap_D}}
\end{table*}

\begin{table*}
\begin{tabular}{cc|cccccccc|c}
 &  & \multicolumn{8}{c|}{BL} & $\SI$\tabularnewline
$\mathcal{K}_{i}''\subseteq\mathcal{K}_{i}'\subseteq\mathcal{K}_{i}$ & Phase & $\mathfrak{p}_{(+,-)}^{(0,0,0)}$ & $\mathfrak{p}_{(+,-)}^{(\pi,0,0)}$ & $\mathfrak{p}_{(+,-)}^{(0,\pi,0)}$ & $\mathfrak{p}_{(+,-)}^{(\pi,\pi,0)}$ & $\mathfrak{p}_{(+,-)}^{(0,0,\pi)}$ & $\mathfrak{p}_{(+,-)}^{(\pi,0,\pi)}$ & $\mathfrak{p}_{(+,-)}^{(0,\pi,\pi)}$ & $\mathfrak{p}_{(+,-)}^{(\pi,\pi,\pi)}$ & $\mathbb{Z}_{2}^{9}$\tabularnewline
\hline 
- & $\vec{x}=(0,0,0),\ \alpha=\begin{cases}
+\\
-
\end{cases}$ & $\begin{array}{c}
(1,0)\\
(0,1)
\end{array}$ & $\begin{array}{c}
(1,0)\\
(0,1)
\end{array}$ & $\begin{array}{c}
(1,0)\\
(0,1)
\end{array}$ & $\begin{array}{c}
(1,0)\\
(0,1)
\end{array}$ & $\begin{array}{c}
(1,0)\\
(0,1)
\end{array}$ & $\begin{array}{c}
(1,0)\\
(0,1)
\end{array}$ & $\begin{array}{c}
(1,0)\\
(0,1)
\end{array}$ & $\begin{array}{c}
(1,0)\\
(0,1)
\end{array}$ & $\begin{array}{c}
\idn\\
\idn
\end{array}$\tabularnewline 
- & $\vec{x}=(\frac{1}{2},0,0),\ \alpha=\begin{cases}
+\\
-
\end{cases}$ & $\begin{array}{c}
(1,0)\\
(0,1)
\end{array}$ & $\begin{array}{c}
(0,1)\\
(1,0)
\end{array}$ & $\begin{array}{c}
(1,0)\\
(0,1)
\end{array}$ & $\begin{array}{c}
(0,1)\\
(1,0)
\end{array}$ & $\begin{array}{c}
(1,0)\\
(0,1)
\end{array}$ & $\begin{array}{c}
(0,1)\\
(1,0)
\end{array}$ & $\begin{array}{c}
(1,0)\\
(0,1)
\end{array}$ & $\begin{array}{c}
(0,1)\\
(1,0)
\end{array}$ & $\begin{array}{c}
\idn\\
\idn
\end{array}$\tabularnewline
\hline 
$\mathbb{Z}_{2}$ & $(1,+;x,z)$ & $(1,0)$ & $(1,0)$ & $(0,0)$ & $(0,0)$ & $(1,0)$ & $(1,0)$ & $(0,0)$ & $(0,0)$ & $\gen_{1,+;x,z}^{(2)}$\tabularnewline
$\mathbb{Z}_{2}$ & $(1,-;x,z)$ & $(0,1)$ & $(0,1)$ & $(0,0)$ & $(0,0)$ & $(0,1)$ & $(0,1)$ & $(0,0)$ & $(0,0)$ & $\gen_{1,-;x,z}^{(2)}$\tabularnewline
$\mathbb{Z}_{2}$ & $(1,+;x,y)$ & $(1,0)$ & $(1,0)$ & $(1,0)$ & $(1,0)$ & $(0,0)$ & $(0,0)$ & $(0,0)$ & $(0,0)$ & $\gen_{1,+;x,y}^{(2)}$\tabularnewline
$\mathbb{Z}_{2}$ & $(1,-;x,y)$ & $(0,1)$ & $(0,1)$ & $(0,1)$ & $(0,1)$ & $(0,0)$ & $(0,0)$ & $(0,0)$ & $(0,0)$ & $\gen_{1,-;x,y}^{(2)}$\tabularnewline
$0\subseteq\mathbb{Z}$ & $(2,+;x)$ & $(1,0)$ & $(1,0)$ & $(0,0)$ & $(0,0)$ & $(0,0)$ & $(0,0)$ & $(0,0)$ & $(0,0)$ & $\gen_{2,+;x}^{(2)}$\tabularnewline
$0\subseteq\mathbb{Z}$ & $(2,-;x)$ & $(0,1)$ & $(0,1)$ & $(0,0)$ & $(0,0)$ & $(0,0)$ & $(0,0)$ & $(0,0)$ & $(0,0)$ & $\gen_{2,-;x}^{(2)}$\tabularnewline
$\mathbb{Z}_{2}\subseteq\mathbb{Z}_{2}$ & $(2;z)^\prime$ & $(1,1)$ & $(0,0)$ & $(0,0)$ & $(0,0)$ & $(1,1)$ & $(0,0)$ & $(0,0)$ & $(0,0)$ & $\gen_{2;z}^{(2)}$\tabularnewline
$\mathbb{Z}_{2}\subseteq\mathbb{Z}_{2}$ & $(2;y)^\prime$ & $(1,1)$ & $(0,0)$ & $(1,1)$ & $(0,0)$ & $(0,0)$ & $(0,0)$ & $(0,0)$ & $(0,0)$ & $\gen_{2;y}^{(2)}$\tabularnewline
$0\subseteq\mathbb{Z}_{2}\subseteq\mathbb{Z}_{2}$ & $(3)^\prime$ & $(1,1)$ & $(0,0)$ & $(0,0)$ & $(0,0)$ & $(0,0)$ & $(0,0)$ & $(0,0)$ & $(0,0)$ & $\gen_{3}^{(2)}$\tabularnewline
\end{tabular}
\caption[]{Topological band labels of three-dimensional superconductors in tenfold-way class D with point group $C_{s}$ and representation $\Theta=A''$.
\label{tab:SM_ex_3d_Cs_App_D}}
\end{table*}

\begin{table*}
\begin{tabular}{cc|cccccccc|c}
 &  & \multicolumn{8}{c|}{BL} & $\SI$\tabularnewline
$\mathcal{K}_{i}''\subseteq\mathcal{K}_{i}'\subseteq\mathcal{K}_{i}$ & Phase & $\mathfrak{N}_{+}^{(0,0,0)}$ & $\mathfrak{N}_{+}^{(\pi,0,0)}$ & $\mathfrak{N}_{+}^{(0,\pi,0)}$ & $\mathfrak{N}_{+}^{(\pi,\pi,0)}$ & $\mathfrak{N}_{+}^{(0,0,\pi)}$ & $\mathfrak{N}_{+}^{(\pi,0,\pi)}$ & $\mathfrak{N}_{+}^{(0,\pi,\pi)}$ & $\mathfrak{N}_{+}^{(\pi,\pi,\pi)}$ & $\mathbb{Z}_{2}^{2}\times \mathbb{Z}_{4}$\tabularnewline
\hline 
- & $\vec{x}=(0,0,0)$ & 1 & 1 & 1 & 1 & 1 & 1 & 1 & 1 & $\idn$\tabularnewline
- & $\vec{x}=(\frac{1}{2},0,0)$ & 1 & -1 & 1 & -1 & 1 & -1 & 1 & -1 & $\idn$\tabularnewline
- & $\vec{x}=(0,\frac{1}{2},0)$ & 1 & 1 & -1 & -1 & 1 & 1 & -1 & -1 & $\idn$\tabularnewline
- & $\vec{x}=(\frac{1}{2},\frac{1}{2},0)$ & 1 & -1 & -1 & 1 & 1 & -1 & -1 & 1 & $\idn$\tabularnewline
\hline 
$\mathbb{Z}_{2}$ & $(1;y,z)$ & 1 & 0 & 1 & 0 & 1 & 0 & 1 & 0 & $\gen_{1;y,z}^{(2)}$\tabularnewline
$\mathbb{Z}_{2}$ & $(1;x,z)$ & 1 & 1 & 0 & 0 & 1 & 1 & 0 & 0 & $\gen_{1;x,z}^{(2)}$\tabularnewline
$\mathbb{Z}_{2}\subseteq\mathbb{Z}_{2}$ & $(2;z)$' & 2 & 0 & 0 & 0 & 2 & 0 & 0 & 0 & $2\gen_{2;z}^{(4)}$\tabularnewline
$0\subseteq\mathbb{Z}$ & $(2;z)$ & 1 & 0 & 0 & 0 & 1 & 0 & 0 & 0 & $\gen_{2;z}^{(4)}$\tabularnewline
\end{tabular}
\caption[]{Topological band labels of three-dimensional superconductors in tenfold-way class D with point group $C_{2}$ and representation $\Theta=A$.  
\label{tab:SM_ex_3d_C2_A_D}}
\end{table*}

\begin{table*}
\begin{tabular}{cc|cccccccc|c}
 &  & \multicolumn{8}{c|}{BL} & $\SI$\tabularnewline
$\mathcal{K}_{i}''\subseteq\mathcal{K}_{i}'\subseteq\mathcal{K}_{i}$ & Phase & $\mathfrak{p}_{(+,-)}^{(0,0,0)}$ & $\mathfrak{p}_{(+,-)}^{(\pi,0,0)}$ & $\mathfrak{p}_{(+,-)}^{(0,\pi,0)}$ & $\mathfrak{p}_{(+,-)}^{(\pi,\pi,0)}$ & $\mathfrak{p}_{(+,-)}^{(0,0,\pi)}$ & $\mathfrak{p}_{(+,-)}^{(\pi,0,\pi)}$ & $\mathfrak{p}_{(+,-)}^{(0,\pi,\pi)}$ & $\mathfrak{p}_{(+,-)}^{(\pi,\pi,\pi)}$ & $\mathbb{Z}_{2}^{6}$\tabularnewline
\hline 
- & $\vec{x}=(0,0,0),\ \alpha=\begin{cases}
+\\
-
\end{cases}$ & $\begin{array}{c}
(1,0)\\
(0,1)
\end{array}$ & $\begin{array}{c}
(1,0)\\
(0,1)
\end{array}$ & $\begin{array}{c}
(1,0)\\
(0,1)
\end{array}$ & $\begin{array}{c}
(1,0)\\
(0,1)
\end{array}$ & $\begin{array}{c}
(1,0)\\
(0,1)
\end{array}$ & $\begin{array}{c}
(1,0)\\
(0,1)
\end{array}$ & $\begin{array}{c}
(1,0)\\
(0,1)
\end{array}$ & $\begin{array}{c}
(1,0)\\
(0,1)
\end{array}$ & $\begin{array}{c}
\idn\\
\idn
\end{array}$\tabularnewline
- & $\vec{x}=(\frac{1}{2},0,0),\ \alpha=\begin{cases}
+\\
-
\end{cases}$ & $\begin{array}{c}
(1,0)\\
(0,1)
\end{array}$ & $\begin{array}{c}
(0,1)\\
(1,0)
\end{array}$ & $\begin{array}{c}
(1,0)\\
(0,1)
\end{array}$ & $\begin{array}{c}
(0,1)\\
(1,0)
\end{array}$ & $\begin{array}{c}
(1,0)\\
(0,1)
\end{array}$ & $\begin{array}{c}
(0,1)\\
(1,0)
\end{array}$ & $\begin{array}{c}
(1,0)\\
(0,1)
\end{array}$ & $\begin{array}{c}
(0,1)\\
(1,0)
\end{array}$ & $\begin{array}{c}
\idn\\
\idn
\end{array}$\tabularnewline
- & $\vec{x}=(0,\frac{1}{2},0),\ \alpha=\begin{cases}
+\\
-
\end{cases}$ & $\begin{array}{c}
(1,0)\\
(0,1)
\end{array}$ & $\begin{array}{c}
(1,0)\\
(0,1)
\end{array}$ & $\begin{array}{c}
(0,1)\\
(1,0)
\end{array}$ & $\begin{array}{c}
(0,1)\\
(1,0)
\end{array}$ & $\begin{array}{c}
(1,0)\\
(0,1)
\end{array}$ & $\begin{array}{c}
(1,0)\\
(0,1)
\end{array}$ & $\begin{array}{c}
(0,1)\\
(1,0)
\end{array}$ & $\begin{array}{c}
(0,1)\\
(1,0)
\end{array}$ & $\begin{array}{c}
\idn\\
\idn
\end{array}$\tabularnewline
- & $\vec{x}=(\frac{1}{2},\frac{1}{2},0),\ \alpha=\begin{cases}
+\\
-
\end{cases}$ & $\begin{array}{c}
(1,0)\\
(0,1)
\end{array}$ & $\begin{array}{c}
(0,1)\\
(1,0)
\end{array}$ & $\begin{array}{c}
(0,1)\\
(1,0)
\end{array}$ & $\begin{array}{c}
(1,0)\\
(0,1)
\end{array}$ & $\begin{array}{c}
(1,0)\\
(0,1)
\end{array}$ & $\begin{array}{c}
(0,1)\\
(1,0)
\end{array}$ & $\begin{array}{c}
(0,1)\\
(1,0)
\end{array}$ & $\begin{array}{c}
(1,0)\\
(0,1)
\end{array}$ & $\begin{array}{c}
\idn\\
\idn
\end{array}$\tabularnewline
\hline 
$\mathbb{Z}_{2}$ & $(1,+;x,y)$ & $(1,0)$ & $(1,0)$ & $(1,0)$ & $(1,0)$ & $(0,0)$ & $(0,0)$ & $(0,0)$ & $(0,0)$ & $\gen_{1,+;x,y}^{(2)}$\tabularnewline
$\mathbb{Z}_{2}$ & $(1,-;x,y)$ & $(0,1)$ & $(0,1)$ & $(0,1)$ & $(0,1)$ & $(0,0)$ & $(0,0)$ & $(0,0)$ & $(0,0)$ & $\gen_{1,-;x,y}^{(2)}$\tabularnewline
$\mathbb{Z}_{2}\subseteq\mathbb{Z}_{2}$ & $(2;x)$' & $(1,1)$ & $(1,1)$ & $(0,0)$ & $(0,0)$ & $(0,0)$ & $(0,0)$ & $(0,0)$ & $(0,0)$ &  $\gen_{2;x}^{(2)}$\tabularnewline
$\mathbb{Z}_{2}\subseteq\mathbb{Z}_{2}$ & $(2;y)$' & $(1,1)$ & $(0,0)$ & $(1,1)$ & $(0,0)$ & $(0,0)$ & $(0,0)$ & $(0,0)$ & $(0,0)$ &  $\gen_{2;y}^{(2)}$\tabularnewline
$0\subseteq\mathbb{Z}$ & $(2;z)$ & $(1,1)$ & $(0,0)$ & $(0,0)$ & $(0,0)$ & $(1,1)$ & $(0,0)$ & $(0,0)$ & $(0,0)$ & $\gen_{2;z}^{(2)}$\tabularnewline
\end{tabular}
\caption[]{Topological band labels of three-dimensional superconductors in tenfold-way class D with point group $C_{2}$ and representation $\Theta=B$. 
\label{tab:SM_ex_3d_C2_B_D}}
\end{table*}

\begin{table*}
\begin{tabular}{cc|cc|c|cc|cc|c|cc|c}
 &  & \multicolumn{10}{c|}{BL} & $\SI$\tabularnewline
 &  & \multicolumn{2}{c|}{$(0,0,0)$} & $(\pi,0,0)$ & \multicolumn{2}{c|}{$(\pi,\pi,0)$} & \multicolumn{2}{c|}{$(0,0,\pi)$} & $(\pi,0,\pi)$ & \multicolumn{2}{c|}{$(\pi,\pi,\pi)$} & \tabularnewline
$\mathcal{K}_{i}''\subseteq\mathcal{K}_{i}'\subseteq\mathcal{K}_{i}$ & Phase & $\mathfrak{N}_{\frac{1}{2}}$ & $\mathfrak{N}_{\frac{5}{2}}$ & $\mathfrak{N}_{\frac{1}{2}}$ & $\mathfrak{N}_{\frac{1}{2}}$ & $\mathfrak{N}_{\frac{5}{2}}$ & $\mathfrak{N}_{\frac{1}{2}}$ & $\mathfrak{N}_{\frac{5}{2}}$ & $\mathfrak{N}_{\frac{1}{2}}$ & $\mathfrak{N}_{\frac{1}{2}}$ & $\mathfrak{N}_{\frac{5}{2}}$ & $\mathbb{Z}_{2}\times \mathbb{Z}_{8}$\tabularnewline
\hline 
- & $\vec{x}=(0,0),\ j=\frac{1}{2}$ & 1 & 0 & 1 & 1 & 0 & 1 & 0 & 1 & 1 & 0 & $\idn$\tabularnewline
- & $\vec{x}=(0,0),\ j=\frac{5}{2}$ & 0 & 1 & 1 & 0 & 1 & 0 & 1 & 1 & 0 & 1 & $\idn$\tabularnewline
- & $\vec{x}=(\frac{1}{2},\frac{1}{2}),\ j=\frac{1}{2}$ & 1 & 0 & -1 & 0 & 1 & 1 & 0 & -1 & 0 & 1 & $\idn$\tabularnewline
- & $\vec{x}=(\frac{1}{2},\frac{1}{2}),\ j=\frac{5}{2}$ & 0 & 1 & -1 & 1 & 0 & 0 & 1 & -1 & 1 & 0 & $\idn$\tabularnewline
- & $\vec{x}=(\frac{1}{2},0),\ j=\frac{1}{2}$ & 1 & 1 & 0 & -1 & -1 & 1 & 1 & 0 & -1 & -1 & $\idn$\tabularnewline
\hline 
$\mathbb{Z}_{2}$ & $(1;x,y,z)$ & 1 & 1 & 1 & 0 & 0 & 1 & 1 & 1 & 0 & 0 & $\gen_{1;x,y,z}^{(2)}$\tabularnewline
$0\subseteq\mathbb{Z}$ & $(2;z)$ & 0 & 0 & 1 & 1 & 0 & 0 & 0 & 1 & 1 & 0 & $\gen_{2;z}^{(8)}$\tabularnewline
$\mathbb{Z}_{2}\subseteq\mathbb{Z}_{2}$ & $(2;z)^\prime$ & 2 & 2 & 0 & 0 & 0 & 2 & 2 & 0 & 0 & 0 & $4\gen_{2;z}^{(8)}$\tabularnewline
\end{tabular}
\caption[]{Band labels of three-dimensional superconductors in tenfold-way class D with point group $C_{4}$ and representations $\Theta=A$ or $\Theta=B$.
\label{tab:SM_ex_3d_C4_A_D}}
\end{table*}

\begin{table*}
\begin{tabular}{cc|ccc|cc|ccc|ccc|cc|ccc|c}
 &  & \multicolumn{16}{c|}{BL} & $\SI$\tabularnewline
 &  & \multicolumn{3}{c|}{$(0,0,0)$} & \multicolumn{2}{c|}{$(\pi,0,0)$} & \multicolumn{3}{c|}{$(\pi,\pi,0)$} & \multicolumn{3}{c|}{$(0,0,\pi)$} & \multicolumn{2}{c|}{$(\pi,0,\pi)$} & \multicolumn{3}{c|}{$(\pi,\pi,\pi)$} & \tabularnewline
$\mathcal{K}_{i}''\subseteq\mathcal{K}_{i}'\subseteq\mathcal{K}_{i}$ & Phase & $\mathfrak{p}_{\frac{1}{2}}$ & $\mathfrak{p}_{\frac{5}{2}}$ & $\mathfrak{N}_{\frac{3}{2}}$ & $\mathfrak{p}_{\frac{1}{2}}$ & $\mathfrak{p}_{\frac{3}{2}}$ & $\mathfrak{p}_{\frac{1}{2}}$ & $\mathfrak{p}_{\frac{5}{2}}$ & $\mathfrak{N}_{\frac{3}{2}}$ & $\mathfrak{p}_{\frac{1}{2}}$ & $\mathfrak{p}_{\frac{5}{2}}$ & $\mathfrak{N}_{\frac{3}{2}}$ & $\mathfrak{p}_{\frac{1}{2}}$ & $\mathfrak{p}_{\frac{3}{2}}$ & $\mathfrak{p}_{\frac{1}{2}}$ & $\mathfrak{p}_{\frac{5}{2}}$ & $\mathfrak{N}_{\frac{3}{2}}$ & $\mathbb{Z}_{2}^{4}\times \mathbb{Z}_{4}$\tabularnewline
\hline 
- & $\vec{x}=(0,0),\ j=\frac{1}{2}$ & 1 & 0 & 0 & 1 & 0 & 1 & 0 & 0 & 1 & 0 & 0 & 1 & 0 & 1 & 0 & 0 & $\idn$\tabularnewline
- & $\vec{x}=(0,0),\ j=\frac{5}{2}$ & 0 & 1 & 0 & 1 & 0 & 0 & 1 & 0 & 0 & 1 & 0 & 1 & 0 & 0 & 1 & 0 & $\idn$\tabularnewline
- & $\vec{x}=(0,0),\ j=\frac{3}{2}$ & 0 & 0 & 1 & 0 & 1 & 0 & 0 & 1 & 0 & 0 & 1 & 0 & 1 & 0 & 0 & 1 & $\idn$\tabularnewline
- & $\vec{x}=(\frac{1}{2},\frac{1}{2}),\ j=\frac{1}{2}$ & 1 & 0 & 0 & 0 & 1 & 0 & 1 & 0 & 1 & 0 & 0 & 0 & 1 & 0 & 1 & 0 & $\idn$\tabularnewline
- & $\vec{x}=(\frac{1}{2},\frac{1}{2}),\ j=\frac{5}{2}$ & 0 & 1 & 0 & 0 & 1 & 1 & 0 & 0 & 0 & 1 & 0 & 0 & 1 & 1 & 0 & 0 & $\idn$\tabularnewline
- & $\vec{x}=(\frac{1}{2},\frac{1}{2}),\ j=\frac{3}{2}$ & 0 & 0 & 1 & 1 & 0 & 0 & 0 & -1 & 0 & 0 & 1 & 1 & 0 & 0 & 0 & -1 & $\idn$\tabularnewline
- & $\vec{x}=(\frac{1}{2},0),\ j=\frac{1}{2}$ & 1 & 1 & 0 & 1 & 1 & 0 & 0 & 0 & 1 & 1 & 0 & 1 & 1 & 0 & 0 & 0 & $\idn$\tabularnewline
- & $\vec{x}=(\frac{1}{2},0),\ j=\frac{3}{2}$ & 0 & 0 & 0 & 1 & 1 & 1 & 1 & 0 & 0 & 0 & 0 & 1 & 1 & 1 & 1 & 0 & $\idn$\tabularnewline
\hline 
$\mathbb{Z}_{2}$ & $(1,\frac{1}{2};x,y)$ & 1 & 0 & 0 & 1 & 0 & 1 & 0 & 0 & 0 & 0 & 0 & 0 & 0 & 0 & 0 & 0 & $\gen_{1,\frac{1}{2};x,y}^{(2)}$\tabularnewline
$\mathbb{Z}_{2}$ & $(1,\frac{5}{2};x,y)$ & 0 & 1 & 0 & 1 & 0 & 0 & 1 & 0 & 0 & 0 & 0 & 0 & 0 & 0 & 0 & 0 & $\gen_{1,\frac{5}{2};x,y}^{(2)}$\tabularnewline
$0\subseteq\mathbb{Z}$ & $(2;z)$ & 0 & 0 & 0 & 1 & 1 & 1 & 0 & 1 & 0 & 0 & 0 & 1 & 1 & 1 & 0 & 1 & $\gen_{2;z}^{(4)}$\tabularnewline
$\mathbb{Z}_{2}\subseteq\mathbb{Z}_{2}$ & $(2;x,y)^\prime$ & 0 & 0 & 1 & 1 & 0 & 1 & 1 & 1 & 1 & 1 & 1 & 0 & 1 & 1 & 1 & 1 & $\gen_{2;x,y}^{(2)}$\tabularnewline
\end{tabular}
\caption[]{Topological band labels of three-dimensional superconductors in tenfold-way class D with point group $C_{4}$ and representation $\Theta=^{1}E$ or $\Theta=^{2}E$.
\label{tab:SM_ex_3d_C4_E_D}}
\end{table*}

\begin{table*}
\begin{tabular}{cc|cccccccc|c}
 & & \multicolumn{8}{c|}{BL} & $\SI$\tabularnewline
$\mathcal{K}_{i}''\subseteq\mathcal{K}_{i}'\subseteq\mathcal{K}_{i}$ & Phase & $\mathfrak{p}^{(0,0,0)}$ & $\mathfrak{p}^{(\pi,0,0)}$ & $\mathfrak{p}^{(0,\pi,0)}$ & $\mathfrak{p}^{(\pi,\pi,0)}$ & $\mathfrak{p}^{(0,0,\pi)}$ & $\mathfrak{p}^{(\pi,0,\pi)}$ & $\mathfrak{p}^{(0,\pi,\pi)}$ & $\mathfrak{p}^{(\pi,\pi,\pi)}$ & $\mathbb{Z}_{2}^{7}$\tabularnewline
\hline 
 & $\vec{x}=(0,0,0)$ & 1 & 1 & 1 & 1 & 1 & 1 & 1 & 1 & $\idn$\tabularnewline
 & $\vec{x}=(\frac{1}{2},0,0)$ & 1 & 1 & 1 & 1 & 1 & 1 & 1 & 1 & $\idn$\tabularnewline
 & $\vec{x}=(0,\frac{1}{2},0)$ & 1 & 1 & 1 & 1 & 1 & 1 & 1 & 1 & $\idn$\tabularnewline
 & $\vec{x}=(\frac{1}{2},\frac{1}{2},0)$ & 1 & 1 & 1 & 1 & 1 & 1 & 1 & 1 & $\idn$\tabularnewline
\hline 
$\mathbb{Z}_{2}$ & $(1;y,z)$   & 1 & 0 & 1 & 0 & 1 & 0 & 1 & 0 & $\gen_{1;y,z}^{(2)}$\tabularnewline
$\mathbb{Z}_{2}$ & $(1;x,z)$   & 1 & 1 & 0 & 0 & 1 & 1 & 0 & 0 & $\gen_{1;x,z}^{(2)}$\tabularnewline
$\mathbb{Z}_{2}$ & $(1;x,y)$   & 1 & 1 & 1 & 1 & 0 & 0 & 0 & 0 & $\gen_{1;x,y}^{(2)}$\tabularnewline
$\mathbb{Z}_{2}\subseteq\mathbb{Z}_{2}$ & $(2,+;z)^\prime$   & 1 & 0 & 0 & 0 & 1 & 0 & 0 & 0 & $\gen_{2;z}^{(2)}$\tabularnewline
$\mathbb{Z}_{2}\subseteq\mathbb{Z}_{2}$ & $(2,-;z)^\prime$   & 1 & 0 & 0 & 0 & 1 & 0 & 0 & 0 & $\gen_{2;z}^{(2)}$\tabularnewline
$0\subseteq\mathbb{Z}$ & $(2;y)$   & 1 & 0 & 1 & 0 & 0 & 0 & 0 & 0 & $\gen_{2;y}^{(2)}$\tabularnewline
$0\subseteq\mathbb{Z}$ & $(2;x)$   & 1 & 1 & 0 & 0 & 0 & 0 & 0 & 0 & $\gen_{2;x}^{(2)}$\tabularnewline
$0\subseteq\mathbb{Z}\subseteq\mathbb{Z}$ & $(3,+)^\prime$   & 1 & 0 & 0 & 0 & 0 & 0 & 0 & 0 & $\gen_{3}^{(2)}$\tabularnewline
$0\subseteq\mathbb{Z}\subseteq\mathbb{Z}$ & $(3,-)^\prime$   & 1 & 0 & 0 & 0 & 0 & 0 & 0 & 0 & $\gen_{3}^{(2)}$\tabularnewline
\end{tabular}
\caption{Topological band labels and symmetry-based indicators for three-dimensional superconductors in tenfold-way class D with point group $C_{2v}$ and representation $\Theta = A_2$. 
\label{tab:SM_ex_3d_C2v_A2_D}}
\end{table*}

\begin{table*}
\begin{tabular}{cc|cccccccc|c}
 &  & \multicolumn{8}{c|}{BL} & $\SI$\tabularnewline
$\mathcal{K}_{i}''\subseteq\mathcal{K}_{i}'\subseteq\mathcal{K}_{i}$ & Phase & $\mathfrak{p}^{(0,0,0)}$ & $\mathfrak{p}^{(\pi,0,0)}$ & $\mathfrak{p}^{(0,\pi,0)}$ & $\mathfrak{p}^{(\pi,\pi,0)}$ & $\mathfrak{p}^{(0,0,\pi)}$ & $\mathfrak{p}^{(\pi,0,\pi)}$ & $\mathfrak{p}^{(0,\pi,\pi)}$ & $\mathfrak{p}^{(\pi,\pi,\pi)}$ & $\mathbb{Z}_{2}^{3}$\tabularnewline
\hline 
 & $\vec{x}=(0,0,0)$ & 1 & 1 & 1 & 1 & 1 & 1 & 1 & 1 & $\idn$\tabularnewline
 & $\vec{x}=(\frac{1}{2},0,0)$ & 1 & 1 & 1 & 1 & 1 & 1 & 1 & 1 & $\idn$\tabularnewline
 & $\vec{x}=(0,\frac{1}{2},0)$ & 1 & 1 & 1 & 1 & 1 & 1 & 1 & 1 & $\idn$\tabularnewline
 & $\vec{x}=(\frac{1}{2},\frac{1}{2},0)$ & 1 & 1 & 1 & 1 & 1 & 1 & 1 & 1 & $\idn$\tabularnewline
\hline 
$\mathbb{Z}_{2}$ & $(1;x,z)$   & 1 & 1 & 0 & 0 & 1 & 1 & 0 & 0 & $\gen_{1;x,z}^{(2)}$\tabularnewline
$\mathbb{Z}_{2}$ & $(1;x,y)$   & 1 & 1 & 1 & 1 & 0 & 0 & 0 & 0 & $\gen_{1;x,y}^{(2)}$\tabularnewline
$\mathbb{Z}_{2}\subseteq\mathbb{Z}_{2}$ & $(2;y)^\prime$   & 0 & 0 & 0 & 0 & 0 & 0 & 0 & 0 & $\idn$\tabularnewline
$0\subseteq\mathbb{Z}$ & $(2;x)$   & 1 & 1 & 0 & 0 & 0 & 0 & 0 & 0 & $\gen_{2;x}^{(2)}$\tabularnewline
$0\subseteq\mathbb{Z}\subseteq\mathbb{Z}$ & $(3)^\prime$   & 0 & 0 & 0 & 0 & 0 & 0 & 0 & 0 & $\idn$\tabularnewline
\end{tabular}
\caption[]{Topological band labels of three-dimensional superconductors in tenfold-way class D with point group $C_{2v}$ and representation $\Theta=B_2$. 
\label{tab:SM_ex_3d_C2v_B2_D}}
\end{table*}

\end{document}